\documentclass[11pt,a4paper,twoside,openright]{book} 

\usepackage[english]{babel}
\usepackage{amsmath}
\usepackage{amssymb}
\usepackage{epsfig}
\usepackage{array}
\usepackage{ifthen}
\usepackage{amsthm}
\usepackage[chapter]{algorithm}
\usepackage{algorithmic}
\usepackage{enumitem}
\usepackage{placeins} 
\usepackage[format=default,font=footnotesize,labelfont=sc,margin=0.5cm]{caption}
\usepackage{subfigure} 
\usepackage{booktabs} 
\usepackage{tikz}
\usepackage{aicescover}

\usepackage[Lenny]{fncychap}

\usepackage{layout}  

% ...

\usepackage[totalwidth=400pt, totalheight=591pt, centering]{geometry}

\makeatletter
\renewcommand{\paragraph}{%
  \@startsection{paragraph}{4}%
  {\z@}{1.25ex \@plus 1ex \@minus .2ex}{-1em}%
  {\normalfont\normalsize\bfseries}%
}
\makeatother

\newtheorem{mythm}[]{Theorem}[section]

\newtheorem{mydef}[]{Definition}[section]
\newtheorem{myguide}{Observation}

\newcommand{\R}{\mathbb{R}}
\newcommand{\C}{\mathbb{C}}

\newcommand{\Rn}{\R^{n}}
\newcommand{\Cn}{\C^{n}}

\newcommand{\Rnn}{\R^{n \times n}}
\newcommand{\Cnn}{\C^{n \times n}}

\newcommand{\gap}{\mbox{gap}}
\newcommand{\relgap}{\mbox{relgap}}
\newcommand{\gaptol}{gaptol}  
\newcommand{\abstol}{abstol}  

\newcommand\norm[1]{\lVert#1\rVert}
\newcommand\order[1]{\mathcal{O}(#1)}
\newcommand\abs[1]{\lvert#1\rvert}

\newcommand{\mrsmp}{{\tt mr3smp}}
\newcommand{\PMRRR}{{\tt PMRRR}}
\newcommand{\SSTEMR}{{\tt SSTEMR}}
\newcommand{\DSTEMR}{{\tt DSTEMR}}

\newcommand{\DSTEDC}{{\tt DSTEDC}}

\newcommand{\DUNN}{{\sc Dunnington}}

\newcommand{\DeltaT}{\Delta \hspace*{-2.4pt} T}

\newcommand{\binaryx}{{\it binary\_\hspace*{1pt}x}}
\newcommand{\binaryy}{{\it binary\_\hspace*{0pt}y}}

\title{\Huge MRRR-based Eigensolvers for Multi-core Processors and
  Supercomputers}
\aicescoverauthor{Matthias Petschow}
\date{\small
Der Fakult\"at f\"ur Mathematik, Informatik und Naturwissenschaften der RWTH
Aachen \\
University vorgelegte Dissertation zur Erlangung des akademischen Grades
eines \\
Doktors der Naturwissenschaften
}

\begin{document}



\usetikzlibrary{shapes,arrows}
\tikzstyle{decision} = [diamond, draw,  
    text width=4.5em, text badly centered, node distance=2.5cm, inner sep=0pt]
\tikzstyle{block} = [rectangle, draw, fill=blue!10, 
    text width=5em, text centered, rounded corners, minimum height=4em,node
    distance=3.5cm]
\tikzstyle{block2} = [rectangle, draw, fill=blue!10, 
    text width=5em, text centered, rounded corners, minimum height=4em,node
    distance=2.1cm]
\tikzstyle{line} = [draw, -latex']

\aicescoverpage
\thispagestyle{empty}


\chapter*{Abstract}
\thispagestyle{empty}

\vspace{-1cm}

The real symmetric tridiagonal eigenproblem is of outstanding importance in numerical
computations; it arises frequently as part of eigensolvers for 
standard and generalized dense Hermitian eigenproblems that are based on a
reduction to tridiagonal form. For its solution, the algorithm
of {\it Multiple Relatively 
  Robust Representations} (MRRR or MR$^3$ in
short) -- introduced in the late 1990s -- is among
the fastest methods. To compute $k$ eigenpairs of tridiagonal $T \in \Rnn$,
MRRR only requires $\order{kn}$ arithmetic operations; in contrast, all the
other practical methods require $\order{k^2 n}$ or $\order{n^3}$ operations in the
worst case. 
This thesis centers around the performance and accuracy of MRRR.

First, we investigate how MRRR can make efficient use of modern multi-core
architectures. We present a parallelization strategy that dynamically divides and 
schedules the work into tasks. This task-based approach is flexible and
produces remarkable workload balancing. In a number of experiments,
comparing our multi-threaded eigensolver, \mrsmp, with the widely used
LAPACK (Linear Algebra PACKage) and Intel's Math Kernel library (MKL), we
show that \mrsmp\ outperforms even the
fastest solvers available.

Second, for massively parallel
distributed/shared-memory systems, we introduce an eigensolver, \PMRRR,  
that merges the task-based approach with a parallelization based on
message-passing. Our design uses non-blocking communications, thus 
allowing processes to proceed computation while waiting to receive
data. Experimentally, we show the importance of such an
overlap of computation and communication for load balancing and
scalability. 
Moreover, with a thorough performance study, we demonstrate that the
new eigensolvers of the Elemental library, which are based on
\PMRRR, are faster and more scalable than the widely used eigensolvers of ScaLAPACK. 

Third, we present a mixed precision variant of MRRR, which improves the
standard algorithm in multiple ways. 
In fact, compared with the best available methods (Divide
\& Conquer and the QR algorithm), the standard MRRR exhibits inferior
accuracy and robustness. Moreover, when confronted with heavily clustered
eigenvalues, its performance and scalability can suffer severely; such
scalability problems especially arise on distributed-memory 
architectures.
Our mixed precision approach makes MRRR at least as accurate as Divide
\& Conquer and the QR algorithm. 
In particular in context of direct methods for large-scale standard and
generalized Hermitian eigenproblems, such an improvement comes with little or no
performance penalty: eigensolvers based on our mixed precision 
MRRR are oftentimes faster than solvers based on Divide \& Conquer and, in some
circumstances, even faster than solvers based on the standard MRRR.
Additionally, the use of mixed precisions considerably 
enhances robustness and parallel scalability. 

\thispagestyle{empty}


\chapter*{Acknowledgments}
\thispagestyle{empty}

\vspace{-1cm}

I am very thankful for the guidance and support of Prof.~Paolo
Bientinesi. Despite my different professional background, he gave me the opportunity
to work in the interesting field of numerical linear algebra and
high-performance computing. 
My work would not have been possible without his technical advise as
well as his constructive suggestions on scientific writing. Besides
technical help, I am equally thankful for his efforts to initiate
group activities outside the work environment and thereby generating a
friendly atmosphere at work as well as outside of work.

I wish to express my sincere gratitude to a number of people and
institutions. In particular, I would like to acknowledge the former and
current members of the HPAC group: Edoardo Di Napoli, Diego Fabregat Traver, 
Roman Iakymchuk, Elmar Peise, Paul Springer, Daniel Tameling, Viola
Wierschem, and Lucas Beyer. I would like to especially express my deep gratitude to
Diego Fabregat Traver for many technical and even more non-technical
discussions. 
I also want to thank all other AICES students and the
AICES technical stuff, particularly, Annette de Haes, Nadine Bachem, Joelle
Janssen, Nicole Faber, and my former office colleague Aravind Balan. 

I would also like to extend my thanks to the people of the Center for
Computing and Communication at RWTH Aachen and the people of the J\"ulich Supercomputing
Center (JSC) for granting access to their computational resources and their
support. From the JSC, I like to explicitly mention Inge Gutheil and
Prof.~Stefan Bl\"ugel. 
Financial support from the Deutsche Forschungsgemeinschaft (German Research Association) through grant GSC 111
is gratefully acknowledged. 

I received valuable and constructive help from Jack Poulson, Prof.~Robert
van de Geijn, and Prof.~Enrique S.~Quintana-Ort\'{i}. I enjoyed my visits to
University Jaime I, Castell\'{o}n, Spain -- enabled by DAAD
(Deutscher Akademischer Austausch Dienst) project 50225798
PARSEMUL. I would also like to thank Prof.~Lars Grasedyck for agreeing to be a
reviewer of this dissertation. 

Last but not least, I want thank my family and friends for making life
worth living. I feel this is not the right place for a long speech 
and I rather convey my feeling to you in person.

\newpage
\mbox{}
\thispagestyle{empty}

\pagenumbering{roman}
\setcounter{page}{0}
\tableofcontents



\clearpage
\pagenumbering{arabic}

\chapter{Motivation \& Contributions}
\label{chapter:ch01}
\thispagestyle{empty}

\section{Motivation}

The algorithm of {\it Multiple Relatively Robust Representations} (MRRR or
MR$^3$ in
short) is a method for computing a set of eigenvalues and eigenvectors (i.e.,
eigenpairs) of a real symmetric tridiagonal matrix~\cite{Dhillon:Diss,Dhillon:2004:Ortvecs,Dhillon:2004:MRRR,Fernando97,Parlett2000121}.  
The algorithm achieves what has been sometimes called the ``{\em holy grail
of numerical linear algebra}''~\cite{handbook}: It is capable of computing
$k$ eigenpairs of tridiagonal $T \in \Rnn$ with $\order{kn}$ arithmetic operations, while
all the other existing 
methods require $\order{k^2 n}$ or $\order{n^3}$ operations in the
worst case. For this reason, its invention by Dhillon and
Parlett~\cite{Dhillon:Diss,Dhillon:2004:Ortvecs,Dhillon:2004:MRRR} in the 
late 1990s was widely acknowledged as a breakthrough in the field. The
method was expected to be (almost always) faster than all existing methods, while being
equally accurate. Furthermore, it promised
to be embarrassingly parallel and thus ideally suited for parallel computers. In light of these
expectations, early after its introduction, it was natural to believe that the
MRRR algorithm would make all the other methods 
obsolete.\footnote{``[...] the new method will make all other tridiagonal
  eigensolvers obsolete''~\cite{LangDirectSolvers}.} 
After roughly 15 years of experience with the algorithm, this has
not been the case for a number of reasons.

\begin{figure}[htb]
  \centering
   \subfigure[Execution time.]{
     \includegraphics[width=.47\textwidth]{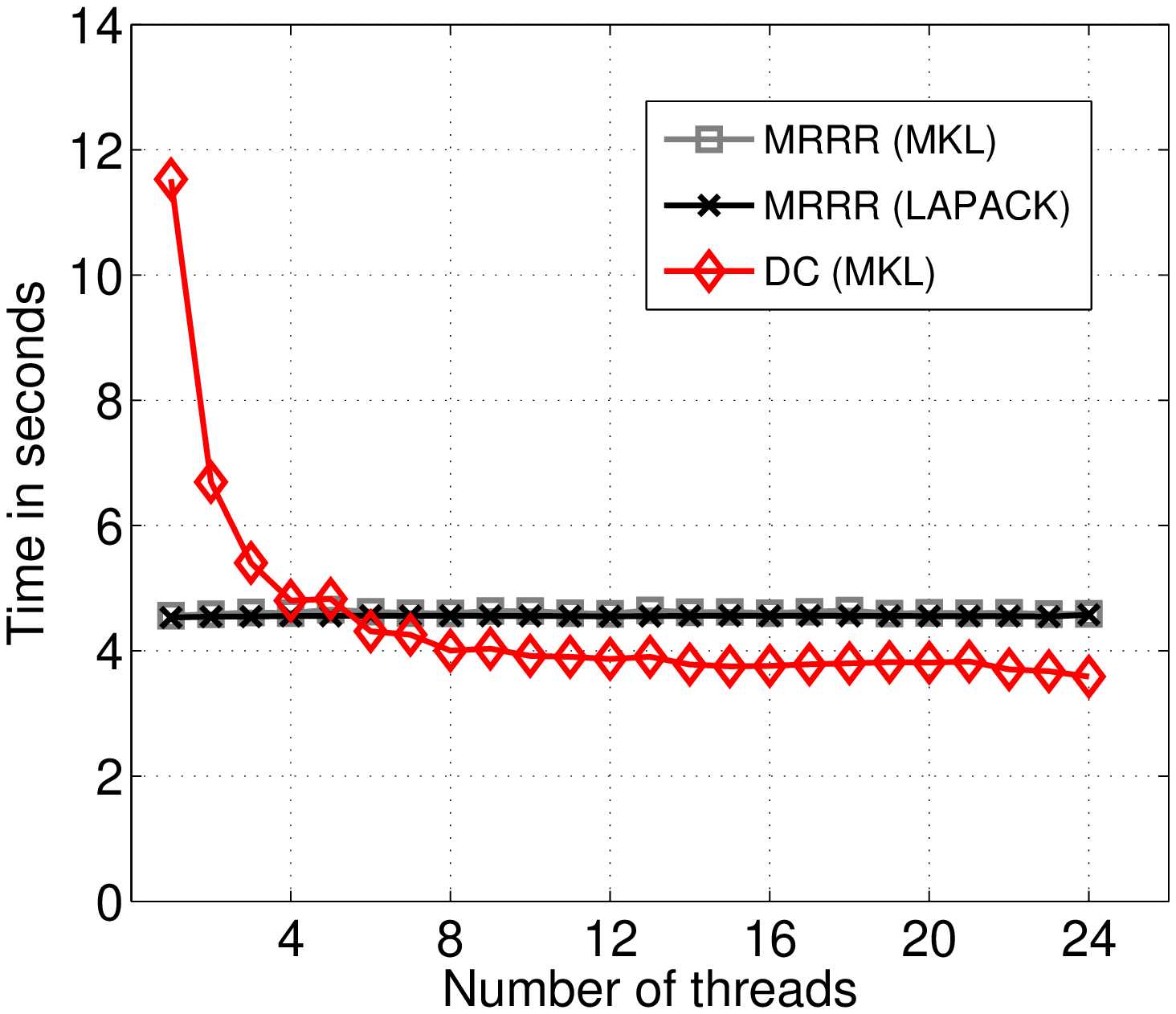} 
     \label{fig1:timinga}
   } \subfigure[Breakdown of time by stages.]{
     \includegraphics[width=.47\textwidth]{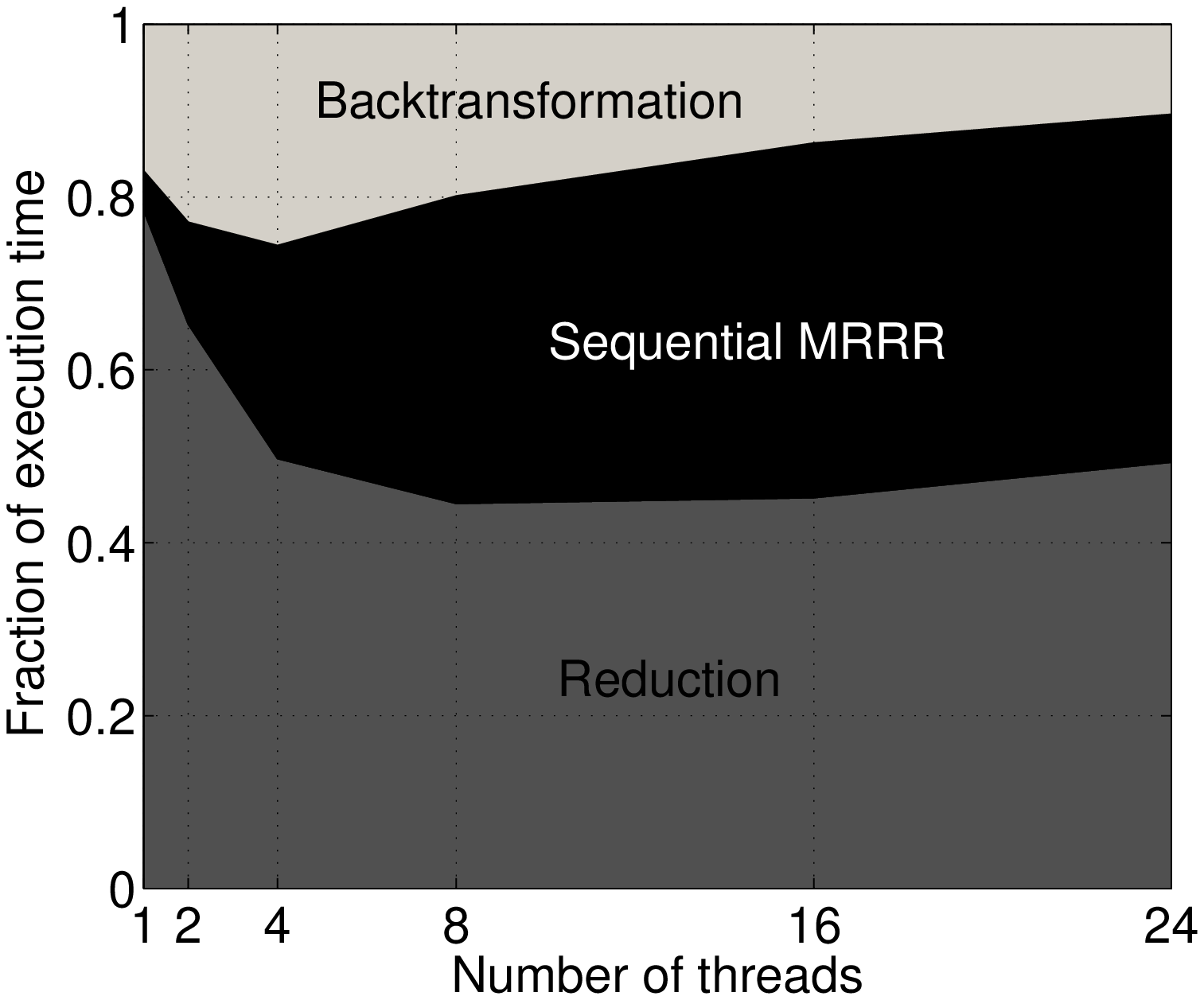}
     \label{fig1:timingb}
   }
  \caption{
    (a) Timings as function of the number of threads
    used in the computation. Qualitatively, the graph is typical for
    the applications matrices that we tested. 
    As the available parallelism increases, DC 
    becomes faster than MRRR.  
    (b) Fraction of time spent in the solution of
    the real symmetric dense eigenproblem for (1)
    reduction to tridiagonal form, (2) tridiagonal eigenproblem, and (3)
    backtransformation of the eigenvectors. For details
    of the experiment, see~\cite{para2010}.} 
  \label{fig1:timing}
\end{figure}

\paragraph{Speed.} A detailed investigation on the performance and accuracy of
  LAPACK's~\cite{lapack} symmetric tridiagonal eigensolvers
  revealed that, although MRRR performs the fewest 
  floating point operations (flops), due the phenomenon of
  numerical deflation and 
  its higher flop-rate (flops per 
  second), the {\em Divide \& Conquer} algorithm (DC) can
  be faster~\cite{perf09}. Not investigated in~\cite{perf09} 
  is how the performance is 
  influenced by the parallelism of modern  
  architectures. 
  For a matrix of size $4{,}289$ and by increasing the number of
  computational threads, Fig.~\ref{fig1:timing} illustrates that,
  although MRRR is the fastest sequential method, as the
  amount of available parallelism increases, DC becomes
  faster than MRRR.  
The reason
  for this behavior is that DC casts most of the work in terms of
  matrix-matrix multiplication and takes advantage of parallelism by
  multi-threaded Basic Linear Algebra Subprograms
(BLAS)~\cite{blas1,blas2,blas3}. In contrast, neither LAPACK's MRRR nor the
version included in Intel's MKL exploit
any parallelism offered by multiple cores.  
  Figure~\ref{fig1:timingb} displays that, since MRRR does not scale, it can become a
  significant portion of the common three stage approach to the dense
  Hermitian eigenproblem.  

\paragraph{Scalability.} 
Today, parallel computing is everywhere. While for many years the usage
of parallel computers was limited to high-end products, the
advent of multi-core processors has revolutionized the world of
computing: 
Performance improvements of existing sequential code will not occur by
simply waiting for the next generation of processors, but by 
explicitly exploiting the available parallelism~\cite{FreeLunchIsOver}. 
Since multi-core processors 
became the standard engines for both commodity computers (desktops and
laptops) and supercomputers, the sequential
computation model became obsolete and replaced by parallel
paradigms~\cite{Reinders:2012}. 
Given the current trends in the development of computer hardware, it is
expected that the number of processing units
(cores) of uniprocessors and supercomputers increase rapidly
in the near future. In light of this development, many algorithms and existing software must be 
reevaluated and rewritten and {\it it becomes increasingly important 
how well algorithms can exploit the ever growing
parallelism}~\cite{Asanovic:EECS-2006-183,Herlihy:2008}.  


A number of software packages (like LAPACK and ScaLAPACK) address the need
for eigensolvers on both uniprocessor and distributed-memory architectures. However, at the
time of beginning this dissertation, the support for multi-core architectures
and mixed distributed/shared-memory architectures was limited. Explicit
exploitation of shared-memory was largely confined to methods that rely
heavily on multi-threaded BLAS. On a uniprocessor, methods that do
not make use of suitable BLAS kernels -- e.g., MRRR -- cannot exploit any parallelism beyond
that at the instruction level and, consequently, do not scale. 

\begin{figure}[htb]
  \centering
   \subfigure[1--2--1 type matrices.]{
     \includegraphics[width=.47\textwidth]{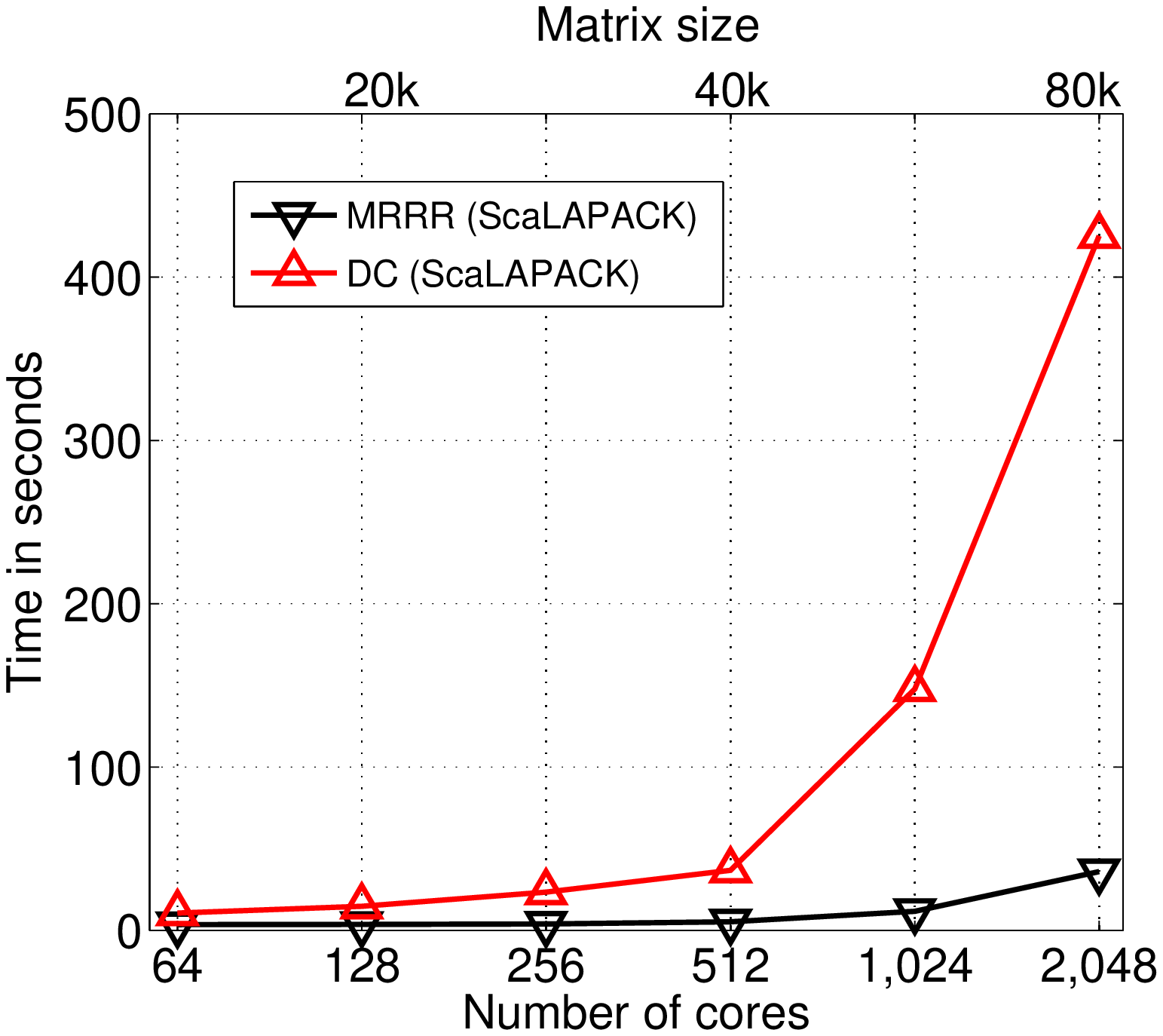} 
     \label{fig:motivationscalingscalapacka}
   } \subfigure[Wilkinson type matrices.]{
     \includegraphics[width=.47\textwidth]{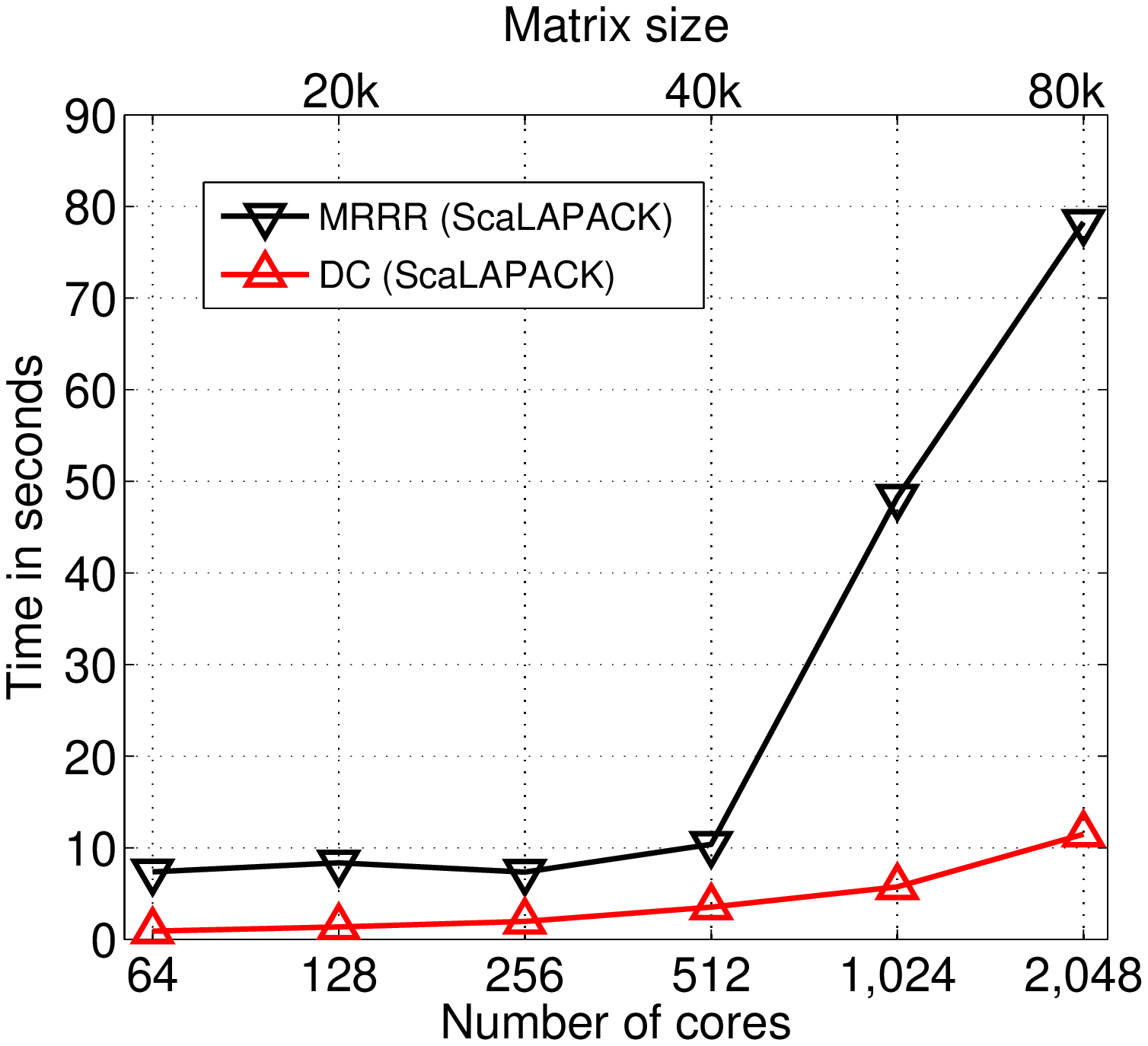}
     \label{fig:motivationscalingscalapackb}
   }
  \caption{Weak scalability (the matrix size is increased according to the
    number of cores) for the computation of all eigenpairs of 
     two different test matrix types. Each type provides an example in which
     one of the two solvers performs rather poorly for highly parallel executions.
     The left and right graphs have
     different scales. For MRRR the execution time should remain roughly
     constant. For details
    of the experiment, see~\cite{EleMRRR}.} 
  \label{fig:motivationscalingscalapack}
\end{figure}

Despite the initial expectations on MRRR, the computation is {\it not} embarrassingly
parallel. However, as demonstrated by Bientinesi et
al.~\cite{Bientinesi:2005:PMR3} and later
V\"omel~\cite{Vomel:2010:ScaLAPACKsMRRR}, MRRR is well suited for parallel
computations. Existing parallel implementations --
ParEig~\cite{Bientinesi:2005:PMR3} and
ScaLAPACK~\cite{scalapack,Vomel:2010:ScaLAPACKsMRRR} -- target
distributed-memory architectures. On multi-core architectures, those
implementations require the availability and initialization of a message-passing library and are
penalized by the redundant computations they perform to avoid costly 
communication. Furthermore, they might not achieve load balancing due to
static workload division. Even on distributed-memory systems, neither
ScaLAPACK's MRRR --  added to the library in 
2011~\cite{Vomel:2010:ScaLAPACKsMRRR} -- nor alternatives like ScaLAPACK's DC
scale perfectly to a large number of processor cores, as illustrated in
Fig.~\ref{fig:motivationscalingscalapack}. 
All these factors motivated 
research that targets multi-core architectures and hybrid
distributed/shared-memory architectures specifically.

\begin{figure}[htb]
  \centering
   \subfigure[Largest residual norm.]{
     \includegraphics[width=.47\textwidth]{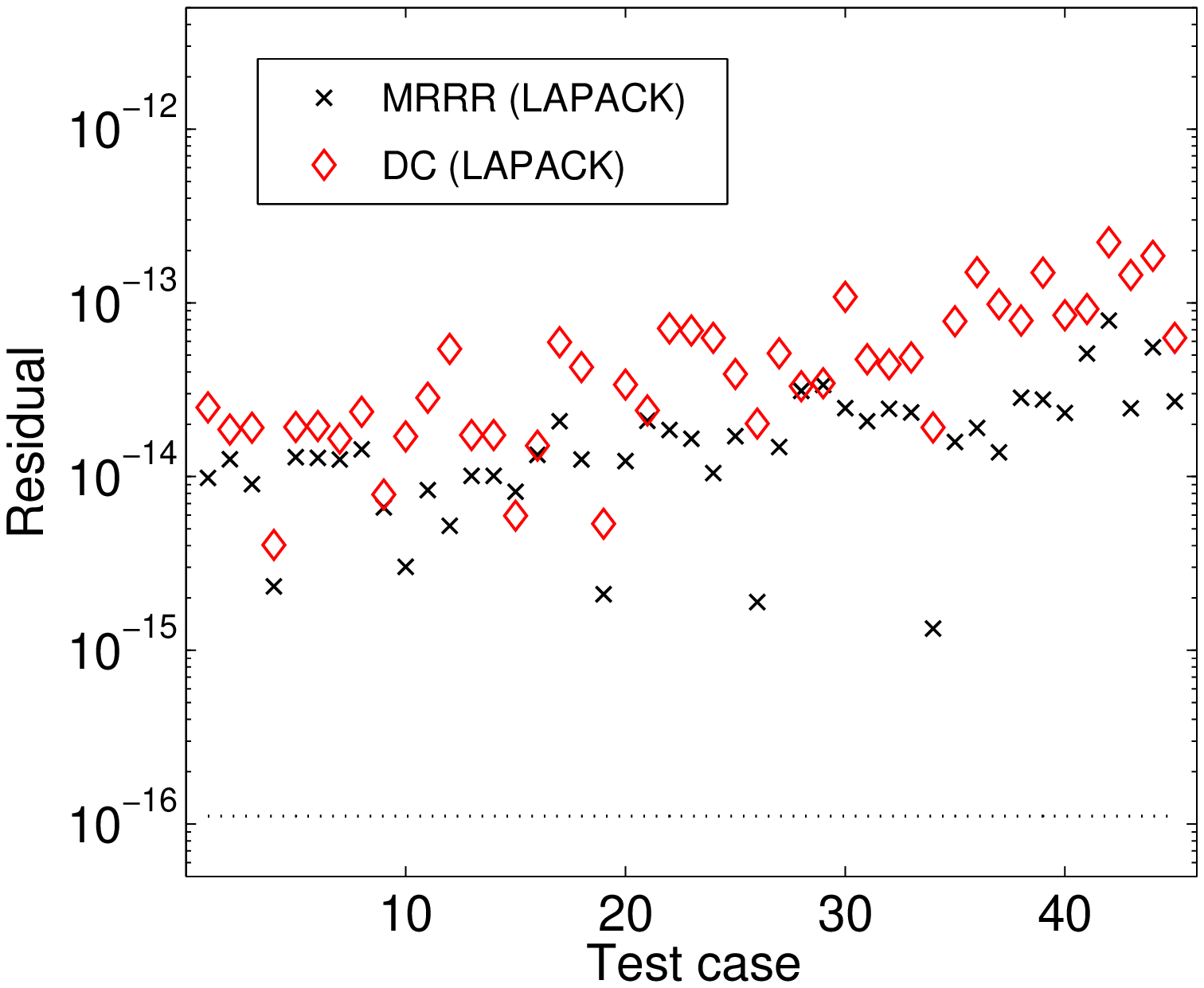} 
     \label{fig:motivationacclapacka}
   } \subfigure[Orthogonality.]{
     \includegraphics[width=.47\textwidth]{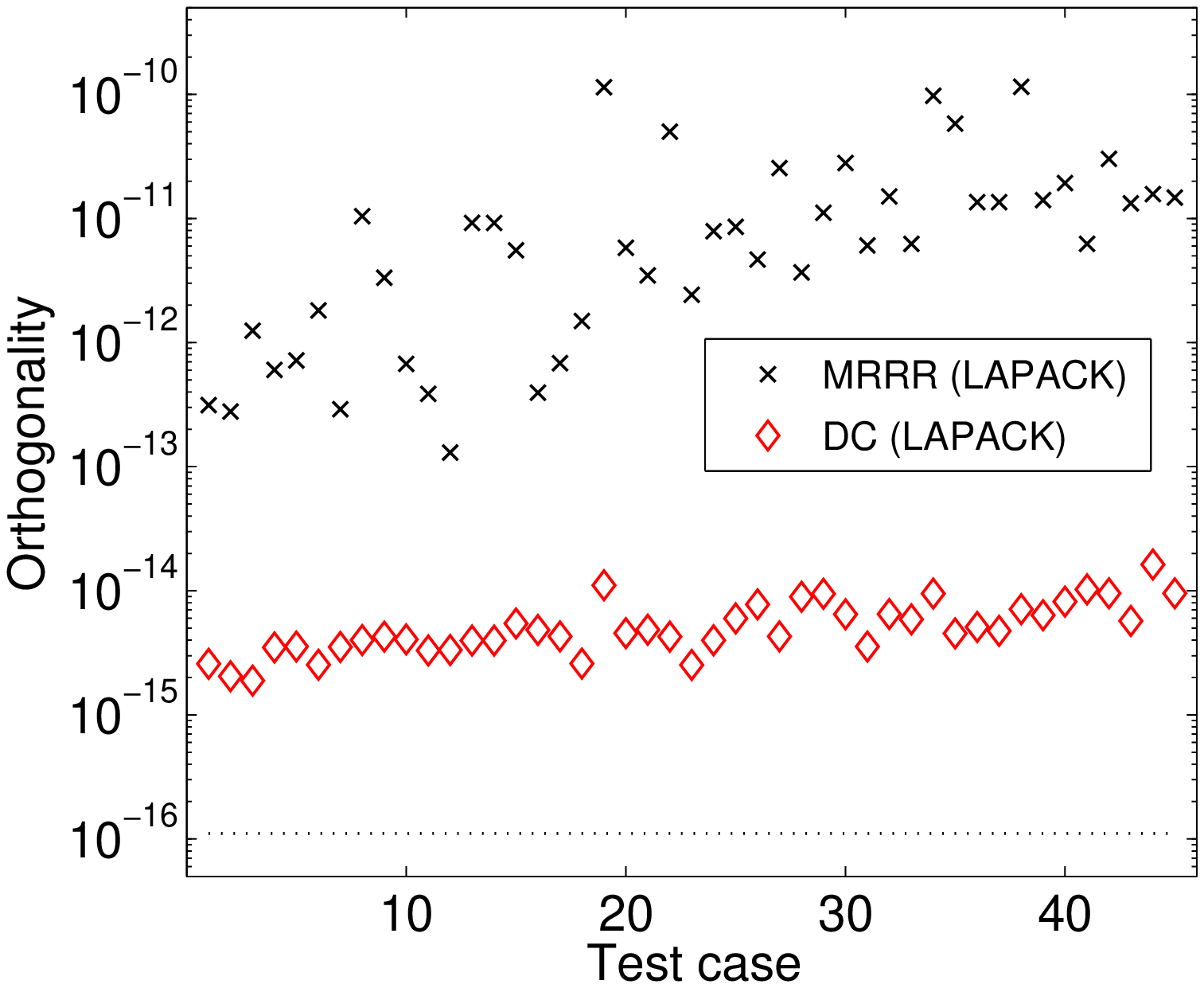}
     \label{fig:motivationacclapackb}
   }
  \caption{
  Accuracy of LAPACK's MRRR and DC for a set of test matrices from the {\sc
    Stetester}~\cite{Marques:2008} test suite, ranging in size
  from $1{,}000$ to about $8{,}000$. The results for QR are similar to
  the ones of DC.    
  } 
  \label{fig:motivationacclapack}
\end{figure}

\paragraph{Accuracy.} The aforementioned study 
also shows that
MRRR is less accurate than DC or 
the QR algorithm (QR)~\cite{perf09}. 
Especially for large-scale problems,
  both residuals and orthogonality of the computed eigenpairs are
  inferior,  
  cf.~\cite[Figure 6.1]{perf09} and \cite[Table
  5.1]{Willems:framework}. In our experience, as depicted in
  Fig.~\ref{fig:motivationacclapack}, the disadvantages are mainly
  confined to the orthogonality. 
  Through increased robustness, Willems and
  Lang~\cite{Willems:Diss,Willems:twisted,Willems:blocked} were 
  able to improve the orthogonality. Nonetheless, their
  analysis~\cite{Willems:framework} shows that the accuracy of any MRRR implementation  
  is inferior to those of DC and QR. For
  instance, ``one must be prepared for orthogonality levels of about
  $\order{1000 n \varepsilon}$, [where $\varepsilon$ denotes
  the unit roundoff,] because they can occur even if all of the 
  requirements [of the algorithm] are fulfilled with very benign
  parameters''~\cite{Willems:framework}. 

\paragraph{Robustness.} The QR algorithm has been analyzed for more than
half a century, which led to extremely robust implementations. 
MRRR -- with a number of novel features -- was only introduced
in the 1990s and, almost naturally, its use on challenging problems brought
forth non-satisfactory results and even failure~\cite{glued}. The causes of
failure were soon identified and mostly eliminated~\cite{glued,Willems:Diss}. In
particular, the research of V\"omel, Willems and
Lang improved the
reliability of the method~\cite{glued,Willems:Diss,Willems:twisted,Willems:blocked}. Even though MRRR usually gives satisfactory results, there is
a problem not quite apparent to a user: The proofs of correctness 
in~\cite{Dhillon:2004:MRRR,Willems:framework} hinge 
on finding so called {\it relatively robust representations} (RRRs) of the
input matrix with shifted spectrum. A representation is accepted as an RRR if it
passes a simple test, which is frequently the case. However, sometimes no
representation is found that passes the test and instead a promising
candidate is selected, which might or might not fulfill the
requirements. In such a situation, accuracy is not 
guaranteed. Willems and Lang reduce -- but not eliminate -- this problem by expanding the forms a
representation can take~\cite{Willems:twisted,Willems:blocked}, changing the
shifting strategy, and using a more sophisticated test for relative
robustness~\cite{localization,Willems:framework}. Nevertheless, for some
input matrices, the accuracy of MRRR cannot be guaranteed.

\section{Contributions}

In this dissertation, we make four main contributions: 
\begin{enumerate}
\item We introduce a strategy, MR$^3$-SMP, specifically
tailored for current multi-core and future many-core processors. Our
design makes efficient use of the on-chip parallelism and 
the low communication overhead thanks to the shared-memory and caches.
Parallelism is achieved by dynamically dividing the computation into
tasks that are executed by multiple threads. 
We show that the task-based approach scales
well on multi-core processors and small shared-memory systems made out
of them; in most cases, it scales better than state-of-the-art eigensolvers for
uniprocessors and distributed-memory systems. Good speedups are 
observed even in the case of relatively small 
matrices. For the example in Fig.~\ref{fig1:timing}, the results of our
solver \mrsmp\ are shown in
Fig.~\ref{fig:mottimingwithmr3smp2}. 
With the exploitation of parallelism, independent of the number of threads used for
the computation, MRRR remains faster than DC. The good scalability of the tridiagonal eigensolver is reflected in the
execution time of the dense Hermitian problem. While previously the fraction of
the tridiagonal stage was up to 40\% of the total execution time
[Fig.~\ref{fig1:timingb}], the fraction spent in the tridiagonal stage is
reduced to less than 7\% [Fig.~\ref{fig:mottimingwithmr3smp2b}].  
\begin{figure}[htb]
  \centering
   \subfigure[Execution time.]{
     \includegraphics[width=.47\textwidth]{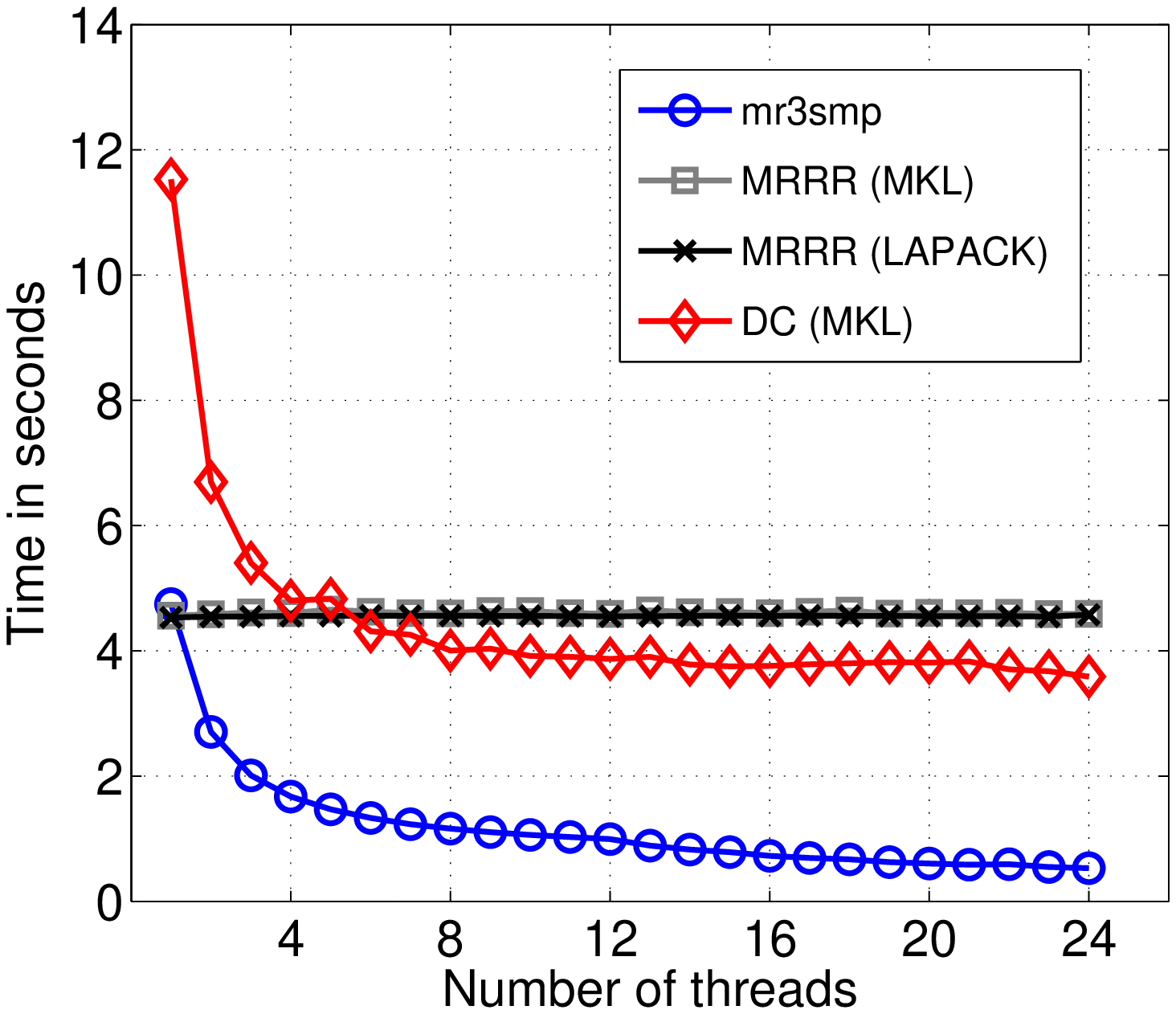} 
     \label{fig:mottimingwithmr3smp2a}
   } \subfigure[Breakdown of time by stages.]{
     \includegraphics[width=.47\textwidth]{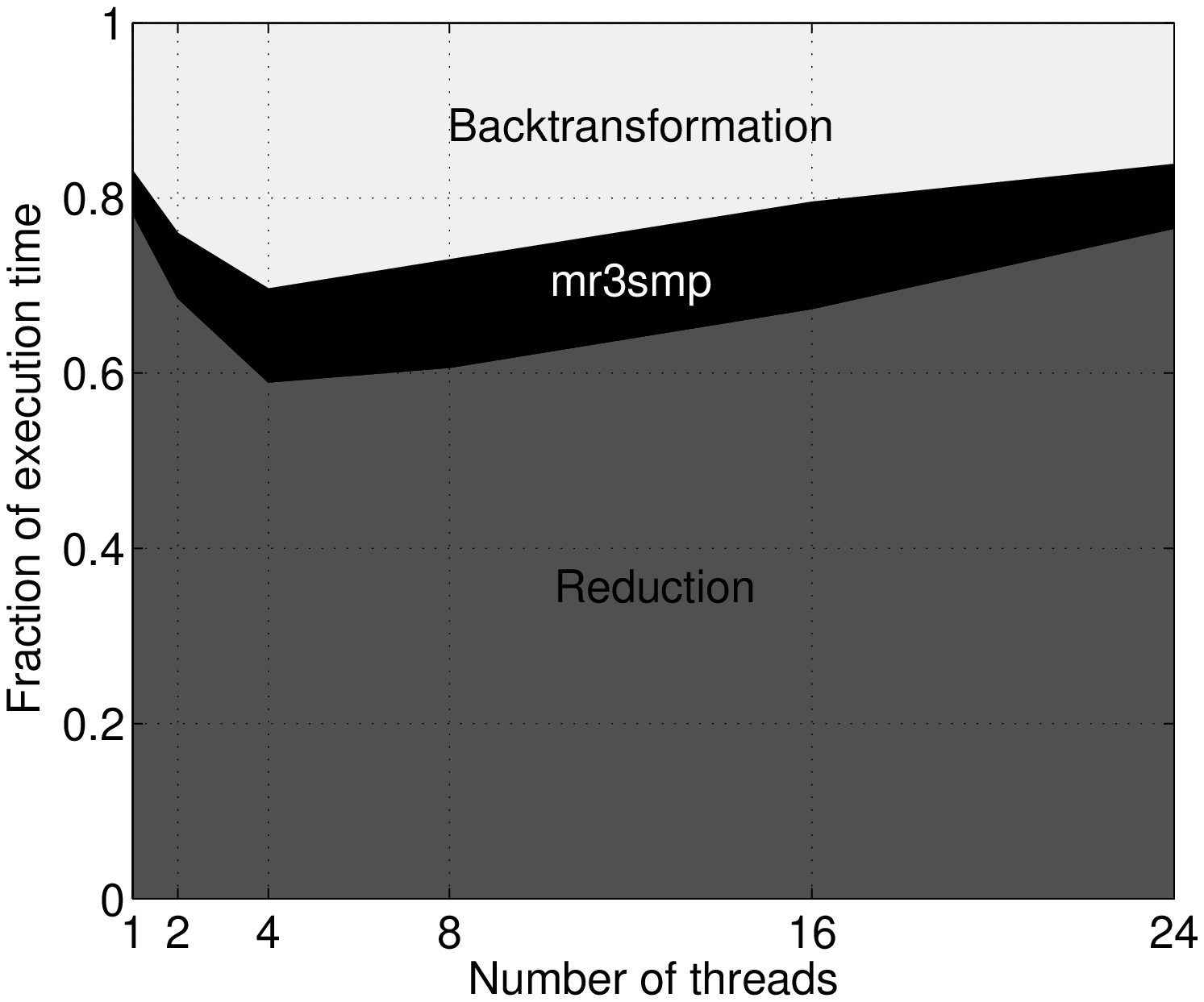}
     \label{fig:mottimingwithmr3smp2b}
   }
  \caption{
    (a) Timings as function of the number of threads
    used in the computation. Qualitatively, the graph is typical for
    the applications matrices that we tested. 
    (b) Fraction of time spent in the solution of
    the corresponding real symmetric dense eigenproblem for (1)
    reduction to tridiagonal form, (2) tridiagonal eigenproblem, and (3)
    backtransformation of the eigenvectors. For details
    of the experiment, see~\cite{para2010}.
  } 
  \label{fig:mottimingwithmr3smp2}
\end{figure} 

\item For massively parallel distributed/shared-memory architectures, we
  develop a variant of MRRR that merges the task-based approach with the parallelization
  strategy proposed by Bientinesi et 
  al.~\cite{Bientinesi:2005:PMR3}. Our new solver, {\tt PMRRR}, can make use of
  messages-passing for inter-node communication and shared-memory for
  intra-node communication. It can also be used in a purely
  message-passing or shared-memory mode, allowing the user to decide which
  programming model to employ.  
  Our design uses {\it non-blocking communications in conjunction with a
    task-based approach}, which enables processes to proceed the computation
  while waiting to receive data. Such an overlap of computation and communication is
  crucial for load balancing and scalability. For the example in
Fig.~\ref{fig:motivationscalingscalapack}, the results are shown in 
Fig.~\ref{fig:motivationscalingscalapack2}. 
\begin{figure}[htb]
  \centering
   \subfigure[1--2--1 type matrices.]{
     \includegraphics[width=.47\textwidth]{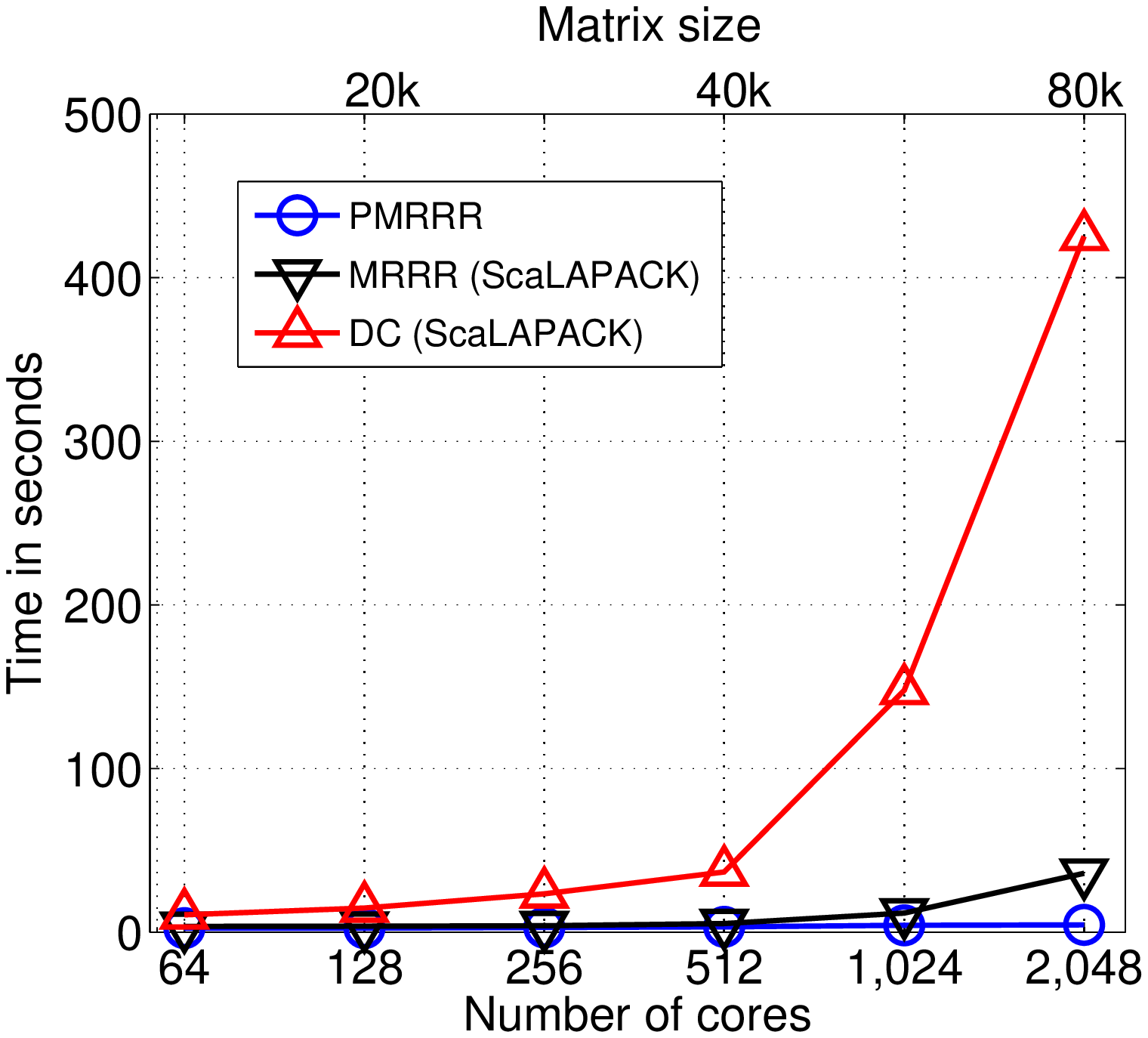} 
     \label{fig:motivationscalingscalapack2a}
   } \subfigure[Wilkinson type matrices.]{
     \includegraphics[width=.47\textwidth]{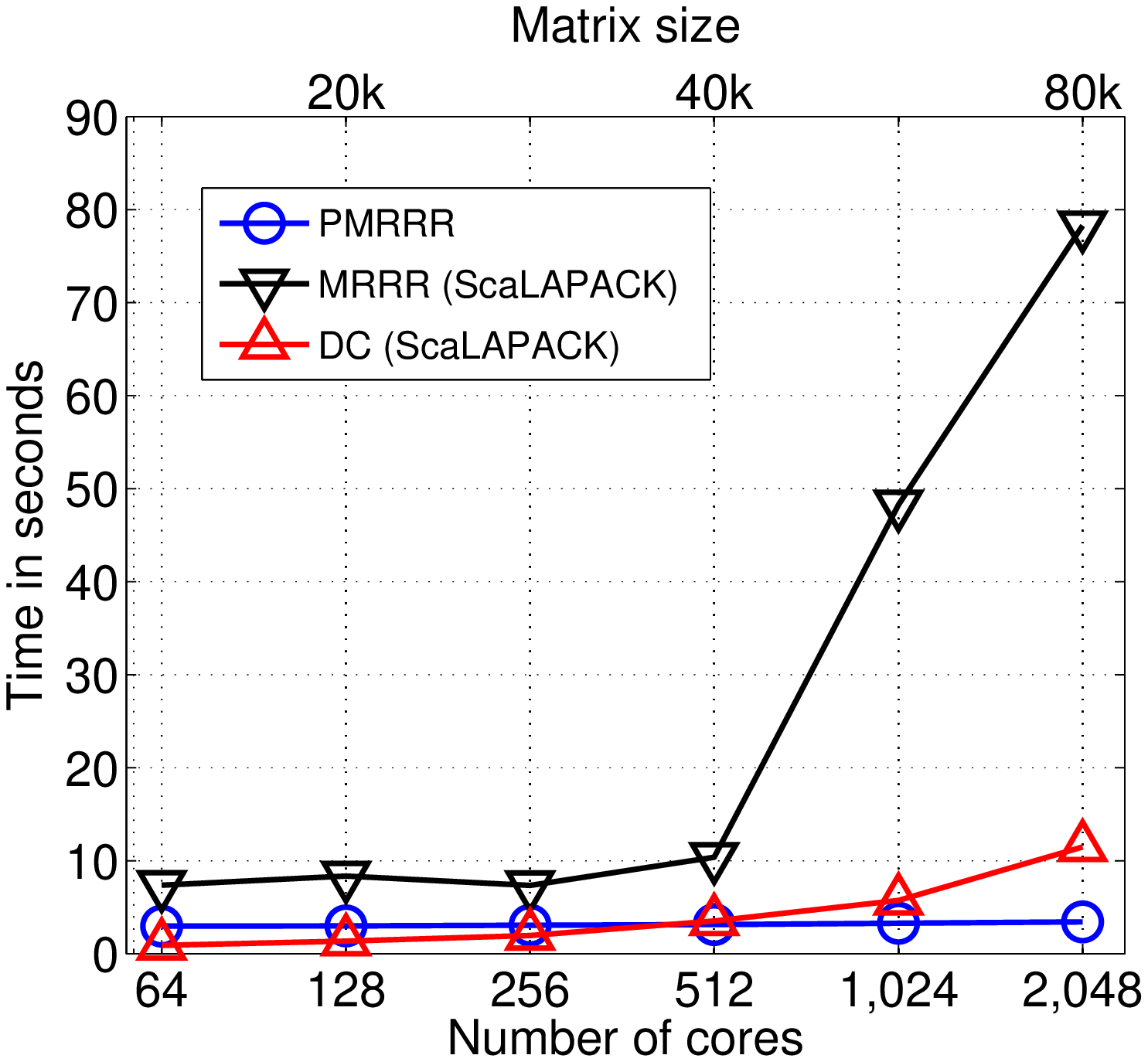}
     \label{fig:motivationscalingscalapack2b}
   }
  \caption{Weak scalability (the matrix size is increased according to the
    number of cores) for the computation of all eigenpairs of 
     two different test matrix types. The left and right graphs have
     different scales. For MRRR the execution time should remain roughly
     constant. For details of the experiment, see~\cite{EleMRRR}.}  
  \label{fig:motivationscalingscalapack2}
\end{figure} 
For instance, in the experiment with a Wilkinson type matrix on 1024 cores, ScaLAPACK's MRRR spends about 30 out of 50
seconds in exposed communication. 
As Fig.~\ref{fig:motivationscalingscalapack2b} demonstrates, even for matrices that strongly
favor DC, due to its superior scalability,
eventually our solver becomes faster than ScaLAPACK's DC.  

\item {\tt PMRRR} is integrated into {\it Elemental}, a development
  framework and library for distributed-memory dense linear algebra, which
  can now be used to solve large-scale dense eigenproblems. 
  We perform a thorough performance study of Elemental's new eigensolvers
  on two high-end computing platforms. A comparison with the widely used
  ScaLAPACK eigensolvers reveals that each ScaLAPACK routine present
  performance penalties that are avoided by calling a different sequence of
  subroutines and choosing suitable settings. We show how to built -- within
  the ScaLAPACK framework -- an eigensolver faster than the existing ones. 
  By comparing Elemental's eigensolvers with the standard ScaLAPACK solvers
  as well as the ones build according to our guidelines, we show that
  Elemental is fast and highly scalable. 
  
\item We present a variant of MRRR based on mixed precisions, which 
addresses the existing weaknesses of the algorithm: ($i$) inferior
accuracy compared with DC or QR;
($ii$) the danger of not finding suitable representations; and ($iii$) for distributed-memory
architectures, load balancing problems and communication overhead for matrices with
large clustering of the eigenvalues. 
Our approach adopts a
new perspective: Given input/output arguments in a \binaryx\ floating
point format, we use a higher precision \binaryy\ arithmetic to
obtain the desired accuracy. An analysis shows that we gain enormous freedom
to choose important parameters of the algorithm. In particular,
we are able to select these parameters to reduce the operation count,
increase robustness, and improve parallelism; at the same time, we meet
more stringent accuracy goals [Fig.~\ref{fig:motivationacclapack2}]. 
\begin{figure}[htb]
  \centering
   \subfigure[Largest residual norm.]{
     \includegraphics[width=.47\textwidth]{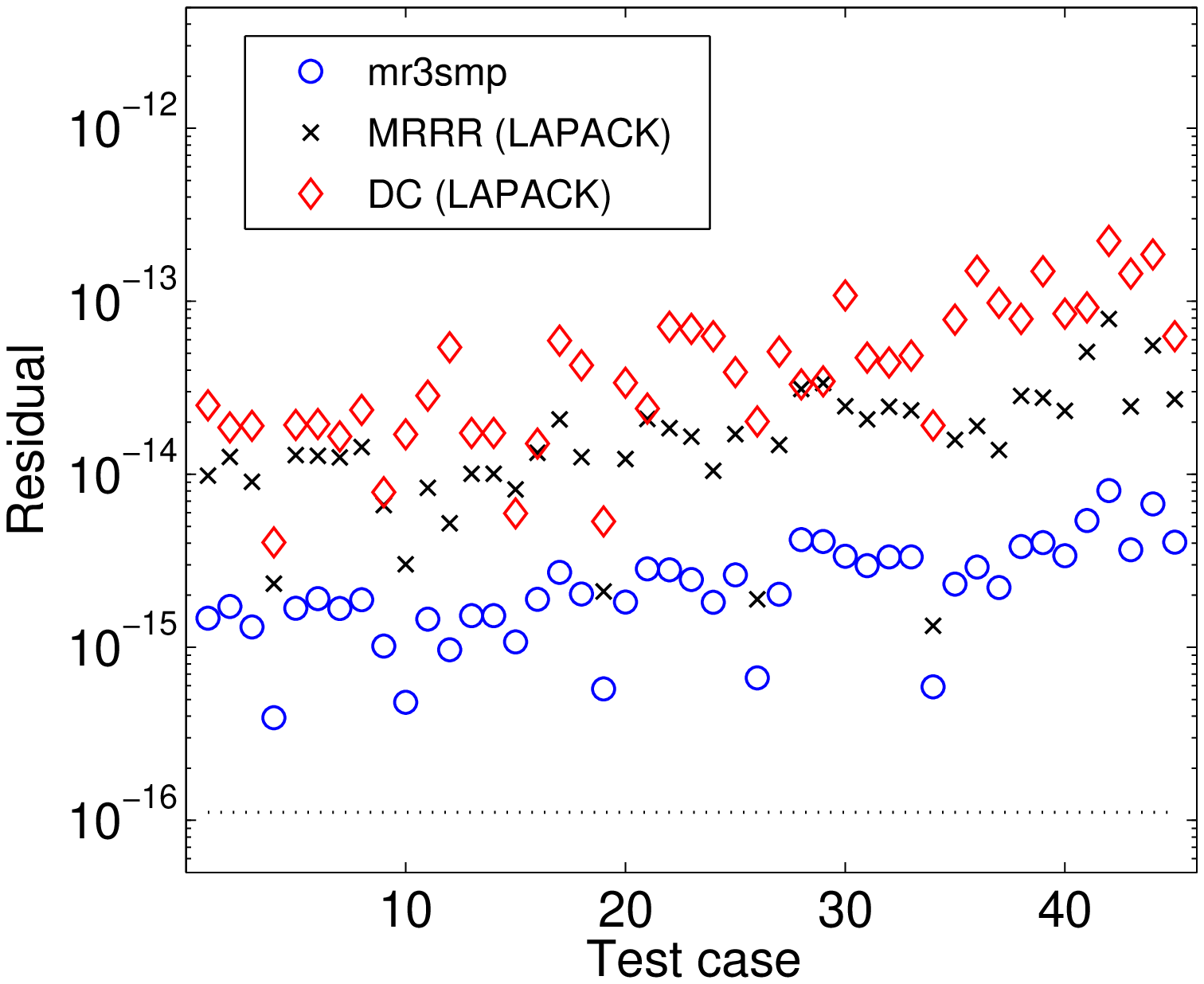} 
     \label{fig:motivationacclapack2a}
   } \subfigure[Orthogonality.]{
     \includegraphics[width=.47\textwidth]{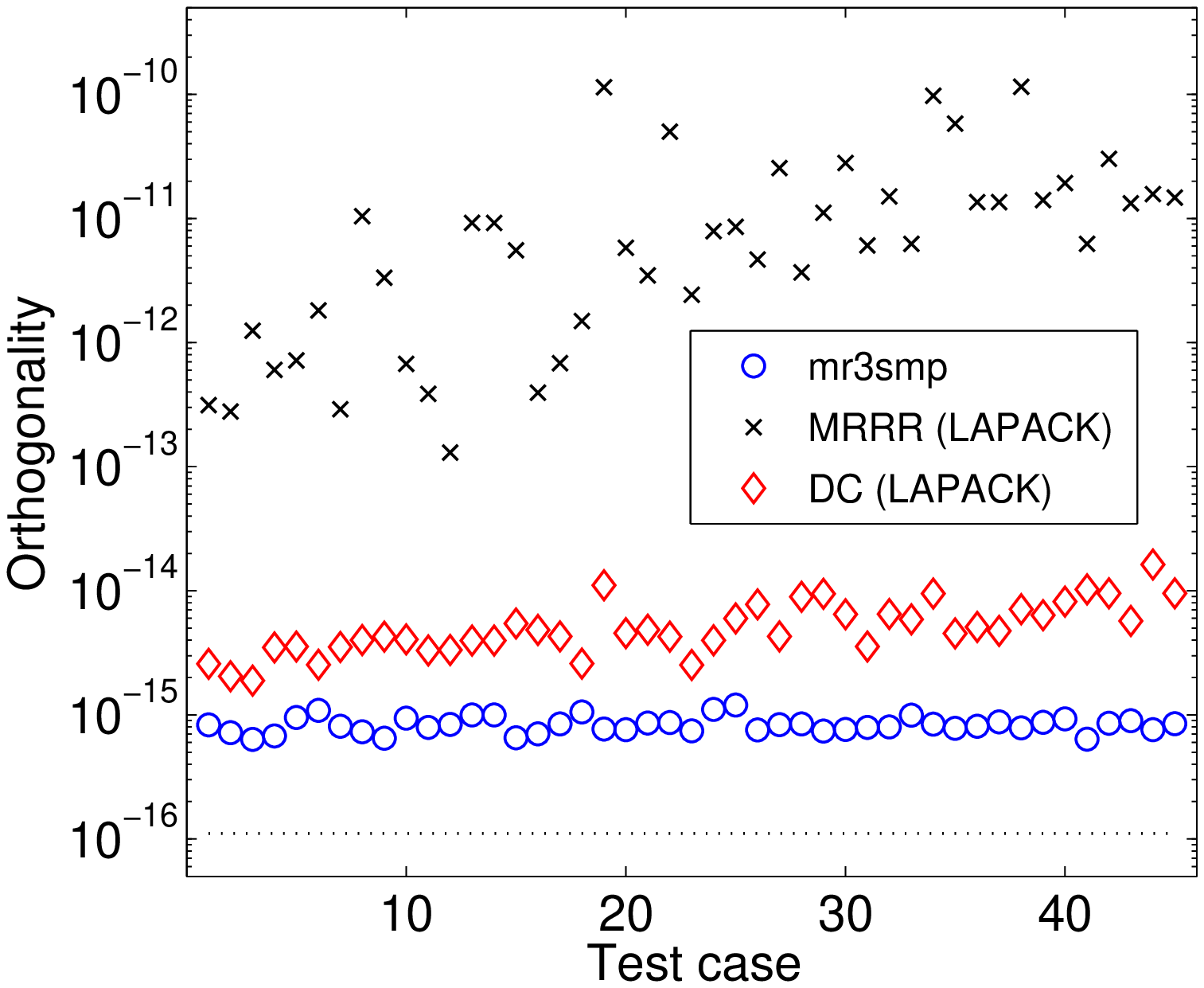}
     \label{fig:motivationacclapack2b}
   }
  \caption{
    For real symmetric tridiagonal matrices with dimension from $1{,}000$ to
    about $8{,}000$,
    accuracy of our mixed precision solver compared with LAPACK's DC and MRRR.
  }
  \label{fig:motivationacclapack2}
\end{figure} 

\begin{figure}[htb]
  \centering
   \subfigure[Execution time: multi-threaded.]{
      \includegraphics[width=.47\textwidth]{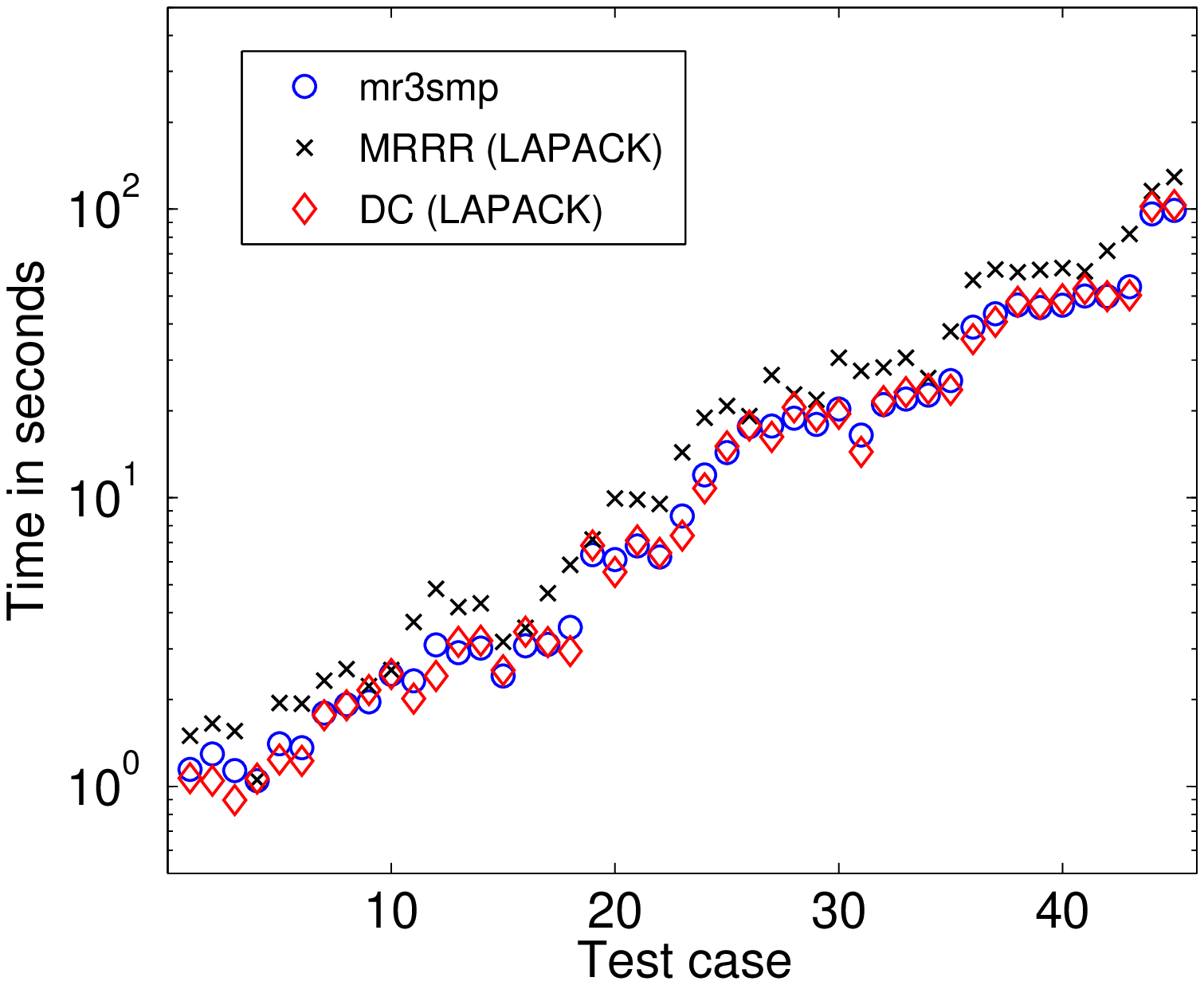}        
  \label{fig:motivationacclapack2densea}
   } \subfigure[Orthogonality.]{
      \includegraphics[width=.47\textwidth]{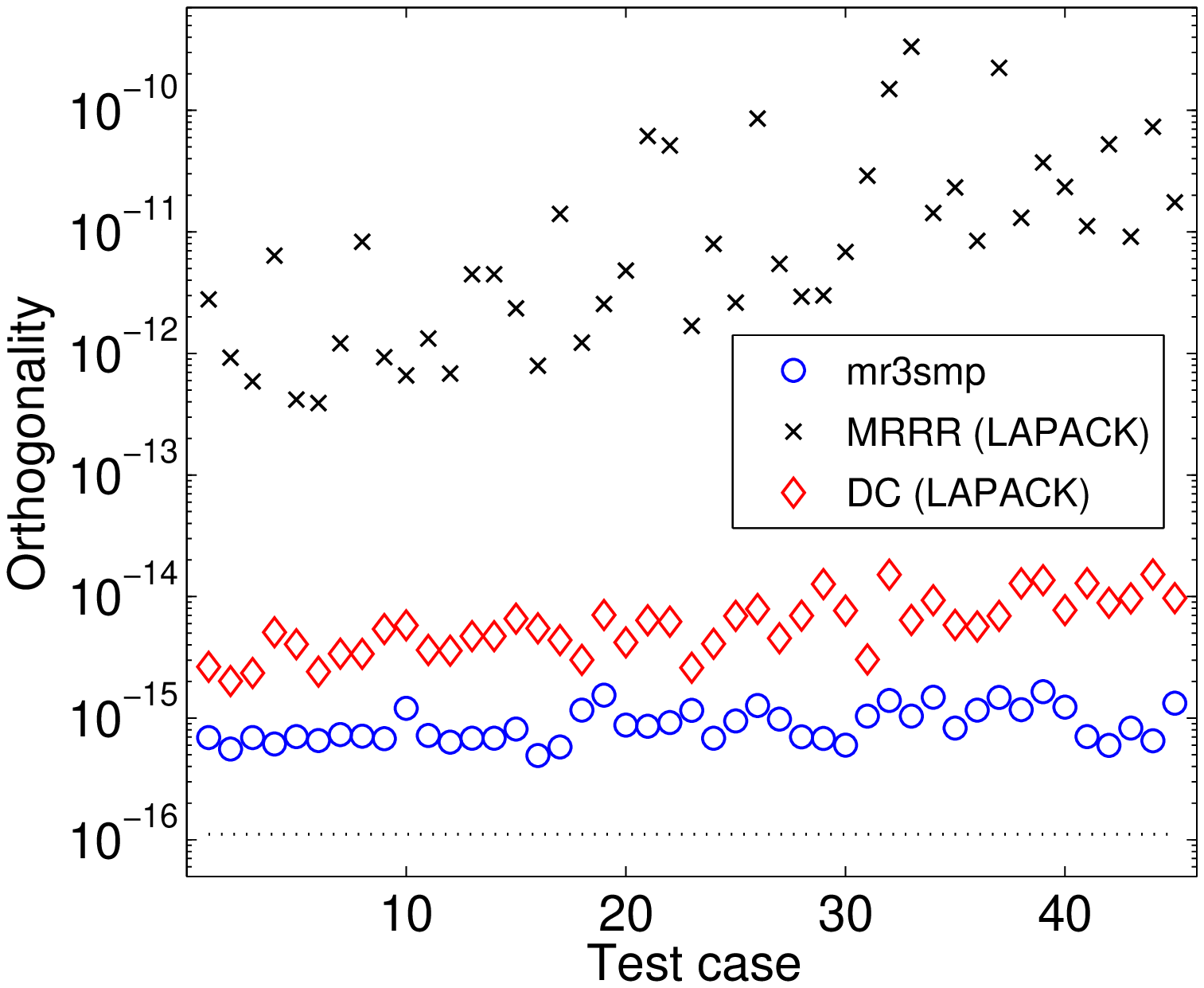}        
  \label{fig:motivationacclapack2denseb}
   }
  \caption{
    For real symmetric dense matrices with dimension from $1{,}000$ to about
    $8{,}000$, time and accuracy of our mixed precision solver compared with LAPACK's DC and
    MRRR. 
  }
  \label{fig:motivationacclapack2dense}
\end{figure} 

This work is mainly motivated by the performance study of Elemental's
eigensolvers. In the context of dense 
eigenproblems, the tridiagonal stage is often completely negligible
in terms of execution time: to compute $k$ eigenpairs of a tridiagonal matrix, it
only requires $\order{kn}$ operations; the reduction to
tridiagonal form requires $\order{n^3}$ operations and is the performance
bottleneck. In terms of accuracy, the 
tridiagonal stage is responsible for most of the loss of orthogonality. The
natural question is whether it is possible to improve the accuracy to the
level of the best methods without sacrificing too much performance. We show
that this is indeed possible. In fact, our mixed precision 
solver is more accurate than the ones based on DC or
QR [Fig.~\ref{fig:motivationacclapack2denseb}], and remains as fast as the classical MRRR
[Fig.~\ref{fig:motivationacclapack2densea}].    
Finally, an important feature of the mixed precision approach is a
considerably increased robustness  
and parallel scalability. 

\end{enumerate}

\section{Outline of the thesis}

In Chapter~\ref{chapter:background}, we give background material concerning
eigenproblems. In particular, we focus on methods for real symmetric tridiagonal
eigenproblems and direct methods for standard and generalized Hermitian
eigenproblems. We introduce basic terminology, existing algorithms, and
available software. We also quantify goals such as accuracy, scalability,
and load balance. A reader familiar with the above issues can safely skip all
or parts of Chapter~\ref{chapter:background}.

In Chapter~\ref{chapter:mrrr}, we focus on the MRRR algorithm and its
most prominent features. We introduce the method to an appropriate level for our
purposes. 
Our goal is neither to provide a rigorous exposition nor all the numerous
details of MRRR. Instead, the chapter serves -- together with
Chapter~\ref{chapter:background} -- as a basis for the later discussion of
Chapters~\ref{chapter:parallel} and \ref{chapter:mixed}. Therefore, we focus
on features that are important for 
parallelization and introduce the factors that influence parallelism,
robustness, and accuracy.  
An expert of MRRR will not find anything new
and can safely skip all or parts of Chapter~\ref{chapter:mrrr}. 

In Chapter~\ref{chapter:parallel}, we introduce our work on parallel versions of MRRR
 targeting modern multi-core and hybrid
distributed/shared-memory architectures. Section~\ref{sec:mr3smp} presents
the task-based parallelization strategy for multi-core architectures. Section~\ref{sec:pmrrr}
 presents \PMRRR, which merges the task-based parallelization with a
 parallelization for distributed-memory architectures via
 message-passing. We further discuss Elemental's new eigensolvers for dense
 Hermitian eigenproblems, which make
 use of \PMRRR, and their performance compared with ScaLAPACK. 

 Finally, in Chapter~\ref{chapter:mixed}, we introduce our mixed precision
 approach for MRRR. We demonstrate how it improves
 accuracy, robustness, and parallelism. Experiments indicate that these benefits come with little or even
 no extra cost in terms of performance.

\chapter{Background \& Related Work}
\label{chapter:background}
\thispagestyle{empty}

In this chapter, we compile background material concerning basic
terminology, existing algorithms, and state-of-the-art software.
The organization is as follows:
Section~\ref{sec:EVP} introduces the {\it Hermitian
eigenproblem} (HEP) and summarizes its basics properties. 
In Section~\ref{sec:notation}, we comment on our notation. 
In Section~\ref{sec:existingmethods}, we give a brief overview of the
existing methods with special emphasis on the 
{\it real symmetric tridiagonal eigenproblem} (STEP), which is the main
focus of this dissertation. We then return to the more general HEP and
introduce existing methods, in particular, direct methods based
on a reduction to tridiagonal form.   
The discussion of the HEP is important for two reasons: first, it
demonstrates that the STEP underlies most methods for the HEP, and
second, in later chapters, we show results of our solvers in the context of
direct methods for the HEP, referring to the material of this chapter
repeatedly. For the same reasons, we briefly discuss the 
{\it generalized Hermitian eigenproblem} (GHEP). 
In Section~\ref{sec:existingsoftware}, we list 
popular software that implements the previously specified algorithms. 
Finally, in Section~\ref{sec:objectives},
we list a set of objectives for any eigensolver. 
These objectives are 
used to compare different implementations throughout later chapters.

Our discussion is far from complete as ``the computation of
eigenvalues of matrices is one of the problems most intensively studied by
numerical analysts, and the amount of understanding incorporated in
state-of-the-art software [...] is [tremendous]''~\cite{trefethenbau}. 
As most of the material can be found in textbooks
(e.g.,~\cite{Parlett:1998:SEP,Demmel97appliednumerical,stewart2001matrix,watkins2010fundamentals,Golub1996,trefethenbau}), 
we limit the content of this chapter to important
aspects for this dissertation. 
Hence, a
reader familiar with the basics of Hermitian eigenproblems can safely skip
the rest this chapter. For the other readers, we
recommend to read this chapter as we use the material and
notation without reference in later chapters.

\section{The Hermitian eigenproblem}
\label{sec:EVP}

The {\it Hermitian eigenproblem} (HEP) is the following: Given an Hermitian
matrix $A \in \Cnn$ (i.e., $A = A^*$, where $A^*$ denotes the
complex-conjugate-transpose of $A$), find solutions to the equation
\begin{equation*}
A x = \lambda x \,,  
\end{equation*}
where $\lambda \in \R$, $x \in \Cn$, and $x \neq 0$. Without loss of
generality, we assume $\norm{x} = \sqrt{x^* x} = 1$ subsequently.
For such a 
solution, $\lambda$ is called an 
\textit{eigenvalue} (of $A$) and $x$ an associated
\textit{eigenvector}. An eigenvalue together with an associated
eigenvector are said to form an \textit{eigenpair}, $(\lambda, x)$. In
matrix form, the computation of $k$ eigenpairs is written
\begin{equation*}
A X = X \Lambda \,,  
\end{equation*}
where the eigenvalues are entries of the diagonal matrix $\Lambda \in \R^{k
  \times k}$ and the associated eigenvectors form the columns of $X \in \R^{n
  \times k}$. 

The HEP is ``one of the best understood`` and ``the most
commonly occurring algebraic''~\cite{templates} eigenproblems;
it has a number of distinct features.
In particular, we make use of the
following well-known result, cf.~\cite[Theorem 4.1.5]{hornjohnson}.

\begin{mythm}[Spectral Theorem for Hermitian matrices]
Let $A \in \Cnn$ be given. Then $A$
is Hermitian if and only if there is a unitary matrix $X \in \Cnn$, $X^* =
X^{-1}$, and a diagonal matrix $\Lambda \in \Rnn$ such that $A = X \Lambda
X^*$. Proof: See~\cite{hornjohnson,Golub1996}.
\label{thm:spectral}
\end{mythm}

The decomposition $A = X \Lambda X^*$ is called an {\it eigendecomposition}
of $A$ and, by the Spectral
Theorem, such a decomposition always exists for an Hermitian matrix. This is
equivalent to saying 
that $n$ distinct eigenpairs exist, i.e., $A x_i =
\lambda_i x_i$ for $1 \leq i \leq n$, where all eigenvalues are
real and the eigenvectors can be chosen to form an orthonormal basis for $\Cn$. Hence, the
eigenvalues can be ordered as
\begin{equation*}
  \lambda_1 \leq \lambda_2 \leq \ldots \leq \lambda_i\leq \ldots \leq \lambda_n  \,,
\end{equation*}
where $\lambda_i$ is the $i$-th smallest eigenvalue of $A$. In
situations where the underlying matrix is not clear, we write
$\lambda_i[A]$ explicitly. 
The set of all eigenvalues is called the {\it spectrum} of $A$ and denoted
by $spec[A]$. 

The eigenvectors can be chosen such that for all $i,j \in
\{1,2,\ldots,n\}$,   
\begin{equation}
  x_i^* x_j = \left\{ 
\begin{array}{l l}
    1 & \quad \text{if $i = j$} \,, \\
    0 & \quad \text{if $i \neq j$} \,. \\
  \end{array} \right.
\end{equation}
For distinct eigenvalues, corresponding eigenvectors are given as
(orthonormal) bases of $\mathcal{N}(A
- \lambda_i I)$ -- the null space of $A - \lambda_i I$. This means that the
eigenvectors are generally not 
unique. However, when an eigenvalue $\lambda_i$ is {\it simple}, i.e.,
distinct from all the other eigenvalues, then $\mbox{dim}\, \mathcal{N}(A -
\lambda_i I) = 1$ and the corresponding eigenvector is unique up to
scaling. 

Besides individual eigenvectors, subspaces spanned by a set of eigenvectors --
called {\it invariant subspaces}\footnote{For every invariant subspace, we
  can chose a set of eigenvectors as a basis, cf.~\cite{Parlett:1998:SEP}.}
-- are of special 
importance. For a given 
index set $\mathcal{I}
\subseteq \{1,2,\ldots,n\}$,  
\begin{equation*}
  \mathcal{X}_{\mathcal{I}} = \mbox{span} \{ x_i : i \in \mathcal{I}\}
\end{equation*}
denotes the invariant subspace associated with $\mathcal{I}$. As with the eigenvalues, we
write $\mathcal{X}_{\mathcal{I}}[A]$ for $\mathcal{X}_{\mathcal{I}}$ whenever the underlying matrix
is not understood from context.

When computing eigenpairs, it is useful to have transformations that change
the problem in a simple prescribed way. Two of these transformations are {\it
  shifts} and {\it similarities}. Transforming $A$ to $A -
\sigma I$ is called {\it shifting} and $\sigma \in \R$ is called a {\it
  shift}. It is easily verified that $(\lambda,x)$ is an eigenpair of $A$ if
and only if $(\lambda -
\sigma,x)$ is an 
eigenpair of  
$A - \sigma I$. Thus, the spectrum is
shifted by $\sigma$ and the eigenvectors are invariant. 
Similarly, given a nonsingular $Q \in \Cnn$, transforming $A$ to $Q^{-1} A Q$ is
called a {\it similarity transformation} or a {\it change of basis}. In this
case, $(\lambda,x)$ is an eigenpair of $A$ if and only if $(\lambda, Q^{-1} x)$ is an
eigenpair of $Q^{-1} A Q$. Thus, the 
spectrum is invariant under similarity transformations, while the
eigenvectors change in a simple way. In particular, if $Q$ is unitary,
$Q^{*} = Q^{-1}$, the transformation is called a {\it unitary similarity
  transformation}. Using this notion, the Spectral Theorem states that every
Hermitian matrix is unitarily similar to a real diagonal matrix. 
 
An important subclass of the HEP is the {\it real symmetric} eigenproblem,
which means $A$ is restricted to be real-valued, $A \in \Rnn$. In this
case, all complex-valued quantities become real-valued  
and the discussion of the HEP holds with the words 'Hermitian' and 
'unitary' respectively replaced by 'symmetric' and 'orthogonal'. 
The focus of this dissertation is on the even more specialized case of real
symmetric {\it 
  tridiagonal} matrices. Before providing an overview of existing methods for
the real symmetric tridiagonal eigenproblem, we give a number of comments
regarding our notation throughout this document.

\section{Notation}
\label{sec:notation}

Generally, matrices are denoted by
upper case Roman or Greek letters, while vectors and scalars are denoted
with lower case Roman or Greek letters. The underlying field ($\R$ or $\C$)
and the dimensions are specified in context. Some letters and symbols are
reserved for special quantities: 
\begin{itemize}[noitemsep,nolistsep]
\item $I$ denotes the identity matrix of appropriate size and $e_j$ denotes its
  $j$-th column, with all elements being zero except the $j$-th, which is one.
\item $\lambda$ is used for eigenvalue and is always real-valued in this document.
\item $A$ of size $n$-by-$n$ denotes a generic Hermitian matrix, possibly real-valued, with
  eigenvalues $\lambda_1 \leq \ldots \leq \lambda_n$ and corresponding
  eigenvectors $x_1, \ldots, x_n$. When computing (a subset of) eigenpairs,
  $\Lambda$ is the diagonal matrix with the eigenvalues as its 
  entries and $X$ contains the corresponding eigenvectors as columns. 
\item $T$ of size $n$-by-$n$ denotes a generic real symmetric tridiagonal
  matrix with 
  eigenvalues $\lambda_1 \leq \ldots \leq \lambda_n$ and corresponding
  eigenvectors $z_1, \ldots, z_n$. When computing (a subset of) eigenpairs,
  $\Lambda$ is the diagonal matrix with the eigenvalues as its 
  entries and $Z$ contains the corresponding eigenvectors as columns. 
\item $\varepsilon$ is used for {\it machine precision} or {\it unit roundoff},
  defined as half the distance between one and the next larger floating
  point number.
\item $spdiam[A]$ denotes the spectral diameter of matrix A, i.e.,
  $spdiam[A] = \lambda_n - \lambda_1$.
\item For $x \in \Rn$,
  $\norm{x} = \sqrt{x^* x}$ is the Euclidean norm; $\norm{x}_1
  = \sum_{i = 1}^n |x(i)|$ is the 1-norm. 
\item For $A \in \Cnn$, $\norm{A} = \max\{|\lambda_1|,|\lambda_n|\}$ denotes
  the spectral norm;  $\norm{A}_1 = \max_{1 \leq j \leq n} \sum_{i=1}^n |A(i,j)|$ denotes the 1-norm.
\item For $t \in \R$, we use the notation $\order{t}$ informally as ``of the
  order of $t$ in magnitude.'' The notion is used to hide moderate constants
  that are of no particular interest. 
\end{itemize}

Despite $\R \subset \C$, we frequently use the term {\it complex-valued
matrix} or $A \in \Cnn$ implying that at least one of its elements is not
real-valued. 
We use $A^*$ to denote the complex-conjugate-transpose of
matrix $A$. (If $A \in \Rnn$, $A^*$ denotes its
transpose.) 
We use the term {\it Hermitian} as the generic term; {\it
  symmetric}, however, is used synonymously with {\it real symmetric}, that is, no
complex symmetric matrices are encountered in this document. Similarly, a
{\it tridiagonal} matrix is implicitly {\it real symmetric}. 

Although eigenvectors are not unique, we talk about {\it the}
eigenvectors and, although we compute {\it approximations} to eigenvalues
and eigenvectors, we simply refer to {\it computed eigenvalues}
and {\it computed eigenvectors}. Such computed quantities are represented by
hatted symbols 
(e.g., $\hat{\lambda}$ or $\hat{z}$). The {\it set} of computed eigenvectors is
{\it numerically} orthogonal. We frequently, omit the terms 'set' and 'numerically' and say
that computed eigenvectors are orthogonal.

The topic of this dissertation is the efficient solution of ``large-scale''
eigenvalue problems. On a single processor, we consider problems as
``small'' if the entire input and output fits into the processors
cache. Similarly, on a distributed-memory system, we consider problems as
``small'' if the entire problem can effectively be solved on a uniprocessor
system. For instance, for today's hardware, computing the 
eigendecomposition of $A \in \C^{5000 \times 5000}$ (in IEEE double
precision) is considered large on a uniprocessor system, but small on a
massively parallel supercomputer.

\section{Existing methods}
\label{sec:existingmethods}

Methods differ in various aspects: the number of floating point operations (flops)
they have to perform (on average, in the worst case), the flop-rate at
which the operations are performed, the amount of memory required, the
possibility of computing subsets 
of eigenpairs at reduced cost, the attainable accuracy (on average, in the
worst case), the simplicity and robustness of an implementation, and the suitability
for parallel computations. In this section, we comment on the main features of different
methods for solving Hermitian eigenproblems in various forms.  

\subsection{The real symmetric tridiagonal eigenproblem}
\label{sec:existingmethodsSTEP}

A number of excellent algorithms for the real symmetric tridiagonal
eigenproblem exist. Among them,
bisection~\cite{wilkinson67,Parlett:1998:SEP,Demmel:bisec} 
and inverse iteration~\cite{wilkinson58,Wilkinsion:InvitReview,Wilkinson:1988,Jessup:1992:Invit,Dhillon98currentinverse,Ipsen:1997:Invit},
the QR algorithm~\cite{qr61a,qr61b,Parlett:1998:SEP},   
the Divide \& Conquer algorithm~\cite{dc81,Dongarra:1987:DC,dc94,dc95}, and
the method
of Multiple 
Relatively Robust
Representations (MRRR)~\cite{Dhillon:Diss,Dhillon:2004:Ortvecs,Dhillon:2004:MRRR,Fernando97,Parlett2000121}.\footnote{We concentrate on these four well-established
  algorithms. Other methods -- often modifications of four mentioned methods
  -- exist
  (e.g.,~\cite{homotopy1,Li92laguerre,Oettli:1999,multisection,Matsekh:2005,TsuboiKTKIN06,weko}).} 
All four methods are 
implemented in publicly available software (see
Section~\ref{sec:existingsoftware}) and a user is required  
to make a selection among the algorithms, each of them with its
strengths and  weaknesses and without a clear winner in all situations. 
The ``best'' algorithm might be influenced by a number of factors: 
the problem (e.g., size, spectrum, subsets), the architecture (e.g., cache sizes, parallelism), external
libraries (e.g., BLAS), and the specific implementation of the algorithm. 
At this point, we collect the main characteristics of each
algorithm; an experimental
comparison of their implementations is 
found in later chapters. 
Further comparisons -- general and experimental --
of the algorithms can be found
in~\cite{perf09,Dhillon:Diss,Bientinesi:2005:PMR3,handbook,Demmel97appliednumerical,templates}. 

We motivate our presentation: The comments
about QR demonstrate why the method is, when eigenvectors are desired, not competitive for
the STEP. The discussion about DC is important as
the method is generally fast, scalable, and
accurate; we therefore frequently compare our results to it. 
We introduce inverse iteration as it is closely
related to MRRR; in fact, ``it is difficult to understand MRRR without appreciating the limitations of
standard inverse iteration''~\cite{Dhillon:DesignMRRR}. 
Finally, MRRR is the main focus of this dissertation and
it is described more detailed in Chapter~\ref{chapter:mrrr}; at this point, we only mention its most salient
features.  

We remark that all four methods are numerically stable and, if
completed successfully, computed eigenpairs of input matrix $T$ fulfill
\begin{equation}
\norm{T \hat{z}_i - \hat{\lambda}_i \hat{z}_i} = \order{n \varepsilon
  \norm{T}} \,,
\label{eq:residualgoalfirst}
\end{equation}
and for $i \neq j$
\begin{equation}
|\hat{z}_i^* \hat{z}_j| = \order{n \varepsilon} \,.
\label{eq:orthogoalfirst}
\end{equation}
MRRR's accuracy and limitations (not for all inputs accuracy is guaranteed)
compared with other methods is the topic of Chapter~\ref{chapter:mixed}. 

\subsubsection{Bisection and inverse iteration (BI)}
Bisection, which is discussed in more detail in
Section~\ref{sec:eigvalscomputation}, may be used to find an approximation
to any eigenvalue with 
$\order{n}$ arithmetic operations~\cite{Demmel97appliednumerical,stewart2001matrix}; 
thus, it computes $k$ eigenvalues requiring only $\order{kn}$ flops. Given an
initial interval $[\alpha,\beta]$ which contains a sought after
eigenvalue, the interval is bisected until sufficiently
small.
Roughly $- 3n \log_2\varepsilon$ flops are required to
compute an eigenvalue to sufficient accuracy~\cite{Dhillon:Diss}.
Convergence is linear and rather slow.\footnote{``Convergence of
  the intervals can be accelerated by using a zero-finder such as {\tt
    zeroin} [...], Newton's method, Rayleigh quotient iteration [...],
  Laguerre's method, or other methods''~\cite{DemmelPNLA}.}
Nonetheless,
bisection has a number of great
features~\cite{Parlett:1998:SEP,Demmel:bisec,Marques:2006:BIF}: 
(1) It can be used to obtain approximations to just a subset of eigenvalues
at reduced cost; (2) It 
can be used to obtain approximations to low accuracy at reduced cost; (3) It
can be used to refine approximate eigenvalues to higher accuracy; (4) It can
be used to compute eigenvalues to high relative accuracy (whenever the data
defines them to such accuracy); (5) all computations are embarrassingly parallel.

After finding approximations to $k$ eigenvalues via bisection, inverse
iteration may be used to compute the corresponding eigenvectors. Inverse
iteration is one of the oldest methods for computing selected eigenvectors
when given approximations to the eigenvalues. According to an interesting
article on the history of the method~\cite{Ipsen95ahistory}, it was
introduced by Wielandt as early as 1944. 
Given an approximate
eigenvalue $\hat{\lambda}_i$ of input matrix $T$ and a
starting vector $\hat{z}_i^{(0)} \in \Rn$, the procedure consists of repeated
solutions of linear systems: for $k \geq 1$, 
\begin{equation*}
  (T - \hat{\lambda}_i I) \hat{z}_i^{(k)} = s^{(k)} \hat{z}_i^{(k-1)} \,,
\end{equation*}
where $s^{(k)} \in \R$ is a positive 
scaling factor, say for simplicity chosen such that
$\norm{\hat{z}_i^{(k)}} = 1$. Under mild assumptions, the sequence
$\hat{z}_i^{(k)}$ converges to an eigenvector approximation $\hat{z}_i$ with
small residual norm, i.e., such that \eqref{eq:residualgoalfirst} is
satisfied. 
In fact, the residual
norm $\norm{r} = \norm{T \hat{z}_i^{(k)} - \hat{\lambda}_i \hat{z}_i^{(k)}}
= s^{(k)}$ is readily available and is used to signal
convergence~\cite{Ipsen:1997:Invit}. A small residual of $\order{n
  \varepsilon \norm{T} }$ implies that
$(\hat{\lambda}_i, \hat{z}_i)$ is an eigenpair of a ``close matrix'', that is,
there exist a perturbation $E$ such that $(\hat{\lambda}_i, \hat{z}_i)$ is
an eigenpair of $A + E$ and $\norm{E} \leq
\norm{r}$~\cite{Ipsen:1997:Invit}. 
However, small residuals (backward
errors) are not sufficient to guarantee orthogonality among independently
computed eigenvectors -- i.e., \eqref{eq:orthogoalfirst}
 might not hold. 
Whenever $\hat{\lambda}_i$ is not 
well-separated (in an absolute sense) from the rest of the spectrum, 
even for a simple eigenvalue, the computed eigenvector
$\hat{z}_i$ might be a poor approximation to the true eigenvector $z_i$.\footnote{Measured by
the acute angle $\angle(\hat{z}_i,z_i) = \arccos{|\hat{z}_i^* z_i |}$,
 see Theorem~\ref{thm:gapthm}
 or~\cite{Ipsen:1997:Invit,Dhillon98currentinverse}.} 
Only if almost all eigenvalues are well-separated, inverse iteration is
efficient and requires $\order{kn}$ 
operations to compute the eigenvectors. In such a scenario, the 
computation of the eigenvectors is also easily performed in parallel. 
In contrast, if eigenvalues are clustered, orthogonality must be enforced,
usually by means of {\it Gram-Schmidt
orthogonalization}~\cite{Björck1994297}. This process is 
potentially costly and requires in the worst case $\order{k^2n}$ flops; 
additionally,
parallelism might be lost almost entirely. For
large-scale problems, the Gram-Schmidt  
procedure almost always increases the computation
time significantly~\cite{Bientinesi:2005:PMR3}. 

Together, bisection and inverse iteration, which is often considered a single method, has the
advantage of being adaptable;  
that is, the method may be used to 
compute a subset of eigenpairs at reduced cost. 
For this reason, it is still
today probably the most commonly used method for computing subsets of eigenpairs.  
Unfortunately, current software can fail to deliver
correct results~\cite{Dhillon98currentinverse,Demmel97appliednumerical}
and, {\it due to the explicit orthogonalization of
eigenvectors, its performance suffers severely 
on matrices with tightly clustered
eigenvalues}~\cite{Demmel97appliednumerical}. 
While BI has been the method of 
choice for computing a subset of eigenpairs for many years, the authors of
\cite{perf09} suggest that today MRRR ``is preferable to BI for subset
computations.''

\subsubsection{The QR algorithm (QR)}  

The QR algorithm\footnote{It is oftentimes called the {\it QR Iteration} and sometimes
  {\it Francis
  algorithm}. We
  do not distinguish between QR and QL algorithms. Excellent
  expositions of the method can be found
  in~\cite{Parlett:1998:SEP,stewart2001matrix,watkins2010fundamentals,Demmel97appliednumerical}.} is 
arguably the most ubiquitous tool for the 
solution of (dense) eigenvalue problems. On that account, it was placed in the
list of the top ten algorithms of the 20th century~\cite{Top10QR}. The
algorithm was independently discovered in the 1950s by Francis~\cite{qr61a} and
Kublanovskaja~\cite{qr61b} and has been studied extensively since
then. While ``for the general, nonsymmetric eigenvalue problem, QR is still
king''~\cite{Watkins:2008:QAR}, it faces serious competition 
in the real symmetric tridiagonal case -- for instance,
Divide \& Conquer, which is usually faster and equally
accurate~\cite{Rutter,Tisseur:1999:PDC}. 

The method generates a sequence of orthogonally similar 
tridiagonals $T_k = Z_k^* T Z_k$ whose off-diagonal entries are driven
rapidly to zero. In other words, $T_k$ converges to $\Lambda$, containing
the eigenvalues, and $Z_k$
converges to 
$Z$, containing
the eigenvectors. For each QR step, $T_{k} = Q_k^* (Z_{k-1}^* T Z_{k-1}) Q_k$, with $k
\geq 1$ and $Z_0 = I$, $Q_k$ is a
product of $n-1$ Givens rotations~\cite{Givens}, which must be accumulated,
$Z_{k} = Z_{k-1} Q_k $, if the eigenvectors are desired.
The process is guaranteed to
converge~\cite{Wilkinson1968409,doi:10.1137/0715060,Parlett:1998:SEP,Wang2001}
and leads to simple, robust and elegant implementations. Crucial for performance
and convergence are the implicit use of 
shifted matrices $T_k - \sigma_kI$ and the exploitation of deflation
whenever an off-diagonal is close to zero~\cite{Parlett:1998:SEP}.  

When only eigenvalues are desired, the rotations need not be
accumulated and the computation can be rearranged such that no square-roots,
which are usually more expensive than other operations, are needed. This
leads to the so called square-root free QR or the PWK
algorithm~\cite{Ortega01011963,Reinsch1971,Parlett:1998:SEP}. This algorithm
is quite efficient and requires only about $9n^2$ operations to compute all
eigenvalues~\cite{Parlett:1998:SEP}. 
The square-root free variant -- with its $\order{n^2}$ costs -- is frequently
used when only eigenvalues are computed on a uniprocessor. On the other
hand, alternative methods like bisection are more amendable on parallel
computing environments~\cite{DemmelPNLA}. 

The cost of QR changes dramatically when also the eigenvectors are
desired. In this case, the majority of the work is performed in accumulating
the Givens rotations. As the accumulation costs about $3 n^2$ operation per
QR step and roughly $2n$ QR steps are necessary for convergence, about $6 n^3$
arithmetic operations are necessary in total. For large problems, due to the
higher cost compared to other methods, QR is usually not competitive for the
STEP when eigenvectors are desired. In contrast to the ``eigenvalues only''
case, if eigenvectors are desired, the computation is parallelized quite
effectively~\cite{Bientinesi:2005:PMR3,DemmelPNLA}.

\subsubsection{Divide \& Conquer (DC)}  

{\it DC is among the fastest and most accurate methods
available}~\cite{perf09,Rutter,Demmel97appliednumerical,Tisseur:1999:PDC}. The
method is 
called by the authors of~\cite{trefethenbau} ``the most important advance in
matrix eigenvalue algorithms since the 1960s.'' 
It was
introduced 
by Cuppen in 1981~\cite{dc81}, but it took more than a decade to find a
stable variant of the algorithm~\cite{dc94,dc95}. 
The divide and conquer
strategy works by dividing the problem into two smaller
subproblems, which are solved recursively and whose solutions are combined
to the 
solution of the original problem. In our case, the tridiagonal is expressed
as rank-one modification of a direct sum of tridiagonals $T_1 = Z_1
\Lambda_1 Z_1^*$ and $T_2 = Z_2 \Lambda_1 Z_2^*$:
\begin{equation*} 
\small
  T = \left( \begin{array}{cc} T_1 & \\ & T_2 \end{array} \right) + \rho u
  u^* = \left( \begin{array}{cc} Z_1 & \\ & Z_2 \end{array}
  \right) \left[ \left( \begin{array}{cc} \Lambda_1 & \\ & \Lambda_2 \end{array}
  \right) + \rho v v^* \right] \left( \begin{array}{cc} Z_1^* & \\ & Z_2^* \end{array} \right) \,,
\end{equation*} 
where $\rho \in \R$ and $u,v \in \Rn$ are readily available without further computation.
The problem is solved by recursively applying the procedure to $T_1$
and $T_2$ to find their eigendecompositions and solving the eigenproblem for
a rank-one update of a diagonal matrix: $(\Lambda_1 \oplus \Lambda_2) +
\rho v v^* = Z_D \Lambda Z_D^*$. To complete the computation of $T$'s
eigendecomposition, $T = Z \Lambda Z^*$, the
eigenvectors are found by the matrix-matrix product $Z = (Z_1 \oplus Z_2)
\cdot Z_D$.  

Both the eigenvalues and the eigenvectors of a rank-one update of a diagonal
matrix are computed efficiently with only $\order{n^2}$ operations,
but the process must be done with great care to be numerically
stable~\cite{Li93solvingsecular,dc94,dc95}. The vast majority of
the computation is spent in the matrix-matrix product to obtain $Z$ --
hence, {\it it is crucial for 
performance of the algorithm that an optimized matrix-matrix multiplication
is available}. Neglecting the cost for computing the
eigendecomposition of $(\Lambda_1 \oplus \Lambda_2) + \rho v v^*$, 
assuming that all the matrices are dense, the subproblems are of size
$n/2$, and a standard matrix multiplication is used, the overall process requires roughly
$\frac{4}{3} n^3$
flops~\cite{Demmel97appliednumerical,stewart2001matrix,Tisseur98parallelDC}. 
However, the   
method frequently does much better than this because of numerical {\it
  deflation}~\cite{Dongarra:1987:DC,Rutter,Tisseur98parallelDC}: Whenever entries of $v$ are sufficiently small in magnitude or
two entries in $(\Lambda_1 \oplus \Lambda_2)$ are almost equal, some columns
of $Z_D$ can be made essentially columns of the identity matrix. 
This means, computations in the matrix-matrix product are
saved and eigenpairs
of the subproblem (padded with zeros) are accepted as eigenpairs of the
larger problem.\footnote{Assuming a fraction of $\delta$ eigenpairs that can
be deflated at any level, the total work becomes $\frac{4}{3} (1-\delta)^2 n^3 $
flops~\cite{Rutter}.} 
For both performance and numerical stability, the
``deflation 
process is essential for the success of the divide and conquer
algorithm''~\cite{Tisseur98parallelDC}. 
The amount of deflation depends on
the eigenvalue distribution and the structure of the
eigenvectors~\cite{Dhillon:Diss}. 

Depending on the sample of test matrices and the deflation criterion,
a different behavior of the algorithm is observed in practice:
in~\cite{Demmel97appliednumerical,Tisseur98parallelDC} it is reported that
for random test matrices ``it appears that [the algorithm takes] only
$\order{n^{2.3}}$ flops on average, and as low as $\order{n^{2}}$ for some
eigenvalue distributions.'' More complete tests in~\cite{perf09} lead to
the conclusion that ``DC is $\order{n^{2.5}}$ measured using time and
$\order{n^{2.8}}$ measured using flop counts'' on average, where the run
time behavior 
is specific to the architecture of the experiment and is explained by an
increasing flop-rate with the size of matrix-matrix multiplications. 

A big advantage of DC lies in the inherent parallelism of the
divide and conquer approach and its reliance on the
(usually available) highly optimized matrix-matrix kernel for computing the
eigenvectors. The main drawback of the method is that {\it it 
requires the largest amount of memory of all
methods} (it requires at least additional $n^2$ floating point numbers of
workspace~\cite{perf09,Rutter,Tisseur98parallelDC}).
Additionally, most
implementations cannot be used to compute a subset of 
eigenpairs at reduced cost. A study in~\cite{Auckenthaler2011} argues that up
to 70\% of the compute time is spent in the final two matrix-matrix
multiplications, unless extreme deflation takes place,
cf.~also~\cite{Rutter}. By only propagating 
the desired subset of eigenvectors in 
these steps, the computation time is reduced by up to a factor
three; see~\cite{Auckenthaler2011,Rutter} for details. While these savings for computing subsets are significant in
practice, ``the adaptation [of Divide \& Conquer] to this case is somewhat
artificial''~\cite{Dhillon:2004:MRRR}. Similarly, if only eigenvalues are
desired, DC requires only $\order{n^2}$ operations, but other methods, like
the square-root free QR Iteration, are often preferred. 

Therefore, as stated in~\cite{Rutter}, {\it ``if [enough memory] is available, if
the full eigendecomposition is 
desired, and if either highly optimized BLAS are available or highly
clustered eigenvalues are possible, we recommend divide and conquer as the
algorithm of choice.''} Since all these conditions are frequently satisfied, DC is one of the
most commonly used method. 

For a pedagogical treatment of the main principles behind DC, we
refer
to~\cite{Demmel97appliednumerical,stewart2001matrix,watkins2010fundamentals}. For 
the sake of completeness, we mention that {\it the}
Divide \& Conquer method does not exist, but a family of 
methods varying in different parts of the described procedure. Also, the operation count
can be lowered to 
$\order{n^2}$ for computing all eigenpairs and $\order{n \log_2 n}$ for
computing only eigenvalues -- with implicit constants too high to be used in
practice today. As these aspects are not
important for our discussion, we simply refer
to~\cite{Demmel97appliednumerical,dc95} and the references therein.

\subsubsection{Multiple Relatively Robust
  Representations (MRRR)} 

To compute $n$ eigenpairs, all the practical methods discussed so far charge $\order{n^3}$
arithmetic operations in the worst case. The MRRR algorithm is the first stable
method that does the job using $\order{n^2}$ flops.
Furthermore, the method is adaptable and computes $k$ eigenpairs in $\order{kn}$
time.\footnote{In practice, usually less than 200$kn$ flops are required.}

The method was introduced by Dhillon and
Parlett in the late 1990s~\cite{Dhillon:Diss}. Proofs of its correctness and a number of
improvements were added at the beginning of this
century~\cite{Dhillon:2004:MRRR,Dhillon:2004:Ortvecs,Willems:Diss};
however, the
method remains the subject of active research. 
It is
a variant of inverse iteration that removes the need for the costly
orthogonalization of the eigenvectors; details of the approach are given in
Chapter~\ref{chapter:mrrr}. While MRRR is a tremendous accomplishment, in
Chapter~\ref{chapter:ch01}, we already discussed remaining limitations and
justified the need for further improvements.

\subsubsection{A brief comparison} 

As mentioned, a number of factors, such as the spectrum of the input matrix
and the underlying hardware, influence the performance and accuracy of the
methods. 
Consequently, their evaluation is a difficult task. In
particular, when experimentally comparing different algorithms, we compare
{\it specific implementations} of the 
algorithms\footnote{On the danger of judging an algorithms based on a
  specific implementation, see~\cite{Stanley94theperformance}.} on a {\it
  specific architecture} and with {\it specific 
  external libraries} (e.g., BLAS or MPI) for a {\it specific task} (e.g.,
20\% of eigenpairs) on a set of {\it specific test matrices}. 

Probably the most complete study is given by Demmel et
al.~\cite{perf09}, comparing implementations of the four
algorithms on a number of architectures and a
wide range of test matrices. 
Here we summarize their main results: (1)~Despite the fact that all methods
deliver results that satisfy \eqref{eq:residualgoalfirst} and
\eqref{eq:orthogoalfirst}, QR and DC are more accurate than BI and
MRRR; (2)~DC requires $O(n^2)$ additional memory and therefore much more than all
the other algorithms, which only require $O(n)$ extra storage; 
(3)~DC and MRRR are {\em much faster} than QR and BI; despite the fact that MRRR
uses the fewest flops, DC is faster on certain classes of
matrices. If the full eigendecomposition is desired, DC is generally the method of
choice, but whether DC or MRRR is faster depends on the spectral distribution
of the input matrix; (4)~If only a subset of eigenpairs is desired, 
MRRR is the method of choice.

All these results are for {\it sequential} executions and do not take
into account how well-suited the algorithms are for parallel
computations. For experimental comparisons of parallel implementations, we
refer to Chapter~\ref{chapter:parallel}
and~\cite{Bientinesi:2005:PMR3,Tisseur:1999:PDC,Vomel:2010:ScaLAPACKsMRRR,para2010,mr3smp,EleMRRR}.

\subsection{The Hermitian eigenproblem}
\label{sec:HEP}

\subsubsection{Direct methods} 

So called {\it direct methods} are usually employed if the input and output
can be stored as full matrices in a computers memory and a significant
portion (say more than 3\%~\cite{DiNapoli2012}) of eigenvalues and
optionally eigenvectors are desired~\cite{templates}. A good
overview targeting the 
non-expert is given by Lang~\cite{LangDirectSolvers}; we 
concentrate on
the importance of the STEP within most methods and keep the discussion of
the other parts to a minimum.

Given an arbitrary (or banded) Hermitian matrix $A$, the most common approach to
solve the eigenproblem consists of three stages: 
\begin{enumerate}[noitemsep,nolistsep] 
  \item Reduction of $A$ to a real symmetric tridiagonal $T = Q^* A Q$ via a
    unitary similarity transformation. 
  \item Solution of the {\it real symmetric tridiagonal eigenproblem}: compute a
    (partial) eigendecomposition $T = Z \Lambda Z^*$, where $\Lambda \in \R^{k
      \times k}$ and $Z \in \R^{n \times k}$.
  \item Backtransformation of the eigenvectors via $X = QZ$.
\end{enumerate} 
Stage 3 becomes unnecessary in case only eigenvalues are desired.
In the following, we give comments and pointers to the literature for
all three stages.

\paragraph{Stage 1: Reduction to tridiagonal form.}

A classical direct reduction to tridiagonal form, the so called {\it Householder
  tridiagonalization}, is covered in most textbooks of
numerical linear algebra, 
e.g.,~\cite{Golub1996,watkins2010fundamentals,stewart2001matrix,trefethenbau,handbook}.\footnote{Alternatively,
the reduction can be performed with rotations~\cite{Givens} or by Lanczos
algorithm~\cite{Parlett:1998:SEP}, which are both less efficient.} Details
on the procedure in general and parallel implementations 
-- as actually used on modern (parallel) architectures -- are given
respectively in~\cite{blockedreduction,VanZee:2012:FAR} and 
\cite{Hendrickson1999,Stanley:Diss,Hendrickson94thetoruswrap,DongarraG92,Chang1988297}. 

The classical tridiagonalization proceeds in $n-2$ steps (unitary
similarity transformations):
\begin{equation*}
  Q_1^* A Q_1 = A_1 \rightarrow Q_2^* A_1 Q_2 = A_2 \rightarrow \ldots
  \rightarrow Q_{n-2}^* A_{n-3} Q_{n-2} = A_{n-2} =: T \,,
\end{equation*}
where $Q_j = I - 2 \frac{u_j u_j^*}{u_j^* u_j}$, $u_j\in \Cn$, are
Householder reflectors responsible for setting the elements of the $j$-th
column below the first
subdiagonal (and by symmetry, the elements of the $j$-th
row above the first super-diagonal) to zero. Due to
the use of unitary matrices the process is numerically
stable~\cite{Wilkinson:1988}. The
overall cost are $\frac{16}{3}n^3$ flops ($\frac{4}{3}n^3$
flops if $A \in \Rnn$). 
It is not necessary to
compute $Q = Q_1 Q_2 \cdots Q_{n-2}$ explicitly, which would require about
the same number of flops as the reduction; instead, the Householder vectors,
which define the transformation, are commonly stored in the original matrix $A$.

Dongarra, Sorensen, and Hammarling~\cite{blockedreduction} showed that
the computation can be restructured for more efficient data reuse; thereby, increasing the
performance on processors with a distinct memory hierarchy. 
The computation is restructured by delaying the application of a ``block'' of $n_b$ transformations
to a part of the matrix and then applying them in an aggregated  
fashion, a level-3 BLAS operation. 
``The performance of the blocked tridiagonalization algorithm depends heavily on an appropriate choice of the 'blocking factor' $n_b$. On the one hand,
increasing $n_b$ will usually increase the performance [...]. On
the other hand, blocking introduces $\order{n_b n^2}$ additional
operations''~\cite{Lang:1999:EES:329112}. For $n_b \ll n$, the effect of
blocking on the overall flop count is negligible, while data
reuse is greatly enhanced. Despite the
improved data reuse, only about half of
the flops are performed as matrix-matrix operations (level-3 BLAS),
while the other half of the flops are in matrix-vector products (level-2
BLAS). The slow flop-rate at which
the matrix-vector products are performed -- not the total number of flops --
makes the tridiagonalization {\it in many situations the 
performance bottleneck} in the three-stage approach. 
However, this is only true
provided the tridiagonal eigensolver is properly chosen; in later chapters,
we show situations in which, due to its inferior scalability, the tridiagonal stage is responsible for a significant fraction of the
overall execution time.\footnote{For instance, see  
Fig.~\ref{fig1:timingb} in Chapter~\ref{chapter:ch01} or
Fig.~\ref{fig:timepzhegvx1b} in Chapter~\ref{chapter:parallel}.}

An alternative to the blocked tridiagonalization 
is {\it successive band reduction} (SBR)~\cite{BischofSBR2000}.  
The idea is to split the reduction in two 
stages.\footnote{In general more than two stages might be
  used~\cite{BischofSBR2000}.} In the  
first stage, the matrix is reduced to 
banded form with bandwidth $b >
1$~\cite{Bischof94paralleltridiag,lang1997effiziente}.  
Unlike the direct reduction to
tridiagonal form, this stage can be cast almost entirely in terms of
matrix-matrix operations, thus attaining high-performance and parallel
scalability.  
The reduction is then completed, in a second stage, with a final band-to-tridiagonal
reduction~\cite{LangReductBandedtoTri,Rajamanickamblockedband}. While this stage
is negligible in terms of arithmetic operations (assuming $b \ll n$), it can
significantly contribute the the overall execution time due to its limited
parallelism~\cite{auckenthaler:developing}.\footnote{If $A$ is banded
  with moderate bandwidth to begin with, the reduction to tridiagonal form
  requires significantly less effort than for a full matrix.}
Compared with the classical reduction, SBR requires only $\order{n^2 b}$
additional flops, which is negligible if $b \ll n$~\cite{Auckenthaler2011}. 
However, achieving optimal 
performance is a balancing game as ``larger $b$ allows BLAS routines to
operate near peak performance and decreases [communication], but it also
increases the run-time of the reduction from banded to tridiagonal
form''~\cite{Auckenthaler2011}. 
The downside of SBR lies in the increased operation count in the
backtransformation stage. 
For this reason, {\it SBR is commonly used
when only the eigenvalues are desired or only a (small) fraction of the
eigenvectors is computed.}\footnote{For more on SBR in general,
  as well as implementations for uniprocessors and distributed-memory systems, we refer
  to respectively~\cite{BischofSBR2000,lang1997effiziente},
  \cite{Bischof:theSBRtoolbox,OurSBR,multicoreSBR,Ballard:2012:CAS}, and
  \cite{Bischof94paralleltridiag,lang1997effiziente,Lang:1999:EES:329112,auckenthaler:developing,Auckenthaler2011}.}

\paragraph{Stage 2: Solution to the STEP.}

Any tridiagonal eigensolver can be used; in particular, one
of the aforementioned methods (BI, QR, DC, MRRR). 
Since the various solvers based on a reduction to tridiagonal form commonly
differ only in this stage,
{\it differences in performance and accuracy are solely
attributed to the tridiagonal eigensolver}. Consequently, the previous comparison of the
methods largely applies to the three-stage approach for the HEP. 

The only exception
is QR: if eigenvectors are desired, it is not competitive for the
STEP. In contrast, using the techniques of~\cite{Lang:1998:RotBLAS3}, it
has been shown that, in the context of direct methods for the HEP, QR can be
(almost) competitive to MRRR or DC~\cite{RestructuredQR}.\footnote{The tests were performed
sequentially on a uniprocessor and it remains to been seen if QR can be equally competitive in
a parallel environment.}
For QR, the three-stage approach is
(usually) modified:\footnote{See~\cite{RestructuredQR}, which also discusses
  an alternative approach.} after Stage 1, matrix $Q$ is build 
explicitly and the rotations arising in the tridiagonal solver are applied to $Q$, i.e.,
we set $Z_0 = Q$ instead of $Z_0 = I$; Stage 3 becomes unnecessary.\footnote{A similar approach can be used for DC, but it
  is not commonly done.}  
If the one-stage reduction is used, building $Q$ requires about
$\frac{16}{3}n^3$ flops ($\frac{4}{3}n^3$ flops if $A \in \Rnn$); 
if SBR is used, it requires additional $8n^3$ flops ($2n^3$ flops if $A \in
\Rnn$)~\cite{RestructuredQR,Lang:1999:EES:329112}.    
Applying the rotations to $Q$ requires about $6 n^2$ flops per QR step ($3
n^2$ flops if
$A \in \Rnn$). Since roughly $2n$ QR steps are necessary for
convergence, about $12 n^3$ flops ($6 n^3$ flops if $A \in \Rnn$) are
required for the accumulation of rotations. As Stage 3 is omitted, if
properly implemented, QR becomes competitive to MRRR and DC in the context of
the standard HEP. 

Furthermore, while QR overwrites input $A$ with the eigenvectors, all the other methods
require additional storage for the eigenvectors of $T$. 
Consequently, if {\it all}
eigenvectors are desired, QR requires the least amount of memory. 
If only $k \ll n$ eigenvectors are computed, no
such savings are observed. 

Independent of the tridiagonal eigensolver, if only eigenvalues are desired,
the $\order{kn}$ or $\order{n^2}$ cost of Stage 2 is usually negligible
compared with the reduction to tridiagonal form. 
Similarly, if MRRR is used to compute $k$ eigenpairs, its $\order{kn}$ cost is
oftentimes insignificant compared with the $\order{n^3}$ cost of the reduction.

\paragraph{Stage 3: Backtransformation.}
The matrix $Q$ of the first stage is applied to the
eigenvectors of $T$. Normally, $Q$ is given implicitly by a sequence of
Householder transformation and the computation is cast almost entirely 
in terms of efficient matrix-matrix multiplications using the compact 
WY or the UT 
representation of a product of Householder
transformations~\cite{Schreiber:compactWY,Joffrain:2006:AHT}. 
Consequently, the backtransformation attains high performance and ``like
any algorithm involving mainly products 
of large matrices, [it is] easily and efficiently
parallelized''~\cite{LangDirectSolvers}.

If the one-stage reduction is used, the cost of the backtransformation is
$8kn^2$ flops ($2kn^2$ flops if $A \in \Rnn$). If the two-stage reduction is used, the
backtransformation stage equally slits into two stages: first, the
eigenvectors of the intermediate banded matrix are computed, and then the
eigenvectors of the input matrix~\cite{Auckenthaler2011}.\footnote{Alternatively, $Q$ is built 
  explicitly during the reduction phase and applied via a matrix-matrix
  multiplication.} The cost is about twice that of the one-stage reduction, which is 
considerable if $k$ is large. As already mentioned, this is the reason why
SBR is usually used if $k$ is sufficiently small compared with the matrix size
or no eigenvectors are desired.

\subsubsection{Methods without tridiagonalization.}
A variation of the described methods is the reduction to banded form
followed by a direct solution of the banded
eigenproblem~\cite{bandedQR,Arbenz19921105,Gansterer98dc1,Gansterer98dc2,tileDC2012}. In 
terms of performance, if eigenvectors are desired, such an approach is currently inferior to methods based on a reduction
to tridiagonal form~\cite{tileDC2012}. 
Methods without any initial reduction to 
condensed (tridiagonal
or banded) form are Jacobi's
method~\cite{Jacobi1846}, spectral divide and conquer
approaches~\cite{SpectralDC,PRISMinfra,Zha:1998,ZhangDC}, and
others~\cite{YauLu,Domas}.  

In terms of execution time, Jacobi's method is generally not competitive to
methods that are based on a reduction to tridiagonal
form~\cite{Demmel97appliednumerical}, but it remains valuable: 
When implemented carefully, it has the advantage of finding eigenvalues to
high relative accuracy (whenever the data defines them to such
accuracy)~\cite{Demmel92jacobivsQR,Parlett:1998:SEP}; furthermore, it is  
    naturally suitable for parallelism~\cite{Golub1996,Sameh71parallelJM}
    and it is fast when used on strongly diagonally dominant matrices.
 
Spectral divide and conquer techniques, such as presented by Nakatsukasa and
Higham in~\cite{SpectralDC},
``have great potential for efficient, numerically stable
computations on computing architectures where the cost of communication
dominates the cost of arithmetic''~\cite{SpectralDC}. 
By casting all computation in terms of QR (or Cholesky) factorizations and
matrix-matrix multiplications, the computation is performed efficiently and
with low communication 
overhead~\cite{SpectralDC}. Furthermore, the natural
parallelism of the divide and conquer approach makes it promising for
parallel 
computations. So far, no parallel implementation of the algorithm presented
in~\cite{SpectralDC} exists and future investigations will show if and under
which 
conditions the approach is superior to a reduction to condensed form.

\subsubsection{Iterative methods.} 
If problem sizes exceed the capability to store input and output in main
memory, sometimes direct methods are still applied in an
out-of-core fashion~\cite{Lang99outofcore,Toledo:1999}. However, very large-scale
problems are usually sparse and sparse matrix storage in
combination with {\it iterative methods} are employed. Commonly, these
methods are used to compute just a few eigenvalues and eigenvectors,
while computing a large number of eigenpairs efficiently remains an open
research question~\cite{Saad:2010}. 
While extremely important, iterative methods are not considered in
this dissertation. 
Instead, we merely note that 
``all subspace-based algorithms [...] need to use dense, tridiagonal, or
banded matrix routines as inner iterations to get Ritz approximations to
subspace eigenvalues''~\cite{templates}. Also, ``the large and dense
eigenvalue problem will gain importance as systems 
become larger. This is because most methods solve a dense eigenvalue problem
[that can] reach a size in the tens of thousands. Because of the
cubic scaling of standard eigenvalue methods for dense matrices, these
calculations may become a bottleneck''~\cite{Saad:2010}.

The only algorithm we mention explicitly is the well-known {\it Lanczos
  method}, as it naturally leads to real symmetric tridiagonal
eigenproblems -- demonstrating once more the importance of the STEP. Lanczos
method and other iterative methods are discussed 
in~\cite{Saad:eig,templates,Parlett:1998:SEP,stewart2001matrix}; a survey of
available software is given in~\cite{SurveySoftware,templates}.

\subsection{The generalized Hermitian eigenproblem} 
\label{sec:GHEPsixstages}

As we consider a specific generalization of the Hermitian
eigenproblem in later chapters, we introduce it briefly.  
A {\em generalized} Hermitian eigenproblem (GHEP) is the following: Given
Hermitian matrices 
$A, B \in \mathbb{C}^{n\times n}$, with $B$ positive definite (i.e.,
  $\lambda_1[B] > 0$), find solutions to the equation
\begin{equation}
  A x = \lambda B x \,,   \label{eq:GenEig}
\end{equation}
where $\lambda \in \R$, $x \in \Cn$, and $x \neq 0$. The
sought after 
scalars $\lambda$ and associated 
vectors $x$ are called  
{\em eigenvalues} and {\em eigenvectors}, respectively. We say that
$(\lambda, x)$ is an {\em eigenpair} of the {\em pencil} $(A,B)$. 
If $B$ is the identity matrix, \eqref{eq:GenEig} reduces to 
the {\em standard} HEP. 

Subsequently, we concentrate on {\it direct methods} for the GHEP, which
make use of the following fact: Given nonsingular matrices $G$ and $F$,
the eigenvalues of the pencil $(A,B)$ are invariant under the equivalence
transformation $(GAF, GBF)$; furthermore, $x$ is an eigenvector of
$(A,B)$ if and only if $F^{-1}x$ is an eigenvector of $(GAF,
GBF)$~\cite{Parlett:1998:SEP}.

The most versatile tool for the generalized eigenproblem, the QZ
algorithm~\cite{Moler73}, uses a sequence of unitary 
equivalence transformations to reduce the original pencil to generalized
(real) Schur form. By design, the  
QZ algorithm is numerically backward stable and imposes no restrictions on the
input matrices; unfortunately, the algorithm does not respect the symmetry
of the Hermitian pencil $(A,B)$ and is computationally rather
costly. The QZ algorithm and other methods -- both direct and iterative --
are discussed 
in~\cite{Parlett:1998:SEP,Golub1996,templates,Chandrasekaran:2000} 
and references therein.

To preserve the symmetry of the problem while reducing the pencil
$(A,B)$ to simpler form, methods are limited to sequences of
congruence transformations -- that is, using $G = F^*$, where $G$ and $F$ are no
longer required to be unitary.   
The traditional approach for computing all or a
    significant fraction of the eigenpairs 
of $(A,B)$ -- and the only one that will be of relevance later -- relies on
a transformation to a HEP~\cite{Martin68}. The HEP is in 
turn solved via a reduction to tridiagonal form.
Overall, the process for solving a generalized eigenproblem -- also known as
the Cholesky-Wilkinson method -- consists
of six stages: 
\begin{enumerate}[noitemsep,nolistsep]
\item {\it Cholesky factorization:}
$B = L L^*$, where $L \in \Cnn$ is lower triangular.

\item 
  {\it Reduction to standard form:} 
The original pencil $(A,B)$ is transformed to $(L^{-1} A
L^{-*},I)$, which takes the form of a standard Hermitian
eigenproblem. With $M = L^{-1} A L^{-*}$, an eigenpair $(\lambda_i, x_i)$ of
the pencil $(A,B)$ is related to an eigenpair $(\lambda_i, y_i)$ of $M$ by
$y_i = L^* x_i$. 
\item 
  {\it Reduction to tridiagonal form:} 
Reduction of $M$ to a real symmetric tridiagonal form via a
    unitary similarity transformation, $T = Q^* M Q$. 
The pencil $(M, I)$ is transformed to $(Q^* M Q, I)$, which takes the form
of a STEP. An eigenpair $(\lambda_i, y_i)$ of $M$ and an eigenpair
$(\lambda_i, z_i)$ of $T$ are related by $z_i = Q^* y_i$.
\item {\it Solution of the tridiagonal eigenproblem:} compute a
    (partial) eigendecomposition $T = Z \Lambda Z^*$, where $\Lambda \in \R^{k
      \times k}$ and $Z \in \R^{n \times k}$.
\item {\it First backtransformation:} In accordance to Stage 3, the 
  eigenvectors of the standard eigenproblem are obtained by computing $Y = Q
  Z$. 
\item {\it Second backtransformation:}
In accordance to Stage 2, the
eigenvectors of the original pencil are obtained by computing $X = L^{-*} Y$.
\end{enumerate}
The above discussion shows that $X^* A X = \Lambda$, $X^* B X = I$, and $A X = B X \Lambda$. 
Furthermore, with slight modifications of Stages 2 and 6, the same six-stage
procedure also applies to eigenproblems in the form $A B x=\lambda x$ and
$ B A x=\lambda x$. In the first case, the reduction to standard form and
the final backtransformation become $ M = L^* A L $ and $ X = L^{-*} Y $,
respectively; in the second case, they become $ M = L^* A L $
and $ X = L Y $. 

The Cholesky-Wilkinson method should only be used if $B$ is sufficiently
well-conditioned with respect to inversion. For a detailed discussion of
problems arising for ill-conditioned $B$, we refer to the standard
literature, including~\cite{Parlett:1998:SEP,stewart-sun:1990,templates,Golub1996}. 
As we have already discussed Stages 3--5, we now give some remarks
on the other three stages. 
\paragraph{Stage 1: Cholesky factorization.}
The factorization requires about $\frac{4}{3}n^3$ flops ($\frac{1}{3}n^3$
flops if $A,B \in \Rnn$), which can be cast
almost entirely in terms of level-3 BLAS. Consequently, it is highly
efficient and scalable. For details, we refer to~\cite{elemental,SuperMatrix,Quintana-Orti:2012}. 
\paragraph{Stage 2: Reduction to standard form.}
A comprehensive exposition for the two-sided triangular solve, $L^{-1} A
L^{-*}$, and the two-sided multiplication, $L^* A L$, is given
in~\cite{Poulson:TwoSided,FLAWN56}. 
The derived algorithms require about
$4 n^3$ flops ($n^3$ flops if $A,B \in \Rnn$). 
Efficient and scalable implementations for distributed-memory architectures
are discussed in \cite{Poulson:TwoSided,FLAWN56,Sears:1998}. 
If $A$ and $B$ are banded with moderate bandwidth the algorithm presented in~\cite{Crawford:1973} can be
  more efficient as it ``reduces the
  generalized problem to an ordinary eigenvalue problem for a symmetric band
  matrix whose bandwidth is the same as $A$ and $B$''.
\paragraph{Stage 6: Backtransformation for the GHEP.}
The triangular solve with multiple right hand sides or the
triangular matrix multiply is a level-3 BLAS operation; consequently, it is
efficient and parallelizable~\cite{ParallelBLAS3}. The computations require
roughly the same number of flops as Stage 2; the flop count is reduced if $B$ is banded.

\begin{figure}[htb]
  \centering
  \includegraphics[scale=.60, trim= 0mm 140mm 0mm 20mm]{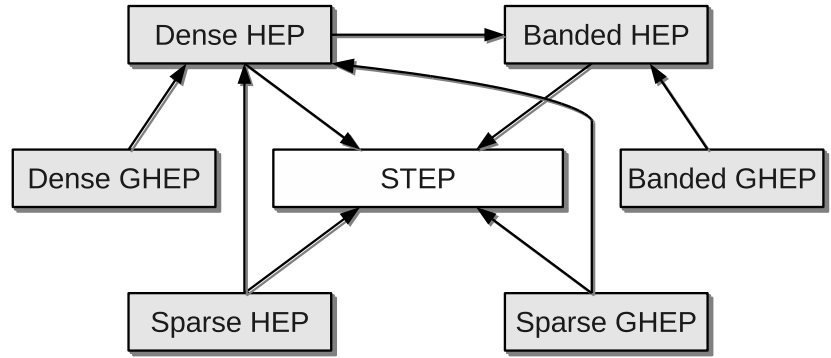}
  \caption{Computational paths leading to the real symmetric tridiagonal
    eigenproblem (STEP). {\it Dense} means that the matrix is treated as a
    dense matrix and a direct method is used for the solution, while {\it sparse} refers to
    the use of an iterative method, which are mainly used for large-scale
    sparse eigenproblems.} 
  \label{fig:steppaths}
\end{figure}

Section~\ref{sec:existingmethods} is summarized in
Fig.~\ref{fig:steppaths}, which emphasizes the 
importance of the STEP in solving dense or sparse Hermitian
eigenproblems. 
As {\em direct methods} do not
exploit any sparsity of the inputs, here and in the following, the term {\em dense
  eigenproblem} implies the use of direct methods for its solution, i.e., the input
is treated as if it were dense. Besides its importance for eigenvalue
problems, the STEP is an integral part in the computation of singular value
decompositions~\cite{Willems:2012:MR3GK}.   

\section{Existing software}
\label{sec:existingsoftware}

Numerous software implements the previously
described methods for the solution of eigenvalue problems. 
In this section, we list a few libraries -- with no intention of being
exhaustive.\footnote{For a list of freely 
  available software for linear algebra computations, see~\cite{FreeSoftware}.} 

In 1972, EISPACK~\cite{eispack} was released --  a collection of Fortran
subroutines based on the ALGOL programs by Wilkinson et
al.~\cite{Wilkinson:handbook}. LAPACK  -- first released in 
1992 -- superseded EISPACK and includes a number of new and improved
algorithms for eigenvalue problems.  
Efficiency and 
(trans)portability is achieved by casting the
computation in terms of common building blocks -- the BLAS (Basic Linear Algebra
Subprograms)~\cite{blas1,blas2,blas3}.\footnote{The term
  ``transportability'' is sometimes used instead of ``portability'' because LAPACK
  relies on a highly optimized BLAS implemented for each machine~\cite{lapack}.}
 BLAS are commonly
provided in optimized, machine-specific form such as included in Intel's MKL. 
Today, LAPACK is the most widely used numerical linear algebra library on serial and
shared-memory machines; it is provided in optimized form by all major
computer vendors (e.g.,
Intel's MKL, IBM's ESSL, AMD's ACML, Sun Performance library) and commonly
used through high-level interfaces (e.g., Matlab, Maple, NAG Numerical
libraries). For the tridiagonal eigenproblem, LAPACK includes
implementations of the four described methods (BI, DC, QR, and MRRR). In
line with the discussion in Section~\ref{sec:existingmethods}, the
tridiagonal eigensolvers form the basis of direct methods for standard and
generalized Hermitian eigenproblems. A list of LAPACK's 
eigensolvers relevant to this dissertation can be found in Appendix~\ref{appendix:routinenames}.\footnote{As a side note: all these libraries are usually designed to be
  efficient for larger problem sizes. For efficient solution of 3x3 problems,
  see~\cite{Kopp}.} 
LAPACK does not support successive band reduction, but the
corresponding routines are provided by the SBR
toolbox~\cite{Bischof:theSBRtoolbox} and are included in some 
vendor-specific LAPACK libraries. 
LAPACK's support for multi-core
architectures is confined to the use of multi-threaded BLAS. 

A number of research efforts are currently devoted to adapt LAPACK's
functionality to multi-core and many-core processors. Among them, 
PLASMA~\cite{plasma}, MAGMA~\cite{plasma}, and
FLAME~\cite{libflame,Zee:2009:LLD}. When  
  this dissertation started, none of those projects included routines for
  the solution of eigenproblems.\footnote{PLASMA User's Guide
    Version 2.0: ``PLASMA does not
    support band matrices and does not solve eigenvalue and singular value
    problems.''} Recent effort changed this
  situation: Today, FLAME includes a solver for the HEP based on the direct
  reduction to tridiagonal form~\cite{VanZee:2012:FAR} and a
  QR~\cite{RestructuredQR} that is significantly faster than its LAPACK
  analog. 
  MAGMA has support for the HEP~\cite{Haidar:2011,MagmaDC} and PLASMA
  added support for the GHEP and the HEP based on SBR~\cite{multicoreSBR}
  and QR. 

  Several libraries address distributed-memory architectures, including
  ScaLAPACK~\cite{scalapack}, PLAPACK~\cite{PLAPACKbook},
  PeIGS~\cite{peigs,dhillonfannparlett1}, PRISM~\cite{PRISMinfra,PRISMeig}, and 
  Elemental~\cite{elemental}. ScaLAPACK was introduced in 
  1995; its goal is to provide
  the same functionality as LAPACK. 
  A list of ScaLAPACK's eigensolvers relevant to this dissertation
  can be found in Appendix~\ref{appendix:routinenames}.  
  Routines for an alternative one-stage reduction and the
  two-stage SBR were released within the ELPA
  project in 2011~\cite{auckenthaler:developing,Auckenthaler2011}. 
  Introduced in the 1990s was also PLAPACK, which is
  ``an [...] infrastructure for rapidly prototyping''~\cite{plapackSBR}
  parallel algorithms and ``resulted from a desire to solve the programmability
  crisis that faced computational scientists in the early days of massively
  parallel computing''~\cite{FLAWN44}. For PLAPACK related to eigenproblems,
  see~\cite{Bientinesi:2005:PMR3,plapackSBR}.
  As the ``major development on
  the PLAPACK project ceased around 2000''~\cite{FLAWN44}, today, it is 
  largely outdated and superseded by Elemental.
  Elemental is equally a framework for dense matrix computations on distributed-memory
architectures. 
Its main objective is to ease the process of
implementing matrix operations without conceding performance or scalability. 
Elemental includes a library for commonly used matrix operations; for
standard and generalized
 Hermitian eigenproblems, the
library includes eigensolvers based on the parallel MRRR presented in
Chapter~\ref{chapter:parallel}. More on Elemental's eigensolvers can be found
in Chapter~\ref{chapter:parallel}. 

\section{Objectives}
\label{sec:objectives}

The ideal eigensolver is accurate, fast, scalable, and
reliable. In this section, we state how to quantify these
attributes. 
Basic definitions such as accuracy, scalability and load balancing are
adopted from~\cite{Parlett:1998:SEP,scott2005scientific}. 


\paragraph{Accuracy.}

Given an Hermitian matrix $A \in \Cnn$ (possibly real-valued and tridiagonal) and a
set of computed eigenpairs 
$\{ (\hat{\lambda}_i,\hat{x}_i) : i \in
\mathcal{I} \}$, $\|\hat{x}_i\| = 1$, we quantify the results by the {\it largest residual norm}
and the {\it orthogonality}, which are defined as  
\begin{equation}
  R = \max_i \frac{\| A \hat{x}_i - \hat{\lambda}_i \hat{x}_i
    \|_1}{\| A \|_1} \quad \mbox{and} \quad
  O = \max_i \max_{j \neq i} | \hat{x}_i^* \hat{x}_j | \label{def:defresortho2}
  \,,
\end{equation}
respectively.\footnote{In the following, we assume that $\|\hat{x}_i\| = 1$ holds exactly.} 
If both $R$ and $O$ are $\order{n \varepsilon}$, with implied constant
in the hundreds, we say that the eigenpairs are computed {\it accurately}.  
In the following, we accept this 
definition of accuracy without further discussion.

The accuracy depends on a number of
factors: the algorithm $\mathcal{A}$ used\footnote{See
  Section~\ref{sec:existingmethods} for the 
  discussion of different algorithms.}, a set of parameters $\mathcal{P}$ of a specific 
implementation of $\mathcal{A}$, and the input matrix $A$ -- in particular
its dimension $n$. 
Consequently, we 
have $R(\mathcal{A}, \mathcal{P},
n,A)$ and $O(\mathcal{A}, \mathcal{P},
n,A)$.  
From each algorithm $\mathcal{A}$, we isolate the set
of parameters $\mathcal{P}$, including convergence criteria and thresholds,
so that if two implementations of the same algorithm only differ in $\mathcal{P}$, then 
they are still considered implementations of the same algorithm. 

To compare a set of algorithms, each with a fixed set of
parameters, we factor in the dependence of the
input matrix by obtaining results for a large set of test matrices. 
Ideally, such a test set would be somewhat standardized and consist of a variety
of application and artificial matrices in a wide range of sizes. 
Taking for each size $n$ the average and worst case accuracy usually represents well the
accuracy of an algorithm and gives some {\it practical} upper bounds on the
residuals and orthogonality. In general, for a stable algorithm it is possible to provide
{\it theoretical} error bounds, independently of all properties of the input
matrix but its size. These theoretical upper bounds are essential for proving
stability and for improving algorithms, but often greatly overestimate the
actual error. Wilkinson wrote~\cite{WilkinsonModernError,Golub1996}, ``a 
priori bounds are not, in general, quantities that 
should be used in practice. Practical error bounds should usually be
determined in some form of a posteriori error analysis, since this takes full
advantage of the statistical distribution of rounding error
[...].'' In this thesis, we will evaluate and compare the accuracy of algorithms by
executing them with a set of test matrices.

\paragraph{Speed.}

Suppose we solve a fixed problem using $p$ processing units, indexed
by $1 \leq i \leq p$, with individual execution times $\tilde{t}_i$. 
The time to solution, $t_p$, is naturally defined as 
\begin{equation}
  t_p = \max\{ \tilde{t}_i : 1 \leq i \leq p\} \,.
\label{eq:timetosolution}
\end{equation}

For different values of $p$, similar to the accuracy assessment, the
average and worst case execution times on a set of test matrices can be used
to compare the performance of different eigensolvers. 
Such an approach, with $p=1$, is taken in \cite{perf09} to evaluate the
performance of LAPACK's symmetric tridiagonal eigensolvers. 
As the performance not only depends on the input
matrices, but also on the architecture and the implementation of
external libraries, an evaluation is a difficult task. 
In many of our experiments, not only the test set is 
inadequate in its generality,
but also the variety of the underlying hardware is not large
enough to draw final conclusions (it never is). However, from our
observations, we can derive some general behavior of the different methods
and their implementations. 

\paragraph{Scalability.}

For a fixed problem, let $t_p$ be as defined in \eqref{eq:timetosolution} and let $t_{ref}$ be a 
reference time using $p_{ref} \leq p$ processing units. The {\it speedup}, $s_p$,
is defined as 
\begin{equation}
 s_p = \frac{t_{ref} \cdot p_{ref}}{t_p} \,.
\end{equation}
Usually $t_{ref} = t_1$ refers to the execution of
the {\it best available} sequential solution and, consequently, $p_{ref} = 1$. 
The corresponding {\it parallel efficiency},
$e_p$, is defined as
\begin{equation}
 e_p = \frac{s_p}{p} = \frac{t_{ref} \cdot p_{ref}}{t_p \cdot p} \,.
\label{eq:paralleleffstrng}
\end{equation}
When investigating the scalability (speedup or efficiency) for a fixed
problem, we refer to {\it strong scalability}.  For practical problems and
values for $p$, the goal is to achieve $s_p \approx p$ and $e_p
\approx 1$. If this is achieved, we say that we obtain {\it perfect speedups} or {\it
  perfect scalability}. 

By {\it Amdahl's law}~\cite{Amdahllaw:1967}, even without taking
synchronization and communication cost into account, perfect speedup is
not always attainable. If a fraction $f$ of a code is inherently sequential,
the speedup and efficiency are limited by  
\begin{equation}
 s_p \leq \frac{1}{f + (1-f)/p} < \frac{1}{f} \quad \mbox{and} \quad e_p \leq \frac{1}{1 + (p-1)f} \,.
\end{equation}
Since always $f > 0$, the efficiency eventually goes to zero as
$p$ increases for any parallel code. 

Besides strong scalability, we are
interested in the so called {\it weak scalability}, where the number of processors
is increased together with the problem size. 
As attested by Fred Gustafson in his article ``Reevaluating Amdahl's Law'',
weak scalability has often a greater significance as in ``practice, the
problem size scales with the number of processors [...] to make use of the
increased facilities''~\cite{Gustafson88}. In particular, we consider the so
called {\it memory-constraint scaling},
where the amount of memory per processor is kept constant while problem size
and number of processors are increased.

For weak scalability, the {\it parallel efficiency}, $\check{e}_p$, is
defined as
\begin{equation}
 \check{e}_p = \frac{e_p \cdot \theta}{\theta_{ref}} = \frac{t_{ref} \cdot p_{ref}
   \cdot \theta}{t_p \cdot p \cdot \theta_{ref}} \,,
\label{eq:paralleleffweak}
\end{equation}
where $\theta$ and $\theta_{ref}$ are measures of the work required to
solve respectively the problem and the reference problem.
For standard and generalized dense eigenvalue problems, we have $\theta/\theta_{ref} =
n^3/n_{ref}^3$ and, for MRRR, we have $\theta/\theta_{ref} \approx n^2/n_{ref}^2$.
If in an experiment $\check{e}_p \approx 1$, we say that we obtain {\it
  perfect scalability}. As with accuracy and timings, 
the average and worst case efficiency on a set of test
matrices can be used to compare the scalability of different eigensolvers. 

\paragraph{Load and memory balancing.}

With $\tilde{t}_i$ as defined above, the average execution time,
$\bar{t}_p$, is given by 
\begin{equation}
  \bar{t}_p = \frac{1}{p} \sum_{i = 1}^p \tilde{t}_i \,.
\end{equation}
The {\it load balance} is quantified by
\begin{equation}
  b_p = \frac{\bar{t}_p}{\max \{\tilde{t}_i : 1 \leq i \leq p\}} \,.
\end{equation}
An execution is called {\it load balanced} if $b_p \approx 1$ and {\it
unbalanced} if $b_p \approx 0$. 
Using \eqref{eq:paralleleffstrng} with $t_{ref} = t_1$ and assuming $t_1
\leq \sum_{i=1}^n \tilde{t}_i$, we have $e_p 
\leq b_p$. Consequently, {\it without load balancing, it is not possible to achieve
good parallel efficiency}. 

Besides a balanced workload, the total memory usage should be equally distributed
among the $p$ processing units. For a specific problem, let $m$ denote the
total memory requirement and let $\tilde{m}_i = \tilde{c}_i m/p$ be the memory
requirement for processing unit $i$. If 
 $c_p = \max \{c_i: 1 \leq i \leq
p\}$ is bounded by a small constant for the tested values of $p$, 
we say that we achieved {\it perfect memory
  balancing}. Again, for a fixed problem, if $p$ increases, eventually memory
balancing is lost for any parallel code. Therefore, as for the weak scaling, we
often are interested in the memory balancing for problems that increase in size as
the number of processors increase. For computing all eigenpairs, we have $m =
\nu n^2$, with $\nu$ being a constant depending on the solver. 
Consequently, memory balancing is achieved when the maximum memory required by any processing
unit, 
\begin{equation}
m_p = \max \{\tilde{m}_i: 1 \leq i \leq
p\}  \,,
\end{equation}
is $\order{ n^2 / p }$, with a small implicit constant.

\paragraph{Robustness.}

In order to quantify the robustness of an eigensolver, we use the
following measure: For a given test set of matrices, {\sc TestSet}, the
robustness $\phi \in [0,1]$ is expressed as 
\begin{equation}
  \phi(\mbox{\sc TestSet}) = 1 - \frac{\mbox{\sc NumFailures}}{|\mbox{\sc TestSet}|} \,
\end{equation}
where {\sc NumFailures} is the number of inputs for which the method
``fails''. We will be more concrete on what to consider failure in
Chapter~\ref{chapter:mixed}.

\chapter{The MRRR Algorithm}
\label{chapter:mrrr}
\thispagestyle{empty}

Since its introduction 
in the 1990s, much has been written about the MRRR algorithm; its 
theoretical foundation is discussed in several
publications~\cite{Dhillon:Diss,Fernando97,Dhillon:2004:MRRR,Dhillon:2004:Ortvecs,Parlett2000121,perturbLDL,localization,Willems:Diss,Willems:framework,Willems:twisted}
and practical aspects of efficient and robust implementations are discussed
in~\cite{Dhillon:DesignMRRR,glued,NLA:NLA493,Marques:2006:BIF,Bientinesi:2005:PMR3,Vomel:2010:ScaLAPACKsMRRR,EleMRRR,mr3smp,Vomel:2012:FineGrain}. 
In particular, Paul Willem's dissertation~\cite{Willems:Diss} as well as
Inderjit Dhillon's 
and Beresford Parlett's two seminal articles~\cite{Dhillon:2004:MRRR,Dhillon:2004:Ortvecs}
provide excellent and essential reading for everyone interested in the algorithm. 
One could say, with an implementation of the
algorithm in the widely used LAPACK library and the
description of (parts of) the algorithm in textbooks such
as~\cite{watkins2010fundamentals}, 
MRRR has become mainstream. 

Most publications however are either concerned about a proof of correctness or only a specific
detail of the algorithm. Our description of the algorithm has a different
purpose: for the non-expert, we describe the factors influencing
performance, parallelism, accuracy, and robustness. We highlight the
main features of the algorithm, the sources of parallelism, and the
existing weaknesses.   
Our exposition, which is largely based
on~\cite{Dhillon:2004:MRRR,Dhillon:2004:Ortvecs,Willems:Diss,Willems:framework},
serves as a basis for the discussion in Chapters~\ref{chapter:parallel} and
\ref{chapter:mixed}. We are not afraid to omit details and
 proofs, which can be found in other places in the literature. 
In contrast to Chapters~\ref{chapter:parallel} and \ref{chapter:mixed}, in
this chapter we are not primarily concerned with efficiency.  

The chapter is organized as follows: In Section~\ref{sec:bigpicture}, we
give a high-level description of MRRR, with the main goal being the
introduction of Algorithm~\ref{alg:mrrr}. Furthermore, we establish the terminology and
notation used in later chapters. For the discussion in Chapter~\ref{chapter:mixed}, we
present in Theorem~\ref{resthm} all the factors that influence
the accuracy of MRRR. By fixing the form to represent tridiagonals,
Section~\ref{sec:closerlook} provides a more concrete description of the 
computation of eigenvalues and eigenvectors. 
An expert of MRRR might safely skip the entire chapter and
continue with Chapter~\ref{chapter:parallel} or~\ref{chapter:mixed}. We
recommend however to read at least Section~\ref{sec:bigpicture}, as it
contains all the required background and notation for the later chapters.

\section{The big picture} 
\label{sec:bigpicture}

Consider the symmetric tridiagonal $T \in \Rnn$ with diagonal $a = (\alpha_1, \ldots, \alpha_n)$ and off-diagonal $b = (\beta_1,
\ldots, \beta_{n-1})$:
\begin{equation}
  T = \left( \begin{array}{cccc}
\alpha_1 & \beta_1    &         &  \\
\beta_1 & \alpha_2    & \ddots  &  \\
    & \ddots & \ddots  & \beta_{n-1} \\
    &        &  \beta_{n-1} & \alpha_n \\
\end{array} \right) \,.
\label{eq:thetridiagonal}
\end{equation}
Recall that the goal is to compute a set of eigenpairs
$\{ (\hat{\lambda}_i,\hat{z}_i) : i \in \mathcal{I} \}$, $\norm{\hat{z}_i} = 1$, 
such that for all $i \in \mathcal{I}$
\begin{equation}
\norm{T \hat{z}_i - \hat{\lambda}_i \hat{z}_i} = \order{n \varepsilon
  \norm{T}} \,,
\label{eq:residualgoal}
\end{equation}
and for all $i,j \in \mathcal{I}$ with $i \neq j$
\begin{equation}
|\hat{z}_i^* \hat{z}_j| = \order{n \varepsilon} \,.
\label{eq:orthogoal}
\end{equation}
For now, without any loss of generality, we assume $\beta_i \neq 0$ for $1 \leq
i \leq n-1$, as otherwise the matrix is the direct sum of tridiagonals and
the eigenpairs are found by inspecting each tridiagonal submatrix
separately~\cite{Wilkinson:1988}. 
As no off-diagonal element is equals
to zero, the matrix $T$ is {\it irreducible} and the following theorem
holds.

\begin{mythm}
  The eigenvalues of an irreducible real symmetric tridiagonal matrix are
  simple, i.e., they are distinct. Proof:
  See~\cite{Parlett:1998:SEP}. 
\end{mythm}

Although the eigenvalues are distinct, they might be equal to
working precision~\cite{Parlett:1998:SEP}. Nonetheless, distinct eigenvalues
imply that each 
normalized eigenvector is uniquely determined up to a factor of
$-1$. Consequently, if we could compute normalized eigenvectors $\hat{z}_i$ and
$\hat{z}_j$, $i \neq j$, such that the (acute) error angle\footnote{Given by
  $\angle(\hat{z}_i,z_i) = \arccos|\hat{z}_i^* z_i|$.} with the true 
eigenvectors are small, that is
\begin{equation}
\sin \angle(\hat{z}_i,z_i) \leq
\order{n \varepsilon} \quad \mbox{and} \quad \sin \angle(\hat{z}_j,z_j) \leq
\order{n \varepsilon} \,,
\label{eq:accurateeigevecs}
\end{equation}
the computed eigenvectors would be numerically orthogonal:
\begin{equation}
| \hat{z}_i^* \hat{z}_j | \leq \sin \angle(\hat{z}_i,z_i) + \sin
\angle(\hat{z}_j,z_j) \leq \order{n \varepsilon} \,.
\label{eq:accurateeigevecs2}
\end{equation}

In general, in finite precision computations, accuracy as in
\eqref{eq:accurateeigevecs} is not achievable as, whenever the eigenvalues are
``close'' to one another, ``small'' perturbation in the data can lead
to ``large'' perturbations in the
eigenvectors~\cite{Wilkinson:1988,Parlett:1998:SEP}. Different algorithms 
therefore have particular means to ensure the 
goal of numerical orthogonality. For example, algorithms
like QR or Jacobi, which obtain eigenvector approximations by accumulating
orthogonal  
transformations, achieve the goal automatically without
\eqref{eq:accurateeigevecs} necessarily being
satisfied~\cite{Dhillon:2004:Ortvecs}. The method of inverse 
iteration addresses the problem by explicitly orthogonalizing the
eigenvectors corresponding to close eigenvalues (in the absolute sense) --
as discussed in Section~\ref{sec:existingmethods}, potentially this is an  
expensive procedure. For Divide \& Conquer and MRRR, more sophisticated
techniques are
used~\cite{dc94,dc95,Dhillon:2004:MRRR,Dhillon:2004:Ortvecs}. The way the
MRRR algorithm computes numerically orthogonal eigenvectors -- without
explicit othogonalization -- is at the heart of this chapter. 

In the next two sections, we illustrate the principles behind MRRR assuming
exact and finite precision arithmetic, respectively. Generally,
\eqref{eq:accurateeigevecs} can only be
achieved in exact arithmetic, while in finite precision more effort is
required to ensure orthogonality 
among eigenvectors.

\subsection{Computing with exact arithmetic} 
\label{sec:exactaritmetic}

In general, even when operating in exact arithmetic, there exists no procedure to
compute eigenvalues of matrices in a finite number of steps,
cf.~\cite[Theorem 25.1]{trefethenbau}.  
As a results, only {\em approximations} to eigenvalues and eigenvectors can
be computed. In this section, we concentrate on the way approximations to
eigenvectors are obtained from given approximations to eigenvalues. 

First however, suppose an eigenvalue $\lambda$
is known exactly. The corresponding
true eigenvector $z$ is given by 
\begin{equation}
  ( T - \lambda I ) z = 0 \,,
\end{equation}
or, using~\eqref{eq:thetridiagonal}, equally by
\begin{subequations}
\label{eq:test}
\begin{align}
(\alpha_1 - \lambda) z(1) + \beta_1 z(2) &= 0 \,,  \\
\beta_{i-1} z(i-1) + (\alpha_i - \lambda) z(i) + \beta_{i} z(i+1) &= 0 \,,  \\
\beta_{n-1} z(n-1) + (\alpha_n - \lambda) z(n) &= 0 \,, 
\end{align}
\end{subequations}
where the second equation holds for $1 < i < n$. Equations~\eqref{eq:test}
imply that $z(1) \neq 0$ and $z(n) \neq 0$ as otherwise $z =
0$. Therefore, setting either $z(1) = 1$ or $z(n) = 1$ and using
respectively the first or last $n-1$ equations yields the sought after
eigenvector. Furthermore, as
$T - \lambda I$ is singular, the unused equation is automatically satisfied. 

Due to finiteness of any approximation procedure, even in exact arithmetic,
the approximation $\hat{\lambda}$ has generally a nonzero error. 
(In practice, whenever we attempt to
compute the corresponding eigenvector, we 
ensure that
$\hat{\lambda}$ 
is closer to $\lambda$ than to any other eigenvalue.) 
Since $\hat{\lambda} \notin
spec[T]$, $T - \hat{\lambda} I$ is nonsingular and solving in 
the described way for an
eigenvector results in $\hat{z}$ that satisfies $(T -
\hat{\lambda} I) \hat{z} = \gamma_n e_n$ 
or $(T - \hat{\lambda} I) \hat{z} = \gamma_1 e_1$, where 
the scaling factor $\gamma_k$ takes
into account that the $k$-th equation in \eqref{eq:test} 
does not hold automatically anymore. In fact, there is nothing special about
omitting the last or the first  
equation: 
we might omit the $k$-th
equation for any $1 \leq k \leq n$ and therefore solve $(T - \hat{\lambda}
I) \hat{z} = \gamma_k e_k$. Those computations are naturally not equivalent:
Omitting the $k$-th equation, $( T - \hat{\lambda} I ) \hat{z} = 
\gamma_k  e_k$, leads to $\hat{z}$ with residual norm $\norm{\bar{r}}$ given by
\begin{equation}
\norm{\bar{r}} = \frac{\norm{T \hat{z} - \hat{\lambda} \hat{z}}}{\norm{\hat{z}}} =
\frac{|\gamma_k|}{\norm{\hat{z}}} \,.
\label{eq:localresidualintermsofgamma}
\end{equation}
It has been known for a long time (see~\cite{Ipsen:1997:Invit}) that 
there exists at least one index $r$ with $|z(r)| \geq n^{-1/2}$
satisfying\footnote{See~\cite[Theorem 11]{Dhillon:2004:Ortvecs},
  \cite{Ipsen:1997:Invit}, and \cite[Theorem 2.19]{Willems:Diss} for a
  proof. A corresponding vector $\hat{z}$ is sometimes 
  called an FP-vector, where FP stands for Fernando and Parlett. Note that
  by \eqref{eq:gapthmeq2} the residual norm of such an eigenpair is within
  a factor $\sqrt{n}$ of the optimal.} 
\begin{equation}
\norm{{\bar{r}}} =
\frac{|\gamma_r|}{\norm{\hat{z}}} \leq \frac{|\hat{\lambda} - \lambda|}{|z(r)|} \leq \sqrt{n} |\hat{\lambda} - \lambda| \,.
\label{eq:localresidual}
\end{equation}
Due to the inability of determining such an index $r$ cheaply, the procedure was
abandoned by Wilkinson and replaced by inverse iteration using a
``random'' starting vector~\cite{Wilkinson:1988}. In the 1990s
however, Dhillon
and Parlett~\cite{Fernando97} showed -- extending the work of Godunov et
al.~\cite{godunov85} and
Fernando~\cite{Fernando95ona,Fernando95oncomputing,Fernando95babe} -- how to
find such 
an index $r$.\footnote{We refer to~\cite{Fernando97,Dhillon:2004:Ortvecs} for a historical review.}   
As a consequence, if $\hat{\lambda}$ is a good
approximation of $\lambda$, say $|\hat{\lambda} - \lambda| = \order{\sqrt{n}
  \varepsilon \norm{T}}$, the method delivers an eigenvector approximation
$\hat{z}$ with a small residual norm
(backward error) that satisfies~\eqref{eq:residualgoal}.
Unfortunately, if the eigenvalue is not
well-separated from the rest of the spectrum, such $\hat{z}$
does not necessarily satisfy the accuracy dictated by~\eqref{eq:accurateeigevecs} as
the following classical theorem reveals.

\begin{mythm}[Gap Theorem]
Given a symmetric matrix $T \in \Rnn$ and an approximation $(\hat{\lambda},
\hat{z})$, $\| \hat{z} \| = 1$, 
to the eigenpair $(\lambda,
z)$, with $\hat{\lambda}$ closer to $\lambda$ than to any other eigenvalue,
let $\bar{r}$ 
be the residual $T \hat{z} -
\hat{\lambda} \hat{z}$; then
\begin{equation}
\sin \angle (\hat{z},z) \leq \frac{\| \bar{r} \|}{\gap(\hat{\lambda})} \,,
\label{eq:gapthmmain}
\end{equation}
with $\gap(\hat{\lambda}) = \min_j \{
|\hat{\lambda} - \lambda_j| : \lambda_j \in spec[T]\,\wedge\,\lambda_j \neq
\lambda \}$. The residual norm is minimized if $\hat{\lambda}$ is
the Rayleigh quotient of $\hat{z}$, $\hat{\lambda} = \hat{z}^* T \hat{z}$. In
this case, 
\begin{equation}
\frac{\| \bar{r} \|}{spdiam[T]} \leq \sin \angle (\hat{z},z)  \quad
\mbox{and} \quad |\hat{\lambda} - \lambda| \leq \min \left\{
 \norm{\bar{r}}, \frac{\norm{\bar{r}}^2}{\gap(\hat{\lambda})} \right\} \,.
\label{eq:gapthmeq2}
 \end{equation}
Proof:
See~\cite{Parlett:1998:SEP,Demmel97appliednumerical,Kahan1970,Willems:Diss}.   
\label{thm:gapthm}
\end{mythm}

Inequality~\eqref{eq:gapthmmain} gives not only an upper bound, but also a good
approximation for the error angle~\cite{ReplaceTwithLDL}. Therefore, by
\eqref{eq:localresidual} and \eqref{eq:gapthmmain}, we require the separation relative to
$\norm{T}$ to be reasonably large for \eqref{eq:accurateeigevecs} to
hold, say $\gap(\hat{\lambda})/\norm{T} \geq
\abstol$ with $\abstol \approx 10^{-3}$.  
On the other hand, if $\hat{\lambda}$ approximates $\lambda$ to high
{\em relative} accuracy, i.e., $|\hat{\lambda} - \lambda| =
\order{n \varepsilon |\lambda|}$, a small error angle to the true
eigenvector is obtained whenever the {\em relative} gap, i.e.,
\begin{equation}
\relgap(\hat{\lambda}) = \gap(\hat{\lambda})/|\lambda|
\end{equation}
is sufficiently large, say 
$\relgap(\hat{\lambda}) \geq \gaptol$ with $gaptol \approx 10^{-3}$.\footnote{If
  $\lambda = 0$, its relative gap is $\infty$. Although 
\eqref{eq:localresidual} delivers $\norm{\bar{r}} = \order{n^{3/2}
  \varepsilon |\lambda|}$ and \eqref{eq:gapthmmain} $\sin \angle (\hat{z},z)
= \order{n^{3/2} \varepsilon / gaptol}$, there exist various reasons to 
ignore the $\sqrt{n}$-factor introduced by \eqref{eq:localresidual}.
Among them, the simple facts that often $\relgap(\hat{\lambda}) \gg gaptol$
and $n^{-1/2} \ll z(r) \leq 1$. A more thorough discussion is found in~\cite{Dhillon:2004:Ortvecs}.
}  
{\it Thus, given approximations to the desired
eigenvalues to high relative accuracy, the above procedure allows to compute
an accurate eigenvector whenever the relative gap of the eigenvalue is large.} 

Assuming we have
approximated the desired eigenvalues $\{\hat{\lambda}_i: i \in \mathcal{I}\}$ to high relative
accuracy, whenever $\relgap(\hat{\lambda}_i) \geq  \gaptol$, we compute
eigenvector $\hat{z}_i$ with a small error angle.  
In practice, we are slightly more restrictive on when to compute an
eigenvector; at this point of the discussion, this detail is not
important and we will return to this matter later. 
If our (slightly adjusted) criterion indicates that we can compute the
eigenvector with a small error angle, $\hat{\lambda}_i$ is said to be
{\it well-separated}, {\it isolated}, or a {\it singleton}.  
{\it For all well-separated eigenvalues, we can independently compute
the corresponding eigenvectors  so that all resulting eigenpairs
satisfy our accuracy goals given by \eqref{eq:residualgoal} and \eqref{eq:orthogoal}.}

\begin{algorithm}[t]
 \small
    {\bf Input:} Irreducible symmetric tridiagonal $T \in \mathbb{R}^{n
    \times n}$; index set $\mathcal{I}_{in}  \subseteq \{1,\ldots,n\}$. \\
    {\bf Output:} Eigenpairs $(\hat{\lambda}_i, \hat{z}_i)$ with
    $i \in \mathcal{I}_{in}$.
    
    \vspace{1mm}

  \algsetup{indent=2em}
  \begin{algorithmic}[1]
    \STATE Compute $\hat{\lambda}_i[T]$ with $i \in \mathcal{I}_{in}$ to high
    relative accuracy. 
    \STATE Form a work queue $Q$ and enqueue task
    $\{T, \mathcal{I}_{in}, 0\}$. 
    \WHILE{$Q$ not empty} 
    \STATE Dequeue a task $\{M, \mathcal{I}, \sigma\}$. \label{line:exactmrrr:associatedindexset}
    \STATE Partition $\mathcal{I} = \bigcup_{s=1}^S \mathcal{I}_s$  according 
    to the separation 
    of the  
    eigenvalues.
    \FOR{$s=1$ {\bf to} $S$}
    \IF{$\mathcal{I}_s = \{i\}$} 
    \STATE // {\it process well-separated eigenvalue associated with
      singleton $\mathcal{I}_s$} //
    \STATE Solve $(M - \hat{\lambda}_i[M] I) \hat{z}_i =
    \gamma_r e_r$ with appropriate index $r$ for $\hat{z}_i$. \label{line:exactmrrr:getvec}
    \STATE Return $\hat{\lambda}_i[T] = \hat{\lambda}_i[M] + \sigma$ and
    normalized $\hat{z}_i$. 
    \ELSE
    \STATE // {\it process cluster associated with $\mathcal{I}_s$} //
    \STATE Select shift $\tau \in
    \mathbb{R}$ and compute $M_{shifted} = M - \tau I$. \label{line:exactmrrr:shifting}
    \STATE Refine $\hat{\lambda}_i[M_{shifted}]$ with $i \in \mathcal{I}_s$ to high relative
    accuracy. 
    \STATE Enqueue $\{M_{shifted}, \mathcal{I}_s, \sigma + \tau\}$. 
    \ENDIF
    \ENDFOR
    \ENDWHILE 
  \end{algorithmic}
  \caption{\ MRRR using exact arithmetic}
  \label{alg:mrrrexact}
\end{algorithm}
If some eigenvalues are non-isolated, they come in collections of two or more
consecutive values, say $\{\hat{\lambda}_p[T], \hat{\lambda}_{p+1}[T]
\ldots, \hat{\lambda}_{q}[T]\}$. These collections are called {\it
  clusters} and the eigenvalues are said to be {\it clustered}. 
To compute the eigenvectors for clusters, the following observation is used:
the eigenvectors are invariant under shifts (i.e., forming $T - \tau I$ for some
$\tau \in \R$), while the relative gaps are not. Indeed, by choosing $\tau$ to be
(close to) one eigenvalue, say $\tau = \hat{\lambda}_p[T]$, one can decrease the
magnitude of this eigenvalue and 
therefore increase its relative gap: 
\begin{equation}
  \relgap(\hat{\lambda}_p[T - \tau I]) = \relgap(\hat{\lambda}_p[T])
  \frac{|\lambda_p[T]|}{|\lambda_p[T] - \tau|} \gg
  \relgap(\hat{\lambda}_p[T]) \,.
\end{equation}
In exact arithmetic, there exists a shift $\tau$ to make the relative gap as large as
desired; in particular, we can make the eigenvalue well-separated with
respect to the shifted matrix $T - \tau I$. 
Using $T - \tau I$, for all those eigenvalues of the original cluster that
are now well-separated, we then compute the eigenvectors with small error
angle to the corresponding true eigenvectors. 
For the eigenvalues that are still clustered, we apply the procedure
recursively until we have computed all eigenvectors. The overall procedure
is summarized in Algorithm~\ref{alg:mrrrexact}.

\subsection{Computing with finite precision arithmetic}
\label{sec:finitearitmetic}

While the procedure described above sounds seemingly simple, there are several
obstacles when applied in finite precision arithmetic. Most notably, the
invariance of the eigenvectors under shifts is lost. Furthermore, rounding
errors could spoil the computation of an eigenvector in
Line~\ref{line:exactmrrr:getvec} of Algorithm~\ref{alg:mrrrexact}. Finally, we 
have not specified how the so called {\it twist index} $r$ is chosen in exact
arithmetic and this task might be impossible in finite precision or
computationally expensive.  In
this section, we give an {\it overview} of how these
problems are addressed. The ideas were developed
in~\cite{Dhillon:Diss,Dhillon:2004:MRRR,Dhillon:2004:Ortvecs,Parlett2000121,Willems:Diss,Willems:twisted,Willems:blocked,Willems:framework} 
and realized the above procedure in floating point arithmetic -- today
known as the algorithm of Multiple Relatively Robust Representations.   

\paragraph{Change of representation.} Small
element-wise relative perturbations of the diagonal and off-diagonal
entries of $T$ can lead to large relative perturbations of small
(in magnitude)
eigenvalues~\cite{Dhillon:2004:Ortvecs,ReplaceTwithLDL}. In such a case, it is 
said that the data does not define these eigenvalues to high relative
  accuracy. As a consequence, using finite precision arithmetic, we cannot
hope to compute eigenvalues to high relative accuracy. In order for the
procedure of Algorithm~\ref{alg:mrrrexact} 
to work, the representation of tridiagonals by their diagonal and
off-diagonal entries must be abandoned and alternative representations must be used. 
To discuss these alternatives, we
give a general definition of the concept of a representation
first. 

\begin{mydef}[Representation] A set of 
  $2n-1$ scalars, called the
  {\em data}, together with a mapping $f: \R^{2n -1} \rightarrow \R^{2n-1}$ to
  define the entries of a symmetric tridiagonal matrix $T \in \Rnn$ is
  called a {\em representation} of $T$~\cite{Willems:framework}.
\end{mydef}

According to the above definition, the set containing the diagonal and off-diagonal entries
together with the identity mapping is a representation of a tridiagonal. Unfortunately,
such a representation generally does not have desirable properties under
small element-wise relative perturbations. 
As {\em all} perturbations in this chapter are element-wise and relative, we
sometimes omit these attributes for the sake of brevity. 

\begin{mydef}[Perturbation of a representation] Let $x_1,\ldots,x_{2n-1}$ be the
  scalars used to represent the symmetric tridiagonal $T \in \Rnn$ and
  $\tilde{x}_i = x_i(1+\xi_i)$ be relative perturbations of them; using the
  same representational mapping as for $T$, they define a matrix $\widetilde{T}$. If $\xi_i
  \leq \xi \ll 1$ for all $1 \leq i \leq 2n-1$, we call $\widetilde{T}$ a small
  perturbation of $T$ bounded by $\xi$~\cite{Willems:framework}.
\end{mydef}

We are interested in representations that have the property
that small perturbations 
cause small perturbations in some of the eigenvalues and eigenvectors. Such a
representation is called {\it relatively robust} and constitutes a {\it relatively robust
  representation} (RRR). 
As the definition of relative robustness -- given below -- requires the notion of a relative gap
connected to an index set $\mathcal{I} \subset \{ 1, \ldots , n\}$, we
define such a relative gap first.
\begin{mydef} Given a symmetric $T \in \Rnn$ with simple
  eigenvalues $\{\lambda_i: 1 \leq i \leq n\}$ and an
  index set $\mathcal{I} \subset \{ 1, \ldots , n\}$, 
  the relative gap connected to $\mathcal{I}$ is defined as
  \begin{equation*}
    \mbox{\em relgap} (\mathcal{I}) = \min \left\{ \frac{|\lambda_j -
    \lambda_i|}{|\lambda_i|} : i \in 
    \mathcal{I}, j \notin \mathcal{I} \right\}
  \end{equation*}
  where $|\lambda_j -
  \lambda_i|/|\lambda_i|$ is $\infty$ if $\lambda_i = 0$~\cite{Dhillon:2004:MRRR}.
\label{def:relgapindexset}
\end{mydef}

\begin{mydef}[Relative robustness] Given a representation of the irreducible
  symmetric tridiagonal $T \in \Rnn$ and an
  index set $\mathcal{I} \subset \{ 1, \ldots , n\}$, we say
  that the representation is relatively robust for $\mathcal{I}$
  if for all small perturbations $\widetilde{T}$ bounded by $\xi$ and $i \in \mathcal{I}$, we
  have 
  \begin{align*}
    |\tilde{\lambda}_i - \lambda_i| &\leq k_{rr} n \xi |\lambda_i| \,,  \\
    \sin \angle (\tilde{\mathcal{Z}}_{\mathcal{I}},\mathcal{Z}_{\mathcal{I}}) &\leq \frac{k_{rr} n \xi}{\mbox{\em relgap}(\mathcal{I})} \,,
  \end{align*}
where $\tilde{\lambda}_i$ and $\tilde{\mathcal{Z}}_{\mathcal{I}}$ denote the
eigenvalues and the corresponding invariant subspaces of the perturbed matrices,
respectively, and $\angle
(\tilde{\mathcal{Z}}_{\mathcal{I}},\mathcal{Z}_{\mathcal{I}})$ the largest
principle angle~\cite{Golub1996}; 
$k_{rr}$ is a moderate constant, say about 10~\cite[Property
I]{Dhillon:2004:MRRR}.\footnote{According to 
  \cite{Willems:Diss,Willems:framework}, the requirement on the perturbation
  of the eigenvalues can be removed.} 
\label{def:RRR}
\end{mydef}


We say that an RRR for $\{i\}$ is relatively robust for the eigenpair
$(\lambda_i, z_i)$, or alternatively, that such an RRR defines the
eigenpair to high relative accuracy. 
As we will see later, when using suitable representations and algorithms,
the matrix shifts performed in Line~\ref{line:exactmrrr:shifting} of
Algorithm~\ref{alg:mrrrexact}, $M_{shifted} = M - \tau I$, introduce small relative 
perturbations in the data of $M$ and $M_{shifted}$. In order for those perturbations
not to influence too much the invariant subspace
$\mathcal{Z}_{\mathcal{I}_{s}}$ associated to a cluster, we must 
modify Algorithm~\ref{alg:mrrrexact} in such a way that 
$M$ and $M_{shifted}$ are RRRs for $\mathcal{I}_s$. 
Similarly, in order to compute a highly accurate eigenpair, we need a
representation that defines the eigenpair to high relative accuracy.
Thus, we have to replace the original $T$ with a suitable {\it
  initial} or {\it root representation}  $M_{root} = T - \mu I$ for some $\mu
\in \R$ and carefully select every intermediate representation,
$M_{shifted}$, that is computed in Line~\ref{line:exactmrrr:shifting} of
Algorithm~\ref{alg:mrrrexact}.  

There are multiple 
candidates -- existence assumed -- for playing the role of RRRs,
which often but not always 
are relatively robust for index sets connected to {\it small} eigenvalues
(in magnitude) and the associated invariant subspaces: 

\begin{enumerate}[noitemsep]
\item {\it Lower bidiagonal factorizations} of the form $T = L D L^*$ and {\it
    upper bidiagonal factorizations} of the form $T = U \Omega U^*$, where $D = 
  \mbox{diag}(d_1,d_2,\ldots,d_n) \in \Rnn$ and $\Omega = 
  \mbox{diag}(\omega_1,\omega_2,\ldots,\omega_n) \in \Rnn$ are diagonal,  
  $L \in \Rnn$ and $U \in \Rnn$ are respectively unit lower bidiagonal and
  unit upper bidiagonal, i.e., 
  \begin{equation*}
  \small
  L = \left( \begin{array}{cccccc}
  1      &        &        &            &        \\
  \ell_1 & 1      &        &            &        \\
         & \ell_2 & 1      &            &        \\
         &        & \ddots & \ddots     &    \\
         &        &        & \ell_{n-1}  & 1  \\
\end{array} \right) \quad \mbox{and} \quad 
  U = \left( \begin{array}{cccccc}
  1      & u_1    &        &            &        \\
         & 1      & u_2    &            &        \\
         &        & 1      & \ddots     &        \\
         &        &        & \ddots     & u_{n-1}   \\
         &        &        &            & 1  \\
\end{array} \right) \,.
\end{equation*}
  Lower bidiagonal factorizations were used to
  represent the intermediate tridiagonal matrices ($M$ in
  Algorithm~\ref{alg:mrrrexact}) in the original implementations of the
  algorithm~\cite{Dhillon:Diss,Dhillon:DesignMRRR}. 
  In the definite case, i.e., $|D| = \pm D$ or $|\Omega| = \pm \Omega$, a
  bidiagonal factorization is an RRR for {\it all}
  eigenpairs~\cite{Demmel90accuratesingular}. Therefore, 
  such a definite factorization is often used as an initial representation
  $M_{root} = T - \mu I$.
\item A generalization of the above are so called {\it twisted
    factorizations} or {\it
    BABE-factorizations}~\cite{Fernando95babe,Fernando97} of
  the form $T = N_k 
  \Delta_k N_k^*$, where 
  $k$ denotes the {\it twist index}. $N_k$ has the form
\begin{equation}
\small
  N_k = \left( \begin{array}{ccccccc}
1   &        &         &   &        &        & \\
\ell_1 & 1      &         &   &        &        & \\
    & \ddots & \ddots  &   &        &        & \\
    &        &  \ell_{k-1} & 1 & u_{k} &        & \\
    &        &         &   & \ddots & \ddots & \\
    &        &         &   &        &  1     & u_{n-1} \\
    &        &         &   &        &        & 1 \\
\end{array} \right) \,,
\end{equation}
and $\Delta_k =
\mbox{diag}(d_1,\ldots,d_{k-1},\gamma_k,\omega_{k+1},\ldots,\omega_n)$ 
is diagonal. The factorizations for all $1 \leq k \leq n$ are almost entirely
defined by elements of the bidiagonal factorizations $T = L D L^* = U \Omega U^*$; only
$\gamma_k$, $1 < k < n$, has to be computed. There are multiple formulations
for $\gamma_k$, see~\cite[Corollary 4]{Fernando97}; one of them is:
$\gamma_k = d_k + \omega_k - \alpha_k$. Twisted 
factorizations are crucial in computing accurate eigenvectors
in Line~\ref{line:exactmrrr:getvec} of  
Algorithm~\ref{alg:mrrrexact}~\cite{Fernando97,Dhillon:2004:Ortvecs}. Although
it was known that these 
factorizations can additionally serve as representations of tridiagonals~\cite{Dhillon:Diss,Dhillon:2004:Ortvecs,perturbLDL,ReplaceTwithLDL}, their
benefits were only demonstrated 
recently~\cite{Willems:Diss,Willems:twisted}. 
Due to their additional degree of freedom in choosing $k$, the use of twisted
factorizations, which includes the bidiagonal factorizations as special
cases, is superior to using only lower bidiagonal factorizations. \\ 
\item {\it Blocked
    factorizations}~\cite{Fang06stablefactorizations,Willems:blocked} are
  further generalizations of bidiagonal and twisted factorizations. The
  quantities $D$, $\Omega$, and  
  $\Delta_k$ are allowed to be block diagonal with blocks of size $1 \times
  1$ or $2 \times 2$. The other factors -- $L$, $U$, and $N_k$ -- are
  partitioned conformally with one or the $2 \times 2$ identity on the diagonal. These class of
  factorizations contain the unblocked bidiagonal and twisted factorizations
  as special cases. With their great flexibility, these factorizations
  have been used very successfully within the MRRR
  algorithm~\cite{Willems:Diss,Willems:blocked}. 
\end{enumerate}

All factorizations are determined by $2n-1$ scalars, the data; the
same number as for $T$ given by its diagonal and off-diagonal
elements. Through the mapping determined by the factorization, they
define a tridiagonal. For instance, for lower bidiagonal factorizations, $2n-1$ floating
point numbers 
$d_1,\ldots,d_n,\ell_1,\ldots,\ell_{n-1}$ determine a tridiagonal that is generally
not exactly representable in the same finite precision. Such representation
by the non-trivial entries of the factorization is called
an $N$-representation~\cite{Willems:twisted}.
Similarly, the floating point
numbers $d_1,\ldots,d_n,\beta_1,\ldots,\beta_{n-1}$ -- with $\beta_i = d_i \ell_i$ being $T$'s
off-diagonal elements --
represent a tridiagonal whose diagonal entries are in general not
representable. Such representation, which includes $T$'s off-diagonal elements, is called
an $e$-representation~\cite{Willems:twisted}.\footnote{The name
  $e$-representation is used as the off-diagonals $\beta_i$ are denoted $e_i$
in~\cite{Willems:twisted}.}
It is important to distinguish
between the tridiagonal -- not necessarily machine representable --
and the finite precision scalars that constitute the representation. The
$2n-1$ scalars constituting the representation are
called the {\it primary} data. Other quantities that are computed using the
primary data are called {\it secondary} or {\it derived} data. 
For instance, the off-diagonal
$\beta_i = d_i \ell_i$ is secondary for an $N$-representation while
being primary for an $e$-representation.\footnote{See \cite{Willems:twisted} for
details and another so called $Z$-representation of the data.} 
Subsequently, we do not distinguish between the representation of a
tridiagonal and the tridiagonal itself; that is, it is always implied that
tridiagonals are represented in one of the above forms. More on different
forms of representing tridiagonals can be found
in~\cite{Dhillon:Diss,ReplaceTwithLDL,Willems:Diss,Willems:twisted,Willems:blocked}. 

It is not at all obvious that these representations are more suitable for
computations than the standard way of representing tridiagonals. This is the
topic of the relative 
perturbation theory covered 
in~\cite{perturbLDL,Parlett2000121,Dhillon:2004:Ortvecs,Dhillon:Diss}, which
shows how sensitive individual eigenvalues and eigenvectors are under relative
component-wise perturbations.

\paragraph{Shifting the spectrum.}

In exact arithmetic, shifting the spectrum, as in
Line~\ref{line:exactmrrr:shifting} of Algorithm~\ref{alg:mrrrexact}, leaves the eigenvectors
unchanged; this invariance is lost in finite precision. An essential
ingredient of MRRR is the use of special forms of Rutishauser's {\it
  Quotienten-Differenzen} (qd)
algorithm~\cite{qd,AccurateSVDandQDtrans,Dhillon:2004:Ortvecs} to
perform the spectrum 
shifts. Given representation $M$, we require that 
$M_{shifted} = M - \tau
I$ is computed in a {\it element-wise mixed relative 
stable} way, i.e., $\widetilde{M}_{shifted} = \widetilde{M} - \tau I$ holds
exactly for small perturbations of $M_{shifted}$ and $M$ bounded by 
$\xi_{\uparrow} = \order{\varepsilon}$ and $\xi_{\,\downarrow} =
\order{\varepsilon}$, respectively. In the following, we assume
$\xi_{\uparrow}$ and $\xi_{\,\downarrow}$ are bounds for all spectrum shifts
performed during an execution of the algorithm.\footnote{The requirement is called {\sc
    Shiftrel} in \cite{Willems:framework}.} 

Since both $M_{shifted}$ and $M$ in Line~\ref{line:exactmrrr:shifting} of
Algorithm~\ref{alg:mrrrexact} 
can take any of 
the discussed forms, 
these leads to a variety of (similar) algorithms with different
characteristics.  
For instance, using lower bidiagonal factorizations to represent
intermediate matrices, we require to perform $L_+ D_+ L_+^* = LDL^* - \tau
I$. This can be accomplished by the so called {\it differential form of the
stationary qd transformation} (dstqds) given
in
Algorithm~\ref{alg:dstqds}~\cite{Dhillon:Diss,Dhillon:2004:Ortvecs}. Similarly,
using twisted 
factorizations, 
we require an algorithm that
is stable in the above sense to compute $N_t \Delta_t N_t^*
= N_s \Delta_s N_s^* - \tau I$ for arbitrary twist indices $s$ and $t$~\cite{Willems:twisted}; the same is true 
for blocked factorizations~\cite{Willems:blocked}. 

In Algorithm~\ref{alg:mrrrexact}, provided that $M_{shifted}$ and $M$
are relatively robust for $\mathcal{I}_{s}$ and
$\relgap(\mathcal{I}_{s}) \geq \gaptol$,  the mixed stability implies that
invariant subspaces connected to clusters are not  
perturbed too much due to rounding
errors, $\sin \angle
(\mathcal{Z}_{\mathcal{I}_s}[M_{shifted}],\mathcal{Z}_{\mathcal{I}_s}[M]) \leq
k_{rr} n (\xi_{\,\downarrow} + \xi_{\uparrow}) /gaptol$.\footnote{The requirement $\relgap(\mathcal{I}_{s}) \geq \gaptol$ is
  called {\sc Relgaps} in \cite{Willems:framework} and corresponds to
  Property III in \cite{Dhillon:2004:MRRR}.} After the shifting, we can
therefore hope to compute an orthonormal 
basis for such a subspace, which is {\em automatically} numerically
orthogonal to the subspace spanned by the other computed 
eigenvectors. 
This is one of the main ideas behind MRRR.

The special computation of the spectrum shifts is essential 
for eigenvectors computed from different representations to be numerically
orthogonal. In order to ensure that the eigenpairs also enjoy small residual norms
with respect to the input matrix, the intermediate representations should additionally
exhibit the so called {\it conditional element growth}. 

\begin{mydef}[Conditional element growth] 
A representation $M$ exhibits conditional element growth for $\mathcal{I}
\subset \{1,2,\ldots,n\}$ if for any small
perturbation $\widetilde{M}$ bounded by $\xi$ and $i \in 
\mathcal{I}$ 
\begin{align*}
  \norm{\widetilde{M} - M} &\leq \mbox{\em spdiam}[M_{root}] \,, \quad
  \mbox{and} \\ 
  \norm{(\widetilde{M} - M) \hat{z}_i} &\leq k_{elg} n \xi \cdot \mbox{\em spdiam}[M_{root}]
  \,, 
\end{align*}
where $\hat{z}_i$ denote the computed eigenvectors and $k_{elg}$ is a
moderate constant, say about 10~\cite{Dhillon:2004:MRRR,Willems:framework}. 
\end{mydef}

In particular, $M_{shifted}$, computed in Line~\ref{line:exactmrrr:shifting} of
Algorithm~\ref{alg:mrrrexact}, needs to exhibit
conditional element growth for $\mathcal{I}_s$. 
At this point, we are not concerned about how to ensure that the involved
representations satisfy 
the requirements; this is the topic
of~\cite{perturbLDL,Parlett2000121,Dhillon:2004:Ortvecs,Willems:framework}.\footnote{It
  is {\it not} necessary to compute the eigenvectors in order 
  to give bounds on the conditional element growth.} 
We 
remark however that there exist the danger that no suitable representation
that passes the test ensuring the requirements is found. In this case,
commonly a promising representation is selected. As such a representation might not fulfill the
requirements, the accuracy of MRRR is not guaranteed anymore.    

Independently of the form to represent the intermediate
tridiagonals, the computation of twisted 
factorizations is essential for finding an
eigenvector. Therefore, the possibility of computing twisted
factorizations $N_k \Delta_k N_k^* =
M - \tau I$ for $1 \leq k \leq n$ must be provided for any representation
used for $M$. This means we must be
able to compute the lower bidiagonal factorization $L_+ D_+ L_+^* = M - \tau
I$, the upper bidiagonal factorization $U_- \Omega_- U_-^* = M - \tau I$ and
the $\gamma_k$ terms in a stable way 
in order for rounding errors not to spoil the computation of an
eigenvector~\cite{Dhillon:2004:Ortvecs,Willems:twisted}.

\paragraph{Using twisted factorizations to find an eigenvector.} At the
moment we attempt to compute an accurate eigenvector approximation
$\hat{z}_i$ in Line~\ref{line:exactmrrr:getvec} of
Algorithm~\ref{alg:mrrrexact}, we have given an RRR for $\{i\}$  
and $\hat{\lambda}_i$ -- an approximation to $\lambda_i$ with high relative
accuracy. Furthermore, the relative gap is sufficiently large,
$\relgap(\hat{\lambda}_i) \geq \gaptol$. This special situation 
is analyzed in \cite{Dhillon:2004:Ortvecs,Fernando97}. 

Suppose $M - \hat{\lambda}_i I = L_+ D_+ L_+^* = U_- \Omega_- U_-^*$ permits
lower and upper bidiagonal factorization. We
can determine all the twisted factorizations $N_k \Delta_k N_k^* = M -
\hat{\lambda}_i I$ cheaply by computing the missing $\gamma_k$ for $1 < k
< n$. The computation must however be done with care, i.e., in a
element-wise mixed relative stable way~\cite{Dhillon:2004:Ortvecs,Willems:twisted}.

When solving $(M - \hat{\lambda}_i I) \hat{z}_i  = N_k \Delta_k N_k^*
\hat{z}_i = \gamma_k e_k$, by~\eqref{eq:localresidual}, in exact arithmetic,
the residual norm is
given by $|\gamma_k|/\norm{\hat{z}_i}$. Thus, a natural choice for the twist
index is $r =
\operatorname*{arg\,min}_k |\gamma_k|$, which is indeed used in
practice. This is justified as follows: 
Since $|\gamma_k|/\norm{\hat{z}_i} \leq |\hat{\lambda}_i -
\lambda_i|/|z_i(k)|$ for all $1 \leq k \leq n$ with $z_i(k) \neq 0$,
see~\cite[Theorem 3.2.3]{Dhillon:Diss}, finding an $|z_i(k)| \geq
n^{-1/2}$, i.e., an entry of the {\it true} eigenvector that is above average in
magnitude, results in the desired bound
of~\eqref{eq:localresidual}. As in
the limiting case $\hat{\lambda}_i \rightarrow \lambda_i$, 
\begin{equation*}
\frac{\gamma_k^{-1}}{\sum_{j=1}^n{\gamma_j^{-1}}} \rightarrow z_i(k)^2 \,,
\end{equation*}
provided $\hat{\lambda}_i$ is an accurate eigenvalue approximation,
$r = \operatorname*{arg\,min}_k |\gamma_k|$ implies $|z_i(r)|$ is above
average, see~\cite[Lemma 3.2.1]{Dhillon:Diss} or~\cite[Lemma 11]{Fernando97}. 

After finding $r$ and its associated twisted factorization, the following
system needs to be solved
\begin{equation}
  N_r \Delta_r N_r^* \hat{z}_i = \gamma_r e_r \Longleftrightarrow N_r^* \hat{z}_i
  = e_r \,,
\end{equation}
where the equivalence stems from the fact that $N_r^{-1} e_r = e_r$ and
$\Delta_r e_r = \gamma_r e_r$. This system
is easily solved by setting $\hat{z}_i(r) = 1$ and
\begin{align*}
  \hat{z}_i(j) =
\left\{ 
\begin{array}{l l}
    - \ell^+_j \hat{z}_i(j + 1) & \quad \text{for $j = r-1,\ldots,1$} \,, \\
    - u_{j-1}^- \hat{z}_i(j - 1)  & \quad \text{for $j = r+1, \ldots, n$} \,. \\
  \end{array} \right.
\end{align*}
Finally, $\hat{\lambda}_i$ might be
improved by the Rayleigh quotient correction term $\gamma_r /
\norm{\hat{z}_i}^2$, and finally $\hat{z}_i$ is normalized and returned. 
A more careful implementation of the procedure, taking into account possible
breakdown due to finite precision arithmetic, is given in Algorithm~\ref{alg:Getvec} and
discussed in the next section. For the sake of brevity, we neglect this issue at
this point of the discussion.  

The above procedure of computing an eigenvector approximation using a twisted
factorization is called {\tt Getvec}. A rigorous analysis of {\tt
  Getvec} in~\cite{Dhillon:2004:Ortvecs}, which takes all effects of finite
precision into account, reveals that the computed eigenvector
$\hat{z}_i$ has a small error angle to the true eigenvector $z_i$. 

\begin{mythm}
Suppose $\hat{z}_i$ is computed by {\tt Getvec} under the
conditions stated above. 
For a small perturbation of $M$ bounded by $\alpha =
\order{\varepsilon}$, $\widetilde{M}$, and a small 
element-wise relative perturbation of $\hat{z}_i$ bounded by $\eta = \order{n\varepsilon}$,
$\tilde{z}_i$, the residual norm satisfies 
\begin{equation}
  \norm{\bar{r}^{(local)}} = \norm{\widetilde{M} \tilde{z}_i - \hat{\lambda}_i
    \tilde{z}_i} \leq k_{rs} gap(\hat{\lambda}_i[\widetilde{M}]) n
  \varepsilon / gaptol \,,
\label{localresbound}
\end{equation}
where $k_{rs} = \order{1}$. In this case, 
\begin{equation*}
\sin \angle(\hat{z}_i,z_i) \leq k_{rr} n \alpha / gaptol + k_{rs} n
\varepsilon / gaptol + \eta = \mathcal{G} n \varepsilon \,,
\end{equation*}
where $k_{rr}$ is given by the relative robustness of $M$ for $\{i\}$ and
$\mathcal{G}$ is defined by the equation. Proof:
See~\cite{Dhillon:2004:Ortvecs}.
\label{thm:Getvec}
\end{mythm}
As we have seen
in~\eqref{eq:accurateeigevecs}  and \eqref{eq:accurateeigevecs2}, the small
error angle, $\sin \angle(\hat{z}_i,z_i) \leq \mathcal{G} n \varepsilon =
\order{n \varepsilon / gaptol}$, is essential for computing numerically
orthogonal eigenvectors without explicit
orthogonalization.\footnote{As \eqref{eq:localresidual}
  shows, the best we can hope for is $\norm{\bar{r}^{(local)}} = \order{n^{3/2}
    \varepsilon |\lambda_i|}$ and $\sin \angle(\hat{z}_i,z_i) =
  \order{n^{3/2} \varepsilon / gaptol}$. However, as the discussion
  of~\cite{Dhillon:2004:Ortvecs} indicates, the bounds in the theorem are
  achieved under mild assumptions.}

\paragraph{A high-level view and the representation tree.}

Although most parts of the procedure deserve a more detailed discussion, we
can now apply all the changes to Algorithm~\ref{alg:mrrrexact} to obtain
a high-level view of the so called
{\it core} MRRR algorithm working on irreducible tridiagonals. The whole
procedure is assembled in Algorithm~\ref{alg:mrrr}.
\begin{algorithm}[th]
 \small
    {\bf Input:} Irreducible symmetric tridiagonal $T \in \mathbb{R}^{n
    \times n}$; index set $\mathcal{I}_{in}  \subseteq \{1,\ldots,n\}$. \\
    {\bf Output:} Eigenpairs $(\hat{\lambda}_i, \hat{z}_i)$ with
    $i \in \mathcal{I}_{in}$.
    
    \vspace{1mm}

  \algsetup{indent=2em}
  \begin{algorithmic}[1]
    \STATE Select shift $\mu \in
    \mathbb{R}$ and compute $M_{root} = T - \mu I$. \label{line:root} 
    \STATE Perturb $M_{root}$ by a ``random'' relative amount bounded by a
    small multiple of $\varepsilon$. \label{line:mrrr:perturb}
    \STATE Compute $\hat{\lambda}_i[M_{root}]$ with $i \in \mathcal{I}_{in}$
    to relative accuracy sufficient for classification. 
    \label{line:mrrr:initialeigvals} 
    \STATE Form a work queue $Q$ and enqueue task
    $\{M_{root}, \mathcal{I}_{in}, \mu\}$. \label{line:mrrr:enddlarre}
    \WHILE{$Q$ not empty} 
    \STATE Dequeue a task $\{M, \mathcal{I}, \sigma\}$.
    \STATE Partition $\mathcal{I} = \bigcup_{s=1}^S \mathcal{I}_s$  according 
    to the separation 
    of the eigenvalues. \label{line:mrrr:initialpartitioning}
    \FOR{$s=1$ {\bf to} $S$}
    \IF{$\mathcal{I}_s = \{i\}$} \label{line:mrrr:ifstatement} 
    \STATE // {\it process well-separated eigenvalue associated with
      singleton $\mathcal{I}_s$} //
    \STATE Refine $\hat{\lambda}_i[M]$ to high relative
    accuracy. \label{line:mrrr:refinesingleton} 
    \STATE 
    Find twisted factorization $N_r \Delta_r N_r^* =
    M - \hat{\lambda}_i[M] I$ and solve $N_r^* \hat{z}_i = e_r$ for
    $\hat{z}_i$. \label{line:mrrr:getvec}  
    \STATE Return $\hat{\lambda}_i[T] := \hat{\lambda}_i[M] + \sigma$ and
    normalized $\hat{z}_i$. \label{line:mrrr:returneigenpair} 
    \ELSE
    \STATE // {\it process cluster associated with $\mathcal{I}_s$} //
    \STATE Select shift $\tau \in
    \mathbb{R}$ and compute $M_{shifted} = M - \tau
    I$. \label{line:mrrr:shifting}  
    \STATE Refine $\hat{\lambda}_i[M_{shifted}]$ with $i \in \mathcal{I}_s$ to
    sufficient relative 
    accuracy. \label{line:mrrr:refine} 
    \STATE Enqueue $\{M_{shifted}, \mathcal{I}_s, \sigma +
    \tau\}$. \label{line:mrrr:partition} 
    \ENDIF
    \ENDFOR
    \ENDWHILE \label{line:mrrr:end}
  \end{algorithmic}
  \caption{\ MRRR}
  \label{alg:mrrr}
\end{algorithm}

Most notably,
tridiagonal matrices are replaced by {\it representations} of tridiagonals,
i.e., the tridiagonals are given only implicitly. These
representations are required to exhibit conditional element growth and be
{\it relatively robust}, which
is reflected in the name of the algorithm. In
Line~\ref{line:mrrr:perturb}, we added a small random perturbation of the root
representation.\footnote{True randomness is not necessary; any
  (fixed) sequence of pseudo-random numbers can be used.} Such a perturbation is crucial to break very tight
clusters~\cite{glued}.


The unfolding of Algorithm~\ref{alg:mrrr} is best described as a rooted
tree, the so called {\it representation
  tree}~\cite{Dhillon:Diss,Dhillon:2004:MRRR,Willems:framework}. Each task
$\{M, \mathcal{I}, \sigma\}$ (or just $\{M, \mathcal{I}\}$) is connected to a
node in the tree; that is, all of the nodes consist of a representation and an
index set. The index set 
corresponds to the indices of all eigenpairs computed from the
representation. The node $\{M_{root}, \mathcal{I}_{in}\}$ is the 
root node (hence the name). 
The other tasks, $\{M, \mathcal{I}\}$, are connected to
ordinary nodes. Each node has a depth -- the number of edges on the
unique path from the root to it. The maximum depth for all
nodes (i.e., the height of the tree) is denoted $d_{max}$. The edges
connecting nodes are associated with the spectrum shifts $\tau$ that are
performed in Line~\ref{line:mrrr:shifting} of Algorithm~\ref{alg:mrrr}. 

\begin{figure}[thb]
  \centering
  \includegraphics[scale=.6, trim= 0mm 150mm 20mm
  25mm]{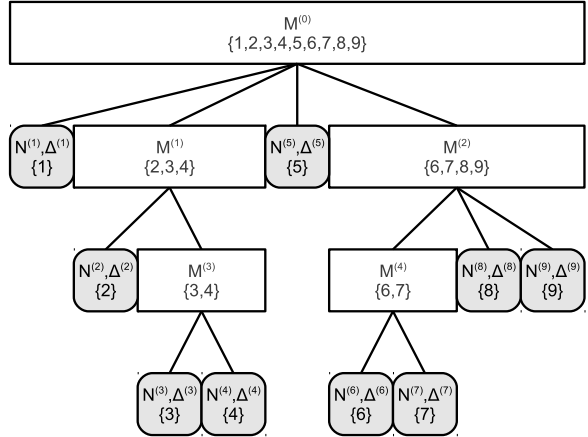}
  \caption{An exemplary representation tree.
}
\label{fig:reptree}
\end{figure}

The concept is best illustrated by an  
example such as in Fig.~\ref{fig:reptree}, where $\mathcal{I}_{in} = \{1,2,\ldots,9\}$. 
The root consists of $M_{root} = M^{(0)}$ together with
$\mathcal{I}_{in}$. 
Ordinary nodes, which are colored in white, correspond to clustered
eigenvalues with respect to the parent representation. 
Eigenvalues
$\{\hat{\lambda}_2[M^{(0)}],\hat{\lambda}_3[M^{(0)}],\hat{\lambda}_4[M^{(0)}]\}$
and 
$\{\hat{\lambda}_6[M^{(0)}],\hat{\lambda}_7[M^{(0)}],\hat{\lambda}_8[M^{(0)}],\hat{\lambda}_9[M^{(0)}]\}$
form clusters and the first invocation of 
Line~\ref{line:mrrr:initialpartitioning} in 
Algorithm~\ref{alg:mrrr} partitions the input index set  
as follows: $\mathcal{I}_{in} = \{1\} \cup \{2,3,4\} \cup \{5\} \cup \{6,7,8,9\}$.
For both clusters, a new representation is computed -- i.e., $M^{(1)} =
M^{(0)} - \tau^{(1)} I$ and $M^{(2)} =
M^{(0)} - \tau^{(2)} I$. 
Although in the definition of $d_{max}$ they are not counted nodes, 
singletons $\{i\}$ (that is, well-separated eigenvalues with
respect the parent 
representation) are associated with {\it leaves} of the tree. 
For each eigenpair there exists exactly one
associated leaf.  
For instance, $\hat{\lambda}_1[M^{(0)}]$ is
well-separated, and the corresponding eigenpair can 
be computed following Lines
\ref{line:mrrr:refinesingleton}--\ref{line:mrrr:returneigenpair} of
Algorithm~\ref{alg:mrrr}, which 
require the twisted factorization $N^{(1)} \Delta^{(1)} (N^{(1)})^{*} = M^{(0)} -
\hat{\lambda}_1[M^{(0)}] I$. According to our definition, which excludes
leaves from the definition of nodes, $d_{max}$ is two. In
\cite{Dhillon:Diss,Dhillon:2004:MRRR} $d_{max}$ is defined 
differently: according to their definition, $d_{max}$ would be three.  
Besides for the definition of $d_{max}$, from now on, we consider {\it
  leaves} as nodes of the representation tree as well. 

The tree of Fig.~\ref{fig:reptree} illustrates sources of natural parallelism: All the
singletons as well as clusters with
the same depth in the tree can be processed independently. As an example, given $M^{(0)}$,
eigenpairs $(\hat{\lambda}_1,\hat{z}_1)$ and $(\hat{\lambda}_5,\hat{z}_5)$ can be
computed in parallel. At the same time, we can independently process the clusters
associated with $\{2,3,4\}$ and $\{6,7,8,9\}$.
When clusters are present, the  
computation of the eigenpairs is {\it not} independent and the amount of
work associated to each eigenpair varies. If the work of computing a set of
eigenpairs is divided statically 
by assigning a subset of indices to each processing unit, 
the presence of large clusters leads to load imbalance and limits 
scalability. 
{\em Only if all
the eigenvalues are well-separated with respect to the root representation,
the computation of the eigenpairs is embarrassingly parallel.}

\paragraph{Accuracy of MRRR.}

As the eigenpairs are computed using possibly different representations, the
main concern of the error analysis given in
\cite{Dhillon:2004:MRRR,Dhillon:2004:Ortvecs,Willems:framework} is if the
resulting eigenpairs enjoy small residual norm with respect to the
  input matrix and are mutually numerically orthogonal. 
The elegant analysis in~\cite{Willems:framework} 
-- a streamlined version of the proofs
in~\cite{Dhillon:2004:MRRR,Dhillon:2004:Ortvecs} --  shows that, provided
the algorithm finds suitable representations, this is indeed the case. 
For our discussion in Chapter~\ref{chapter:mixed}, in following theorem we
state the upper bounds on the residual norm and the orthogonality.

\begin{mythm}[Accuracy] 
Let $\hat{\lambda}_i[M_{root}]$ be computed (exactly) by applying the
spectrum shifts to $\hat{\lambda}_i[M]$ obtained in
Line~\ref{line:mrrr:refinesingleton} of Algorithm~\ref{alg:mrrr}.   
Provided all the requirements of the MRRR algorithm are satisfied (in
particular, that suitable representations are found), we have
\begin{equation}
\norm{M_{root}\, \hat{z}_i - \hat{\lambda}_i[M_{root}]  \, \hat{z}_i} \leq
\left( \norm{\bar{r}^{(local)}} + \gamma \, spdiam[M_{root}] \right)
\frac{1+\eta}{1-\eta}
\label{residualbound}
\end{equation}
with $\norm{\bar{r}^{(local)}}$ being bounded by Theorem~\ref{thm:Getvec}
and $\gamma = k_{elg} n \left( d_{max}(\xi_{\,\downarrow} + \xi_{\uparrow})
  +  \alpha \right) + 2(d_{max} +1) \eta$. 
Furthermore, we have for any computed eigenvectors $\hat{z}_i$ and $\hat{z}_j$, $i \neq j$, 
\begin{equation}
|\hat{z}_i^* \hat{z}_j| \leq 2 \left( \mathcal{G} n \varepsilon +  \frac{k_{rr}
  n (\xi_{\,\downarrow} + \xi_{\uparrow}) d_{max}}{gaptol} \right) \,.
\label{orthogonalitybound}
\end{equation}
where $\mathcal{G} n \varepsilon = k_{rr} n \alpha / gaptol + k_{rs} n
\varepsilon/gaptol + \eta$. Proof: See~\cite{Willems:Diss,Willems:framework} or \cite{Dhillon:2004:MRRR,Dhillon:2004:Ortvecs}.
\label{resthm}
\end{mythm}

We give a number of remarks regarding the theorem:
\begin{enumerate}[noitemsep,nolistsep]
\item The theorem hinges on the fact that suitable representations are
  found. If we accept one or more representations for which conditional
  element growth and relative robustness (i.e., being an RRR) is not
  verified, the accuracy of the result is not guaranteed.
\item For a reasonable implementation, we have $\alpha =
  \order{\varepsilon}$, $\eta = \order{n \varepsilon}$, $\xi_{\,\downarrow} =
  \order{\varepsilon}$, and  $\xi_{\uparrow} =
  \order{\varepsilon}$. Furthermore, $k_{rr}$, $k_{rs}$, and $k_{elg}$ can be
  bounded by a small constant, say 10.
\item The assumption that the
accumulation of the shifts is done in exact arithmetic is not crucial; we
simply stated the theorem as in~\cite{Willems:Diss,Willems:framework}.
\item Provided the computation $M_{root} = T - \mu
I$ is performed in a backward stable manner, 
i.e., $M_{root} = T + \DeltaT - \mu I$ with $\norm{\DeltaT} = \order{n
  \epsilon \norm{T}}$, small residual norms with respect to the root representation, as
given in \eqref{residualbound}, imply small residual norms to the level dictated
by~\eqref{eq:residualgoal}~\cite{Dhillon:Diss}.
\item Instead of  accumulating the shifts to obtain the eigenvalues, the Rayleigh quotient
of $\hat{z}_i$ might be returned as the corresponding eigenvalue $\hat{\lambda}_i$.
\item Element-wise mixed relative stability for the shifts are key to success and imply that 
$\xi_{\,\downarrow} = \order{\varepsilon}$ and  $\xi_{\uparrow} = \order{\varepsilon}$.
\item Relative robustness of the representations is essential and is exposed
  by the multiple appearances of $k_{rr}$ in \eqref{residualbound} (together
  with Theorem~\ref{thm:Getvec}) and
  \eqref{orthogonalitybound}. The first appearance in
  \eqref{orthogonalitybound}, together with  
  the stable spectrum shifts, indicates that invariant subspaces connected
  to clusters are not perturbed too much. The second appearance in
  \eqref{orthogonalitybound} allows the
  computation of an eigenvector with small error angle to the true eigenvector.
\item If $\relgap(\hat{\lambda})$ is almost as small as
  $gaptol$, high relative accuracy of the eigenvalue approximation 
  is necessary. If $\relgap(\hat{\lambda}) \gg gaptol$, the accuracy of the eigenvalue
  approximation can be relaxed~\cite{Dhillon:DesignMRRR}.
\item Large relative gaps of the clusters and well-separated
  eigenvalues are crucial for obtaining orthogonal 
  eigenvectors; this is reflected in the dependence on $gaptol$ in
  \eqref{residualbound} and \eqref{orthogonalitybound}.
\item Shallow representation trees (that is, small values of $d_{max}$) are
preferable over deep trees. In the optimal scenario of $d_{max} = 0$,
many terms in \eqref{residualbound} and \eqref{orthogonalitybound}
cancel. In such a scenario, all eigenvalues are well-separated with 
respect to the root representation, and consequently, not
only is the computation embarrassingly parallel, but the danger of not finding suitable
representations is also entirely removed.\footnote{A root representation can 
always be found, for instance, by making $T - \mu I$ definite.}
\item The bound \eqref{orthogonalitybound} is a quite realistic estimate of the result. The
  observed worst case orthogonality grows as 
$n \varepsilon /\gaptol$. As oftentimes $\gaptol = 10^{-3}$, this explains the
quote in Chapter~\ref{chapter:ch01} that one needs to be prepared of orthogonality levels of
about $\order{1000 n \varepsilon}$, even if all requirements of the algorithm
are fulfilled. 
\end{enumerate}

\section{A closer look}
\label{sec:closerlook}

In the previous section, we discussed the basics of MRRR independently of the 
 form to represent tridiagonals. In this section, we want to be
more concrete: We use of
$N$-representations of lower bidiagonal factorizations and assume full
support for IEEE arithmetic. In such settings, we 
detail the preprocessing stage, the computation and refinement of eigenvalues as
well as the computation of eigenvectors.  

\subsection{Preprocessing}
\label{sec:preprocessing}

The preprocessing of the input matrix $T \in \Rnn$, given by its diagonal
and its off-diagonal entries as 
in~\eqref{eq:thetridiagonal}, includes the {\it scaling} of the entries and
the so called {\it splitting} of the matrix into principal submatrices if off-diagonal entries are
sufficiently small in magnitude.  

Proper scaling strategies are discussed in~\cite{Demmel:bisec,Kahan:1966}
and are of no special importance for our discussion. After the scaling, we
usually set element $\beta_i$ to zero whenever   
\begin{equation}
  |\beta_i| \leq \mbox{tol}_{split} \| T \| \,,
\label{eq:tolsplit}
\end{equation}
and thereby reduce the problem to smaller (numerically irreducible)
subproblems for which we invoke
Algorithm~\ref{alg:mrrr}~\cite{Demmel:bisec,Parlett:1998:SEP,kahan1966neglect}. A
common choice for $\mbox{tol}_{split}$ is a small multiple of  
unit roundoff $\varepsilon$ or even $n \varepsilon$; specifically, we might
use $\mbox{tol}_{split} = \varepsilon \sqrt{n}$.

{\sc Remarks:} (1) If the tridiagonal is known to define its eigenvalues to
high relative accuracy and it is desired to compute them to such accuracy, $\|
T \|$ in \eqref{eq:tolsplit} has to be replaced by for example 
$\sqrt{| \alpha_k \alpha_{k+1} |}$; (2) If all
eigenpairs or a subset with eigenvalues in the interval $[v_{\ell},v_u)$ are
requested, the same is true for each subproblem; on the other hand,
whenever a subset of eigenpairs with indices $i_{\ell}$ to $i_u$ is
requested, it requires to find 
the eigenpairs that need to be computed for each subproblem;
(3) Normally, we assume that the
preprocessing has been done and each subproblem is treated
independently (possibly in parallel) by invoking Algorithm~\ref{alg:mrrr}. 
This justifies our previous assumption that $\beta_i \neq 0$ -- in fact,
$|\beta_i| > \mbox{tol}_{split} \| T \|$. As we did previously and continue
to do, whenever we refer to input matrix
$T$, it is assumed to be (numerically) irreducible; whenever we reference
the matrix size $n$, it refers to the size of the processed block. 

\subsection{Eigenvalues of symmetric tridiagonals}
\label{sec:eigvalscomputation}

Algorithm~\ref{alg:mrrr} requires at several stages (Lines
\ref{line:mrrr:initialeigvals}, \ref{line:mrrr:refinesingleton}, and
\ref{line:mrrr:refine}) to either compute 
eigenvalues of tridiagonals, or refine them to some prescribed relative
accuracy. A natural choice for these tasks (in particular in a parallel environment) is the
method of bisection, whose 
main features were already listed in
Section~\ref{sec:existingmethodsSTEP}.

The bisection procedure relies on the ability to count the number of
eigenvalues smaller than given value $\sigma \in
\mathbb{R}$. Assuming this is accomplished by the function $NegCount(T, \sigma)$, 
Algorithm~\ref{alg:Bisection} bisects a given interval
$[\underline{\lambda}_k,\overline{\lambda}_k]$, which is known to contain
the eigenvalue
$\lambda_k$, until sufficiently small. 
\begin{algorithm}[ht]
 \small
    {\bf Input:} Tridiagonal $T$, function NegCount, index $k$, initial
    interval $[w_{\ell},w_u]$, stopping criteria $rtol$ and $atol$ \\
    {\bf Output:} Interval $[w_{\ell},w_u]$ containing the eigenvalue with
    index $k$ \\
    {\bf Require:} NegCount($T, w_{\ell}$) $< k$ $\wedge$ NegCount($T, w_u$) $\geq k$
    
    \vspace{1mm}
    \algsetup{indent=2em}
    \begin{algorithmic}[1]
      \WHILE{$|w_u - w_{\ell}| > \mbox{rtol} \cdot \max\{|w_{\ell}|,|w_u|\}$
        $\vee$ $|w_u - w_{\ell}| > \mbox{atol}$}
      \STATE $w_{err} := (w_u - w_{\ell})/2$ 
      \STATE $w := w_{\ell} + w_{err}$
      \IF{$\mbox{NegCount}(T, w) < k$}
      \STATE $w_{\ell} := w$
      \ELSE
      \STATE $w_{u} := w$
      \ENDIF
      \ENDWHILE
      \RETURN $[w_{\ell},w_u]$
    \end{algorithmic}
  \caption{Bisection}
  \label{alg:Bisection}
\end{algorithm}
After convergence, we consider $\hat{\lambda}_k =
(\overline{\lambda}_k + \underline{\lambda}_k)/2$ as the approximation of
$\lambda_k$ with error $\hat{\lambda}_{k,err} =
(\overline{\lambda}_k - \underline{\lambda}_k)/2$. Since we are generally
interested in relative accuracy, $atol$ is fixed to a value of
$\order{\omega}$, say $atol = 2\omega$, where $\omega$ denotes the underflow
threshold. Quantity $rtol$ reflects the required accuracy. If the
hypothesis of Algorithm~\ref{alg:Bisection} is not satisfied, an actual
implementation should inflate the input interval until the condition is met. 

At each iteration, the width of the interval is reduced by a factor two and,
consequently, convergence is linear and rather slow. Several schemes might be
used to accelerate the procedure. Bisection is mainly used to
classify eigenvalues into well-separated and clustered, for which only
limited accuracy in the approximations is necessary. To obtain eigenvalues to 
full precision (i.e., in Line \ref{line:mrrr:refinesingleton} of Algorithm~\ref{alg:mrrr}) a form of
Rayleigh quotient iteration (RQI) is commonly used. Bisection to full precision is
only used if the RQI fails to converge to the correct eigenvalue,
 and to compute the extremal eigenvalues before selecting the shifts $\mu$ and
$\tau$ in Lines~\ref{line:root} and \ref{line:mrrr:shifting} of Algorithm~\ref{alg:mrrr}. 

To complete the eigenvalue computation via Algorithm~\ref{alg:Bisection}, we require 
to specify the function 
$NegCount$ for tridiagonals given either in form of
\eqref{eq:thetridiagonal} or in factored form, $LDL^*$, and we need to specify initial
intervals containing the desired eigenvalues.

\paragraph{Counting eigenvalues.} Recall that two symmetric matrices $A, H \in \Rnn$ are congruent if there exist a
nonsingular matrix $S \in \Rnn$ such that $A = S H S^*$. 
The function $NegCount$, which is required by
Algorithm~\ref{alg:Bisection}, is then derived by Sylvester's law of
inertia. 

\begin{mythm}[Sylvester]
Two real symmetric matrices are congruent if and only if they have
the same inertia, that is, the same number of positive, negative, and zero
eigenvalues. Proof: See~\cite{Demmel97appliednumerical,Golub1996}.
\label{thm:sylvester}
\end{mythm}

Given the lower bidiagonal factorization $L D L^* = T - \sigma
I$, matrices $D$ and $T - \sigma I$ have the same inertia, i.e.,
$NegCount(D,0) = NegCount(T, \sigma)$. Hence, all that is needed to determine
$NegCount(T, \sigma)$ is to count
the number of negative entries in $D$. A similar approach is taken if
the tridiagonal is given in factored form, $L D L^*$. In this case, $L_{+}
D_{+} L_{+}^* = L D L^* - \sigma I$, and $D_+$ and $L D L^* - \sigma I$ have the same inertia.
These considerations lead to Algorithm~\ref{alg:negcountT} and
Algorithm~\ref{alg:negcountLDL} in Appendix
\ref{appendix:somealgorithms}. 

\paragraph{Initial intervals.}

If we want to use bisection to compute either the extremal eigenvalues of input matrix $T$ to
full accuracy or a first approximation to the desired eigenvalues, we need a
starting interval. Such an interval that contains all eigenvalues (and therefore the $k$-th
eigenvalue) might be found by means of the Gershgorin Theorem. 

\begin{mythm}[Gershgorin]
Let $A \in \Cnn$ be an arbitrary matrix. The eigenvalues of $A$ are located in the union of
$n$ disks:
\begin{equation*} 
\bigcup_{i=1}^n \{ z \in \C : |z - a_{ii}| \leq \sum_{\substack{j = 1 \\ j
    \neq i}}^n |a_{ij}| \} \,.
\end{equation*} 
Proof: See~\cite{hornjohnson,Golub1996}.
\end{mythm}

Using Gershgorin's theorem, the initial interval is found by Algorithm
\ref{alg:gershgorin} in Appendix \ref{appendix:somealgorithms}. Bisection
is then used to find any eigenvalue of $T$ to the accuracy warranted by the
data. For instance, the statements

\begin{algorithmic}
\small
  \STATE $[g_{\ell}, g_u] = Gershgorin(T)$
  \FOR{$i = 1$ {\bf to} $n$}  
  \STATE $\left[\underline{\lambda}_i[T],\overline{\lambda}_i[T]\right] = Bisection(T
  , NegCount, i, [g_{\ell}, g_u], rtol = 10^{-8}, atol= 2 \omega)$
  \ENDFOR
\end{algorithmic}
compute approximations $\hat{\lambda}_i[T]$ to about 8 digits of accuracy (if
the data defines the eigenvalues to such accuracy). 

In later stages of the algorithm, we generally have approximate
intervals ($\left[\underline{\lambda}_i[T] - \mu,\overline{\lambda}_i[T] - \mu\right]$ or
$\left[\underline{\lambda}_i[M] - \tau,\overline{\lambda}_i[M] -
  \tau\right]$), which can be inflated and used as a starting point 
for limited bisection to refine the eigenvalues to a prescribed
accuracy.  

{\sc Remarks:} (1) Although the eigenvalue computation as well as
their refinement by limited bisection is embarrassingly 
parallel, it is often beneficial to not keep an interval for each eigenvalue
separately. By starting from an initial interval containing all desired
eigenvalues, bisection is performed to obtain nonoverlapping intervals
containing one or more eigenvalues. The computation is described by an
unbalanced binary tree. 
(2) As the work associated with an eigenvalue depends on
its value (and possibly the neighboring eigenvalues), a static division of
work can lead to load imbalance.  
(3) In Lines \ref{line:mrrr:initialeigvals} and
\ref{line:mrrr:refine} of Algorithm~\ref{alg:mrrr}, only sufficient accuracy
to classify eigenvalues into well-separated and clustered is needed. 
For that purpose, $rtol = 10^{-2} \cdot \gaptol$ might be used. 
Depending on the absolute gap 
of the eigenvalue, the refinement can be stopped
earlier on~\cite{Dhillon:DesignMRRR}. 
(4) In Line~\ref{line:mrrr:initialeigvals} of Algorithm~\ref{alg:mrrr}, any
algorithm that computes eigenvalues to sufficient relative accuracy can be
employed. If $M_{root}$ is positive/negative definite and (almost) all
eigenvalues are desired, the best choice
is often the {\it dqds  
algorithm}~\cite{AccurateSVDandQDtrans,dqds99}, as it can be significantly faster than
bisection. Then Lines~\ref{line:root}--\ref{line:mrrr:initialeigvals} of
Algorithm~\ref{alg:mrrr} become similar to  
Algorithm~\ref{alg:initialeigvals}.\footnote{Cf.~\cite[Algorithm
5]{Dhillon:DesignMRRR}.}
 (5) For implementations targeting highly parallel systems, the dqds
algorithm is usually not used. 

\begin{algorithm}[ht]
  \small
  {\bf Input:} Irreducible symmetric tridiagonal $T \in \mathbb{R}^{n
    \times n}$; index set $\mathcal{I}_{in}  \subseteq \{1,2,\ldots,n\}$. \\
    {\bf Output:} Root representation $M_{root}$ and shift $\mu$; eigenvalues
    $\hat{\lambda}_i[M_{root}]$ with $i \in \mathcal{I}_{in}$.
    
    \vspace{1mm}
    
    \algsetup{indent=2em}
    \begin{algorithmic}[1]
      \IF{only a subset of eigenpairs desired or enough parallelism available}  \label{line:startinitialbisectionT}
      \STATE Compute crude approximations $\hat{\lambda}_i[T]$
      for $i \in \mathcal{I}_{in}$ via bisection. 
      \ENDIF \label{line:endinitialbisectionT}
      \STATE Select shift $\mu \in
      \mathbb{R}$ and compute $M_{root} = T - \mu I$. 
      \STATE Perturb $M_{root}$ by a ``random'' relative amount bounded by a
      small multiple of $\varepsilon$.
      \IF{already computed initial eigenvalue approximations}    
      \STATE Refine $\hat{\lambda}_i[M_{root}]$ via bisection to
       accuracy sufficient for classification. 
      \ELSE
      \STATE Compute eigenvalues $\hat{\lambda}_i[M_{root}]$ for $i
      \in \{1,2,\ldots,n\}$ via the dqds algorithm.
      \STATE Discard $\hat{\lambda}_i[M_{root}]$ if $i \in \{1,2,\ldots,n\}
      \setminus \mathcal{I}_{in}$.
      \ENDIF
    \end{algorithmic}
    \caption{\ Initial eigenvalue approximation}
    \label{alg:initialeigvals}
\end{algorithm}

\subsection{Eigenvectors of symmetric tridiagonals}
\label{sec:closerlookeigvecs}

For each irreducible subblock, after computing a root representation and
initial approximations to the eigenvalues, in
Line~\ref{line:mrrr:initialpartitioning} of Algorithm~\ref{alg:mrrr},
the index set is partitioned
according to the separation of the eigenvalues. For well-separated
eigenvalues, the eigenpair is computed directly, while for clustered
eigenvalues more work is necessary. In the
following, we detail the classification, the computation of eigenvectors for
well-separated eigenvalues, and the computation of eigenvectors for
clustered eigenvalues.

\paragraph{Classification.}

For all $i \in \mathcal{I}$, let $\hat{\lambda}_i$ denote the midpoint point
of a computed interval of uncertainty
$[\underline{\lambda}_i,\overline{\lambda}_i]$ containing the eigenvalue
$\lambda_i$. We have to partition the index set, $\mathcal{I} =
\bigcup_{s=1}^S \mathcal{I}_s$, 
such that the resulting subsets have $\relgap(\mathcal{I}_s) \geq gaptol$ and,
whenever $\mathcal{I}_s = \{i\}$, additionally $\relgap(\hat{\lambda}_i) \geq
gaptol$. To achieve the desired partitioning of $\mathcal{I}$, let $j, j+1
\in \mathcal{I}$ and define 
\begin{equation*}
  reldist(j,j+1) = \frac{\underline{\lambda}_{j+1} -
    \overline{\lambda}_j}{\max\{ |\underline{\lambda}_j| ,
    |\overline{\lambda}_j|, |\underline{\lambda}_{j+1}| ,
    |\overline{\lambda}_{j+1}| \}}
\end{equation*}
as a measure of the relative gap. If $reldist(j,j+1) \geq gaptol$, then $j$ and
$j+1$ belong to different subsets of the partition. Additionally, this criterion
based on the relative separation can be amended by a criterion based on the
absolute separation of the eigenvalues~\cite{VoemelRefinedTree2007tr}.

\paragraph{Computation of an eigenvector for a well-separated eigenvalue.}
We briefly discussed Algorithm {\tt Getvec} in
Section~\ref{sec:finitearitmetic}. In this section, by Algorithm~\ref{alg:Getvec},
we give a concrete example for lower
bidiagonal factorizations, $LDL^*$. 
As in Section~\ref{sec:finitearitmetic}, we assume $LDL^*$ is an RRR for
eigenpair $(\lambda,z)$ and
$\hat{\lambda}$ is well-separated as well as an approximation with high relative
accuracy. 

The first ingredient of Algorithm {\tt Getvec} is the computation of a suitable
twisted factorization, which is done by determining the lower bidiagonal
factorization $L_+ D_+ L_+^* = LDL^* - \hat{\lambda} I$, the upper
bidiagonal factorization $U_- \Omega_- U_-^* = LDL^* - \hat{\lambda}
I$ and quantities $\gamma_k$, $1 \leq k \leq n$. 
The bidiagonal factorizations are
determined by the {\it differential form of the stationary
  qd transformation with shift} (dstqds) and the {\it differential form of the progressive qd
  transformation with shift} (dqds), which are given in
Algorithm~\ref{alg:dstqds} and Algorithm~\ref{alg:dqds}, respectively. We
use the following notation: $d \in \Rn$ denotes the diagonal elements of $D$, and
$\ell \in \R^{n-1}$ denotes the off-diagonal elements of $L$.\footnote{In
Matlab notation, $d = \mbox{diag}(D)$ and $\ell = \mbox{diag}(L,-1)$.}
Similarly, $d^+,\omega^{-} \in \Rn$ and $\ell^+, u^- \in \R^{n-1}$
denote the non-trivial entries of respectively $L_+D_+L_+^*$ and
$U_-\Omega_-U_-^*$.

\begin{figure*}
  \begin{minipage}[t]{.48\textwidth}
    \begin{algorithm}[H]
      \caption{dstqds transform}
      \label{alg:dstqds}
      
      \footnotesize
      {\bf Input:} Non-trivial entries of $LDL^*$ given by $d \in \Rn$
      and $\ell \in \R^{n-1}$; shift $\tau \in \R$. \\
      {\bf Output:} Non-trivial entries of $L_+D_+L_+^* = LDL^* - \tau I$,
      $d^+ \in \Rn$ and $\ell^+ \in \R^{n-1}$; auxiliary quantities $s \in \Rn$.
      
      \vspace{1mm}
      \algsetup{indent=2em}
      \begin{algorithmic}
        \STATE $s(1) := -\tau$
        \FOR{$i = 1, \ldots, n-1$}
        \STATE $d^+(i) := s(i) + d(i)$
        \IF{$|s(i)| = \infty \wedge |d^+(i)| = \infty$}
        \STATE $q := 1$
        \ELSE
        \STATE $q := s(i)/d^+(i)$ 
        \ENDIF
        \STATE $\ell^+(i) := d(i)\ell(i) /d^+(i)$
        \STATE $s(i+1) := q \cdot d(i) \ell(i) \ell(i) - \tau$      
        \ENDFOR
        \STATE $d^+(n) := s(n) + d(n)$
        \RETURN $d^+, \ell^+, s$
      \end{algorithmic}
    \end{algorithm}
  \end{minipage}
  %
  \begin{minipage}[t]{0.03\textwidth}
    \begin{center}
    \makebox{\begin{minipage}{0.03\textwidth}
      \hspace{0.01in}
      \vspace{0.03in}
      \end{minipage}
     }
    \end{center}
  \end{minipage}
  \begin{minipage}[t]{.48\textwidth}
    \begin{algorithm}[H]
      \caption{dqds transform}
      \label{alg:dqds}
   
      \footnotesize
      {\bf Input:} Non-trivial entries of $LDL^*$ given by $d \in \Rn$
      and $\ell \in \R^{n-1}$; shift $\tau \in \R$. \\
      {\bf Output:} Non-trivial entries of $U_-\Omega_-U_-^* = LDL^* - \tau I$,
      $\omega^- \in \Rn$ and $u^- \in \R^{n-1}$; auxiliary quantities $p \in \Rn$.
      
      \vspace{1mm}
      \algsetup{indent=2em}
      \begin{algorithmic}
        \STATE $p(n) := d(n) - \tau$
        \FOR{$i = n-1, \ldots, 1$}
        \STATE $u^{-}(i+1) := p(i + 1) + d(i) \ell(i) \ell(i)$
        \IF{$|p(i+1)| = \infty \wedge |\omega^{-}(i+1)| = ·\infty$}
        \STATE $q := 1$
        \ELSE 
        \STATE $q := p(i+1)/\omega^{-}(i+1)$
        \ENDIF 
        \STATE $u^{-}(i) :=  d(i) \ell(i) / \omega^{-}(i+1)$
        \STATE $p(i) := q \cdot d(i) - \tau$      
        \ENDFOR
        \STATE $\omega^{-}(1) := p(1)$
        \RETURN $\omega^{-}, u^{-}, p$
      \end{algorithmic}
    \end{algorithm}
  \end{minipage}
\end{figure*}

We give a number of remarks regarding the transformations, of which the
first three are discussed in~\cite{Marques:2006:BIF}: (1) IEEE
arithmetic is exploited to handle possible breakdown of the algorithms; (2)
Other ways can be employed to prevent breakdown that do not require IEEE
arithmetic; (3) The computation can be accelerated by
removing the if statement in the loop and checking if a NaN (``Not a Number'') value is
produced in course of the computation; (4) An error analysis as well as
alternative formulations can be found in~\cite{Dhillon:2004:Ortvecs} and~\cite{Willems:twisted}. 

In Algorithm~\ref{alg:Getvec}, after determining the appropriate twist index
$r$, the linear system is solved as described in
Section~\ref{sec:finitearitmetic}, only this time we include the
handling of zero pivots that might have occurred in the factorizations. Under
the above assumptions on the inputs, Algorithm~\ref{alg:Getvec} returns an
eigenvector approximation that satisfies Theorem~\ref{thm:Getvec}.

\begin{algorithm}[ht]
  \small
    {\bf Input:} Non-trivial entries of $LDL^*$ given by $d \in \Rn$
    and $\ell \in \R^{n-1}$; eigenvalue $\hat{\lambda} \in \R$. \\
    {\bf Output:} Eigenvector $\hat{z}$.  
    
    \vspace{1mm}
    \algsetup{indent=2em}
    \begin{algorithmic}[1]
      \STATE $[d^{+},\ell^{+},s] = dstqds(d,\ell,\hat{\lambda})$
      \STATE $[\omega^{-},u^{-},p] = dqds(d,\ell,\hat{\lambda})$
      \FOR{$k=1, \ldots, n$}
      \IF{$k = n$}
      \STATE $\gamma_k := s(n) + d(n)$
      \ELSE
      \STATE $\gamma_k := s(k) +  \frac{d(k)}{\omega^{-}(k+1)} \cdot p(k + 1)$
      \ENDIF
      \ENDFOR
      \STATE $r := \operatorname*{arg\,min}_k |\gamma_k|$
      \STATE $\hat{z}(r) := 1$
      \FOR{$i=r-1, \ldots, 1$}
      \IF{$d^{+}(i) \neq 0$}
      \STATE $\hat{z}(i) := - \ell^{+}(i) \cdot \hat{z}(i+1)$
      \ELSE 
      \STATE $\hat{z}(i) := - \frac{d(i+1)\ell(i+1)}{d(i)\ell(i)} \cdot \hat{z}(i+2)$
      \ENDIF
      \ENDFOR
      \FOR{$i=r, \ldots, n-1$}
      \IF{$\omega^{-}(i+1) \neq 0$}
      \STATE $\hat{z}(i+1) := -u^{-}(i) \cdot \hat{z}(i)$
      \ELSE 
      \STATE $\hat{z}(i+1) := -\frac{d(i-1)\ell(i-1)}{d(i)\ell(i)} \cdot \hat{z}(i-1)$
      \ENDIF
      \ENDFOR
      \RETURN $\hat{z} := \hat{z}/\norm{\hat{z}}$ 
    \end{algorithmic}
  \caption{Getvec}
  \label{alg:Getvec}
\end{algorithm}

{\sc Remarks:}
(1) An elaborate version of {\tt Getvec} is implemented in LAPACK as
{\tt xLAR1V}.
(2) Often the computed eigenvector has small numerical support, that is, for
all $i < i_{first}$ and for all $i > i_{last}$, $\hat{z}(i)$ can be set to
zero, where $i_{last} - i_{first} +1 \ll n$. Such phenomenon is easily
detected and exploited. For details on the numerical support as well as {\tt
Getvec} in general, we refer to~\cite{Dhillon:2004:Ortvecs}.
(3) {\tt Getvec} is commonly used to improve $\hat{\lambda} =
\hat{\lambda}^{(0)}$ by Rayleigh quotient iteration (RQI): in the $j$-th
iteration, compute $\hat{z}^{(j)}$ via {\tt Getvec} 
and use the Rayleigh quotient correction term to update the eigenvalue
$\hat{\lambda}^{(j+1)} = \hat{\lambda}^{(j)} + \gamma_r/\norm{\hat{z}^{(j)}}^2$.
The process is stopped if the residual norm is sufficiently
  small~\cite{NLA:NLA493}. During the iteration, it is not needed to recompute
  index $r$ at each iteration and the amount of work in {\tt Getvec} can be
  reduced as only the twisted factorization with index $r$ needs to be
  computed. 
  For performance reasons, RQI is commonly used instead of 
  Lines~\ref{line:mrrr:refinesingleton}--\ref{line:mrrr:returneigenpair} in
  Algorithm~\ref{alg:mrrr}. As $\Delta_r^{(j)}$ (from the 
  twisted factorization) has the same 
  inertia as $LDL^* - \hat{\lambda}^{(j)} I$, 
  $NegCount(LDL^*,\hat{\lambda}^{(j)})$ can be used to monitor if the RQI
  converges to the desired eigenvalue~\cite{Dhillon:DesignMRRR}. 
  If not, bisection is used to compute the eigenvalue to full accuracy, followed by
  one invocation of {\tt Getvec}.  
 
\paragraph{Computation of eigenvectors for clustered eigenvalues.}
If eigenvalues are clustered, in Line~\ref{line:mrrr:shifting} of
Algorithm~\ref{alg:mrrr}, we need to find a new
representation with shifted spectrum. When using lower bidiagonal
factorizations, the spectrum shifts are performed by the 
dstqds transform such as in Algorithm~\ref{alg:dstqds}, $L_+ D_+ L_+^* = LDL^* - \tau
I$. Shift $\tau$ must be selected such that the new representation fulfills
two requirements: relative robustness and conditional element growth. As the
entire algorithm depends on finding such representations, the spectrum shifts
are one of the most crucial steps (and the only possible source of
failure). The topic of how to ensure that the requirements are
satisfied is beyond the scope of our discussion, we refer to~\cite{Parlett2000121,perturbLDL,Dhillon:2004:Ortvecs,Dhillon:Diss,Willems:framework,Willems:Diss} for an
in-depth treatment. However, in Algorithm~\ref{alg:spectrumshift}, we give
a high-level description of the process.\footnote{See \cite[Algorithms 9 and 10]{Dhillon:DesignMRRR} for a similar discussion.}

\begin{algorithm}[ht]
  \small
  {\bf Input:} A representation $LDL^*$; clustered eigenvalues $\{\hat{\lambda}_p,\hat{\lambda}_{p+1},\ldots,\hat{\lambda}_{q}\}$. \\
    {\bf Output:} A representation $L_+D_+L_+^* = LDL^* - \tau I$.
    
    \vspace{1mm}
    
    \algsetup{indent=2em}
    \begin{algorithmic}[1]
      \STATE Refine the extremal eigenvalues to full accuracy by
      bisection to obtain intervals $[\underline{\lambda}_p,\overline{\lambda}_p]$ and
      $[\underline{\lambda}_q,\overline{\lambda}_q]$ with midpoints
      $\hat{\lambda}_p$ and $\hat{\lambda}_q$, respectively. 
      \STATE Select shifts close to  $\hat{\lambda}_p$ and
      $\hat{\lambda}_{q}$; e.g., $\tau_{\ell} = \underline{\lambda}_p - 100
      \varepsilon |\underline{\lambda}_p|$ and $\tau_{r} =
      \overline{\lambda}_q + 100\varepsilon |\overline{\lambda}_q|$.
      \FOR{$i = 1$ {\bf to} a maximal number of iterations}
      \STATE Use shifts $\tau_{\ell}$ and $\tau_{r}$ to compute a shifted
      representation via the dstqds transform, $L_+D_+L_+^* = LDL^* -
      \tau_{\ell}I$ and $L_+D_+L_+^* = LDL^* - \tau_{r}I$. 

      \IF{not both factorizations exist}
      \STATE $\tau_{\ell} := \tau_{\ell} - \delta_{\ell}$ and $\tau_{r} := \tau_{r} +
      \delta_{r}$ for some small $\delta_{\ell}$, $\delta_{r}$
      \STATE {\bf continue}
      \ELSIF{both factorizations exist}
      \STATE Select the one with smaller element growth, i.e., minimizing
      $\norm{D_+}$.
      \ELSE
      \STATE Select the existing one.
      \ENDIF      
      \IF{the element growth of selected $L_+D_+L_+^*$ is below a
        threshold}
      \RETURN the representation, $L_+ D_+ L_+^*$.
      \ELSIF{$L_+D_+L_+^*$ passes a more sophisticated test} 
      \RETURN the representation, $L_+ D_+ L_+^*$.
      \ENDIF
      \STATE $\tau_{\ell} := \tau_{\ell} - \delta_{\ell}$ and $\tau_{r} := \tau_{r} +
      \delta_{r}$ for some small $\delta_{\ell}$, $\delta_{r}$      
      \ENDFOR
      \RETURN $L_+ D_+ L_+^*$ with the smallest element
      growth of all computed representations. \label{line:justaccept}
    \end{algorithmic}
    \caption{\ Shifting the spectrum}
    \label{alg:spectrumshift}
\end{algorithm}

{\sc Remarks:} (1) If a factorization does not exist, nonnumerical values
are produced in its computations and such a factorizations must be discarded. 
(2) As it defines all eigenpairs to high relative 
accuracy, a factorization with essentially no element growth
is accepted immediately~\cite{Dhillon:DesignMRRR}.\footnote{If $\norm{D_+}$ is
  $\order{spdiam[M_{root}]}$, the factorization is 
  accepted; see~\cite[Section 2.5]{Willems:Diss} for comments and more
  thorough tests.}  
(3) A more sophisticated test consists of an approximation of the conditioning
of the relevant eigenpairs as well as the so called envelope of the invariant
subspace to obtain information on where all eigenvectors of the invariant
subspace have small
entries~\cite{Parlett2000121,perturbLDL,localization,Dhillon:2004:Ortvecs,Willems:framework};
with this information, it can be checked if the representation is relatively
robust and has conditional element growth for the relevant eigenpairs.
(4) The values for $\delta_{\ell}$ and $\delta_{r}$ must be chosen with
care, as ``backing off too far even might, in an extreme case, make the
algorithm fail [as tight clusters are not broken, while] backing off too
little will not reduce the element growth''~\cite{Dhillon:DesignMRRR}; 
(5) Shifting inside the cluster is possible, but often
avoided~\cite{Dhillon:2004:MRRR,Willems:framework}.
(6) If no suitable representation is 
found, the candidate with the least element growth is returned. Such a
strategy is potentially dangerous as accuracy might be jeopardized (``code may fail but should never
lie'')~\cite{Dhillon98currentinverse,Dhillon:Diss}. Instead, an implementation could
explicitly check the accuracy of the eigenpairs connected to the cluster or fall back to another
method like the submatrix
method~\cite{Parlett:1996:Submatrices,Parlett:1996:BIT}. 

After we found a new RRR for a cluster,
$L_+D_+L_+^* = LDL^* - \tau I$, the eigenvalues are refined to the desired accuracy via
bisection, and then reclassified. If
$[\underline{\lambda}_i,\overline{\lambda}_i]$ denote the approximation of
an eigenvalue of $LDL^*$, we might use as a starting interval 
$[\underline{\lambda}_i^+,\overline{\lambda}_i^+]$ for limited
bisection: 
\begin{align*}
\underline{\lambda}_i^+ &:= (\underline{\lambda}_i (1 \pm \nu) - \tau)(1
\pm \nu) \,, \\
\overline{\lambda}_i^+ &:= (\overline{\lambda}_i (1 \pm \nu) - \tau)(1 \pm
\nu) \,,
\end{align*}
with $\nu = 10 k_{rr} n\varepsilon$ and the appropriate signs to enlarge the
interval~\cite{Willems:framework}.

\chapter{Parallel MRRR-based Eigensolvers}
\label{chapter:parallel}
\thispagestyle{empty}

In this chapter, we introduce our work on parallel versions of the MRRR
algorithm targeting modern multi-core architectures and distributed-memory systems.
In Section~\ref{sec:mr3smp}, we present a task queue-based parallelization
strategy that is especially designed to take advantage of the features
of multi-core and symmetric multiprocessor (SMP) systems. 
We provide a detailed discussion of the new parallel eigensolver {\tt
  mr3smp}. Experiments give ample evidence that the task
queue-based approach leads to  
remarkable results in terms of scalability and execution time:
Compared with the fastest solvers available on both artificial and
application matrices, {\tt mr3smp} consistently ranks as the
fastest. All results of Section~\ref{sec:mr3smp} are published
in~\cite{para2010,mr3smp}.  

When problem sizes become too large for a typical SMP system, or the execution times are
too high, scientists place their hopes on massively
parallel supercomputers. In Section~\ref{sec:pmrrr}, we focus on modern 
distributed-memory architectures, which themselves use multi-core
processors as their building blocks. These hybrid shared/distributed-memory
systems are often briefly denoted as hybrid architectures.\footnote{The term
  {\it hybrid architecture} is also used if two or more types of processors are part
  of the system.}
We present a novel tridiagonal solver,
\PMRRR, which merges the task queue-based approach introduced
previously in Section~\ref{sec:mr3smp}, and the distributed-memory 
parallelization strategy of Bientinesi et al.~\cite{Bientinesi:2005:PMR3}.

\PMRRR\ was integrated into the Elemental library for the solution of
large-scale standard and generalized dense Hermitian eigenproblems.\footnote{See Section~\ref{sec:existingsoftware} for a brief
  introduction of Elemental.}  
Our study of these problems is motivated by performance issues of 
commonly used routines in the ScaLAPACK library. After identifying
those issues, we provide 
clear guidelines on how to circumvent them: By invoking suitable routines
with carefully chosen settings, users can assemble solvers that perform
considerably better than those included in ScaLAPACK. 
In a thorough performance study on two state-of-the-art
supercomputers, we compare Elemental with the solvers built within the ScaLAPACK
framework according to our guidelines.
For a modest amount of parallelism and provided the fastest routines with suitable
settings are invoked, ScaLAPACK's solvers obtain results comparable to Elemental.  
In general, Elemental attains the best performance and obtains the best
scalability of all solvers. In particular, compared to the most widely used
ScaLAPACK routines, the performance improvements are quite significant. 
Most results of Section~\ref{sec:pmrrr} are published in~\cite{EleMRRR}.

\section{MRRR for multi-core architectures}
\label{sec:mr3smp}

In this section, we present a task-based design of the MRRR algorithm specifically tailored to
multi-core processors and shared-memory architectures. We call our
task-based approach MR$^3$-SMP. The rationale behind MR$^3$-SMP is
that the unfolding of the algorithm, which is summarized (in slightly
altered form than before) in Algorithm~\ref{alg:mrrr2}, is only determined in 
course of the computation. 
\begin{algorithm}[ht]
 \small
    {\bf Input:} Irreducible symmetric tridiagonal $T \in \mathbb{R}^{n
    \times n}$; index set $\mathcal{I}_{in}  \subseteq \{1,\ldots,n\}$. \\
    {\bf Output:} Eigenpairs $(\hat{\lambda}_i, \hat{z}_i)$ with
    $i \in \mathcal{I}_{in}$.
    
    \vspace{1mm}

  \algsetup{indent=2em}
  \begin{algorithmic}[1]
    \STATE Select shift $\mu \in
    \mathbb{R}$ and compute $M_{root} = T - \mu I$. \label{line:mrrr2:root} 
    \STATE Perturb $M_{root}$ by a ``random'' relative amount bounded by a
    small multiple of $\varepsilon$. \label{line:mrrr2:perturb}
    \STATE Compute $\hat{\lambda}_i[M_{root}]$ with $i \in \mathcal{I}_{in}$
    to relative accuracy sufficient for classification. 
    \label{line:mrrr2:initialeigvals} 
    \STATE Partition $\mathcal{I}_{in} = \bigcup_{s=1}^S \mathcal{I}_s$  according 
    to the separation 
    of the eigenvalues. \label{line:mrrr2:initialpartitioning}
    \STATE Form a work queue $Q$ and enqueue each task
    $\{M_{root}, \mathcal{I}_s, \mu\}$. \label{line:mrrr2:enddlarre}
    \WHILE{$Q$ not empty} 
    \STATE Dequeue a task $\{M, \mathcal{I}, \sigma\}$.
    \IF{$\mathcal{I} = \{i\}$} \label{line:mrrr2:ifstatement} 
    \STATE // {\it process well-separated eigenvalue associated with
      singleton $\mathcal{I}$} //
    \STATE Perform Rayleigh quotient iteration (guarded by bisection) to
    obtain eigenpair $(\hat{\lambda}_i[M],\hat{z}_i)$ with sufficiently
    small residual norm, $\norm{M \hat{z}_i - \hat{\lambda}_i[M]
      \,\hat{z}_i}/\norm{\hat{z}_i}$. \label{line:mrrr2:getvec}
    \STATE Return $\hat{\lambda}_i[T] := \hat{\lambda}_i[M] + \sigma$ and
    normalized $\hat{z}_i$. \label{line:mrrr2:returneigenpair} 
    \ELSE
    \STATE // {\it process cluster associated with $\mathcal{I}$} //
    \STATE Select shift $\tau \in
    \mathbb{R}$ and compute $M_{shifted} = M - \tau
    I$. \label{line:mrrr2:shifting}  
    \STATE Refine $\hat{\lambda}_i[M_{shifted}]$ with $i \in \mathcal{I}$ to
    sufficient relative 
    accuracy. \label{line:mrrr2:refine} 
    \STATE Partition $\mathcal{I} = \bigcup_{s=1}^S \mathcal{I}_s$
    according to the separation 
    of the eigenvalues. 
    \STATE Enqueue each 
    $\{M_{shifted}, \mathcal{I}_s, \sigma + \tau\}$. \label{line:mrrr2:partition}
    \ENDIF
    \ENDWHILE \label{line:mrrr2:end}
  \end{algorithmic}
  \caption{\ MRRR}
  \label{alg:mrrr2}
\end{algorithm}
As the work associated with each
eigenpair is unknown a priori, any static assignment of eigenpairs to
computational threads is likely to result in poor load balancing, which in
turn negatively effects parallel efficiency.\footnote{See
  Section~\ref{sec:objectives} for the connection of load balance and
  scalability.} 
In order to
achieve perfect load balancing, computational tasks are 
created and scheduled dynamically. Naturally, the tasks can be executed by
multiple compute threads. Our C implementation of this concept, which we
call {\tt mr3smp}, is based on 
LAPACK's {\tt DSTEMR} version 
3.2, and makes use of {\em POSIX threads} (IEEE POSIX 1003.1c). The use
of POSIX threads is motivated by the desire to have a maximal control about
the threading behavior. Other threading environments supporting task-based
parallelism (such as OpenMP, Threading Building Blocks, or Cilk) might be
used instead. 

During and after our work on
the parallel solver, several
refinements to the MRRR algorithm have been proposed to improve its
robustness~\cite{Willems:Diss,Willems:twisted,Willems:framework,Willems:blocked}.  
Among these improvements is the use of alternative forms of representing intermediate
tridiagonals.\footnote{See Chapter~\ref{chapter:mrrr} for a discussion on
  different form of representing tridiagonals.}
LAPACK, and therefore 
\mrsmp, uses the $N$-representation of lower bidiagonal factorizations;
that is, in the following, any relatively robust representation (RRR) of a
tridiagonal is associated with a lower bidiagonal factorization, $LDL^*$.
We have designed {\tt mr3smp} in a 
modular fashion to be able to incorporate algorithmic changes with
minimal efforts.

\subsection{A brief motivation}

We already gave a motivation for investigating how MRRR can make efficient
use of modern multi-core architectures in Chapter~\ref{chapter:ch01}; here
we partly repeat this argument. Considering the four methods (BI, QR, DC, MRRR) introduced in
Section~\ref{sec:existingmethodsSTEP}, exhaustive experiments of LAPACK's
implementations indicate that for sufficiently large matrices DC and MRRR are the fastest~\cite{perf09}. Whether DC or MRRR is faster  depends on the spectral
distribution of the input matrix.  

\begin{figure}[thb]
   \centering
   \subfigure[Execution time.]{
     \includegraphics[width=.47\textwidth]{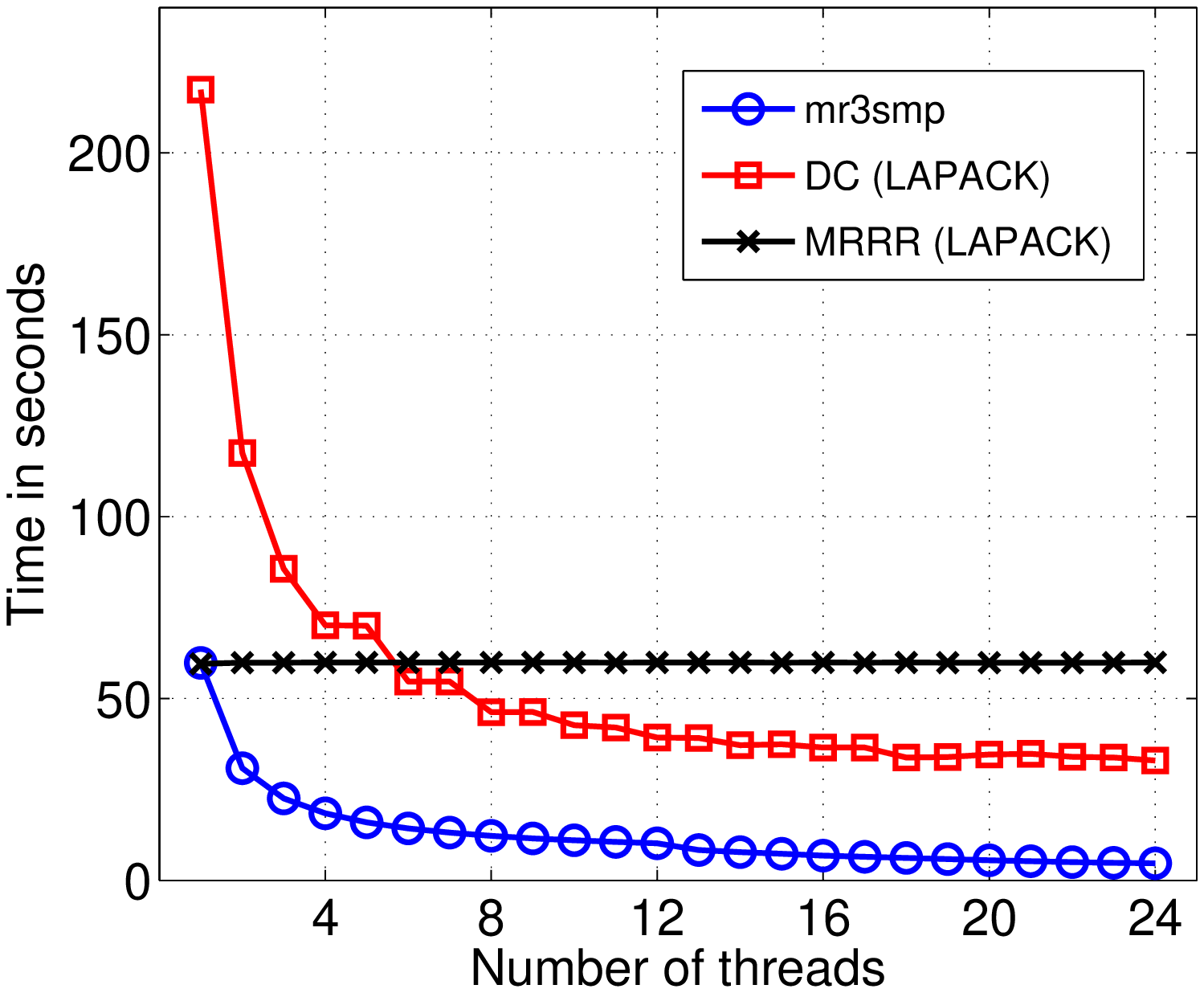} 
     \label{fig:timingmotivationmr3smpa}
   } \subfigure[Speedup.]{
     \includegraphics[width=.47\textwidth]{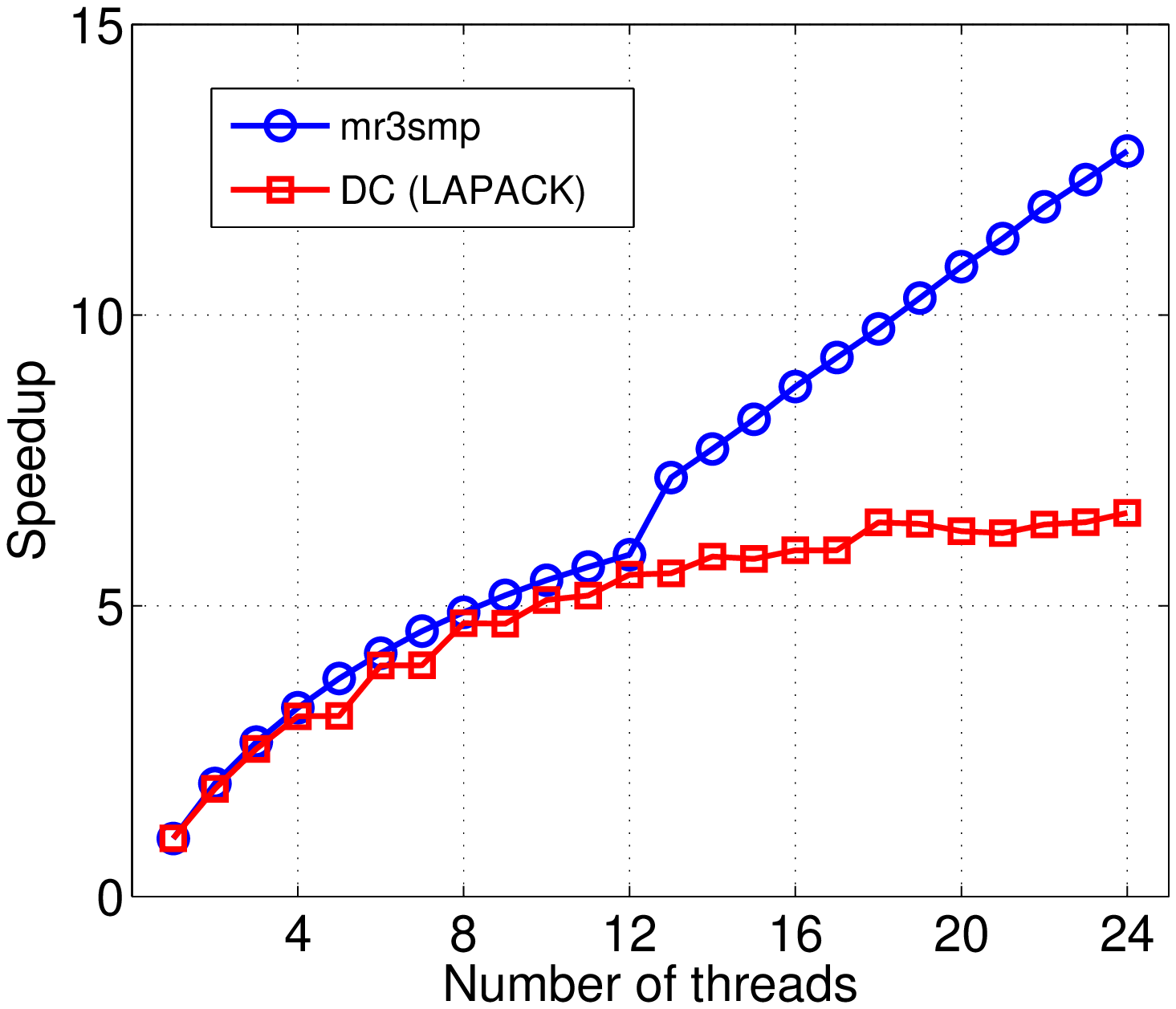}
     \label{fig:timingmotivationmr3smpb}
   }
   \caption{
     Execution of {\tt mr3smp} and LAPACK's DC and
     MRRR for a matrix of size $12{,}387$. The experiment is performed with
     LAPACK 3.2.2 on
     \DUNN\ (see Appendix~\ref{appendix:hardware}).
     The routines are linked to a multi-threaded MKL BLAS. 
     The slope of \mrsmp's speedup curve at 24
     threads is still positive, indicating that more available hardware
     parallelism will yield higher speedups.}
   \label{fig:timingmotivationmr3smp}
\end{figure}

When executed on multi-core architectures, which algorithm is faster
additionally depends on the amount of available parallelism; indeed, if
many cores are available, DC using multi-threaded BLAS becomes faster than MRRR.
In Fig.~\ref{fig:timingmotivationmr3smp}, we report representative timings for the
computation of all eigenpairs as a function of
the number of threads used; 
the tridiagonal input matrix of size $12{,}387$  comes from a finite-element model of 
an automobile body (see~\cite{Bientinesi:2005:PMR3} for more details).
Shown are LAPACK's DC and MRRR implementations, as
well as the new {\tt mr3smp}. LAPACK's BI with about 2 {\em hours} and QR 
with more than 6 {\em hours} are much slower and not shown.

While DC takes advantage of parallelism by using a multi-threaded BLAS library, 
the other LAPACK routines do not exploit multi-core architectures. 
Once enough hardware parallelism is available, DC becomes faster than
the sequential MRRR. Our parallel {\tt mr3smp} on the other hand is
specifically designed for multi-core processors, and results to be faster
and, as Fig.~\ref{fig:timingmotivationmr3smp} suggests, more scalable than all the
other LAPACK routines.\footnote{Only LAPACK's DC takes advantage of the
  multiple cores. In contrast, Intel MKL's QR is equally multi-threaded.} 
As experiments in~\cite{para2010} revealed, similar results hold
for comparisons with the vendor-tuned Intel MKL.  

\subsection{Parallelization strategy}
\label{parallelstrategy}

Existing distributed-memory versions of MRRR aim at minimizing
communication among processors while attaining perfect {\it
  memory}
balancing~\cite{Bientinesi:2005:PMR3,Vomel:2010:ScaLAPACKsMRRR}.\footnote{See
  Section \ref{sec:objectives} for a definition of memory balancing.} As the
division of work is performed statically, this   
approach comes at the expense of load balancing. Since on multi-core and shared-memory
architectures the memory balance is not a concern, our objective is
instead to identify the right computational granularity to distribute the
workload perfectly.  In 
this section, we illustrate how this is accomplished by dividing and
scheduling the work dynamically. 

The MRRR algorithm, as given in Algorithm~\ref{alg:mrrr2},
conceptually splits into two parts:   
(1)~Computation of a root representation and initial approximation of eigenvalues; 
(2)~Computation of eigenvectors, which includes the final refinement of eigenvalues. 
The two parts correspond to
Lines~\ref{line:mrrr2:root}--\ref{line:mrrr2:initialeigvals} and 
\ref{line:mrrr2:initialpartitioning}--\ref{line:mrrr2:end} in
Algorithm~\ref{alg:mrrr2}, respectively. 

The first part is detailed in Algorithm~\ref{alg:initialeigvals} in
Section~\ref{sec:closerlook} and is, 
like all the other computations, performed independently for each irreducible
subblock. The computation of a root representation (as for each RRR
subsequently) only costs $O(n)$ flops and is performed sequentially.
The initial approximation of eigenvalues instead costs $O(n)$ flops per
eigenvalue, and is performed in parallel using bisection;
alternatively, whenever faster, the sequential dqds algorithm
is used. Assuming a user requests $k$ eigenpairs, 
bisection is
preferable over the dqds algorithm if the number of processors is larger
than or equal to $12 \cdot k/n$~\cite{Bientinesi:2005:PMR3}. 
In the parallel case,
the computation of a subset of eigenvalues is encapsulated as an
independent computational task. The 
granularity of the tasks ensures load balancing among the computational
threads.\footnote{We used a simple work division strategy in \mrsmp\ as the
  resulting load imbalance was negligible in the total run time.} 

In the second part, we also divide the computation into tasks, which are
executed by multiple threads in parallel.   
As many different strategies for dividing and scheduling the
computation can be employed, we discuss possible variants in more
detail. Although the eigenvalues are refined in the second part of the
algorithm, for simplicity, we refer to {\em the eigenvalue computation} and
{\em the eigenvector computation} as the first and the second part,
respectively. 

\subsection{Dividing the computation into tasks}

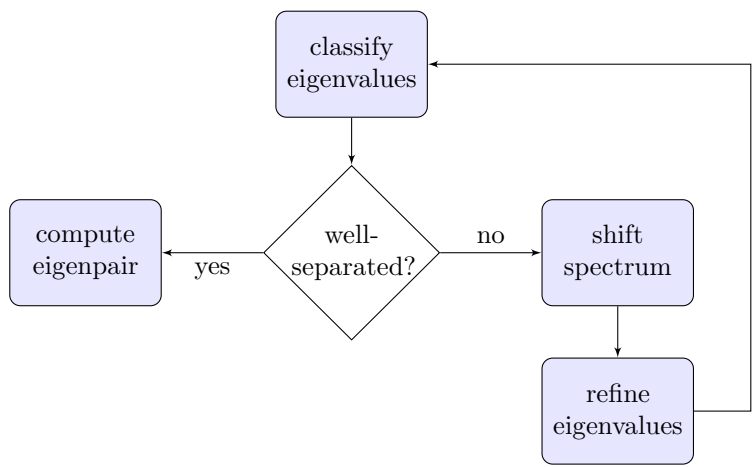
\begin{figure}[bht]
  \small
  \centering
  \begin{tikzpicture}[node distance = 2cm, auto]
    \node [block] (classify) {classify eigenvalues};
    \node [decision, below of=classify] (decide) {well-separated?};
    \node [block, left of=decide] (singleton) {compute eigenpair};
    \node [block, right of=decide] (shift) {shift spectrum};
    \node [block2, below of=shift] (refine) {refine eigenvalues};
    \path [line] (classify) -- (decide);
    \path [line] (decide) -- node {yes}(singleton);
    \path [line] (shift) -- (refine);
    \path [line] (decide) -- node {no}(shift);
    \path [line] (refine) --+(1.75,0) |- (classify);
  \end{tikzpicture}
  \caption{The computational flow of MRRR. An S-task
    performs to the final computation of eigenpairs. A C-task
    performs the spectrum shift, refinement, and reclassification
    of the eigenvalues.}
  \label{fig:flowchart}
\end{figure}

We now concentrate on the second -- more costly --  part of the computation. 
Since the representation tree, which is introduced in
Section~\ref{sec:finitearitmetic}, 
 characterizes the 
unfolding of the algorithm, it is natural to associate 
each node in the tree with a 
unit of computation. Because of this one-to-one correspondence, from
now on, we use the notion of task and node interchangeably.
Corresponding to interior nodes and leaf nodes, we introduce two types
of tasks, {\em C-tasks} and {\em S-tasks}.  C-tasks deal with the
processing of clusters and create other tasks, while S-tasks are
responsible for the final eigenpair computation. C-tasks and S-tasks
embody the two branches of the {\tt if} statement in Algorithm~\ref{alg:mrrr2}:
Lines~\ref{line:mrrr2:shifting}--\ref{line:mrrr2:partition} and
\ref{line:mrrr2:getvec}--\ref{line:mrrr2:returneigenpair},
respectively. The flow of the computation is schematically displayed in
Fig.~\ref{fig:flowchart}.

Irrespective of scheduling strategies, the granularity of this first
approach prevents a parallel execution of tasks: In the presence of a
large cluster, the threads might run out of executable tasks well
before the large cluster is processed and broken up into smaller
 tasks.  
As an example, consider the computation
of the eigenpairs with indices $\{1,2,\ldots,9\}$ of an irreducible input
matrix. Assume eigenvalues $\hat{\lambda}_2$ to $\hat{\lambda}_9$ are clustered,
and that we want to solve the problem on a 4-core processor; in this
case, one of the four cores will tackle the singleton $\{1\}$, a second
core will process the cluster, and the third and forth cores
will sit idle until the cluster is decomposed. Since the cluster
computation is more expensive than the processing of a singleton, even the
first core will have idle time, waiting for new available tasks.

Corresponding to statements~\ref{line:mrrr2:shifting}--\ref{line:mrrr2:partition} in
Algorithm~\ref{alg:mrrr2}, the computation associated with a C-task is summarized in
Algorithm~\ref{alg:ctask}. 
\begin{algorithm}[htb]
 \small
  \textbf{Input:} An RRR and the index set of a cluster $\mathcal{I}_c$,
  $\{\mbox{RRR}, \mathcal{I}_c\}$. \\
  \textbf{Output:} S-tasks and C-tasks associated with the children of
  the cluster.

  \vspace{1mm} 
  \algsetup{indent=2em}
  \begin{algorithmic}[1]
    \STATE Call subroutine {\tt ComputeNewRRR}.
    \STATE Call subroutine {\tt RefineEigenvalues}.
    \STATE Call subroutine {\tt ClassifyEigenvalues}.
  \end{algorithmic}
  \vspace{-2mm}
  \rule{\textwidth}{.1mm}
 
  {\sc Subroutine:} {\tt ComputeNewRRR} \\
  \textbf{Input:} An RRR and the index set of a cluster $\mathcal{I}_c$,
  $\{\mbox{RRR}, \mathcal{I}_c\}$. 
  
  \vspace{1mm}
  
  \algsetup{indent=2em}
  \begin{algorithmic}[1]
    \STATE Compute RRR$_{shifted}$ using {\tt DLARRF}.
  \end{algorithmic}
  \vspace{-2mm}
  \rule{\textwidth}{.1mm}

  {\sc Subroutine:} {\tt RefineEigenvalues} \\
  \textbf{Input:} An RRR and the index set of a cluster $\mathcal{I}_c$,
  $\{\mbox{RRR}_{shifted}, \mathcal{I}_c\}$. 
  
  \vspace{1mm}
  
  \algsetup{indent=2em}
  \begin{algorithmic}[1]
    \IF{$|\mathcal{I}_c| > \lceil nleft/nthreads \rceil$}
    \STATE Decompose the refinement into R-tasks.
    \ELSE
    \STATE Refine eigenvalues $\{\hat{\lambda}_i[\mbox{RRR}_{shifted}] : i
    \in \mathcal{I}_c\}$ using {\tt DLARRB}.
    \ENDIF
  \end{algorithmic}
  \vspace{-2mm}
  \rule{\textwidth}{.1mm}

  {\sc Subroutine:} {\tt ClassifyEigenvalues} \\
  \textbf{Input:} An RRR and the index set of a cluster $\mathcal{I}_c$,
  $\{\mbox{RRR}_{shifted}, \mathcal{I}_c\}$.
  
  \vspace{1mm}
  
  \algsetup{indent=2em}
  \begin{algorithmic}[1]
    \STATE Partition $\mathcal{I}_c$ into subsets 
    $\mathcal{I}_c = \bigcup_{s=1}^{S} \mathcal{I}_s$.
    \FOR{$s=1$ {\bf to} $S$}
    \IF{$|\mathcal{I}_s| > 1$}
    \STATE Create and enqueue C-task $\{\mbox{RRR}_{shifted}, \mathcal{I}_s\}$. 
    \ELSE
    \STATE Create and enqueue S-task $\{\mbox{RRR}_{shifted}, \mathcal{I}_s\}$. 
    \ENDIF
    \ENDFOR
  \end{algorithmic}

  \caption{\ C-task}
  \label{alg:ctask}
\end{algorithm}
Eigenvalues and eigenvectors are specified by their index and
generally shared among all threads. 
Quantities $nleft$ and $nthreads$ denote the
number of eigenpairs not yet computed and the number of threads used,
respectively. In our implementation, an RRR is a reference counted object,
which contains information such as the data of the representation, the
accumulated shift, the depth in the representation tree, and a counter that indicates the
number of tasks that depend on the RRR. Whenever tasks are executed that make
use of an RRR, its counter is decremented; if the counter
becomes zero the memory of the RRR is released. 
To reduce the granularity of the computation, we introduce a
third type of tasks, the {\em R-task}, that allows the decomposition
of the eigenvalue refinement -- the most expensive step in C-tasks --
originating immediately executable tasks for all threads even if large
clusters 
are encountered.  An R-task takes as 
input an RRR and a list of eigenvalues. As output, it returns the refined
eigenvalues. 
As a guiding principle for load balance, we decompose C-tasks into subtasks
whenever they involve more than $s_{max} = \lceil nleft/nthreads \rceil$ eigenpairs; 
the rationale being that at any moment, each thread is conceptually
``responsible'' for the computation of  not more than $s_{max}$
eigenpairs.\footnote{In practice, a minimum threshold 
prevents the splitting of small clusters.}
The algorithm directly invokes LAPACK's routines
{\tt DLARRF} and {\tt DLARRB}; we point out that if these are amended,
our parallelization of MRRR still stands, without modifications.

Vice versa, to avoid too fine a granularity, we extend S-tasks to
compute eigenpairs of multiple singletons from the same RRR; we
call this strategy ``bundling''.  This form of bundling
might be seen as widening the base case (by joining leaf nodes) of the
recursive algorithm.  
Bundling brings several advantages: among them, (1) better usage of cached data
through temporal and spatial locality; (2) fewer tasks and less contention for the work
queues; (3) reduced false sharing; and (4) the possibility of lowering the
overall memory requirement.
For similar reasons, we execute C-tasks without creating
additional tasks, provided they are sufficiently small.

Corresponding to
Lines~\ref{line:mrrr2:getvec}--\ref{line:mrrr2:returneigenpair} in
Algorithm~\ref{alg:mrrr2}, the computation performed by an 
S-task is summarized by Algorithm~\ref{alg:stask}. 
\begin{algorithm}[htb]
 \small
  \textbf{Input:} An RRR and an index set $\mathcal{I}_s$ 
   of well-separated eigenvalues, $\{\mbox{RRR}, \mathcal{I}_s\}$. \\
  \textbf{Output:} Local eigenpairs $\{(\hat{\lambda}_i, \hat{z}_i): i \in
  \mathcal{I}_s\}$. 
  
  \vspace{1mm}
  
  \algsetup{indent=2em}
  \begin{algorithmic}[1]
    \FOR{each index $i \in \mathcal{I}_s$} 
    \WHILE{eigenpair $(\hat{\lambda}_i,\hat{z}_i)$ is not accepted}
    \STATE Invoke {\tt DLAR1V} to solve $(L D L^T - \hat{\lambda}_i I)
    \hat{z}_i = N \Delta N \hat{z}_i = \gamma_r e_r$.
    \STATE Record the residual norm $|\gamma_r|/\|\hat{z}_i\|$ and the RQC
    $ \gamma_r / \|\hat{z}_i\|^2$.
    \IF{$|\gamma_r|/\|\hat{z}_i\| < \mbox{tol}_1\cdot \mbox{gap}(\hat{\lambda}_i)$
      {\bf or} $|\gamma_r| / \|\hat{z}_i\|^2 < \mbox{tol}_{2}\cdot |\hat{\lambda}_i|$}
    \STATE Normalize $\hat{z}_i$ and accept the eigenpair $(\hat{\lambda}_i, \hat{z}_i)$.
    \ENDIF
    \IF{RQC improves $\hat{\lambda}_i$}
    \STATE $ \hat{\lambda}_i := \hat{\lambda}_i + \gamma_r / \|\hat{z}_i\|^2$
    \ELSE
    \STATE Use bisection to refine $\hat{\lambda}_i$ to full accuracy.
    \ENDIF
    \ENDWHILE
    \ENDFOR
  \end{algorithmic}
  
  \caption{\ S-task}
  \label{alg:stask}
  
\end{algorithm}
Instead of bisection followed by one
step of inverse iteration, it uses Rayleigh quotient iteration for performance reasons. 
The Rayleigh quotient correction (RQC) is used
to improve the eigenvalue if the following applies: the sign
of the correction is correct and it leads to a value within the
uncertainty interval $[\hat{\lambda}_i - \hat{\lambda}_{i,err},
\hat{\lambda}_i + \hat{\lambda}_{i,err}]$~\cite{Dhillon:DesignMRRR}. 
When the eigenpair is accepted, a final RQC could be added to reduce the residual.
For the eigenvector computation, the algorithm invokes LAPACK's 
auxiliary routine {\tt DLAR1V}; the same 
comment as for Algorithm~\ref{alg:ctask} applies here. 
Quantities $tol_1 = \order{n \varepsilon}$ and
$tol_2 = \order{\varepsilon}$ denote appropriate constants to signal
convergence.

\paragraph{An example.}

\begin{figure}[tbh]
  \centering
  \includegraphics[scale=.2, trim= 0mm 0mm 0mm 0mm]{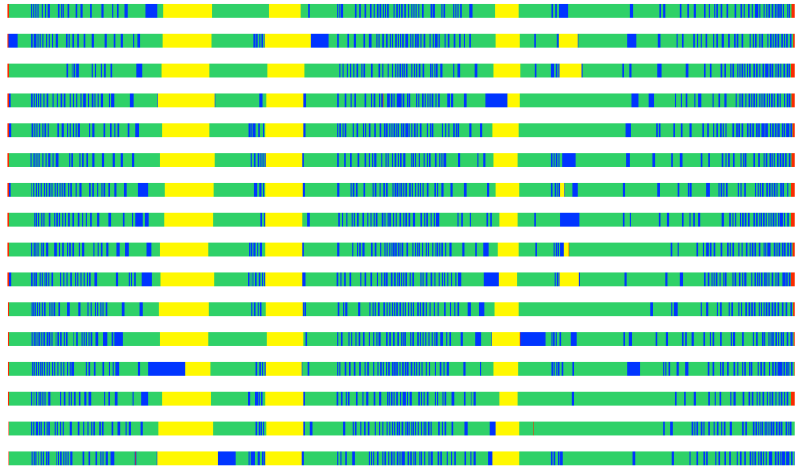}
  \caption{Exemplary execution trace for a matrix of size $12{,}387$
    using 16 threads on \DUNN. The colors
    green, blue, and yellow represent time spent in the execution of S-tasks,
    C-tasks, and R-tasks, respectively.}
  \label{fig:trace12387}
\end{figure}

Figure~\ref{fig:trace12387} shows the execution
trace of an exemplary eigenvector computation. 
The examined matrix is the one for which timings were presented in
Fig.~\ref{fig:timingmotivationmr3smp}. Computing the  
eigenvectors took about 49.3 seconds sequentially and about 3.3 seconds with
16 threads. In the time-line graph, 
the green, blue and yellow sections correspond to the processing of S-tasks,
C-tasks, and R-tasks, respectively. 
Everything considered as parallelization overhead is colored in red.
On average, each thread spends about 66\% of the execution time in
computing the eigenvectors of singletons, 34\% in computing new RRRs of
clusters and to refine the associated eigenvalues. Due to the dynamic task
scheduling, load balancing is achieved and almost no overhead computations
occurs. 

We want to point out important details of the trace. 
The first yellow bar corresponds to a refinement of eigenvalues that is split via R-tasks. A
cluster of size $8{,}871$ is encountered by the bottommost thread. Since the
cluster contains a large part of the 
eigenvectors that are still to be computed, the refinement of its eigenvalues is
split among all the threads. The procedure of splitting the refinement
among all threads is repeated two more times during the computation. Later
during the computation there are yellow regions that provide examples where
the refinement of eigenvalues is split into R-tasks but not distributed
among all threads anymore.

\subsection{The work queues and task scheduling}

We analyze data dependencies among
tasks, and discuss how they may be scheduled for execution.
To differentiate the task priority, we form three work queues for {\em
  high}, {\em medium}, and {\em low} priority tasks, respectively. 
Each active thread polls the queues, from the highest to the lowest in
priority, until an available task is found; the task then is dequeued
and executed. The process is repeated until all eigenpairs are
computed.

\subsubsection{Task data dependencies}
\label{subsec:datadepend}

Since the index sets associated to C-tasks
and S-tasks are disjoint, no data
dependencies are introduced. 
Dependencies among tasks only result from the management of the RRRs.
Each task in the work queues requires its parent RRR, which serves
as input in Algorithms~\ref{alg:ctask} and~\ref{alg:stask}; as a consequence,
an RRR must be available until all its children tasks are processed.
A possible solution -- but not the one used -- is to dynamically allocate memory for each new
RRR, pass the memory address to the children tasks, and keep the RRR
in memory until all the children are executed. 
An example is given by
node $\{6,7,8,9\}$ in Fig.~\ref{fig:reptree} in Section~\ref{sec:bigpicture}. $M^{(2)}$ must 
be available as long as the children nodes $\{6,7\}$, $\{8\}$, and $\{9\}$ have
not been processed.  This solution offers the following advantages:
When multiple threads are executing the children of a C-task, only one
copy of the parent RRR resides in memory and is shared among all threads;
similarly, the auxiliary quantities $d\ell(i) = D(i,i) \cdot L(i+1,i)$
and $d\ell\ell(i) = D(i,i) \cdot L(i+1,i)^2$, for all $1 \leq i
\leq n-1$, might
be evaluated once and shared; in
architectures with shared caches, additional performance improvements might arise from
reuse of cached data.  
On the downside, this approach needs an extra $\order{n^2}$ workspace in
the worst case.  

In order to reduce the memory requirement, we make
the following observation.  Each task can associate the portion of
eigenvector matrix $Z$
corresponding to $\mathcal{I}_c$ in
Algorithm~\ref{alg:ctask} or $\mathcal{I}_s$ in 
Algorithm~\ref{alg:stask} with local workspace, i.e., a section of memory
not accessed by any other task in the work queues. 
Continuing with the
example of node $\{6,7,8,9\}$ at depth one of the tree in
Fig.~\ref{fig:reptree}, the columns 6 to 9 are regarded as local storage of
the corresponding task. 
If the RRR required as input is stored in the
task's local workspace, then we call such a task {\em
  independent}. If all 
the children of a task are independent, then the task's RRR is not needed
anymore and can be discarded. Additionally, all the children tasks can be
executed in any order, without any data dependency. We now illustrate that
all the C-tasks can be made independent and that practically all S-tasks can
be made independent too.

{\em C-tasks:} As a cluster consists of at least two
eigenvalues, the corresponding portion of $Z$ -- used as local
workspace -- is always large enough to contain a representation.
We therefore use $Z$ to store the parent RRR's data.\footnote{An alternative, used
  by {\tt DSTEMR}, is to compute and store the new RRR of the cluster. We use
  the other approach to generate tasks more quickly at the beginning.}
Unfortunately, rendering the tasks independent comes with an overhead due to storing
the parent RRR into $Z$ and retrieving the parent RRR from $Z$.


{\em S-tasks:} The same approach is feasible for S-tasks whenever at
least two singletons are bundled. Conversely, the approach cannot be
applied in the extreme scenario in which a cluster is decomposed into
smaller clusters plus only one singleton.
The leaf node $\{2\}$ in
Fig.~\ref{fig:reptree} represents such an exception. 
All the other
S-tasks may be bundled in groups of two or more, and therefore can be
made independent. 
One drawback of this approach is that when several S-tasks children of the
same node are processed simultaneously, multiple copies of the same
RRR reside in memory, preventing the reuse of cached data.
In the example of node $\{6,7,8,9\}$ in Fig.~\ref{fig:reptree}, storing
$M^{(2)}$ into columns 6 and 7, as well as columns 8 and 9, creates an 
independent C-task for the node $\{6,7\}$, and an independent S-task $\{8,9\}$
in which the two singletons $\{8\}$ and $\{9\}$ are bundled.
We point out that while working with independent tasks introduces some
overhead, it brings great flexibility in the scheduling, as tasks
can now be scheduled {\em in any order}.

\subsubsection{Task scheduling}

Many strategies can be employed for the dynamic scheduling of
tasks. As a general guideline, in a shared memory environment, having
enough tasks to feed all the computational cores is paramount to
achieve high-performance. In this section, we discuss the
implementation of two simple but effective strategies that balance
task-granularity, memory requirement, and creation of tasks.  In both
cases, we implemented three separate FIFO queues, to avoid contention when
many worker threads are used. Both approaches schedule R-tasks with
high priority, because they arise as part of the decomposition of
large clusters, and originate work in the form of tasks.

{\em C-tasks before S-tasks:} All enqueued C-tasks and S-tasks are made
independent. (In the rare event that a S-task cannot be made independent, the
task is executed immediately without being enqueued.) Consequently, all
C-tasks and S-tasks in the work queues can be scheduled in any order. Since 
processing C-tasks creates new tasks that can be executed in parallel, we
impose that C-tasks (medium priority) are dequeued before S-tasks (low
priority). This ordering is a special case of many different strategies in
which the execution of C-tasks and S-tasks are interleaved. No other
ordering offers more executable tasks in the work queues; thus, for the
scenario that all tasks are independent, we expect
that the strategy attains the best performance. 

{\em S-tasks before C-tasks:} No S-task is made independent. In
order to obtain a similar memory requirement as in the first approach, we
are forced to schedule the S-tasks before the C-tasks. The reason is
that for each cluster an RRR must be kept in memory until all its children
S-tasks are executed. The ordering guarantees that at any time only S-tasks originating no
more than $nthreads$ clusters are in the queue, and we limit
the number of RRRs to be kept in memory to $nthreads$.
While the flexibility in  
scheduling tasks is reduced, the overhead from making the S-tasks
independent is avoided.  In practice, this second approach is slightly
faster -- about 5--8\% in our tests; it is used for all timings in
Section~\ref{section:mrsmp:experiments}. 

\subsection{Memory requirement}

Routines \mrsmp\ and LAPACK's \DSTEMR\ make efficient use of memory by using the
eigenvector matrix $Z$ as temporary storage. Additionally, \DSTEMR\
overwrites the input matrix $T$ to store only the RRR of the currently
processed node. This approach is not feasible in a parallel setting, as the
simultaneous processing of multiple nodes requires storing the associated RRRs
separately. Moreover, in our multi-threaded implementation, each thread
requires its own workspace. As a consequence, the parallelization and its
resulting performance gain come at the cost of a slightly higher memory
requirement than in the sequential case.

While \DSTEMR\ requires extra workspace to hold $18n$ double precision
numbers and $10n$ integers~\cite{Dhillon:DesignMRRR}, \mrsmp\ requires extra 
storage of about 
$(12n + 6n \cdot nthreads)$ double precision numbers and $(10n + 5n \cdot
nthreads)$ integers. 
For a single thread the total workspace requirement is roughly the
same as for \DSTEMR, and the requirement increases linearly with the number of
threads participating in the computation. We remark that thanks to shared data, the
required memory is less than $nthreads$ times the memory needed for
the sequential routine.
\begin{figure}[bht]
   \centering
   \subfigure[Using a fixed number of 12 threads.]{
     \includegraphics[width=.47\textwidth]{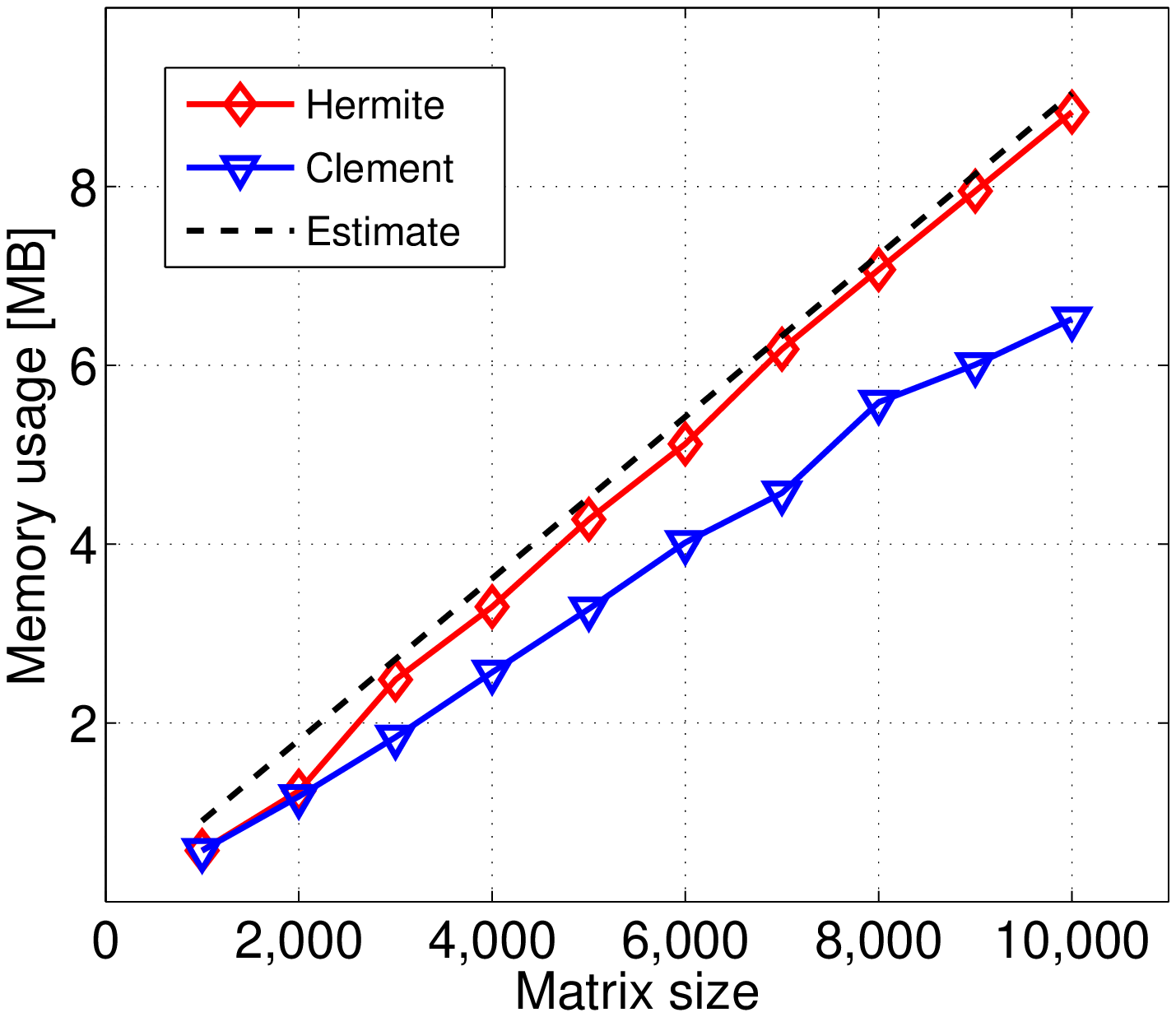}
     \label{fig:memoryofmr3smpa}
   } \subfigure[Using a fixed matrix size of $5{,}000$.]{
     \includegraphics[width=.47\textwidth]{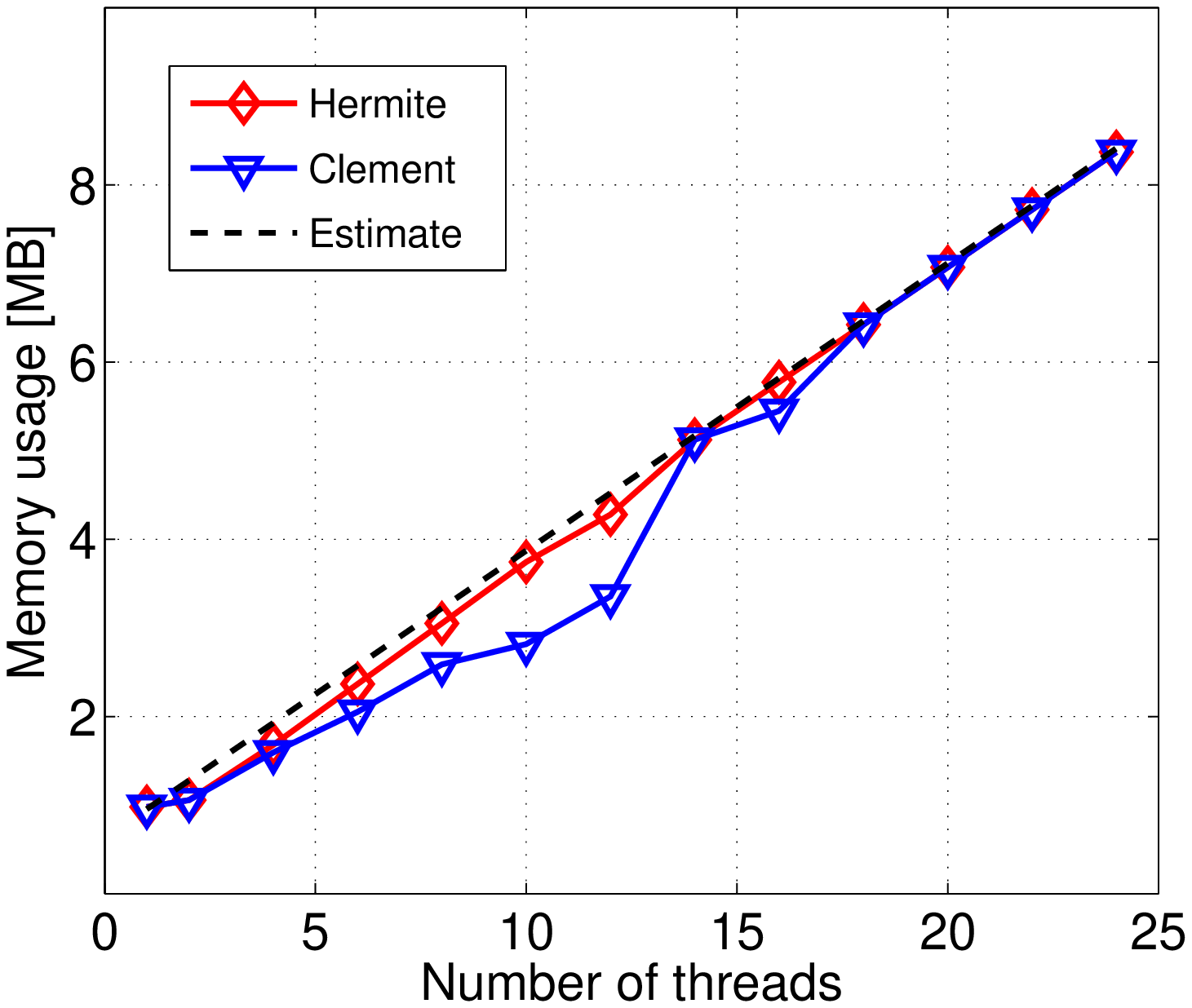}
     \label{fig:memoryofmr3smpb}
   }
   \caption{
     Peak memory requirement of \mrsmp. The
     workspace requirement has to be compared with the $n^2$ double
     precision numbers of the   
     output matrix. For instance, for matrices of size 5{,}000
     the output requires about 190 MB memory. 
   }
   \label{fig:memoryofmr3smp}
\end{figure}
In Fig.~\ref{fig:memoryofmr3smp} we show for two different kind of matrices,
namely the so called Hermite 
and symmetrized Clement matrices, the measured peak memory usage by
\mrsmp.\footnote{See Appendix~\ref{appendix:matrices} for
a description of test matrices.} 
By setting $nthreads$ to 12, Fig.~\ref{fig:memoryofmr3smpa} shows that the
extra memory requirement is linear in the matrix size. In
Fig.~\ref{fig:memoryofmr3smpb}, the matrix size has a fixed value 
of $5{,}000$ and the dependence of the required memory with the number of
threads is shown. Although the memory requirement increases with the number
of threads, even if 24 threads are used, the required extra storage only makes
up for less than 5\% of the memory required by the output matrix $Z$.
Since the DC algorithm requires $\order{n^2}$ double precision workspace, the
advantage of MRRR in requiring much less memory than DC is maintained.

\subsection{Experimental results}
\label{section:mrsmp:experiments}

We now turn the attention on timing and accuracy results of \mrsmp. We
used similar settings for thresholds, convergence 
  parameters, and classification criteria as {\tt DSTEMR} to clearly 
  identify the effects of the parallel execution.  
  
We compare our solver to the fastest solvers available in
LAPACK and the parallel ParEig (a solver for a distributed-memory
systems).\footnote{See~\cite{para2010} for a comparison with Intel's MKL.} As not
all solvers allow for the computation of a subset of
eigenpairs at reduced cost, we consider this scenario for \mrsmp\ first. 
All tests were run on \DUNN, with the settings described in
Appendix~\ref{appendix:hardware}.  We 
used LAPACK version~$3.2.2$ in conjunction 
with Intel's MKL BLAS version
10.2, and ParEig version~$2.0$ together with OpenMPI
version~$1.4.2$. Descriptions of the LAPACK routines and test matrices
are found in Appendix~\ref{appendix:routinenames} and
\ref{appendix:matrices}, respectively. We also refer to
\cite{para2010,mr3smp} for further results and additional information on the
experiments.

\subsubsection{Subset Computations}

\begin{figure}[thb]
  \centering
   \subfigure[Execution time.]{
     \includegraphics[width=.47\textwidth]{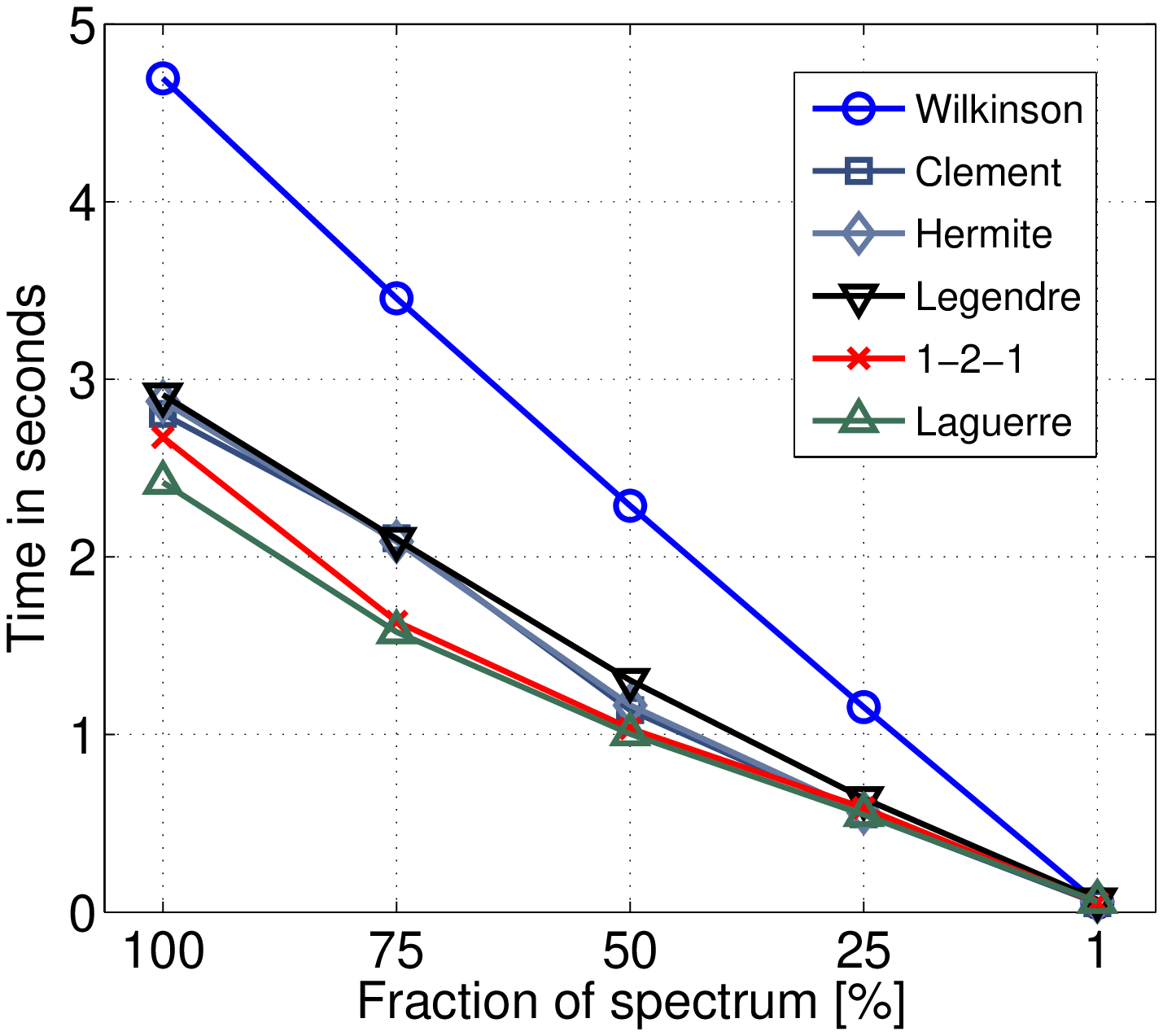} 
     \label{fig:subseta}
   } \subfigure[Speedup with respect
    to {\tt DSTEMR}.]{
     \includegraphics[width=.47\textwidth]{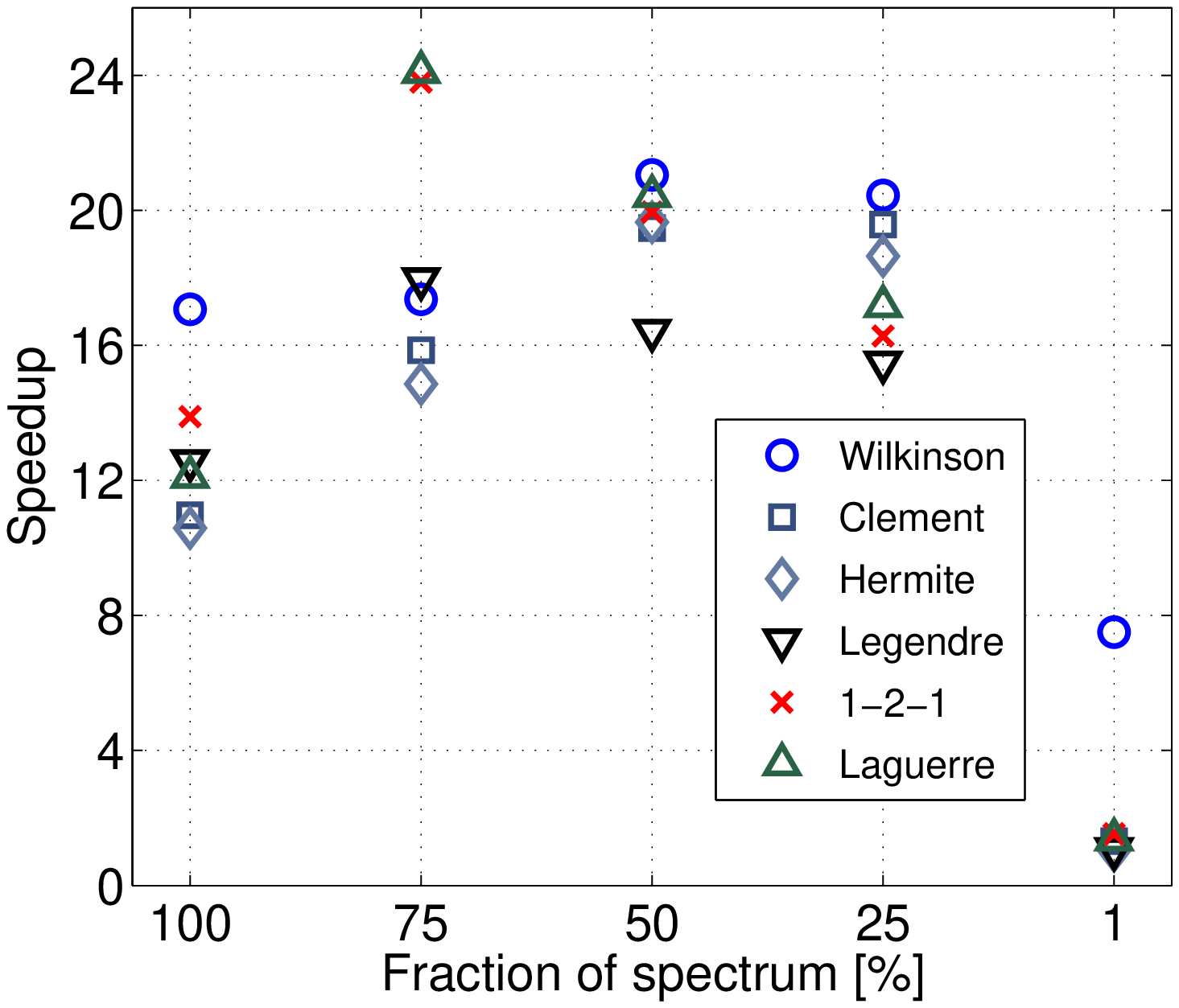}
     \label{fig:subsetb}
   }
  \caption{
    Computation of a subset of eigenpairs using \mrsmp\ with 24
    threads on \DUNN.}
  \label{fig:subset}
\end{figure}

An important feature of MRRR is the ability of computing a 
subset of eigenpairs at reduced cost. We show that with our task-based 
parallelization, the same is true also in the multi-threaded case. In 
Fig.~\ref{fig:subseta}, we show \mrsmp 's total execution time
against the fraction of requested eigenpairs. The test matrices are of
size 10{,}001, and the computed subsets are on the left end of the spectrum.
Additionally, Fig.~\ref{fig:subsetb} shows the obtained speedup with respect
to {\tt DSTEMR}. Intuitively, the speedup is limited if very few
eigenpairs are computed. However, it is often higher for computing
subsets than for finding all eigenpairs. 

The experiments indicate that \mrsmp\ is especially well 
suited for subsets computation. However, in order to compare timings with other
solvers that cannot compute subsets of eigenpairs at reduced cost, we show results for
computing all eigenpairs subsequently. {\it It is important to notice that
the execution time of MRRR is proportional to the number of
eigenpairs, which is not true for other methods.} 
This makes MRRR the method of choice for computing subsets of
eigenpairs, but often MRRR is the fastest method even when all eigenpairs are
desired~\cite{perf09,EleMRRR,Bientinesi:2005:PMR3}. 

As a final note: if MRRR
is used to compute small subsets of eigenpairs of dense matrices after a
reduction to tridiagonal form, it might be advantageous to use successive
band reduction (SBR) instead of the direct reduction to tridiagonal form
provided by LAPACK.\footnote{See Chapter~\ref{chapter:background}.} Routines for SBR
are found in the SBR toolbox~\cite{Bischof:theSBRtoolbox}, and in
vendor-tuned libraries such as Intel's MKL.  

\subsubsection{MR$^{\it 3}$\!--SMP vs.~LAPACK}

Arguably, the most important test matrices come from applications.
In the following, we concentrate on timing and accuracy 
results for a set of tridiagonal matrices arising in different scientific
and engineering disciplines.  

In Table~\ref{tab_appl} we show timings for QR ({\tt DSTEQR}), BI ({\tt
  DSTEVX}), MRRR ({\tt DSTEMR}), as well as for DC ({\tt DSTEDC}) and
MR$^3$-SMP (\mrsmp).  Since they cannot
achieve parallelism through multi-threaded BLAS, the execution time for the
first three routines is independent of the number of threads. 
Conversely, both DC
and \mrsmp\ take advantage of multiple threads, through
multi-threaded BLAS and the task-based approach, respectively.
All the timings refer to the execution with 24 threads, that is, as
many threads as cores available. The execution times for
QR and BI are significantly higher than for the other algorithms; thus, we
omit the results for the largest matrices. In all cases, even if
parallelized and achieving perfect speedup, the routines would still be
noticeably slower than the other methods.
In all tests, MR$^3$-SMP outperforms the other solvers considerably.
\begin{table}[htb]
\small
\centering
\begin{tabular}[htb]{l@{\quad}c@{\quad\quad}c@{\quad}c@{\quad}c@{\quad}c@{\quad}c@{\quad}}
\toprule
 Matrix            &   Size & QR & BI & MRRR & DC & MR$^3$--SMP \\
\midrule
{\it ZSM-5}    & $2{,}053$ &  $68.4$  & $6.24$  &  $0.92$ &  $0.70$ & {\bf 0.15} \\ 
{\it Juel-k2b} & $4{,}289$ &  $921$   & $382$   &  $4.41$ &  $3.39$ & {\bf 0.52} \\ 
{\it Auto-a}   & $7{,}923$ & $6{,}014$&$2{,}286$& $18.8$  & $12.2$  & {\bf 1.88} \\ 
{\it Auto-b}   &$12{,}387$ &$22{,}434$&$7{,}137$& $59.5$  & $32.9$  & {\bf 4.65} \\ 
{\it Auto-c}   &$13{,}786$ &  --    &$9{,}474$& $56.6$  & $35.4$  & {\bf 5.34} \\ 
{\it Auto-d}   &$16{,}023$ &  --    &  --   & $87.8$  & $45.8$  & {\bf 7.76} \\ 
\bottomrule
\end{tabular}
\caption{Execution times in seconds for a set of application matrices.}
\label{tab_appl}
\end{table}

\begin{figure}[thb]
  \centering
   \subfigure[Speedup.]{
     \includegraphics[width=.47\textwidth]{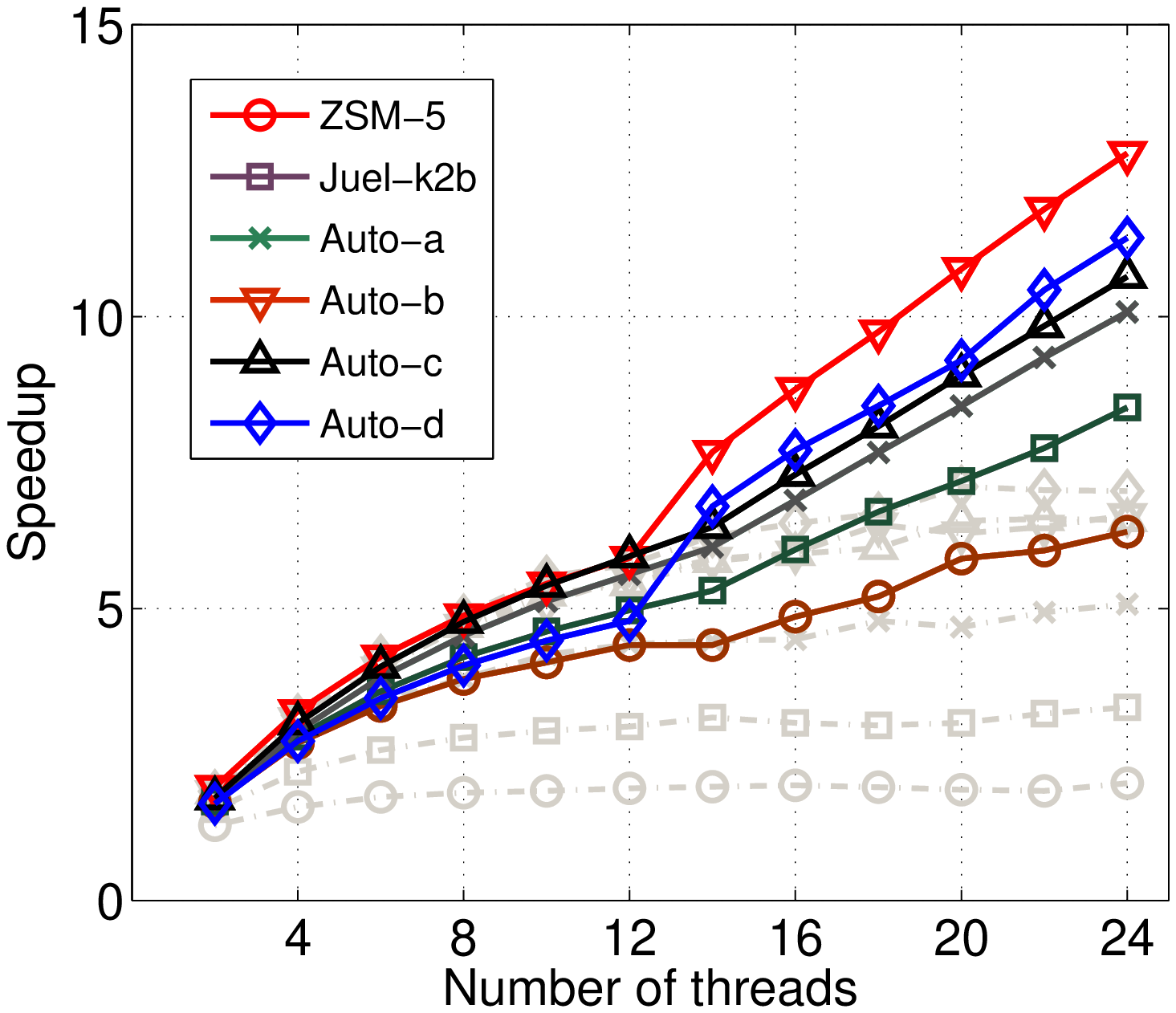} 
     \label{fig:mcspeedupappla}
   } \subfigure[Execution time.]{
     \includegraphics[width=.47\textwidth]{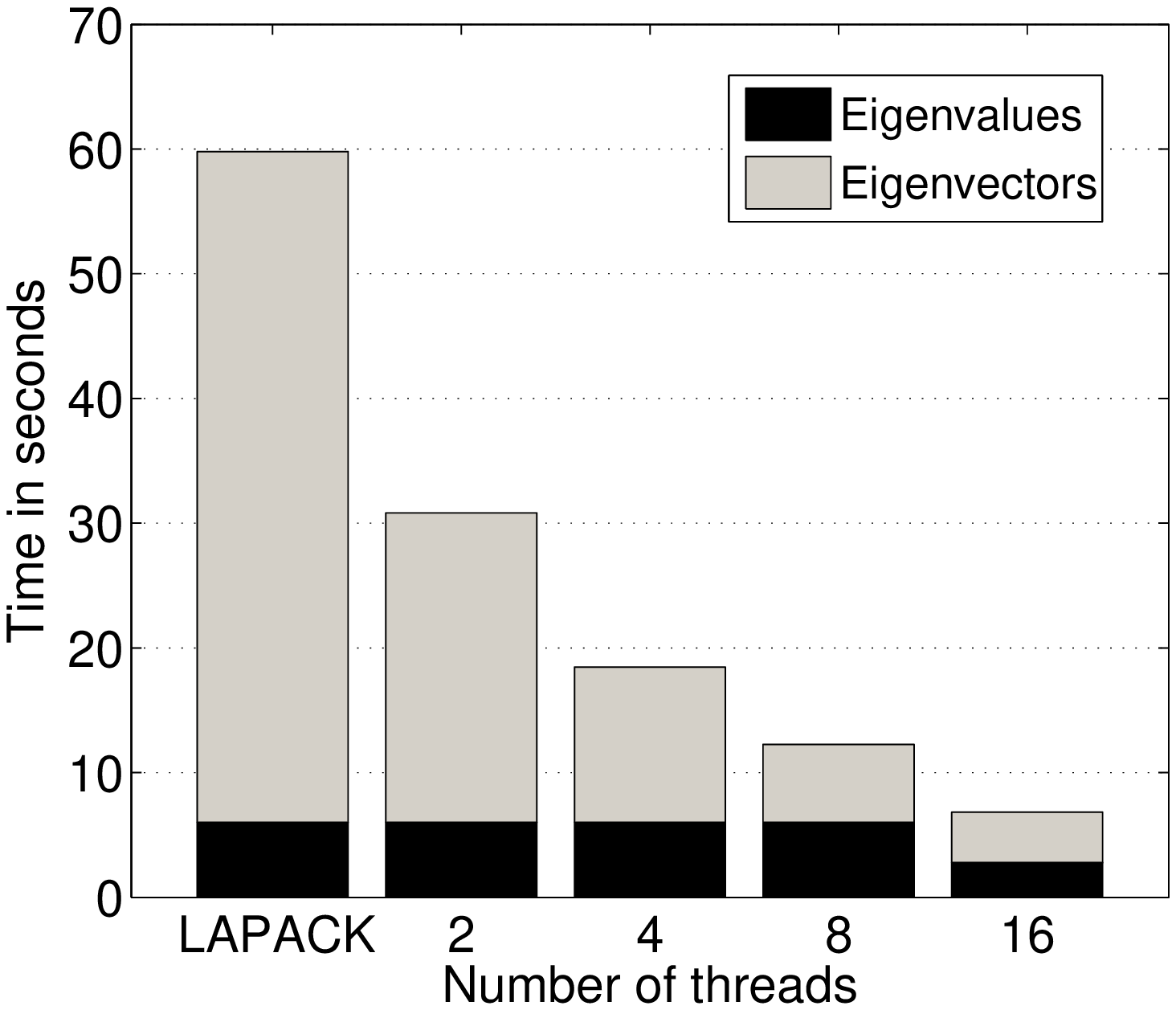}
     \label{fig:mcspeedupapplb}
   }
 \caption{Speedup for the total execution time of \mrsmp\ and \DSTEDC. {\it
     (a)} \mrsmp's speedup is with respect to \DSTEMR, while speedups for
   \DSTEDC\ (shown in light gray) are with respect to its sequential
   execution. {\it (b)} For input matrix {\it Auto-b}, separate execution
   time for the initial eigenvalue approximation and for the subsequent
   eigenvector computation. 
    }
  \label{fig:mcspeedupappl}
\end{figure}

Fig.~\ref{fig:mcspeedupappla} shows \mrsmp's and \DSTEDC's (in light gray) total speedup. 
All the other routines do not attain any 
speedups at all.  Regarding \mrsmp, every line up to 12 threads
converges to a constant value, a phenomenon which is better
understood by looking at Fig.~\ref{fig:mcspeedupapplb}.
For matrix {\it Auto-b}, we plot 
the execution time for the initial eigenvalue approximations and the
eigenvectors computation separately. When all the eigenpairs are
requested, the sequential dqds algorithm is used in a single-threaded
execution to approximate the 
eigenvalues. Because of the sequential nature of the dqds algorithm, by Amdahl's law,
the maximal speedup of \mrsmp\ is limited to about 7.6.  When enough
parallelism is available, bisection becomes faster than dqds; according to the criterion
discussed in Section~\ref{parallelstrategy}, if more than 12 threads are
used, we switch to bisection. Such a strategy is essential to achieve
scalability: using 24 threads, the speedup is 
about 13, that is almost twice of the limit dictated by Amdahl's law.   
Moreover, all the speedup curves in Fig.~\ref{fig:mcspeedupappla}
have positive slope at 24 threads, thus indicating that further
speedups should be expected as the amount of available cores increases. 


\subsubsection{MR$^{\it 3}$\!--SMP vs.~ParEig}

In Table~\ref{tab_appl_pareig} we compare the timings of \mrsmp\ and 
ParEig. ParEig is designed for distributed-memory
architectures, and uses the {\em Message-Passing-Interface} (MPI) 
for communication. Since ParEig was at the time of writing the fastest
distributed-memory parallel symmetric tridiagonal eigensolver, we omit a
comparison to other routines such as ScaLAPACK's {\tt PDSTEDC}. Instead, we
refer to~\cite{Bientinesi:2005:PMR3,Vomel:2010:ScaLAPACKsMRRR} for
comparisons of ParEig to other eigensolvers. 
In the experiments, we present ParEig's timings for 22 processes 
as they were consistently faster than the execution with 24
processes. Timings of ParEig do not include the overhead of initializing the MPI library.
\begin{table}[t]
\small
\centering
\begin{tabular}[htb]{l@{\quad\quad}c@{\quad\quad\quad}c@{\quad\quad}c@{\quad\quad}c@{\quad\quad}c@{\quad\quad}c}

\toprule
Matrix &  Size & MR$^3$-SMP & ParEig \\
\midrule

{\it ZSM-5}    & $2{,}053$ & $0.15$ \ $(0.16)$ & $0.13$ \\ 
{\it Juel-k2b} & $4{,}289$ & $0.52$ \ $(0.49)$ & $1.29$ \\ 
{\it Auto-a}   & $7{,}923$ & $1.88$ \ $(1.62)$ & $2.89$ \\ 
{\it Auto-b}   &$12{,}387$ & $4.65$ \ $(3.93)$ & $5.48$ \\ 
{\it Auto-c}   &$13{,}786$ & $5.34$ \ $(3.02)$ & $5.98$ \\ 
{\it Auto-d}   &$16{,}023$ & $7.76$ \ $(5.69)$ & $7.99$ \\ 
\bottomrule
\end{tabular}
\caption{Execution times in seconds for a set of matrices 
  arising in applications. The timings in brackets are achieved if \mrsmp\
  uses the same splitting criterion as ParEig.} 
\label{tab_appl_pareig}
\end{table}
The performance of \mrsmp\ matches and even surpasses that
of ParEig. 
For a direct comparison of \mrsmp\ with ParEig,
it is important to stress that, even though both routines implement
the MRRR algorithm, several internal parameters are different. Most
importantly, the minimum relative gap, $gaptol$, which determines when an
eigenvalue is to be considered well-separated, is set to
$\min\{10^{-2},n^{-1}\}$ and $10^{-3}$ in ParEig and \mrsmp, respectively.  In
order to make a fair comparison, in 
brackets we show the execution time of \mrsmp\ when using the
parameter $gaptol$ as set in ParEig.\footnote{ 
Despite a slight loss of performance, in \mrsmp\  
we conservatively set $tol = 10^{-3}$ for accuracy
reasons~\cite{Vomel:2010:ScaLAPACKsMRRR}.} 
The numbers indicate that with
similar tolerances, in the shared-memory environment, \mrsmp\
outperforms ParEig.

\subsubsection{MR$^{\it 3}$\!--SMP's accuracy}

\begin{figure}[thb]
  \centering
   \subfigure[Largest residual norm.]{
     \includegraphics[width=.47\textwidth]{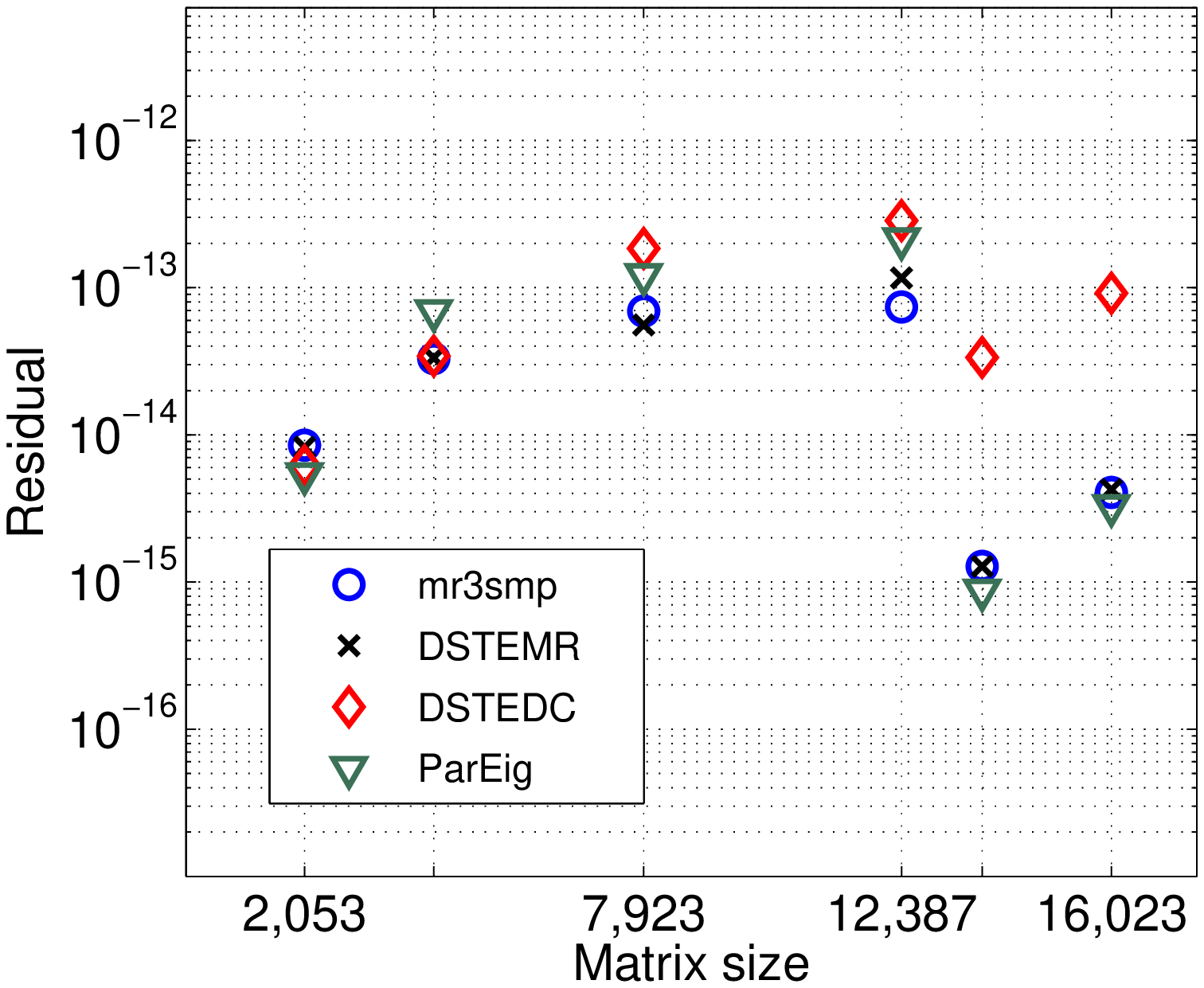} 
     \label{fig:mrsmpaccuracya}
   } \subfigure[Orthogonality.]{
     \includegraphics[width=.47\textwidth]{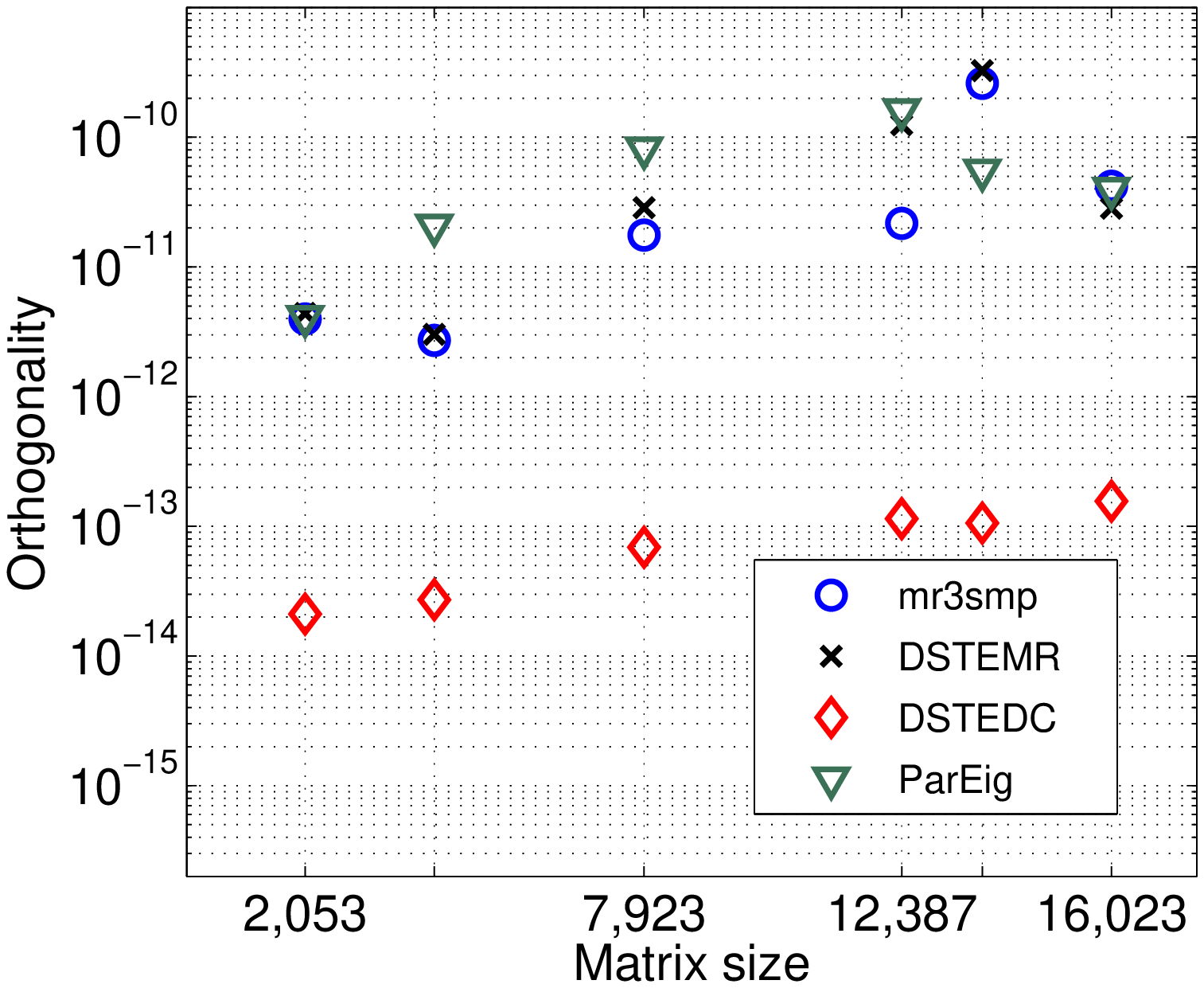}
     \label{fig:mrsmpaccuracyb}
   }
   \caption{ 
    Accuracy of \mrsmp, LAPACK's \DSTEMR\ and \DSTEDC, and ParEig
    for a set of application matrices. The largest residual norm and the
orthogonality are defined in~\eqref{def:defresortho2}. 
  }
  \label{fig:mrsmpaccuracy}
\end{figure}

In Fig.~\ref{fig:mrsmpaccuracy}, we present accuracy results for \mrsmp\ and
ParEig, as well as LAPACK's \DSTEMR\ and \DSTEDC.
For all test matrices, the routines attain equally good 
residuals, while in terms of orthogonality DC is more accurate than
the three MRRR-based routines. The
results are underpinned by the tests of Demmel et al.~\cite{perf09}, which
shows a similar behavior for a large test set of artificial and application
test matrices. In general, {\em our parallel MRRR obtains
  the same accuracy as its sequential counterpart}. 

While the
accuracy of MRRR is sufficient for many users, it might be a concern to
others. 
It is therefore natural to ask whether the accuracy of the MRRR-based routines can be
improved to levels of DC. 
Unfortunately, this is not
an easy task as Theorem~\ref{resthm} shows that one needs to be prepared
of orthogonality levels of about 
$\order{1000 n \varepsilon}$ even if all requirements of the algorithm are
fulfilled. Our workaround to this dilemma resorts to the use of higher
precision arithmetic and is the topic of Chapter~\ref{chapter:mixed}. 
\section{MRRR for modern supercomputers}
\label{sec:pmrrr}

In this section we present a tridiagonal solver, \PMRRR, which 
merges the task-based approach introduced in the previous section and the
distributed-memory parallelization of Bientinesi et
al.~\cite{Bientinesi:2005:PMR3}.\footnote{\PMRRR\ should not be confused with the
  solver introduced
  in~\cite{Bientinesi:2005:PMR3}.} Our target architecture are supercomputers with hybrid
shared/distributed-memory. For these architectures, parallelism is
achieved by executing multiple processes that communicate via message
passing. Each of these processes in turn might execute multiple threads that
communicate via shared memory. In this way, \PMRRR\ is well-suited for both single node 
and large-scale massively parallel computations. 

\PMRRR\ was integrated to the Elemental library~\cite{elemental} for the
solution of large-scale standard and generalized dense Hermitian
eigenproblems (in short, GHEP and HEP, respectively). 
In Section~\ref{sec:elementalssolver}, we introduce EleMRRR (from
Elemental and \PMRRR), a set of distributed-memory 
MRRR-based eigensolvers within Elemental.
EleMRRR provides full support for hybrid message-passing and multi-threading parallelism. If
multi-threading is not desired, EleMRRR can be used in purely message-passing
mode. 

A thorough performance study comparing EleMRRR with ScaLAPACK's eigensolvers
on two high-end computing platforms is provided. 
This study accomplishes two objectives: First, it reveals that
the commonly\footnote{E.g., see
  \cite{dai2008,doi:10.1002/jcc.20549,kent2008,doi:10.1021/ct900539m,tomic2006}.}
used ScaLAPACK routines ({\tt PxHEGVX}, {\tt PxSYGVX}, {\tt PxHEEVD}, {\tt
  PxSYEVD}) present 
performance penalties 
that are avoided by calling a different
sequence of subroutines and by choosing suitable settings. 
Second, it indicates that EleMRRR is scalable -- both strongly and weakly
-- to a large number of processors, and outperforms the ScaLAPACK solvers
even when used according to our guidelines.

\subsection{PMRRR and its parallelization strategy}
\label{elemrrr:parallelstrategy}

\PMRRR\ is built on top of the multi-threaded \mrsmp\ introduced
in the previous section. Our parallelization strategy consists of two
layers, a global and a local
one. At the global level, the $k$ desired eigenpairs are
  statically divided into equal parts and assigned to the processes. 
  Since the unfolding of the
  algorithm depends on the spectrum, it is still possible that the workload
  is not perfectly balanced among the processes; this deficit is accepted in order
  to achieve the more important memory balancing.\footnote{See
    Section~\ref{sec:objectives}.} 
  At the local level (within each
  process), the computation is decomposed into tasks, which are equal to
  the ones introduced in Section~\ref{sec:mr3smp}. The tasks can be executed in
  parallel by multiple threads and lead to the
  dynamic generation of new tasks. The new feature is that some tasks
  involve explicit communication with other processes via messages. 

When executed with $p$ processes, the algorithm starts by
broadcasting the input matrix and by redundantly computing the root
representation $M_{root}$.\footnote{In the discussion, we assume the input
  is numerically irreducible. In fact, after splitting the matrix, a root
  representation is computed for each submatrix.} Once this is
available, the computation of the approximations $\hat{\lambda}_i[M_{root}]$
via bisection is embarrassingly parallel: Each process is
responsible for at most 
$k_{pp} = \lceil k/p \rceil$ eigenvalues.
Within a process however, to obtain workload balance, the assignment of
eigenvalues to threads is done dynamically. 

Once the eigenvalues are computed locally, each process gathers all
eigenvalues, and the corresponding eigenpairs are assigned  
as desired. At this point, each process redundantly classifies the
eigenvalues into singletons and clusters and only enqueues tasks that involve
eigenpairs assigned to the process. To increase workload balance, 
the first classification of eigenvalues is amended 
with a criterion based on the absolute
gaps of the eigenvalues~\cite{VoemelRefinedTree2007tr,Vomel:2010:ScaLAPACKsMRRR,Vomel:2012:FineGrain}.  

Locally, the calculation of the eigenpairs is split into computational
tasks of three types as in MR$^3$-SMP: S-tasks, C-tasks, and R-tasks. 
Multi-threading support within each process is easily obtained by
having multiple threads dequeue and execute tasks.   
In contrast to the multi-core parallelization, the
C-tasks must deal with two scenarios: ($i$) no 
communication is required to process the task, or ($ii$) communication with
other processes is required for its execution.
The computation associated with each of the three tasks is detailed below. 
  \begin{enumerate}[noitemsep,nolistsep]
    \item {\it S-task:} The corresponding
      eigenpairs are computed locally as in Algorithm~\ref{alg:stask}. No
      further communication among processes is necessary.
    \item {\it C-task:} When a cluster 
      contains eigenvalues assigned to only one process, no
      cooperation among processes  
      is needed. The necessary steps are the same as those in Algorithm~\ref{alg:ctask}.
      When a cluster contains a set of eigenvalues which spans multiple processes,
      inter-process communication is needed.
      In this case, all involved processes redundantly compute a new RRR,
      the local set of eigenvalues is refined, and the 
      eigenvalues of the cluster are communicated among the processes. At
      this point, the eigenvalues of the cluster are reclassified and the
      corresponding tasks created and enqueued. 
    \item {\it R-task:} Exactly as in MR$^3$-SMP, the task is
      used to split the work for refining a set of eigenvalue among
      threads. 
    \end{enumerate}

    In order to deal with the two different flavors of C-tasks,
    Algorithm~\ref{alg:ctask} is modified: for each C-task it checks
    whether communication with other processes is required. If communication
    is required, the refinement is limited to the portion of eigenvalues
    assigned to the process and a communication of refined eigenvalues among
    the involved processes is added. If no
    communication is required, the task corresponds to
    Algorithm~\ref{alg:ctask}.   
  
    While the overall execution time depends on the spectral distribution, the
    memory requirement is matrix independent (with $\mathcal{O}(nk/p)$ 
    floating point numbers per process), and perfect memory balance is
    achieved~\cite{Bientinesi:2005:PMR3,Vomel:2010:ScaLAPACKsMRRR}. To
    appreciate this feature, we mention that such a memory balance is not
    guaranteed by ScaLAPACK's bisection and inverse iteration
    implementation. In this case, clusters are processed within a single
    process, effectively leading (in the worst case) to a requirement of $\mathcal{O}(nk^2)$
    operations and $\mathcal{O}(nk)$ floating point numbers memory for a
    single process.

Our tests show that the hybrid parallelization approach is equally fast than
the one purely based on MPI. This is generally true for
architectures with a high degree of inter-node parallelism and limited
intra-node parallelism. 
By contrast, on architectures with a small degree
of inter-node parallelism and high degree of intra-node parallelism, we
expect the hybrid execution of  
\PMRRR\ to be preferable to pure MPI.

  We stress that even when no multi-threading is used, the task-based
  design of \PMRRR\ is advantageous: By scheduling tasks that require
  inter-process communication with priority and using non-blocking
  communication, processes continue executing tasks while waiting to
  receive data. This strategy often leads to a perfect overlap of
  communication and computation.
As an example, the scalability advantage of \PMRRR\ compared with
ScaLAPACK's {\tt PDSTEMR} in Fig.~\ref{fig:timetrdeigb} is the result of
non-blocking communications; in the 
experiment with $1{,}024$ cores, ScaLAPACK's MRRR spends 
about 30 out of 50 seconds in exposed communication.

\subsection{Elemental's eigensolvers}
\label{sec:elementalssolver}

Elemental's eigensolvers, {\tt HermitianGenDefiniteEig} and {\tt
  HermitianEig}, follow the classical reduction and 
backtransformation approach described in Sections~\ref{sec:HEP}
and \ref{sec:GHEPsixstages}.\footnote{Detailed discussions the reduction and
  backtransformation stages, both in general and within the Elemental
  environment, can be found
  in~\cite{StanleyDiss97,Hendrickson1999,Poulson:TwoSided,Sears:1998}.}
As Elemental's solvers are equivalent to their sequential
counterparts in terms of accuracy, we concentrate on their performance in
later experiments. In terms of memory, Elemental's solvers are quite efficient. In
Table~\ref{tab:memoryusgae}, we report approximate total memory
requirements for computing $k$ eigenpairs of generalized and standard
eigenproblems. As a reference, we provide the same numbers for the solvers
we later built from ScaLAPACK routines. (In
Section~\ref{sec:elemrrr:experiments}, we define the meaning of ``ScaLAPACK 
DC'' and ``ScaLAPACK MRRR'' precisely.)  The numbers in
the table are expressed in units of the size of a 
single complex and real floating point number, depending on the required
arithmetic. The memory requirement per process can be obtained by dividing
by the total number of processes.

\begin{table}[t]
\centering
\footnotesize
\begin{tabular}{ c c c c c} 
\toprule
    & \multicolumn{2}{c}{Complex}        & \multicolumn{2}{c}{Real} \\ 
    & GHEP             & HEP             & GHEP         &  HEP \\
\midrule
      ScaLAPACK DC    & $4 n^2$          & $3 n^2$  & $5 n^2$        & $4 n^2$  \\[1mm]
      ScaLAPACK MRRR  & $2 n^2 + 1.5 nk$ & $n^2 + 1.5nk$  & $2 n^2 + 2 nk$ &
      $n^2 + 2nk$  \\[1mm]
      Elemental         & $2 n^2 + nk$     & $n^2 + nk$  & $2 n^2 + nk$   & $n^2 + nk$ \\[1mm]
 \bottomrule
\end{tabular}
\caption{Total memory requirements in units of complex and real
  floating point numbers for the computation of $k$ eigenpairs. 
}
\label{tab:memoryusgae}
\end{table}

For performance reasons, as we discuss later, both Elemental and ScaLAPACK
are best executed with processes that are logically organized in a
two-dimensional grid. For square process grids, Elemental's memory
requirement is between $0.5n^2$ 
and $2n^2$ floating point numbers lower than that of the ScaLAPACK-based
solvers.
If non-square grids are
used, a user concerned about memory usage can make use of the non-square
reduction routines -- at the cost of suboptimal performance. The
  reduction routines for square process grids would otherwise need a data
  redistribution that adds a $n^2$ term to
Elemental and ScaLAPACK MRRR, but not to DC. However, in this
situation, ScaLAPACK
MRRR can save workspace to perform its redistribution of the eigenvectors
from a one-dimensional to a two-dimensional block-cyclic data layout,
reducing its terms by $nk$ real 
floating point numbers. This eigenvector redistribution is performed in
Elemental in-place, resulting in a smaller memory footprint than possible for
ScaLAPACK's routines. 

Subsequently, we call Elemental's eigensolvers based on \PMRRR\ briefly EleMRRR.
Before we show experimental results of EleMRRR, we make a little detour
discussing ScaLAPACK's solvers. The reason being that the solvers that are
included in the library each present
performance penalties, which are easily avoided. In order to have a fair
comparison, and to provide our findings to a large group of ScaLAPACK
users, we build eigensolvers from a sequences of ScaLAPACK
routines with optimal settings, which are generally faster than the commonly
used solvers.

\subsection{A study of ScaLAPACK's eigensolvers}
\label{sec:studyscalapackissues}

Highly parallel computers (e.g., {\sc Juropa}, which is used for the
experiments below and described in
Appendix~\ref{appendix:hardware})  are 
frequently used when the execution time and/or the memory 
requirement of a simulation become limiting factors. With respect to
execution time, the use of more processors would ideally result in 
faster solutions. When memory is the limiting factor, additional resources from large
distributed-memory environments should enable the solution of larger problems. We
study the performance of eigensolvers for both situations: increasing the number of
processors while keeping the problem size constant ({\it strong scaling}), and increasing
the number of processors while keeping the memory per processors constant
({\it weak scaling}). 

ScaLAPACK is a widely used library for the solution of large-scale dense 
eigenproblems on distributed-memory systems.
Version 1.8 of the library, at the time of performing the experiments the
latest release,  is used within this section.
Later versions, such as 2.0
(released in Nov.~2011), presents no significant changes in the tested
routines. From version 2.0 on, the MRRR-based routines for the HEP are added to
the library. The available routines for the GHEP and HEP and their
functionality are described in Appendix~\ref{appendix:routinenames}. 
Without loss of generality, we
concentrate on the case of double precision complex-valued
input.

\subsubsection{Generalized eigenproblems}

\begin{figure}[thb]
   \centering
   \subfigure[Execution time.]{
     \includegraphics[width=.47\textwidth]{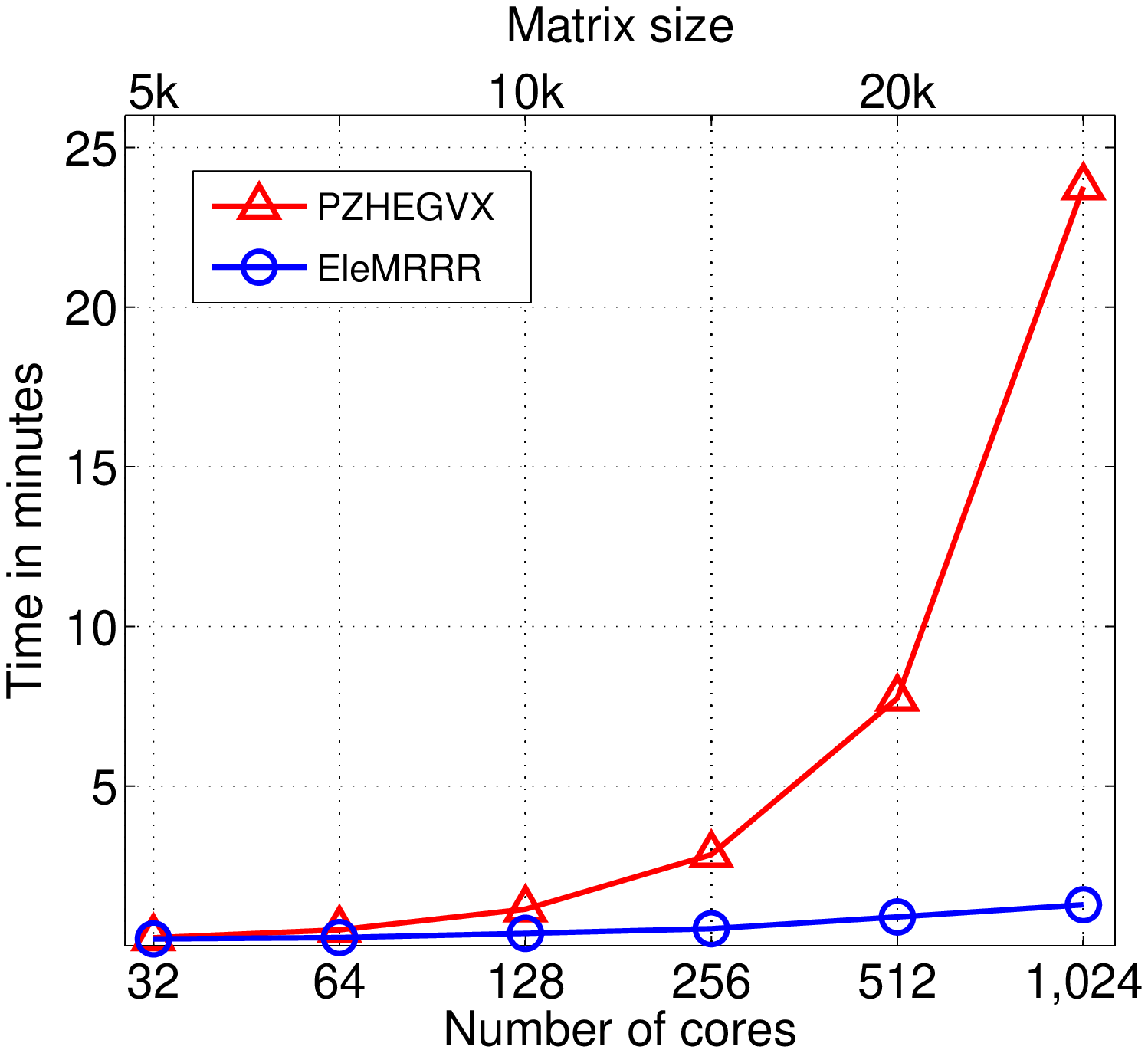}
     \label{fig:timepzhegvx1a}
   } \subfigure[Breakdown of time by stages.]{
     \includegraphics[width=.47\textwidth]{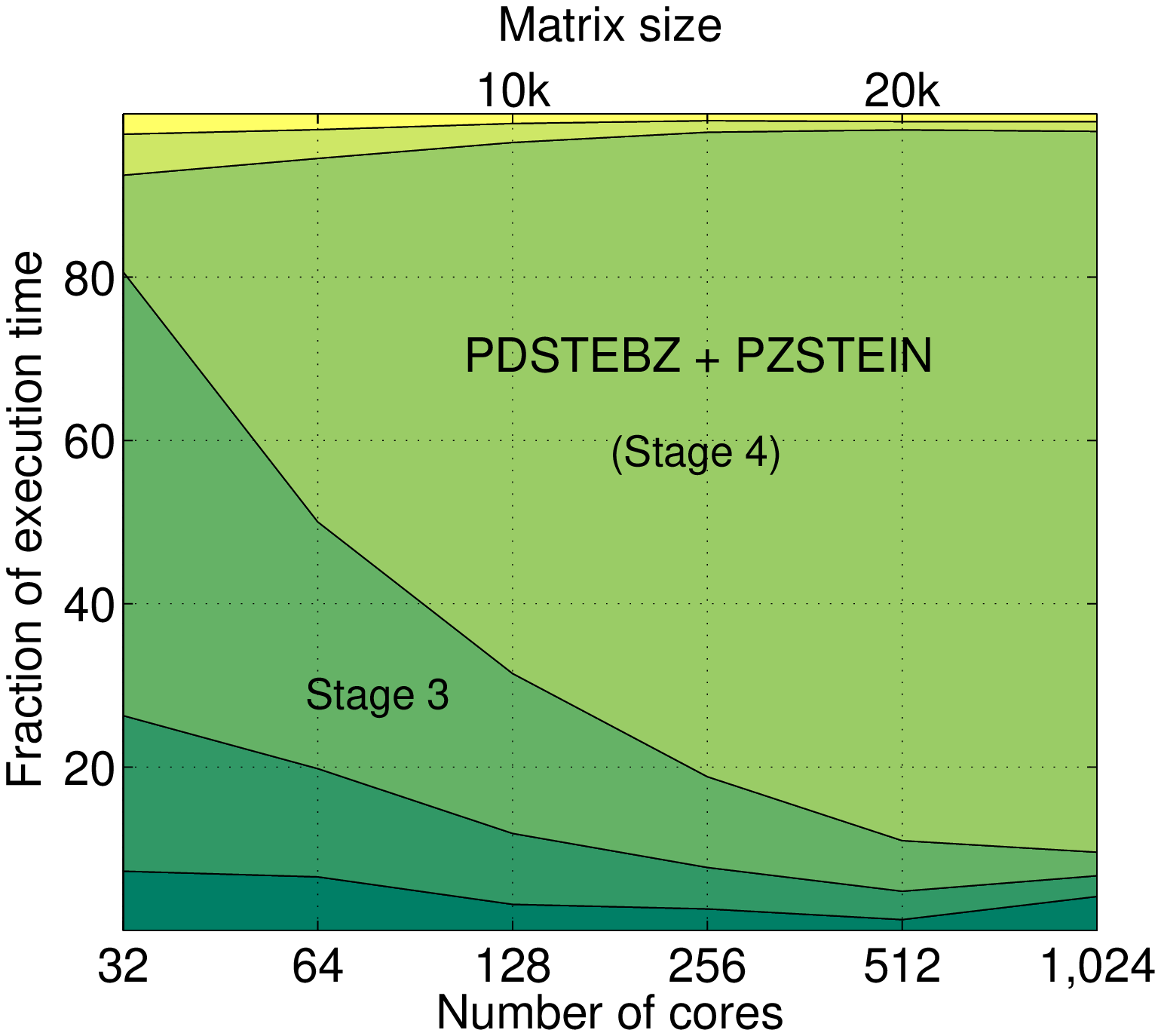}
     \label{fig:timepzhegvx1b}
   }
   \caption{
     Weak scalability for the computation of 15\% of the eigenpairs.
     As commonly done in practice, the
     eigenvectors are requested to be 
     numerically orthogonal~\cite{doi:10.1021/ct900539m}. 
   }
   \label{fig:timepzhegvx1}
\end{figure}

As presented in Table~\ref{tab:scalapackroutinesGHEP} of
Appendix~\ref{appendix:routinenames}, {\tt PZHEGVX} is
ScaLAPACK's only routine for Hermitian generalized eigenproblems. The routine
implements the six-stage procedure described in
Section~\ref{sec:GHEPsixstages} and is based on bisection and inverse
iteration. 
In Fig.~\ref{fig:timepzhegvx1a}, we report the weak
scalability of
 {\tt PZHEGVX} for computing 15\% of the eigenpairs of $Ax = \lambda
B x$ associated with
the smallest eigenvalues.\footnote{We used all default parameters.
  In particular, the parameter
  {\it orfac}, which indicates which eigenvectors should be orthogonalized
  during inverse iteration, has the default value $10^{-3}$. 
  In order to exploit ScaLAPACK's fastest reduction routines, the lower triangular part of the matrices is stored and referenced. 
}
This task frequently arises in electronic structure
calculations via {\it density functional theory}
(DFT);  some methods require between 1/6 and 1/3 of the
eigenpairs associated with the smallest eigenvalues of a large number
matrices whose dimensions can be in the tens of thousands~\cite{Sears:1998}.
Fig.~\ref{fig:timepzhegvx1a} indicates that as the
problem size and the number of processors increase, {\tt PZHEGVX} does
not scale as well as EleMRRR, which we included as a
reference. 

By Fig.~\ref{fig:timepzhegvx1b}, it is evident that the routines {\tt
  PDSTEBZ} and {\tt PZSTEIN}, which 
implement BI for the tridiagonal stage, are the main cause
for the poor performance of {\tt PZHEGVX}. 
For instance in the problem of
size $20{,}000$, these routines are responsible for 
almost 90\% of the compute time.  BI's poor performance is a well-understood
phenomenon (e.g., see the comments in~\cite{Choi19961} and 
Section~\ref{sec:existingmethods}), directly related to the effort necessary
to orthogonalize eigenvectors corresponding to 
clustered eigenvalues. This issue led to the development of MRRR, which avoids
the orthogonalization entirely. In addition to the performance issue, {\tt
  PZHEGVX} suffers from memory imbalances, as all the eigenvalues
belonging to a cluster are computed on a single processor. 

\begin{myguide}
    In light of the above considerations, the use of ScaLAPACK's
    routines based on bisection and inverse iteration is not
    recommended.
\end{myguide}

Consequently, we do not provide further
comparisons between EleMRRR and {\tt PZHEGVX} or {\tt PDSYGVX}.
Instead, we illustrate how the performance of
these drivers changes when BI for the tridiagonal
eigensolver is replaced with other -- faster -- methods, namely 
DC and MRRR.  

\subsubsection{Standard eigenproblems}
\label{sec:scalapacksHEP}

Among the solvers available in ScaLAPACK, only {\tt PZHEEVX} (BI) and 
{\tt PZHEEVR} (MRRR) offer the possibility of computing a subset of
eigenpairs. {\tt PZHEEVX} is widely used, even though, as highlighted in
the previous section, it is highly non-scalable. 
Similarly, if eigenvectors are computed, the QR algorithm is known to be
slower than DC for large
problems~\cite{Bientinesi:2005:PMR3}.
Therefore, we omit 
comparisons with routines that are based on the QR algorithm. However, as
discussed Section~\ref{sec:HEP}, QR requires significantly less memory than DC. 

\begin{myguide}
  Provided enough memory is available, ScaLAPACK's DC is preferable over QR.
\end{myguide}

\begin{figure}[thb]
   \centering
   \subfigure[Execution time.]{
     \includegraphics[width=.47\textwidth]{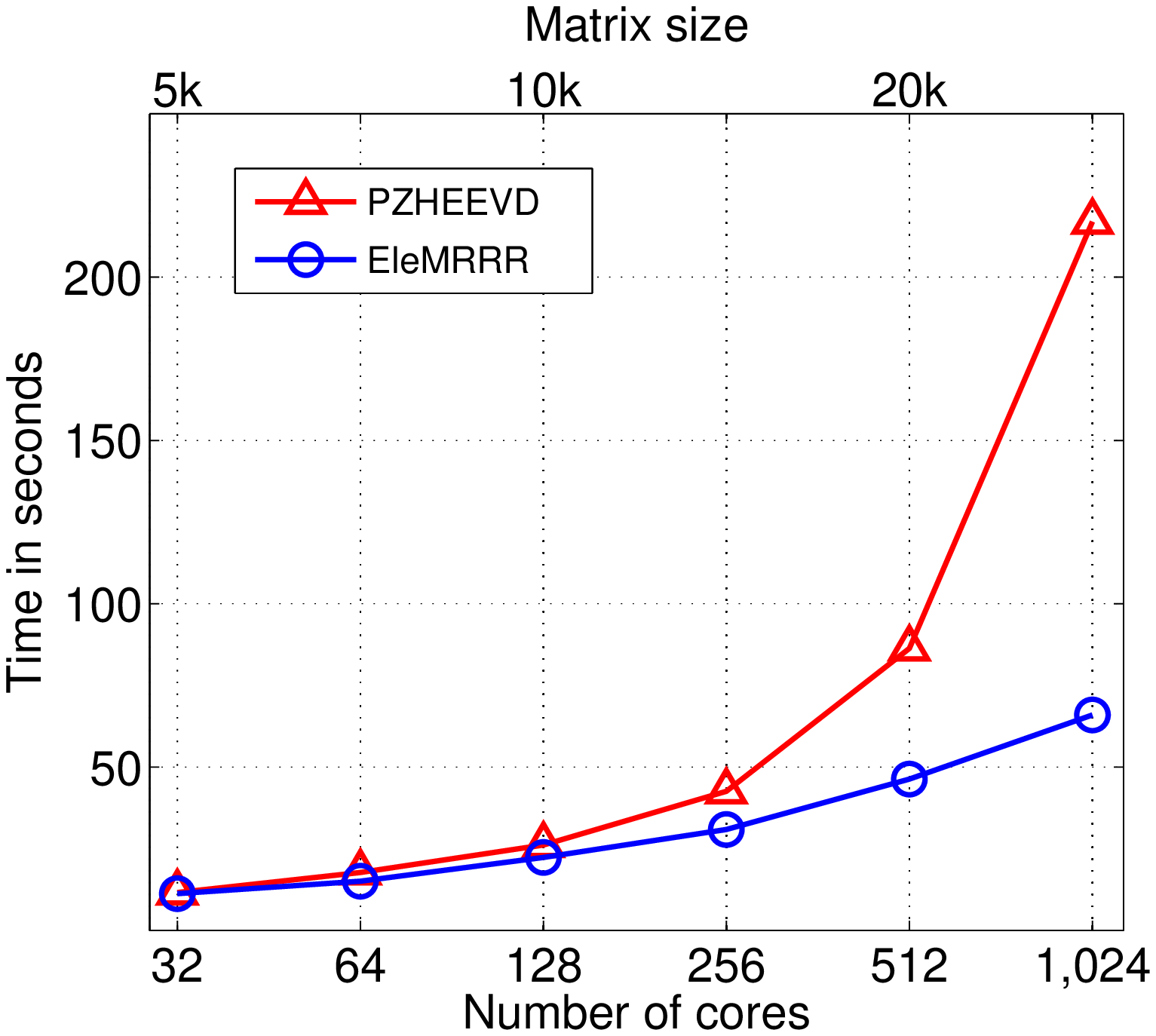}
     \label{fig:timepzheevd2a}
   } \subfigure[Execution time of the first stage.]{
     \includegraphics[width=.47\textwidth]{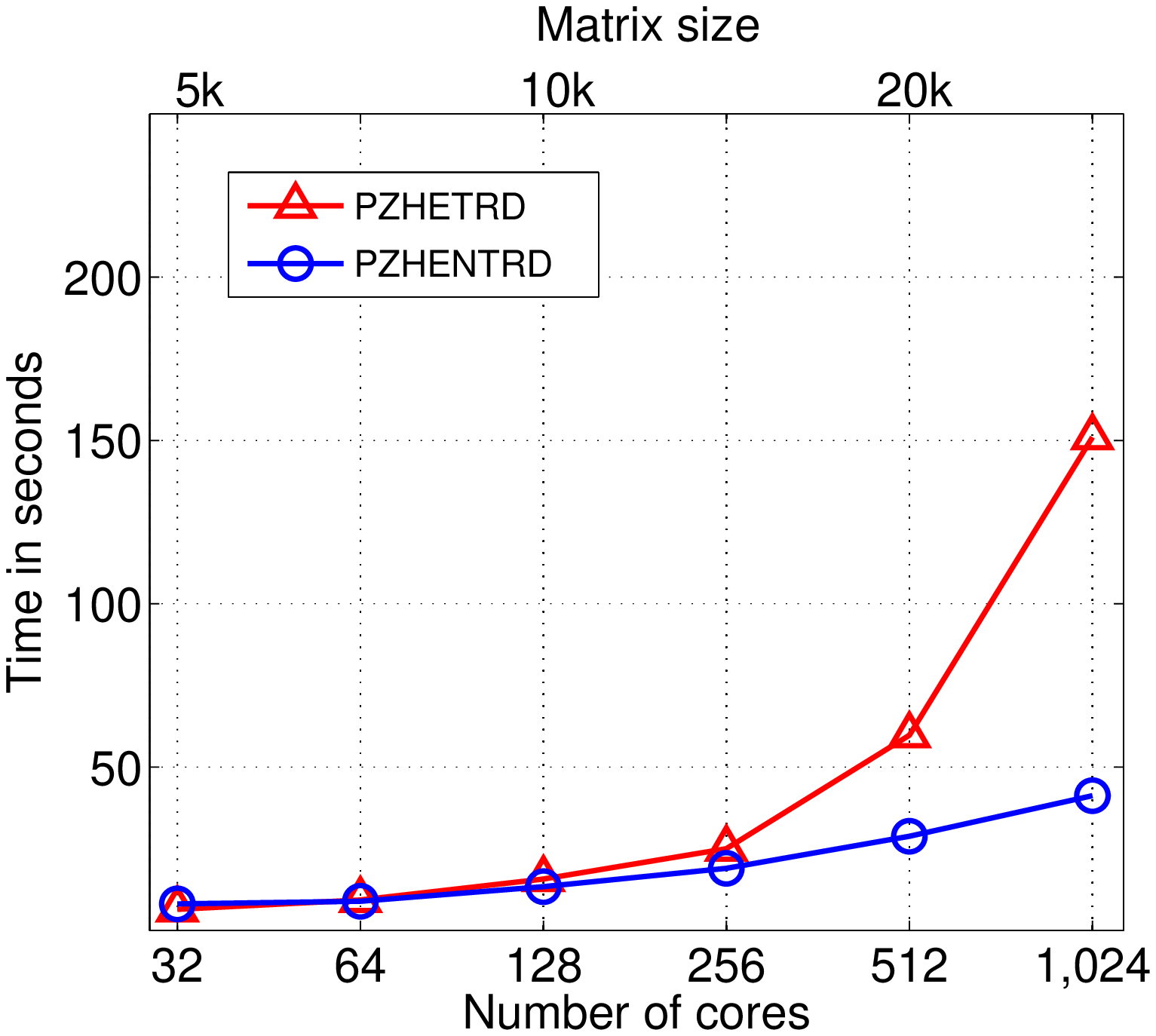}
     \label{fig:timepzheevd2b}
   }
   \caption{
     Weak scalability for the computation of all eigenpairs using DC.
     (a)  Total execution time of {\tt PZHEEVD} and EleMRRR.
     (b) Execution time for ScaLAPACK's routines {\tt PZHETRD} and
     {\tt PZHENTRD}, which are responsible for the reduction to tridiagonal
     form. The former,
     used within the routine {\tt PZHEEVD}, 
     causes a performance penalty and accounts for much of the time
     difference compared with EleMRRR.
   }
   \label{fig:timepzheevd2}
\end{figure}

We now focus on the weak scalability of {\tt
  PZHEEVD}, which uses DC for the tridiagonal eigenproblem. 
For an experiment similar to that of
Fig.~\ref{fig:timepzhegvx1}, we show the results for {\tt PZHEEVD} and
EleMRRR in Fig.~\ref{fig:timepzheevd2a}.
Note that all eigenpairs were computed, since {\tt PZHEEVD} does not allow
for subset computation.
While BI might dominate the
run time of the entire eigenproblem, DC required less
than 10\% of the total execution time. 
Instead, as the matrix size increases, the reduction to tridiagonal
form ({\tt PZHETRD}) becomes the performance bottleneck, requiring
up to 70\% of the total time. A comparison of
Fig.~\ref{fig:timepzheevd2a} and Fig.~\ref{fig:timepzheevd2b} 
reveals that, for large problems, using {\tt PZHETRD} for the reduction to
tridiagonal form is more time consuming than the complete solution with
EleMRRR. 

ScaLAPACK also includes {\tt PZHENTRD}, a routine for the reduction to
tridiagonal form especially optimized for square processor grids. The performance
improvement with respect to {\tt PZHETRD} can be 
so dramatic that, for this stage, it is
  preferable
to limit the computation to a square number of
processors and redistribute the matrix
accordingly~\cite{Hendrickson1999}.\footnote{See also
  the results in Appendix~\ref{appendix:resultsjugene}.} 
Provided enough workspace is made available,  a necessary redistribution is 
automatically done within {\tt
  PZHENTRD}. 
In any case, it is important to note
that the performance benefit of {\tt PZHENTRD} is only exploited if the
lower triangle of the input matrix is stored,
otherwise the slower routine, {\tt PZHETRD}, is invoked.\footnote{Similar
  considerations apply for the reduction to  
  standard form via the routines {\tt PZHEGST} and {\tt PZHENGST},
  see~\cite{Poulson:TwoSided}.}  

\begin{myguide}
ScaLAPACK's reduction routines 
optimized for square grids of processors are to be preferred 
    over the regular reduction 
  routines, even when non-square process grids are used;
  moreover, only the lower triangle of implicitly Hermitian matrices should
  be referenced. 
\end{myguide}

For performance and scalability reasons, we
 use the reduction routines that are optimized for square grids to build the
 fastest solver within the ScaLAPACK framework.

\subsubsection{Tridiagonal eigenproblems}

\begin{figure}[thb]
   \centering
   \subfigure[1--2--1 type.]{
     \includegraphics[width=.47\textwidth]{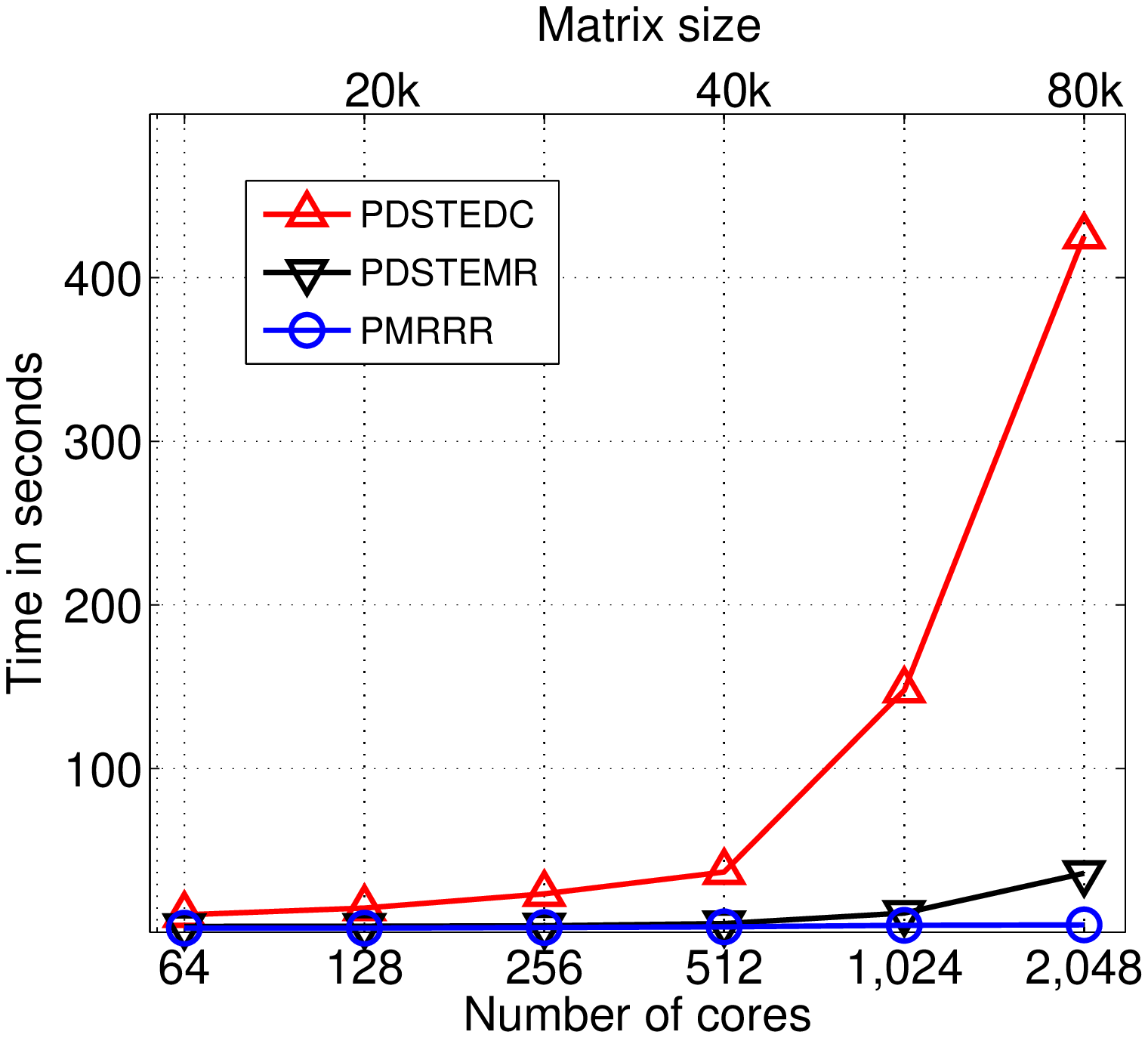} 
     \label{fig:timetrdeiga}
   } \subfigure[Wilkinson type.]{
     \includegraphics[width=.47\textwidth]{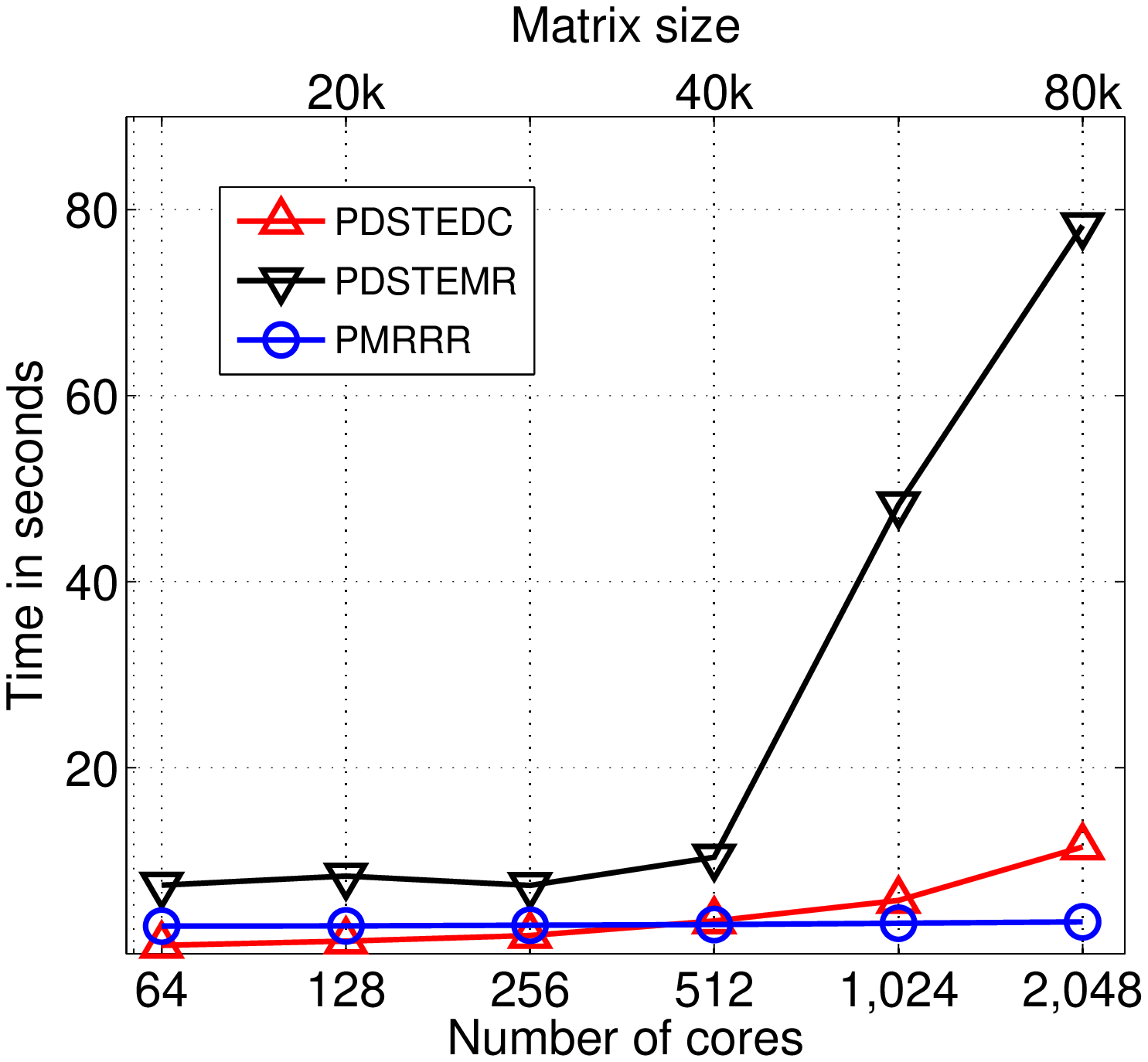}
     \label{fig:timetrdeigb}
   }
   \caption{
     Weak scalability for the computation of all eigenpairs of 
     two different test matrix types. The left and right graphs have
     different scales. 
     The execution time of the MRRR routines should remain roughly constant
     as the number of cores is increased. 
     In contrast to the results reported
     in~\cite{Vomel:2010:ScaLAPACKsMRRR}, where a similar experiment to 
     comparing {\tt PDSTEDC} and {\tt
     PDSTEMR} is performed,  
     even when the matrices offer an opportunity for heavy deflation,
     eventually our \PMRRR\ becomes faster than DC due to its superior
      scalability. 
   }
   \label{fig:timetrdeig}
\end{figure}

In contrast to all other stages of generalized and
standard eigenproblems, the number 
of arithmetic operations of the tridiagonal eigensolver depends on the input
data. When all eigenpairs are desired, depending on the  
matrix entries, either DC or MRRR may be faster. 
Fig.~\ref{fig:timetrdeig} provides an example of how performance is
influenced by the input data. 
We already justified why we
only consider DC and MRRR in our experiments. DC is implemented as
{\tt PDSTEDC}~\cite{Tisseur:1999:PDC}. The MRRR routine
corresponds to the tridiagonal stage of {\tt
  PDSYEVR}~\cite{Vomel:2010:ScaLAPACKsMRRR} and it is not available 
as a separate routine; nonetheless, we call it {\tt PDSTEMR}
subsequently. Both routines are compared on two types of test matrices: 
``1--2--1'' and ``Wilkinson''. Due to deflation, the Wilkinson matrices are
known to strongly favor the DC~\cite{Marques:2008}. 
For both matrix types, the ScaLAPACK codes do
not scale to a
large number of cores and \PMRRR\ eventually becomes the fastest solver. 

\begin{myguide}
ScaLAPACK's tridiagonal eigensolvers based on DC and MRRR are generally 
    fast and reasonably scalable; depending on the target architecture and
    specific requirements from the application, either one may be
    used. Specifically, if only a small 
subset of the spectrum has to be computed, 
in terms of performance, the MRRR-based solver is to be
preferred to DC. In terms of accuracy, DC is to be preferred.
\end{myguide}

\begin{figure}[thb]
   \centering
   \subfigure[Breakdown of time by stages.]{
     \includegraphics[width=.47\textwidth]{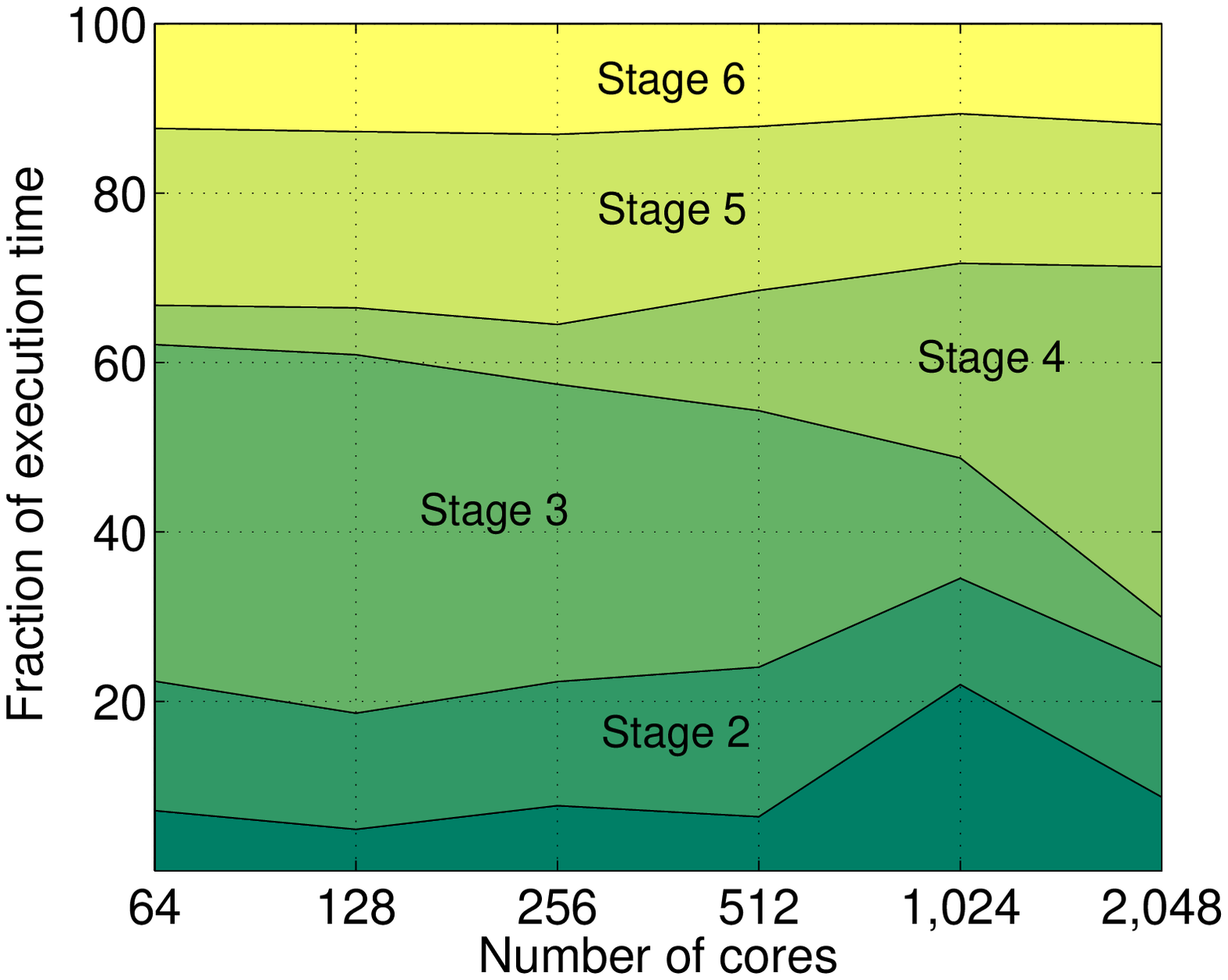} 
     \label{fig:strngscaldcfraca}
   } \subfigure[Parallel efficiency as in \eqref{eq:paralleleffstrng}.]{
     \includegraphics[width=.47\textwidth]{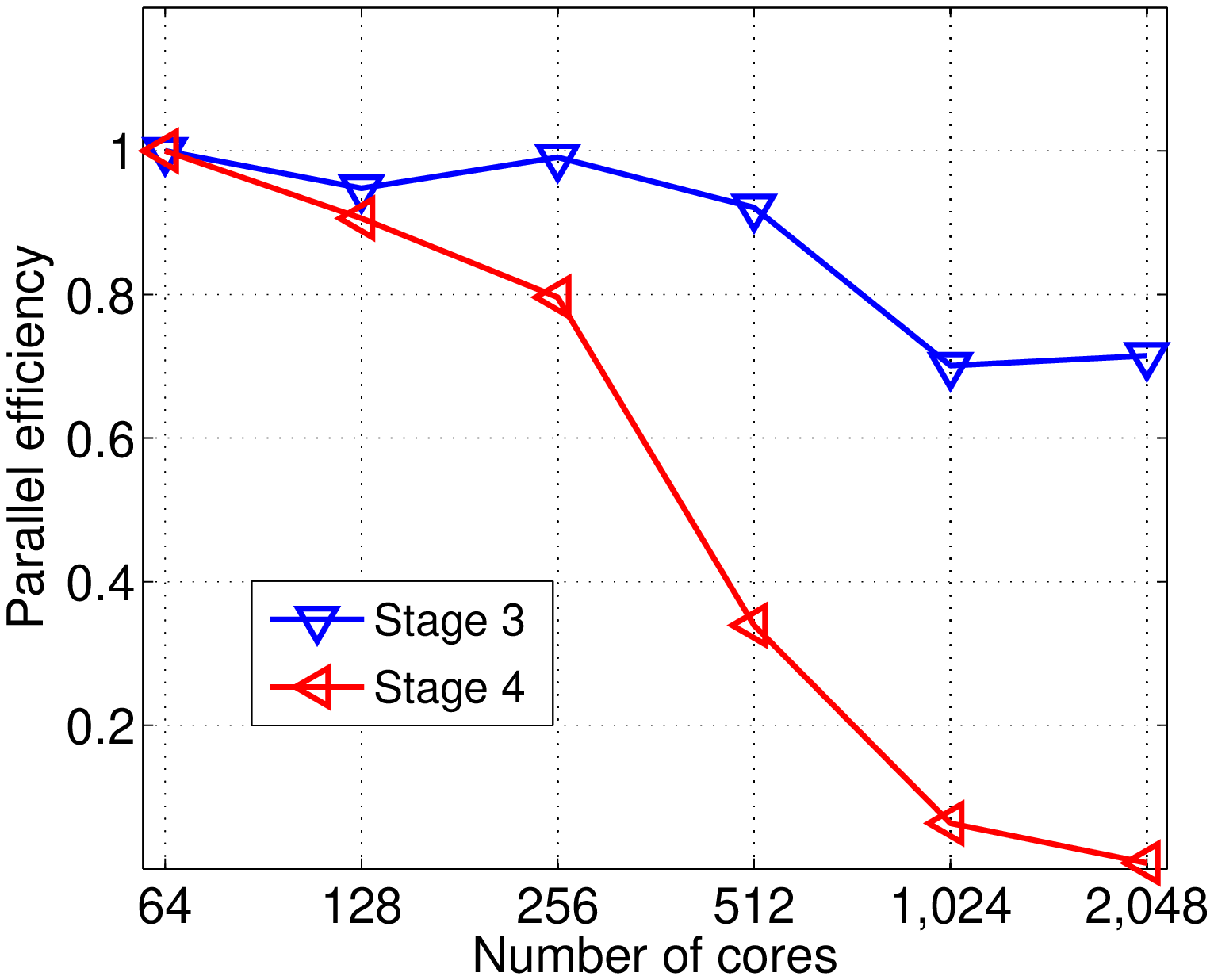}
     \label{fig:strngscaldcfracb}
   }
   \caption{
     Scalability of the computation of all eigenpairs for a GHEP of size
     20{,}000 using ScaLAPACK's DC.
     The details of the experiment can be found in~\cite{EleMRRR}.
   }
   \label{fig:strngscaldcfrac}
\end{figure}

Later, we include experimental data on generalized
eigenproblems for both DC and MRRR.  
One of the challenges in building a scalable solver is that every stage must be scalable.  
This is illustrated in
Fig.~\ref{fig:strngscaldcfrac}, which shows the results for ScaLAPACK's DC
for a GHEP of size $20{,}000$.  
While the reduction to tridiagonal form
and the tridiagonal eigensolver
are respectively the most and the least expensive stages on $64$ cores,
on $2{,}048$ cores the situation is reversed.
The behavior is explained by the parallel efficiencies shown in Fig.~\ref{fig:strngscaldcfracb}. 

The ScaLAPACK experiments demonstrate that a comparison with the commonly
used routines would be misleading. In fact, by calling the suitable set of
routines, we can build much
faster solvers within the ScaLAPACK framework. We use these fast solvers to compare EleMRRR with. However, we
stress that what we call ``ScaLAPACK's DC'' and ``ScaLAPACK's MRRR'' do
not correspond to routines in ScaLAPACK and are not
(yet) frequently used in practice.

\subsection{Experimental results}
\label{sec:elemrrr:experiments}

We present experimental results for the execution
 on two state-of-art supercomputers at the
Research Center J\"ulich, Germany: {\sc Juropa} and {\sc Jugene}. In this
section we concentrate on {\sc Juropa}; results on {\sc Jugene} are
similar and outsourced to Appendix~\ref{appendix:resultsjugene}. We
limit ourselves to generalized eigenproblems of the form $Ax=\lambda Bx$; the
results for standard eigenproblems are given implicitly by stages three to five.

All tested routines were compiled using the {\it Intel compilers}
(ver.~11.1) with the flag {\tt -O3} and linked to the {\it
  ParTec's ParaStation MPI}  
library 
(ver.~5.0.23).\footnote{Version 5.0.24 was used when support for
  multi-threading was needed.} 
Generally, we used a two-dimensional
process grid $P_r \times P_c$ (number of rows $\times$ number of columns)
with $P_r = P_c$ whenever possible, and 
$P_c = 2P_r$ otherwise.\footnote{As discussed in Sections~\ref{sec:scalapacksHEP} and
  \ref{sec:elementalssolver}, $P_c
  \approx P_r$ or the 
  largest square grid possible should be
  preferred. These choices do
  not affect the qualitative behavior of our performance results.}
If not
stated otherwise, one process per core was employed.

The ScaLAPACK library (ver.~1.8) was used in conjunction with
Intel's MKL BLAS (ver.~10.2).
From extensive testing, we identified that in
all cases the optimal block size was close to 32;
therefore, we carried out the ScaLAPACK experiments only with
block sizes of 16, 32, 48; the best result
out of this pool is then reported.
Since no driver for the generalized eigenproblem that makes use of DC is
available,   
we refer to ScaLAPACK's DC as the following sequence of routines: 
{\tt PZPOTRF}--{\tt PZHENGST}--{\tt PZHENTRD}--{\tt PDSTEDC}--{\tt
  PZUNMTR}--{\tt PZTRSM}. 
Similarly, ScaLAPACK's MRRR corresponds to the same 
sequence with {\tt PDSTEDC} replaced by {\tt PDSTEMR}.\footnote{As {\tt
    PDSTEMR} is not contained in ScaLAPACK, it corresponds to the sequence {\tt PZPOTRF}--{\tt PZHENGST}--{\tt PZHEEVR}--{\tt PZTRSM}.} 
We do not use routines {\tt PZHEGST} and {\tt  PZHETRD}
for the reduction to standard and tridiagonal form, respectively.  
Instead, we replaced them (when necessary) by the faster  
{\tt PZHENGST} and {\tt PZHENTRD}.
Furthermore, in order to make
use the fast reduction routines, only the lower 
triangular part of the matrices is referenced, and enough memory for a possible
redistribution of the data is provided. 

Elemental (ver.~0.6) -- incorporating \PMRRR\ (ver.~0.6) -- was used for the
EleMRRR timings. In general, since Elemental does not tie 
the algorithmic block size to the distribution block size, 
different block sizes could be used for each of the stages. 
We do not exploit this fact in the reported timings. 
Instead the same block size is used for all stages.  
A block size of around 96 was optimal in all cases, therefore, 
experiments were carried out for block sizes of 64, 96, and
128, but only the best timings are reported.\footnote{
The block size for matrix vector products were fixed to 32 in all cases. 
For the biggest matrices in the weak scaling experiment only the block 
size of 32 and 96 were used for ScaLAPACK and EleMRRR,
respectively.}   

Since the timings of the tridiagonal eigenproblem depend on the input data,
so does the 
overall solver. In order to compare fairly different solvers, we fixed
the spectral distribution: for $1 \leq k \leq n$, $\lambda_k = 2-2 \cos(
     \pi k / (n + 1))$.
The
performance of every other stage is data independent.
Moreover, since the output of the tridiagonal solvers has to be in a
format suitable for the backtransformation, the MRRR-based routines
have to undergo a data redistribution;  in all the experiments, the timings
for Stage 4 include the cost of such a redistribution.

\subsubsection{Strong Scaling}
\label{sec:strngscal}

\begin{figure}[h!tb]
   \centering
   \subfigure[Execution time.]{
     \includegraphics[width=.47\textwidth]{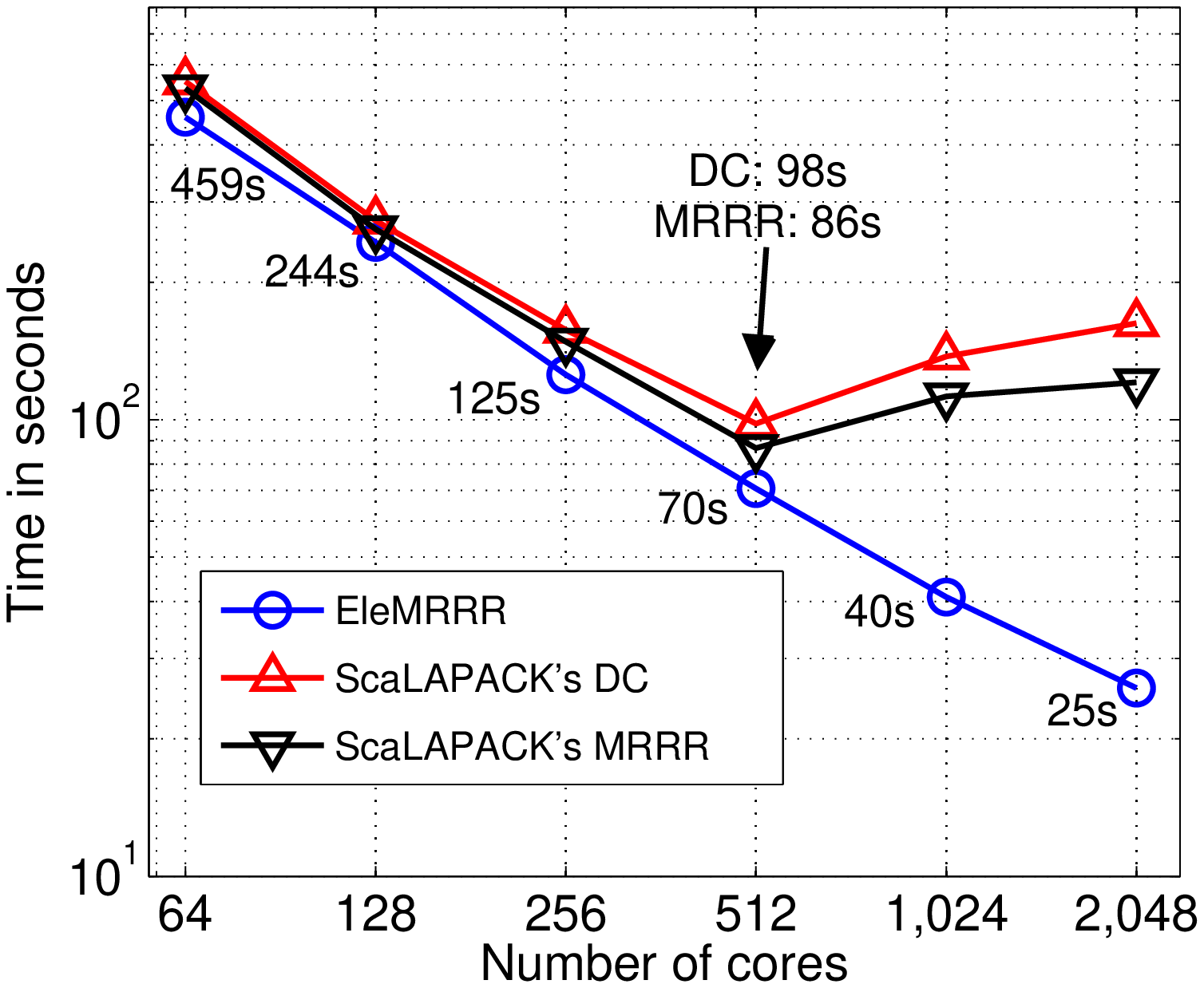}
     \label{fig:time_juropa_strngscala}
   } \subfigure[Parallel efficiency.]{
     \includegraphics[width=.47\textwidth]{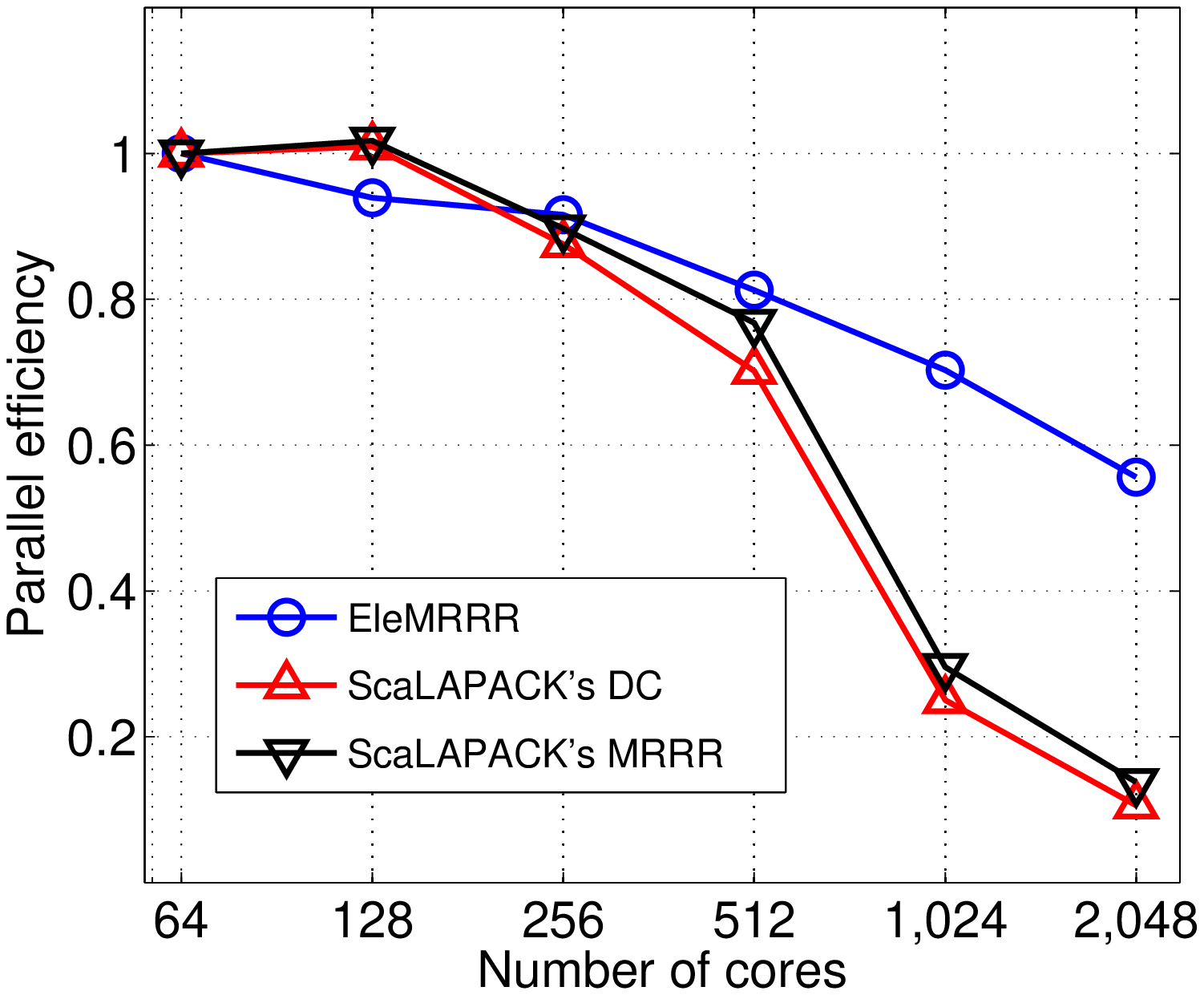}
     \label{fig:time_juropa_strngscalb}
   }
   \subfigure[Breakdown of time by stages.]{
     \includegraphics[width=.47\textwidth]{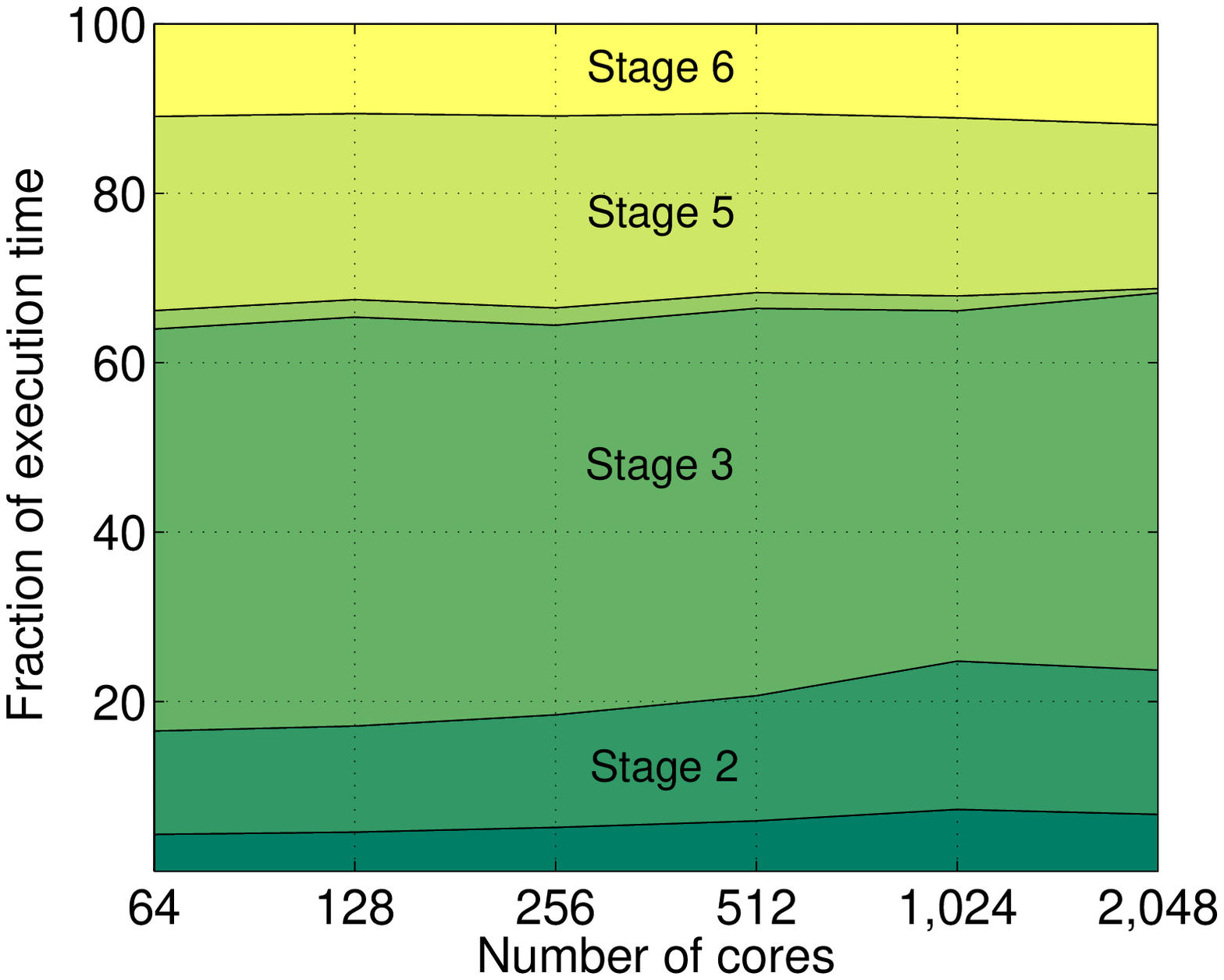} 
     \label{fig:frac_juropa_strngscala}
   } \subfigure[Parallel efficiency.]{
     \includegraphics[width=.47\textwidth]{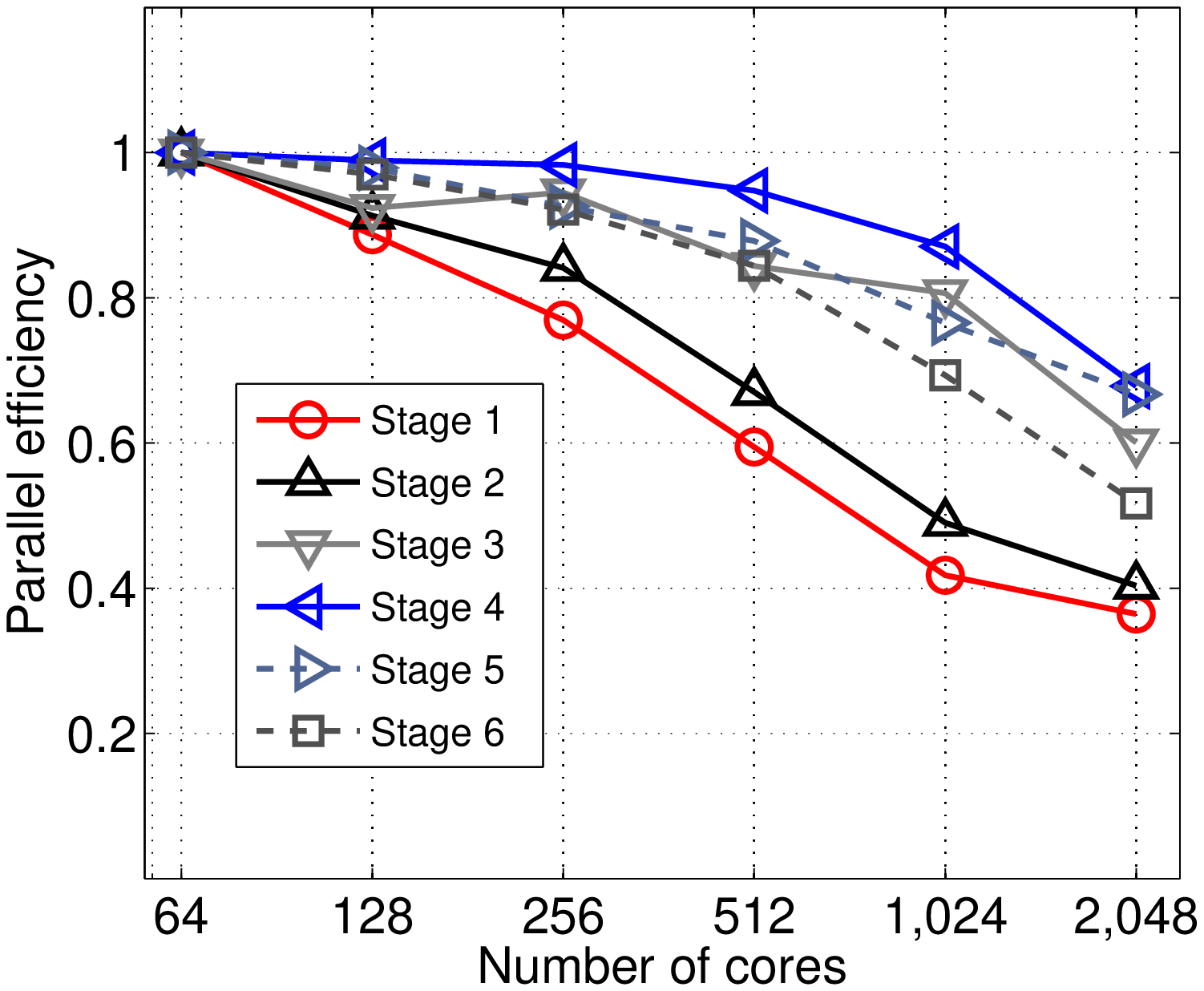}
     \label{fig:frac_juropa_strngscalb}
   }
   \caption{
     Strong scalability for the computation of all eigenpairs, $Ax = \lambda
     B x$.  
     Matrices $A$ and $B$ are of size $20{,}000$. 
     (a) Total execution time in a log-log scale. (b) Parallel
     efficiency as defined in~\eqref{eq:paralleleffstrng}; normalized to the
     execution using 64 cores. 
     (c) Fraction of time
     spent in all six stages of the computation. The time spent in the last three
     stages is proportional to the number of eigenpairs computed. (d) Parallel efficiency
     for all six stages of the computation. 
   }
   \label{fig:time_juropa_strngscal}
\end{figure}

In Fig.~\ref{fig:time_juropa_strngscala}, we present timings of EleMRRR for fixed problem
of size $20{,}000$.\footnote{We did not investigate the cause for the 
  increased 
  run time of ScaLAPACK using $1{,}024$ and $2{,}048$ cores. We observed
  that while most subroutines in the
  sequence are slower compared with the run time using 512 cores, {\tt PZHENTRD}
  scales well up to the tested $2{,}048$ cores -- see also
  Fig.~\ref{fig:timepzheevd2b}.} 
Fig.~\ref{fig:time_juropa_strngscalb} shows the parallel efficiency as defined
in~\eqref{eq:paralleleffstrng}; the reference is the execution on 64 cores.

Once the proper sequence of routines is rectified, the performance of
ScaLAPACK is comparable to that of EleMRRR, up to 512 
cores (see Fig.~\ref{fig:time_juropa_strngscal}).
For 64 to 512 cores, ScaLAPACK's DC is about 10\% to 40\% slower than EleMRRR, while
ScaLAPACK's MRRR is about 7\% to 20\% slower.  The advantage of EleMRRR mainly
comes from the Stages 1, 2, and 6, i.e., those related to the generalized
eigenproblem.  The
timings for the standard problem (Stages 3--5) are nearly identical for all solvers,
with DC slightly slower than both MRRR-based solutions.

The story for $1{,}024$ and $2{,}048$ cores changes; the performance of
ScaLAPACK's routines for the generalized eigenproblem drops dramatically.
Compared to DC, 
EleMRRR is about 3.3 and 6.3 times faster; with respect to ScaLAPACK's MRRR
instead, EleMRRR is 2.7 and 4.7 times faster.  The same holds for the
standard eigenproblem, where EleMRRR is about 2.9 and 6.2 times faster than
DC and 1.9 and 3.7 times faster than MRRR.

In Figs.~\ref{fig:frac_juropa_strngscala} and
~\ref{fig:frac_juropa_strngscalb},  we take a closer look at the six 
different stages of EleMRRR. Fig.~\ref{fig:frac_juropa_strngscala} tells us that
roughly one third of EleMRRR's execution time -- corresponding to Stages
4, 5, and 
6 -- is proportional to the fraction of computed eigenpairs. Computing a
small fraction of eigenpairs would therefore require about two thirds of
computing the complete decomposition. 
In Fig.~\ref{fig:frac_juropa_strngscalb}, we report
the parallel efficiency for all six stages separately.  
When analyzed in conjunction with the left figure, this figure indicates if
and when a routine becomes a bottleneck due to bad scaling.

The most time consuming stages
     are the reduction to tridiagonal form, the reduction to standard form,
     and the first backtransformation; the
     tridiagonal eigensolver (Stage 4) is negligible. Stage 4 attains
the highest parallel efficiency and 
contributes for less than 2.2\% of the overall run time. 

Up to $1{,}024$ cores, ScaLAPACK's MRRR shows a similar behavior: 
the tridiagonal stage
makes up for less than 6\% of the execution time. With $2{,}048$ cores instead, the
percentage increases to 21. 
The situation is even more severe for DC, as the fraction spent in the
tridiagonal 
stage increases from about 4.5\% with $64$ cores to 41\% with $2{,}048$
cores [Fig.~\ref{fig:strngscaldcfraca}]. 
The experiment illustrates that the tridiagonal stage,
unless as scalable as the other stages,
will eventually account for a
significant portion of the execution time.

\subsubsection{Weak Scaling}

\begin{figure}[hbt]
   \centering
   \subfigure[Execution time.]{
     \includegraphics[width=.47\textwidth]{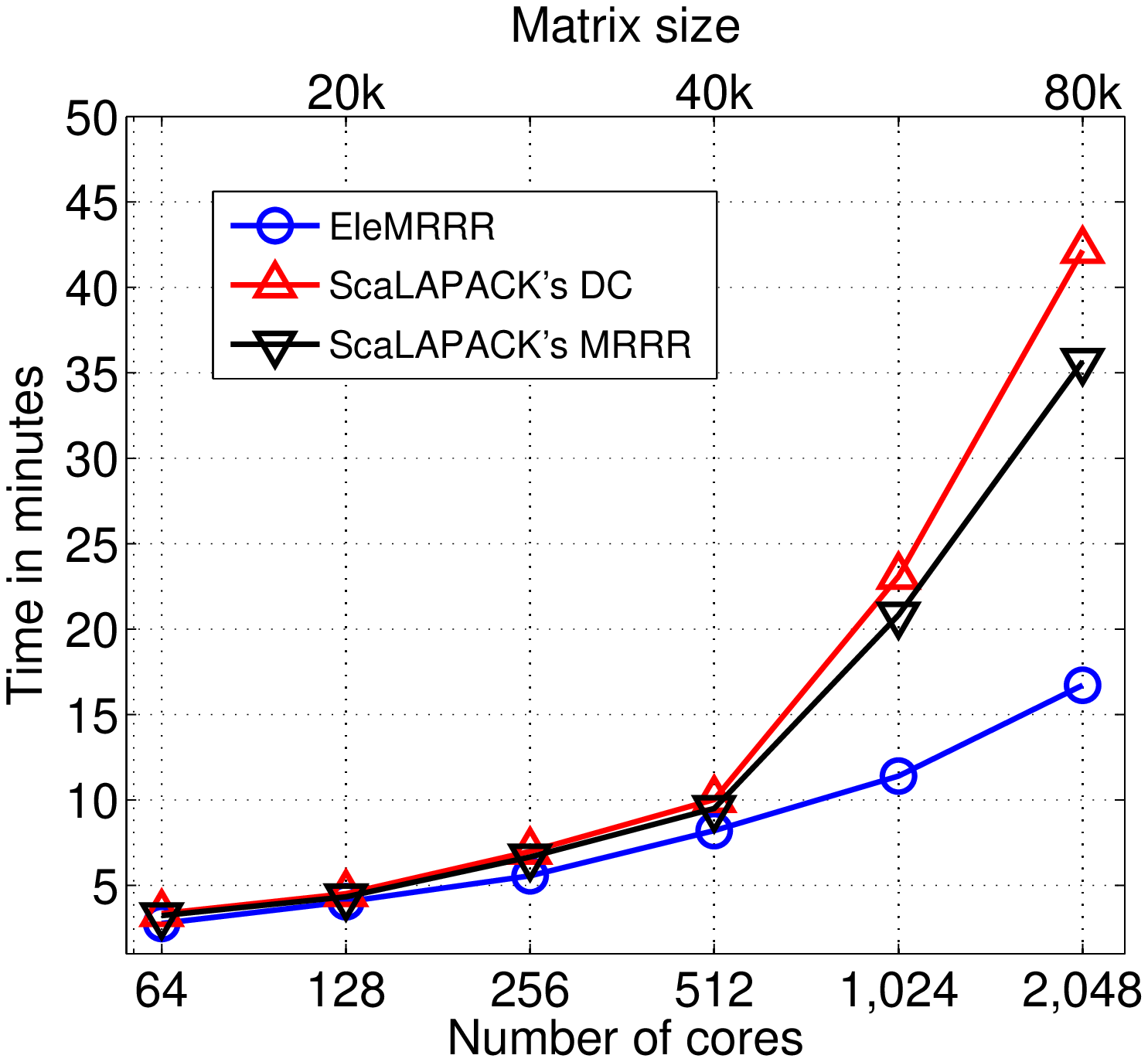} 
     \label{fig:time_juropa_weakscala}
   } \subfigure[Parallel efficiency.]{
     \includegraphics[width=.47\textwidth]{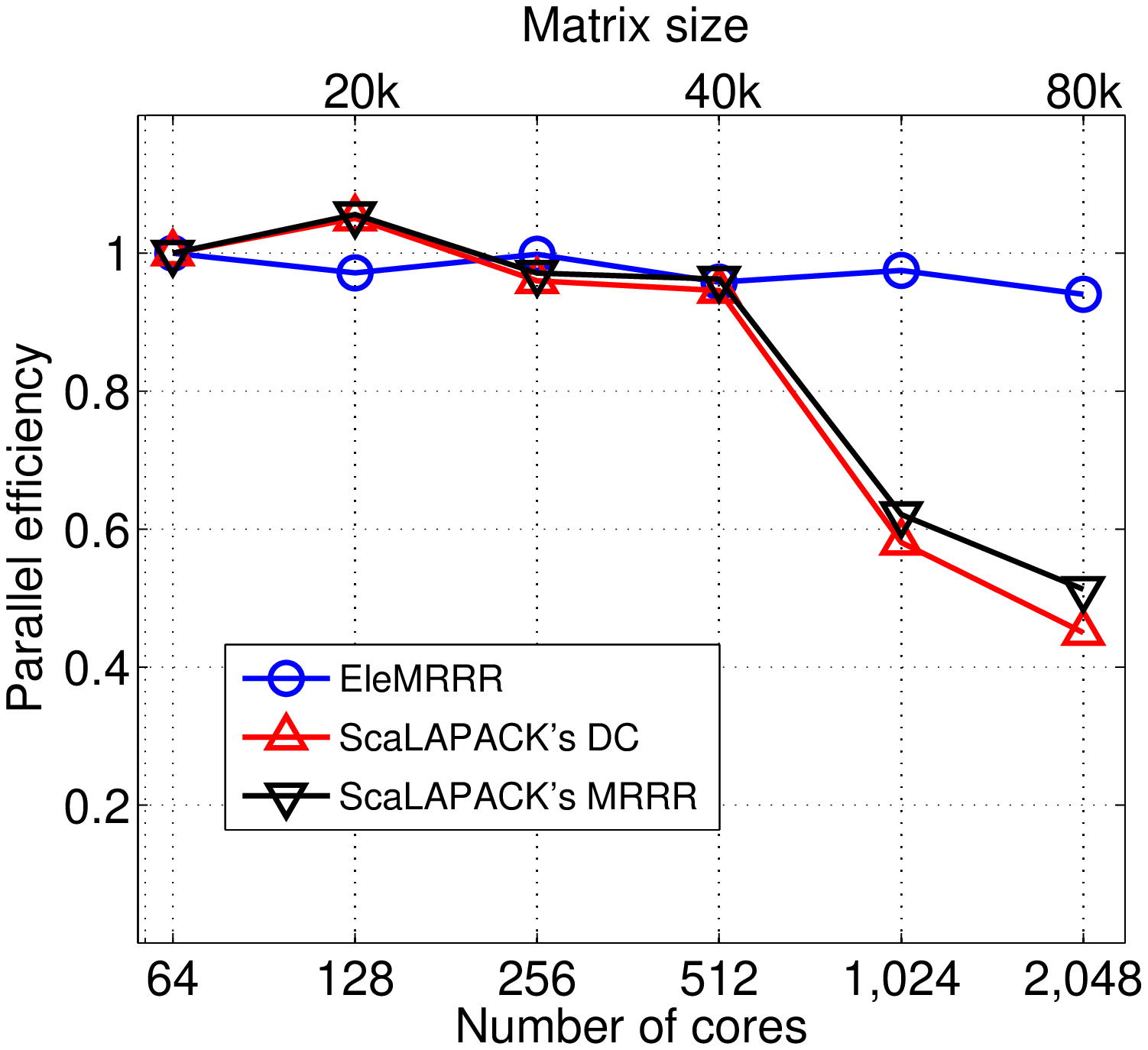} 
     \label{fig:time_juropa_weakscalb}
   }
   \caption{
     Weak scalability for the computation of all eigenpairs, $Ax = \lambda B
     x$. Matrices $A$
     and $B$ are varied in size such that the memory requirement per core remains constant. 
}
   \label{fig:time_juropa_weakscal}
\end{figure}

Fig.~\ref{fig:time_juropa_weakscal} shows EleMRRR's timings, when 
the matrix size increases 
(from $14{,}142$ to $80{,}000$)
 together with the number of cores (from $64$ to $2{,}048$).
 Fig.~\ref{fig:time_juropa_weakscala} contains the parallel
efficiency as defined in~\eqref{eq:paralleleffweak}, where the
reference configuration is the same as in the previous section (the
reference matrix size is $14{,}142$). 

In the tests using 512 cores and less, EleMRRR outperforms ScaLAPACK
only by a small margin, while using $1{,}024$ cores and more, the difference
becomes significant. 
The right graph indicates that EleMRRR scales well to large problem sizes
and high number of processes, with parallel efficiency close to one. 
Thanks to its better scalability, for the biggest problem, EleMRRR is 2.1
and 2.5 times faster than ScaLAPACK's MRRR and DC, respectively.

\begin{figure}[hbt]
   \centering
   \subfigure[Breakdown of time by stages.]{
     \includegraphics[width=.47\textwidth]{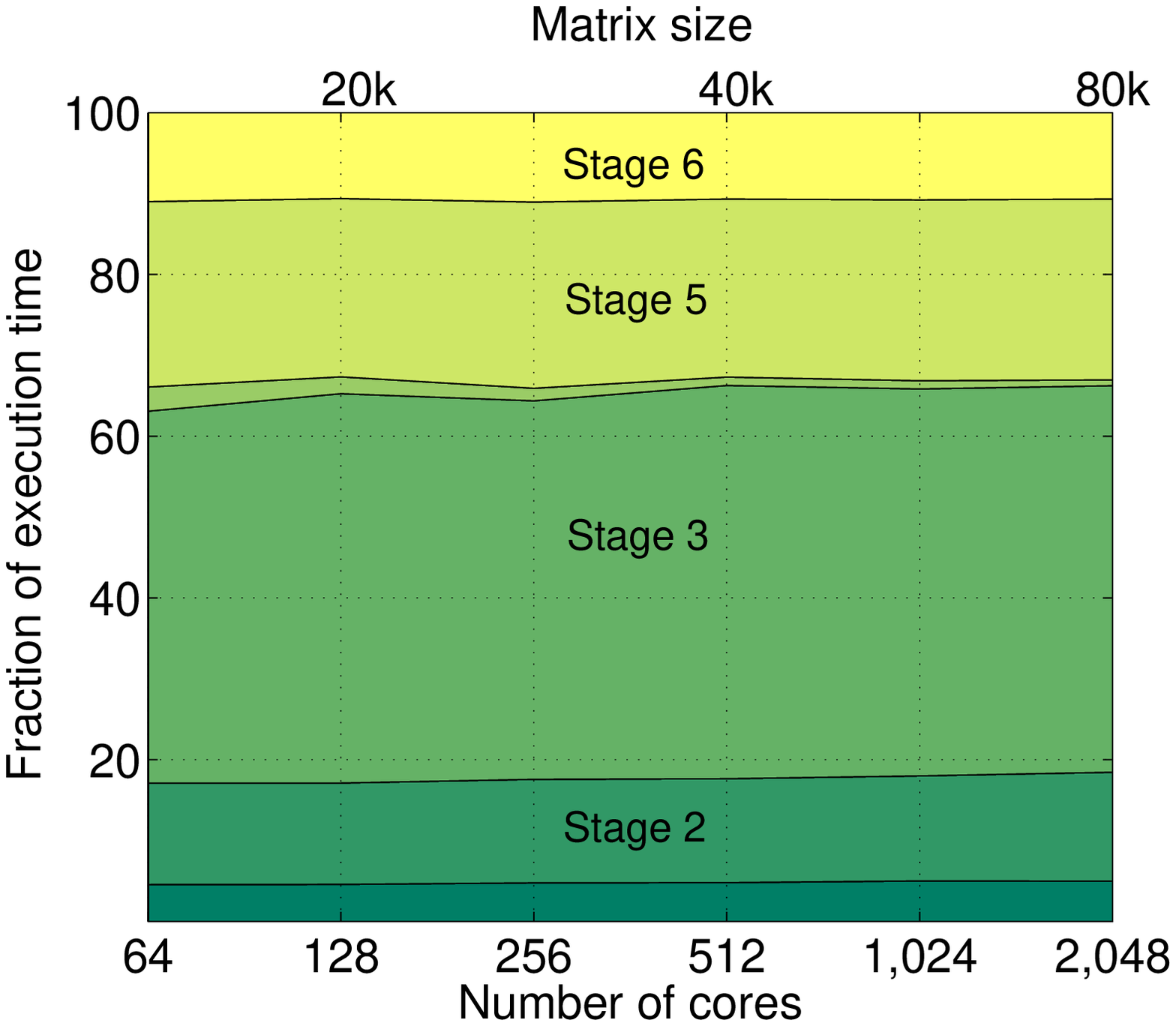} 
     \label{fig:frac_juropa_weakscala}
   } \subfigure[Hybrid mode execution.]{
     \includegraphics[width=.47\textwidth]{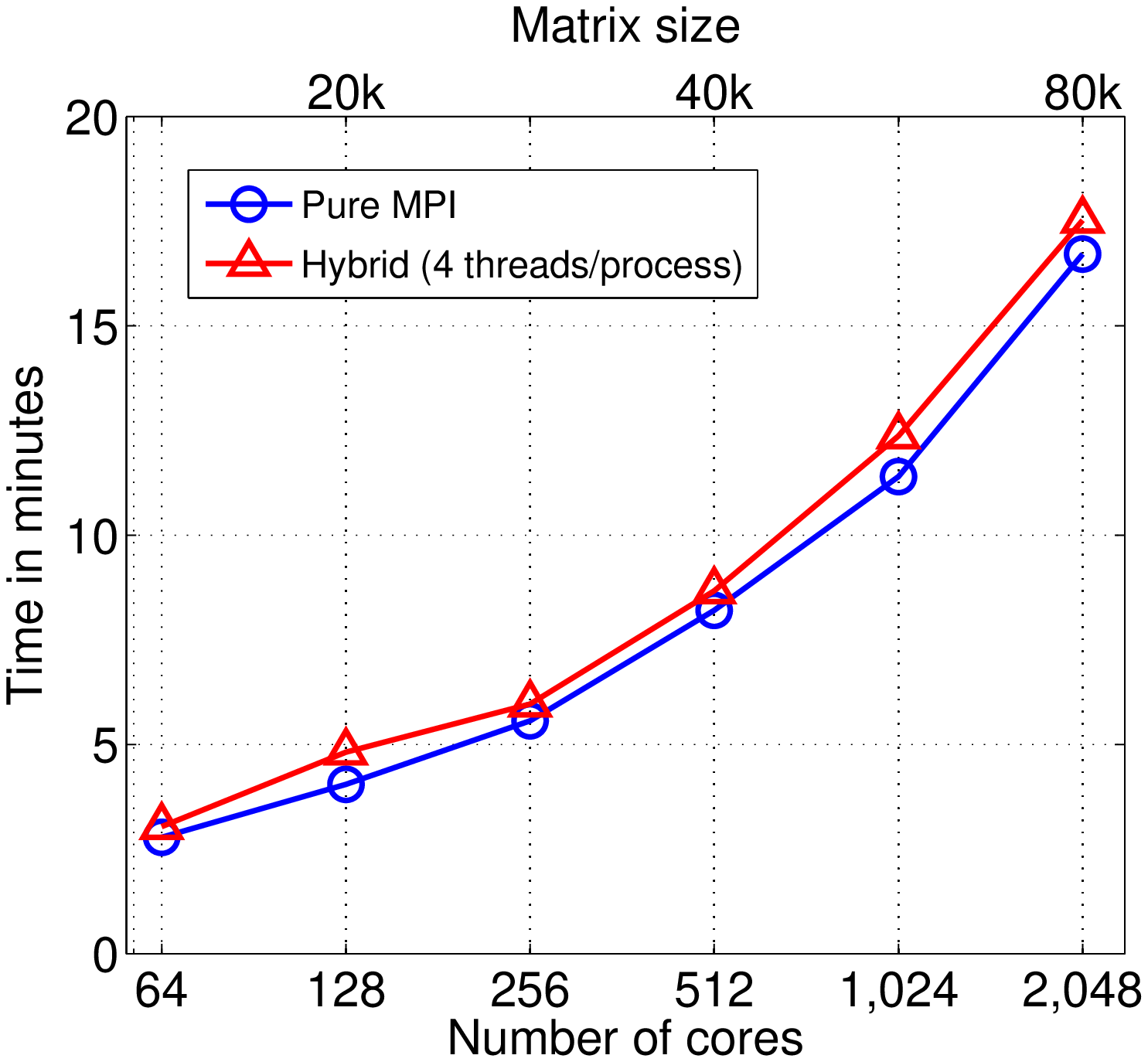} 
     \label{fig:frac_juropa_weakscalb}
   }
   \caption{
     EleMRRR's weak scalability for the computation of all eigenpairs. 
     (a) Fraction of the execution time spent in the six stages, 
     from bottom to top. 
     (b) Comparison between a pure MPI execution and a hybrid execution
     using one process per socket with four threads per process.
   }
   \label{fig:frac_juropa_weakscal}
\end{figure}

The execution time is broken down into stages in 
Fig.~\ref{fig:frac_juropa_weakscala}.
Four comments follow: 
(1) The time spent in \PMRRR\ (Stage 4) is in the range of 
2.5\% to 0.7\% and it is completely negligible, especially for 
large problem sizes. The timings relative to only Stage 4 are detailed in
Fig.~\ref{fig:timetrdeiga}. 
(2) The
timings corresponding to the standard eigenproblem (Stages 3--5)
account for about 72\% of the 
generalized problem's execution time. 
(3) The part of the solver whose execution time is roughly proportional to the
fraction of desired eigenpairs (Stages 4--6)
makes up 32\%--37\% of the 
execution for both the GHEP and HEP. 
(4) No one stage in EleMRRR becomes a performance bottleneck, as all of them 
scale equally well.

Fig.~\ref{fig:frac_juropa_weakscalb} shows the
execution of EleMRRR using one process per socket with four threads per process. 
The resulting execution time is roughly the same as for the pure MPI execution, 
highlighting the potential of Elemental's hybrid mode. Further experimental
results can be found in Appendix~\ref{appendix:resultsjugene}.

\subsection{Remaining limitations}
\label{sec:remaininglimitations}

\begin{figure}[htb]
   \centering
   \subfigure[Execution time.]{
     \includegraphics[width=.47\textwidth]{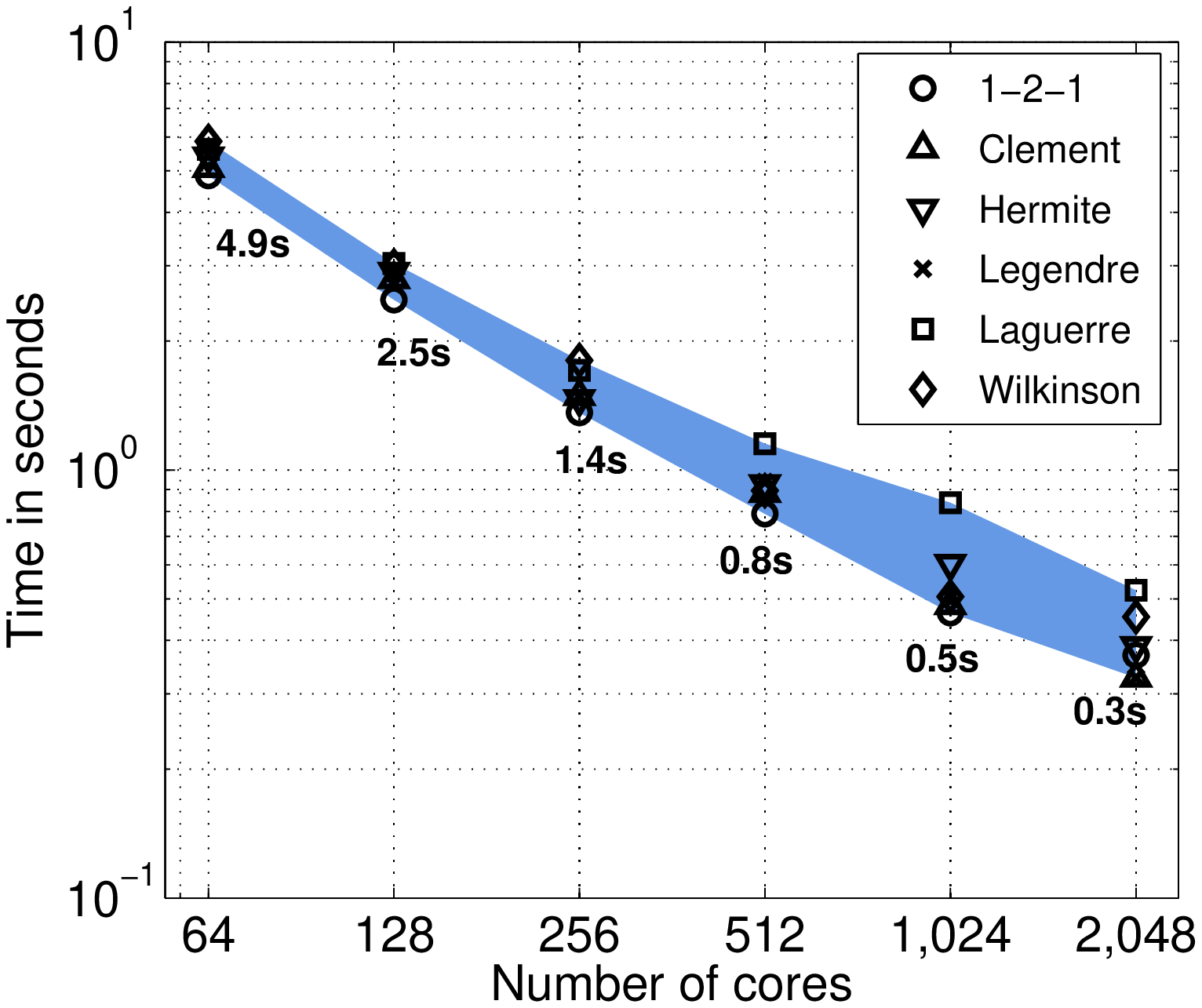} 
     \label{fig:timetrdeigstrng2a}
   } \subfigure[Parallel efficiency.]{
     \includegraphics[width=.47\textwidth]{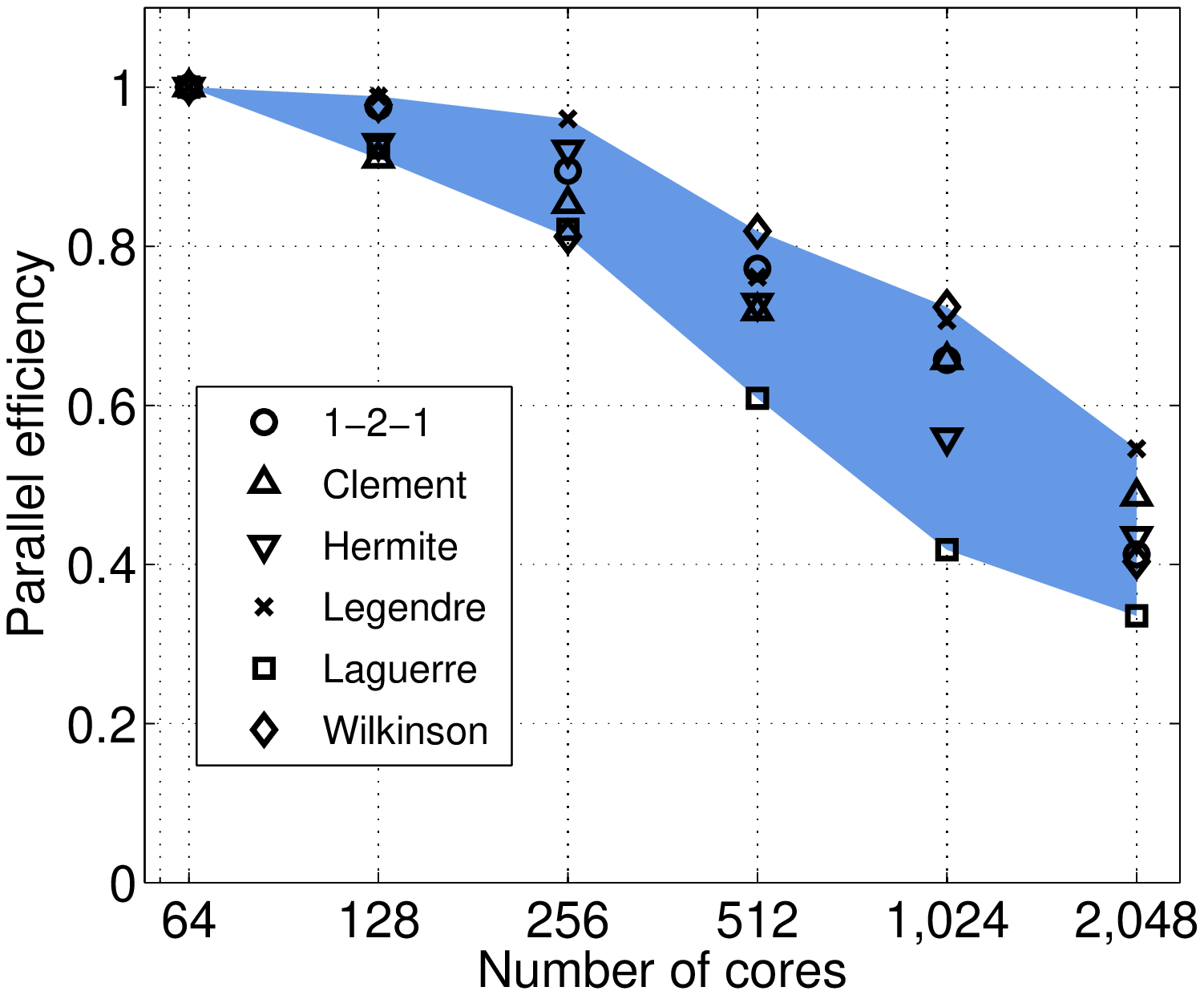}
     \label{fig:timetrdeigstrng2b}
   }
   \caption{
     Strong scalability for the computation of all eigenpairs. 
   }
   \label{fig:timetrdeigstrng2}
\end{figure}

We have shown that, in context of direct methods for generalized and standard Hermitian
eigenproblems, the tridiagonal stage is often negligible. 
However, because
of the lower complexity of MRRR, inefficiencies in the tridiagonal stage do
not become visible. 
For \PMRRR, there are three major issues remaining:
\begin{itemize}[noitemsep,nolistsep]
\item Although mostly negligible in terms of execution time, \PMRRR\ is
  the primary source of ``loss of orthogonality'' in the overall solution
  (see Section~\ref{section:densemixedexperiments}).
\item When matrices possess large clusters of eigenvalues, the work is
  increased, load balance issues arise, and communication among processes is
  introduced. As a consequence, parallel scalability is limited (see
  experiments below).  
\item For some inputs, accuracy is not guaranteed as one or more
  representation is accepted without passing the test for its requirements
  (see Line~\ref{line:justaccept} of Algorithm~\ref{alg:spectrumshift}). 
\end{itemize}
While the last point is discussed in the next chapter, at this point, we
illustrate the remaining performance and accuracy issues of \PMRRR.  
As the
performance is matrix depended, we use a set of artificial matrices for our
experiment. In Fig.~\ref{fig:timetrdeigstrng2}, we show strong
scaling results for matrices of size 20{,}001. The experiment corresponds
to the one displayed in Fig.~\ref{fig:time_juropa_strngscal}, but
differs in two respects. First, we used
a hybrid execution with one process per node and eight threads per process;
the pure MPI results however are very similar. Second, while in the dense case the
redistribution from one-dimensional to two-dimensional matrix layout is
considered part of the tridiagonal stage, here it is excluded from the timings. 

For all test matrices, the execution time of \PMRRR\ is
negligible in the dense case. For instance, EleMRRR timings for the generalized problem
are 459 seconds and 25 seconds on 64 cores and 2{,}048 cores,
respectively. Also, \PMRRR's parallel efficiency is in the same ballpark as for
the other stages of the dense problem. 
Although timings and scalability are sufficient in context of dense
problems, the parallel efficiency drops to only about 0.5 on 2{,}048
cores. Consequently, if we further increase the number of cores, we cannot
expect an adequate reduction in run time.  

\begin{figure}[thb]
   \centering
   \subfigure[Execution time.]{
     \includegraphics[width=.47\textwidth]{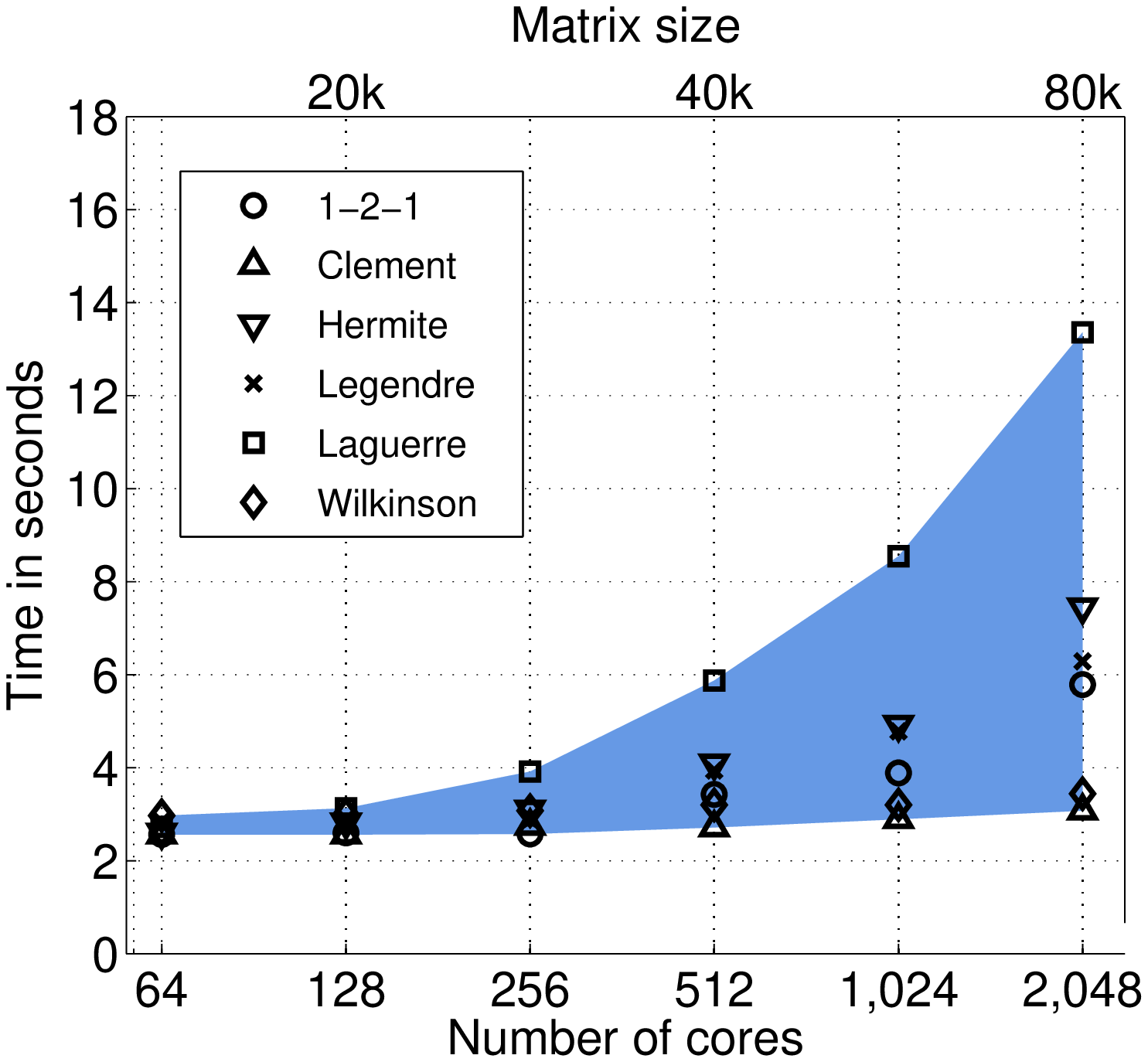} 
     \label{fig:timepmrrrweaka}
   } \subfigure[Parallel efficiency.]{
     \includegraphics[width=.47\textwidth]{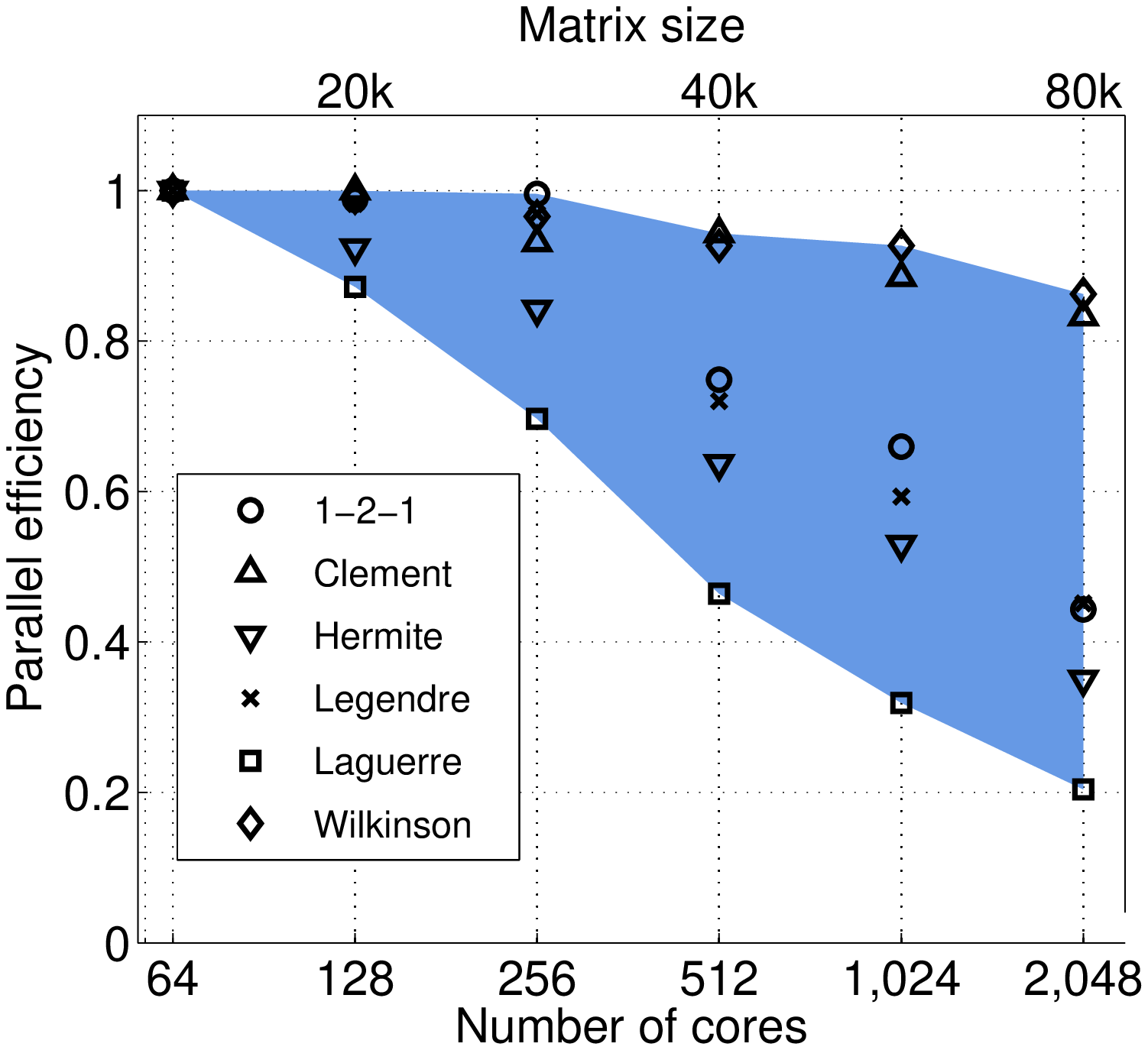}
     \label{fig:timepmrrrweakb}
   }
   \caption{
     Weak scalability for the computation of all eigenpairs. 
   }
   \label{fig:timepmrrrweak}
\end{figure}

In Fig.~\ref{fig:timepmrrrweak}, we detail the weak scaling results of
Figs.~\ref{fig:timetrdeig}, \ref{fig:time_juropa_weakscal} and
\ref{fig:frac_juropa_weakscal}. We
use a hybrid execution mode, but the results for pure MPI are very similar.  
For 1-2-1 and Wilkinson type matrices,
Fig.~\ref{fig:timetrdeig} shows similar results, but with a different scale. 
According to Figs.~\ref{fig:time_juropa_weakscal} and \ref{fig:frac_juropa_weakscal}, for the 
largest problem of size $80{,}000$ on 2{,}048 cores, Elemental requires for the generalized
eigenproblem about 1000 seconds and for the standard eigenproblem about 700
seconds. For all the test matrices, \PMRRR\ contributes less than 15 
seconds to the execution time; and this only if {\it all} eigenpairs
are computed. However, the parallel efficiency drops dramatically for some
matrices, while it behaves well for others. Ideally, the
execution time would remain constant as the number of cores are increased. This
is roughly observed for the Clement and Wilkinson matrices. For the Laguerre
matrices however, the parallel efficiency clearly degenerates.  

The loss in efficiency is directly related to the
clustering of the eigenvalues. To formalize clustering, we define {\it clustering}
$\rho \in [1/n, 1]$ to be the largest encountered cluster divided by the
matrix size.
For two extreme case, we show clustering and parallel efficiency in
Table~\ref{tab:loadbalance}. 
High clustering has two negative
impacts: (1) The overall work is increased by $\order{\rho n^2}$ flops. 
As clustering tends to increase for large matrices, in practice, MRRR does not quite
performs work proportional to $n^2$~\cite{perf09}; (2) The static
assignment of eigenpairs to processes leads to workload imbalance.  
\begin{table}[htb]
\centering
\footnotesize
\begin{tabular}{ l c@{\quad\quad} c c c c c} \toprule
 Metric   & Matrix & \multicolumn{5}{c}{Matrix size}  \\ 
\cmidrule(r){3-7}
    & & $20{,}001$             & $28{,}285$         &  $40{,}001$ & $56{,}569$ & $80{,}001$ \\
\midrule
 Clustering   &  Wilkinson & $1.5e{-4}$ &
 $1.1e{-4}$ & $7.5e{-5}$ &  $5.3e{-5}$ & $3.8e{-5}$ \\
     & Laguerre   & $0.46$  & $0.49$ & $0.51$ & $0.52$ & $0.53$  \\
   \midrule
 Efficiency   &  Wilkinson  & $0.98$  & $0.96$ & $0.94$ & $0.91$ & $0.88$  \\
    &  Laguerre    & $0.87$  & $0.70$ & $0.46$ & $0.32$ & $0.20$  \\
      \bottomrule 
\end{tabular}
\caption{Clustering and parallel efficiency for the two
  extreme types of test matrices. If $\rho$ is close to $1/n$, the parallel
  efficiency remains good.   
}
\label{tab:loadbalance}
\end{table}

The available {\it parallelism} is conservatively approximated by
$\rho^{-1}$. The measure is pessimistic as it assumes that clusters are 
processed sequentially. In reality, the bulk of the work in processing
clusters is parallelized: (1) the refinement of the eigenvalues via R-tasks in
the shared-memory environment and C-tasks with communication in the
distributed-memory environment; (2) the final computation of the
eigenpairs. However, {\it significant clustering poses limitations on
scalability, while small clustering implies great potential for parallelism}.

While \PMRRR's performance is quite satisfactory for most practical
purposes (especially in context of dense eigenproblems), its accuracy can
be problematic. \PMRRR\ generally obtains similar accuracy to its sequential
and multi-core counterparts, which was already discussed in
Section~\ref{section:mrsmp:experiments}: For all MRRR implementations, we
must be prepared for orthogonality of about $1000n\varepsilon$.

\begin{figure}[t!hb]
   \centering
   \subfigure[Largest residual norm.]{
     \includegraphics[width=.47\textwidth]{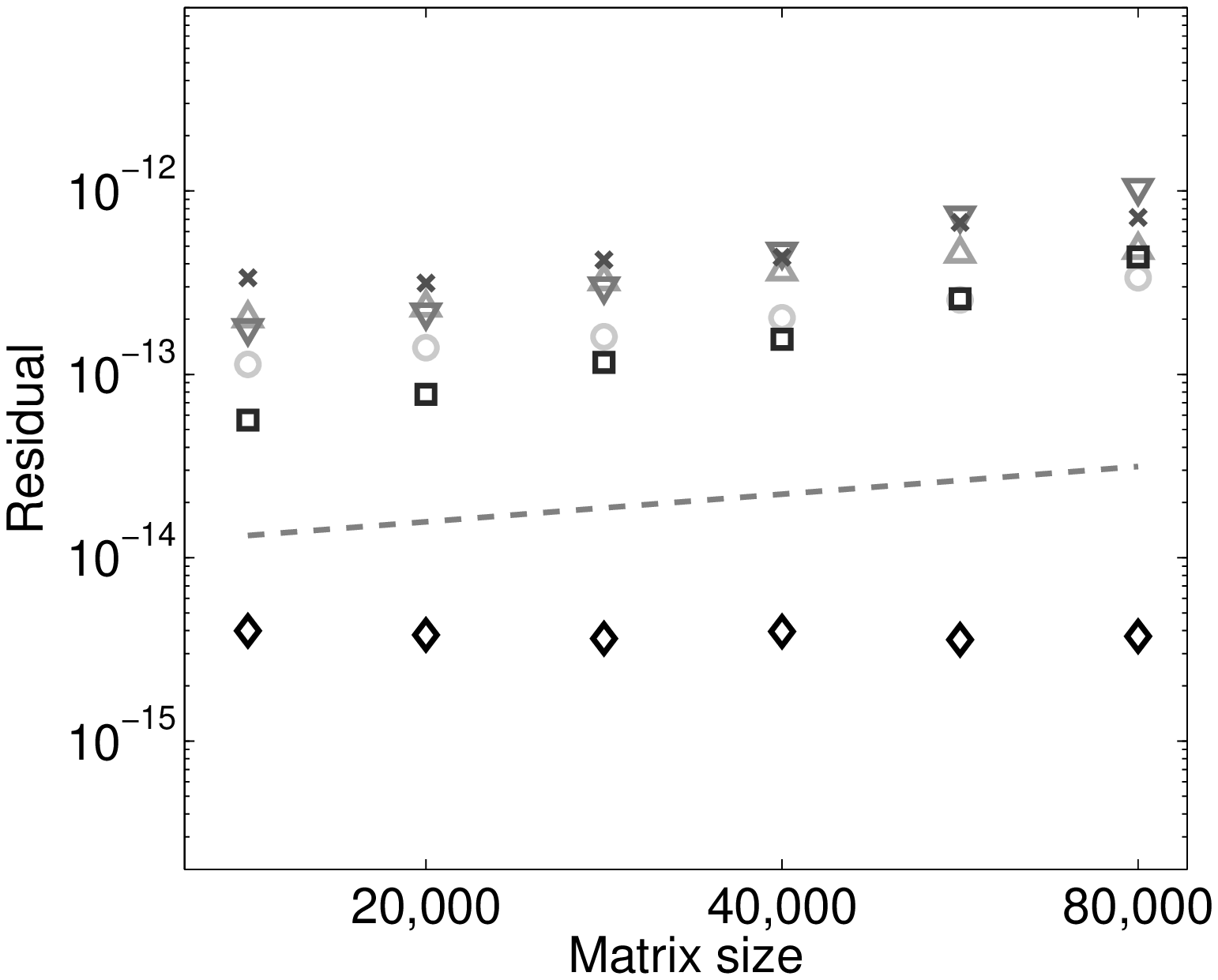} 
   } \subfigure[Orthogonality.]{
     \includegraphics[width=.47\textwidth]{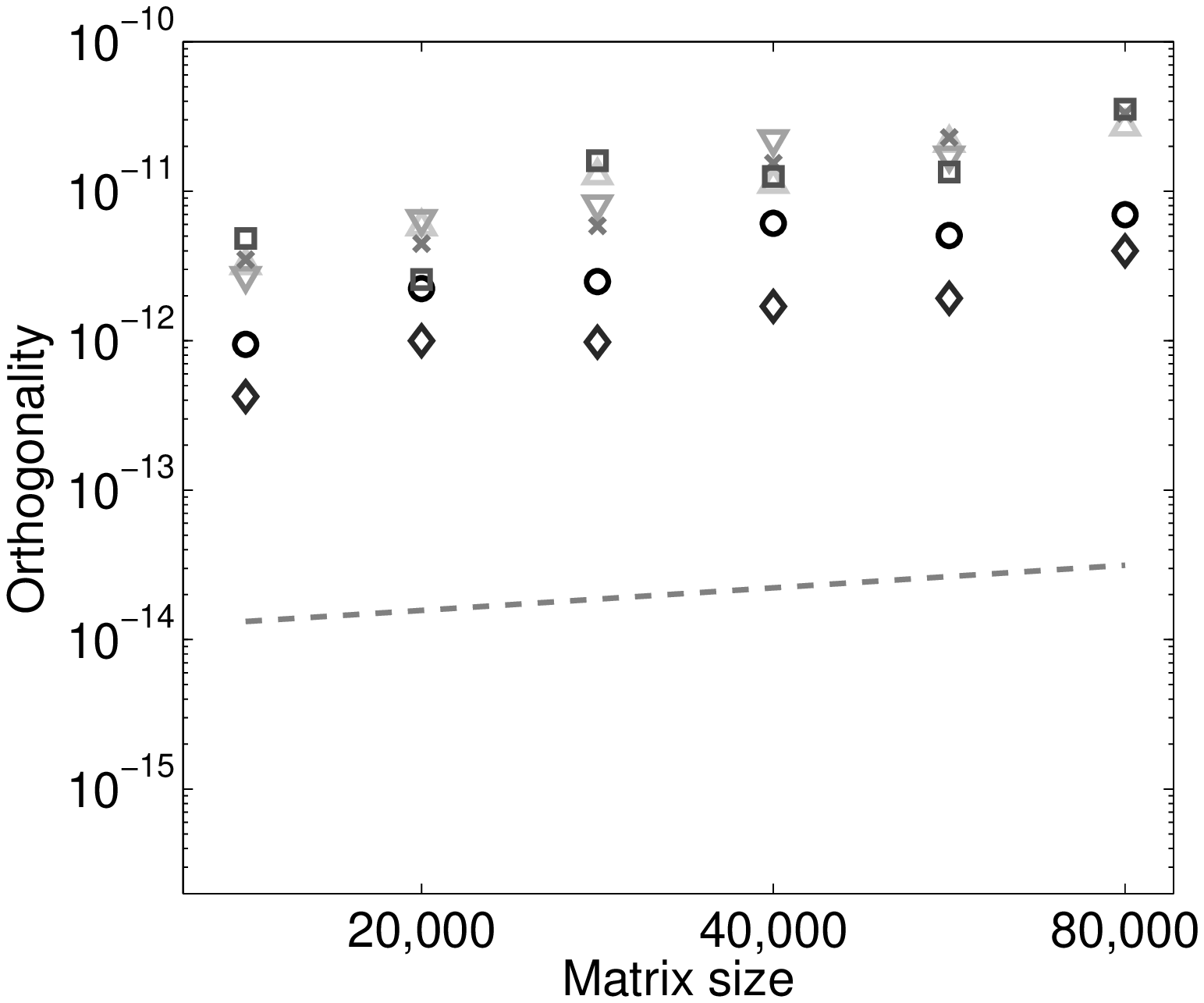}
   }
   \caption{
     Accuracy of \PMRRR\ for the following matrices: ($\bigcirc$) 1--2--1, ($\vartriangle$) Clement,
     ($\triangledown$) Hermite, ($\times$) Legendre, ($\Box$) Laguerre, and 
     ($\Diamond$) Wilkinson.  
   }
   \label{fig:accpmrrr}
\end{figure}
In Fig.~\ref{fig:accpmrrr}, we show the largest residual norm and the
orthogonality as defined in~\eqref{def:defresortho2}.  
Both quantities growth linearly with
the matrix size. The orthogonality results are bounded by $n
\varepsilon$, which is better than the worst case expected for other
matrices. Nevertheless, the accuracy is not at the level of the most
accurate methods, which we approximated by $\varepsilon \sqrt{n}$ (the dashed line).

\chapter{Mixed Precision MRRR}
\label{chapter:mixed}
\thispagestyle{empty}

In the previous chapter, we showed that ($i$) MRRR is oftentimes used in
context of direct methods  for Hermitian 
eigenproblems; ($ii$) MRRR is frequently the fastest
algorithm; and, ($iii$)
MRRR is less accurate than other methods. In particular for dense
eigenproblems, MRRR is responsible for much of  
the loss of orthogonality. However, it has a lower
computational complexity than the reduction to tridiagonal form, which
requires $\order{n^3}$ arithmetic operations.
These observations raise the question whether it is  
possible to trade (some) of MRRR's performance to obtain better accuracy. 

For any MRRR-based solver, we present how the use of mixed
precisions leads to more accurate results at very little or even no extra
costs in terms of performance. 
In this way, {\it MRRR-based solvers are not only among 
of the fastest but also among the most accurate methods}. 
An important feature of our approach is that leads to more scalable and more
robust implementations.

At this point, we assume that the reader is familiar with the content of
Chapter~\ref{chapter:mrrr}. In particular, Algorithm~\ref{alg:mrrr} and
Theorem~\ref{resthm} serve as the basis of
the following discussion.\footnote{The results in this chapter have been
published in form of~\cite{mixedsisc}.}

\section{A mixed precision approach}
\label{sec:mixed:asolver}

The technique is simple, yet powerful: Inside the algorithm, 
we use a precision higher than of the input/output in order to improve accuracy. 
To this end, we build a tridiagonal eigensolver that 
differentiates between two precisions: (1)~the  
input/output precision, say \binaryx, and (2)~the working precision,
\binaryy, with $y \geq x$. 
If $y = x$, we have the original 
situation of a solver based on one precision; in this case, the following analysis
is easily adapted to situations in which we are satisfied with {\em less} accuracy
than achievable by MRRR in $x$-bit arithmetic.\footnote{A similar idea was
  already mentioned in~\cite{Dhillon:Diss}, in relation to a preliminary
  version of the MRRR algorithm, but was never pursued further.}
Since we are interested in accuracy that cannot be accomplished in
$x$-bit arithmetic, we
restrict ourselves to the case $y > x$. Provided the unit roundoff of the $y$-bit format is
sufficiently smaller than the unit roundoff of the $x$-bit format, say four
or five orders of magnitude, we show how to obtain, for practical matrix sizes, improved
accuracy to the desired level. 

Although any $x$-bit and $y$-bit floating point format might be chosen, in
practice, only those shown in
Table~\ref{tab:precisions} are used in high-performance libraries. 
For example, for {\it binary32} input/output (single precision), we
might use a {\it binary64} working format (double precision). Similarly, for
{\it binary64} input/output, we might use a {\it binary80} or {\it binary128} working
format (extended or quadruple precision). For these three configurations, we use the terms
{\it single/double}, {\it double/extended}, and {\it double/quadruple}; 
practical issues of their implementation are 
discussed in Section~\ref{sec:implementation}. In this section, however, we
concentrate on the generic case of \binaryx/\binaryy\ and only use the
concrete cases to illustrate our arguments.\footnote{When we
refer to \binaryx, we mean both the $x$-bit data type and its unit
roundoff $\varepsilon_x$.}  
\begin{table}[htb]
  \begin{center}
\begin{tabular}[thb]{l@{\quad}l@{\quad}l@{\quad}l@{\quad}l@{\quad}}
\toprule
{\bf Name} & {\bf IEEE-754} & {\bf Precision} & {\bf Support}  \\
\midrule
single    & binary32   & $\varepsilon_s = 2^{-24}$ &  Hardware       \\ 
double    & binary64   & $\varepsilon_d = 2^{-53}$ &  Hardware       \\ 
extended  & binary80   & $\varepsilon_e = 2^{-64}$ &  Hardware       \\ 
quadruple & binary128  & $\varepsilon_q = 2^{-113}$ & Software       \\ 
\bottomrule\noalign{\smallskip}
\end{tabular}
  \end{center}
  \caption{The various floating point formats used and their support on common
    hardware. The $\varepsilon$-terms denote the unit roundoff error (for 
    rounding to nearest). We use the letters $s$, $d$, $e$ and
    $q$ synonymously with 32, 64, 80, and 128. For instance,
    $\varepsilon_{_{32}} = \varepsilon_s$.}
  \label{tab:precisions}
\end{table}

In principle, we could perform the entire computation in $y$-bit arithmetic
and, at the end, cast the results to form the $x$-bit output; for 
all practical purposes, we would obtain the desired accuracy.
This naive approach is not satisfactory for two reasons:
First, since the eigenvectors need to be
stored explicitly in the
\binaryy\ format, the memory requirement is increased; and second, if the
$y$-bit floating point arithmetic is much slower than the $x$-bit one, the
performance suffers severely. While the first issue is addressed rather
easily (as discussed Section~\ref{sec:memcost}), the latter requires more
care. 
The key insight is that it is unnecessary to compute eigenpairs with
residual norms and orthogonality bounded by say $1000 n \varepsilon_y$; instead,
these bounds are relaxed to $\varepsilon_x \sqrt{n}$ (for
example, think of $\varepsilon_x \approx 10^{-16}$, $\varepsilon_y
\approx 10^{-34}$, and $n \approx 10{,}000$).
While in the standard MRRR the choice of 
algorithmic parameters is very restricted, as we show below, we gain
enormous freedom in their choice. In 
particular, while meeting more demanding accuracy goals, we are able to select values such
that the amount of necessary computation is reduced, the 
robustness is increased, and parallelism is improved. 

\begin{figure}[thb]
   \centering
   \subfigure[Relative execution time.]{
     \includegraphics[width=.47\textwidth]{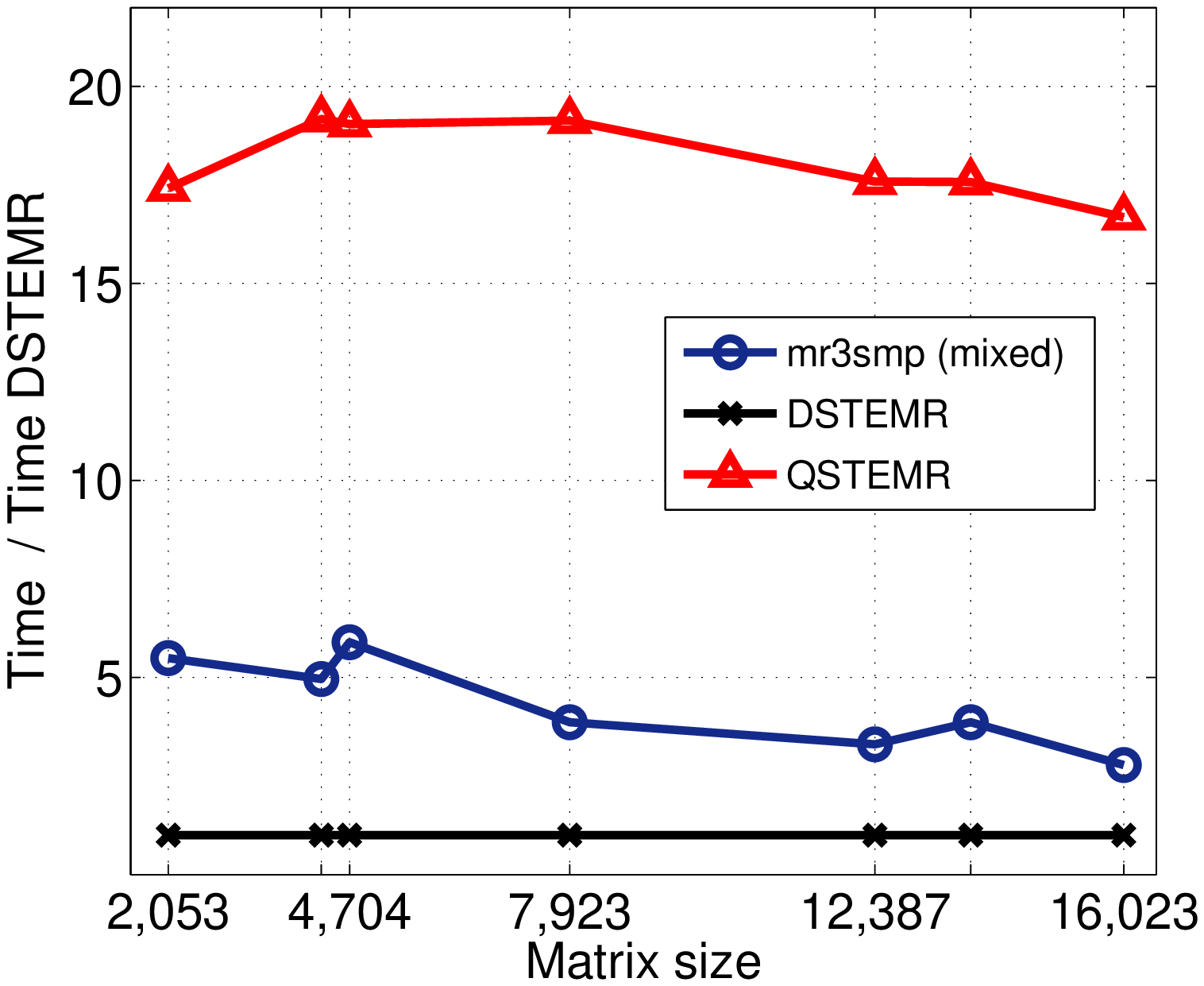}
     \label{fig:optimzedtimetridiaga}
   } \subfigure[Accuracy.]{
     \includegraphics[width=.47\textwidth]{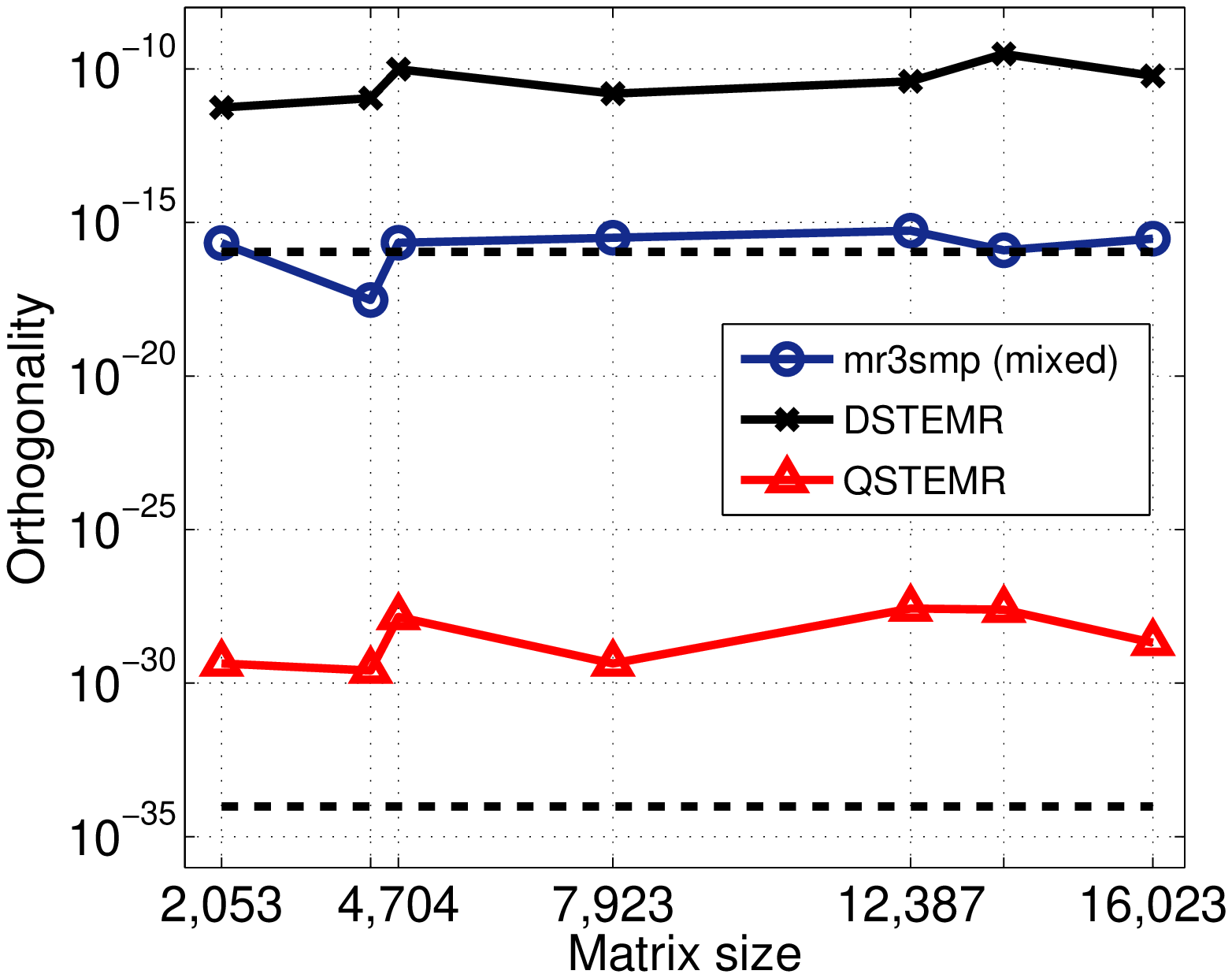}
     \label{fig:optimzedtimetridiagb}
   }
   \caption{
     (a) Execution time of the mixed precision solver relative to
     the standard MRRR and a naive approach to improve accuracy. 
     In the experiment, all eigenpairs were computed and all solvers executed
     {\em sequentially}. The experiment was performed on
     {\sc Westmere} with the application matrices from
     Table~\ref{tab:applmatrices} of Appendix~\ref{sec:firstexperimentsmixed}. (b) 
     Accuracy in form of orthogonality of the eigenvectors. As a reference,
     we added $\varepsilon_d$ and $\varepsilon_q$ as dashed lines.  
   }
   \label{fig:optimzedtimetridiag}
\end{figure}

To illustrate our goal of trading performance for accuracy, we use the
double/quadruple case as an example. 
As depicted
in Fig.~\ref{fig:optimzedtimetridiagb}, the standard MRRR, represented by
LAPACK's {\tt DSTEMR}, computes eigenpairs with accuracy achievable using double
precision arithmetic; with a significant performance penalty,
Fig.~\ref{fig:optimzedtimetridiaga}, {\tt QSTEMR}, which is an adaptation of
{\tt DSTEMR} for quadruple precision, naturally computes more accurate results. In a
naive approach, we use such a solver to achieve better accuracy and cast the
result to double precision. However, all extra accuracy provided by
{\tt QSTEMR} is lost once the result is transformed.

Instead, as we merely require improved accuracy with respect to {\tt DSTEMR}, we speed up a 
solver that uses quadruple precision arithmetic internally. The results of
such an approach is presented in form of \mrsmp\ (with mixed
precision).\footnote{In order to compute the
  orthogonality, we kept the eigenvectors in the
  quadruple format.}  
While computing sufficiently accurate
results, Fig.~\ref{fig:optimzedtimetridiagb}, the reward for the relaxed
accuracy requirement 
is a remarkable up to five-fold speedup compared with the naive
approach, Fig.~\ref{fig:optimzedtimetridiaga}. Although, in a sequential
execution, our solver is slower than the standard double precision solver, it comes with
two important advantages: robustness and 
parallelism are increased significantly. Furthermore, the strategy
can be adapted to any \binaryx/\binaryy\ solver; in case $y$-bit arithmetic
is not that much slower than $x$-bit 
arithmetic, the mixed precision approach leads to {\em faster}
executions compared with the standard MRRR. In the next section, we describe
our strategy in detail.

\subsection{Adjusting the algorithm} 

At this point, it is necessary to recall Theorem~\ref{resthm} in
Section~\ref{sec:finitearitmetic}, which specifies the accuracy of any MRRR implementation. 
 The exact values of the parameters in the theorem differ
slightly for various implementations of the algorithm and need not to be
known in the following analysis. 
The bounds on the residual norm and
orthogonality are {\it theoretical}, that is, for a worst case scenario. 
In practice, the results are smaller than the bounds
suggest: with common parameters, realistic {\it practical} bounds on the residual
norm and on the 
orthogonality are $n \varepsilon$ and $1000 n \varepsilon$, respectively. 
In order to obtain accuracy similar to that of the best available methods,
we need to transform the 
dependence on $n$ by one on $\sqrt{n}$. Furthermore, it is
necessary to reduce the orthogonality by about three orders of
magnitude.\footnote{We will achieve this by transforming
  $n\varepsilon_x/gaptol$ terms (with $gaptol \approx 10^{-3}$) in the
  bound into $\varepsilon_x \sqrt{n}$.} 

Consider the input/output being in a
$x$-bit format and the entire computation being performed in
$y$-bit arithmetic. Starting from this configuration, we expose the new
freedom in the parameter space and justify changes
that we make to the algorithm.
For example, we identify parts that can be executed in $x$-bit
arithmetic, which might be considerably faster.

Assuming $\varepsilon_y \ll \varepsilon_x$ (again, think of $\varepsilon_x
\approx 10^{-16}$ and $\varepsilon_y \approx 10^{-34}$), we simplify
Theorem~\ref{resthm} by canceling terms that do not contribute significantly  
even with adjusted parameters (i.e., terms that are comparable to
$\varepsilon_y$ in magnitude; in particular, we require that $n \varepsilon_y \leq
  \varepsilon_x \sqrt{n}$.\footnote{Commonly, such an assumption does not introduce
    any further restriction on the matrix size, as commonly $n \varepsilon_x
    < 1$ is assumed for any error analysis.}). In our
  argumentation, we hide all constants, 
which anyway correspond to the bounds attainable for a solver purely based
on \binaryy.  For any reasonable implementation of the algorithm, we have the following:
$\alpha = \order{\varepsilon_y}$, $\eta = \order{n \varepsilon_y}$,
$\xi_{\,\downarrow} = \order{\varepsilon_y}$, $\xi_{\uparrow} = \order{\varepsilon_y}$. 
Thus, the orthogonality of the final result is given by 
\begin{equation}
|\hat{z}_i^* \hat{z}_j| = \mathcal{O}\left( k_{rs} \frac{n
    \varepsilon_y}{gaptol} + k_{rr} \, d_{max}\, \frac{n \varepsilon_y}{gaptol} \right) \,.
\label{simpleorthobound}
\end{equation}
Similarly, for the bound on the residual norm, we get 
\begin{equation}
\norm{M_{root}\, \hat{z}_i - \hat{\lambda}_i[M_{root}]  \, \hat{z}_i} =
\mathcal{O}\left(\norm{\bar{r}^{(local)}} + \gamma \, spdiam[M_{root}]\right)
\label{simpleresbound}
\end{equation}
with $\norm{\bar{r}^{(local)}} \leq k_{rs} \, gap\left(\hat{\lambda}_i[M]\right) \frac{n
\varepsilon_y}{gaptol}$
 and $\gamma = \order{k_{elg} \, d_{max} \, n \varepsilon_y}$.

We now provide a list of changes that can be done to the algorithm. 
We discuss their effects on performance, parallelism, and memory
requirement.  

\paragraph{Preprocessing.}
We assume scaling and splitting is done as in a solver purely based on $x$-bit
floating point arithmetic, see Section~\ref{sec:preprocessing}. In particular, off-diagonal
element $\beta_i$ of the input $T$ is set to zero whenever
\begin{equation*} 
|\beta_i| \leq \varepsilon_x \sqrt{n} \norm{T} \,,
\end{equation*}  
where $n$ and $T$ refer to the {\it unreduced} input.\footnote{In our implementation, we
   used $|\beta_i| \leq \varepsilon_x \norm{T}$~\cite{mixedtr}.} 
We remark that this criterion is less strict than setting elements to zero whenever $|\beta_i|
 \leq \varepsilon_y \sqrt{n} \norm{T}$. 
Splitting the input matrix into submatrices is
beneficial for both performance and accuracy as these are mainly determined
by the largest submatrix.
Throughout this
section, we assume that the preprocessing has been done and each subproblem
is treated independently by invoking Algorithm~\ref{alg:mrrr}. In particular, whenever
we refer to matrix $T$, it is assumed to be irreducible; whenever we
reference the matrix size $n$ in the context of parameter settings, it refers
to the size of the processed block. 

\paragraph{Choice of form to represent tridiagonals.} 
For the various forms to represent tridiagonals (e.g., bidiagonal, twisted,
or blocked factorizations) and their 
data (e.g., $N$-, $e$-, or $Z$-representation), different algorithms
implement the shift operation in Line~\ref{line:mrrr:shifting} of
Algorithm~\ref{alg:mrrr}: $M_{shifted} = M - \tau I$. All these 
algorithms are stable in the sense that the relation holds 
exactly if the data for $M_{shifted}$ and $M$ are perturbed element-wise by a
relative amount bounded by $\order{\varepsilon_y}$. The implied constants
for the perturbation bounds vary slightly. As $\varepsilon_y <
\varepsilon_x$, instead of concentrating on accuracy issues, we can make our
choice based on robustness and performance. A discussion of performance
issues related to different forms to represent tridiagonals can be found
in~\cite{Willems:twisted,Willems:Diss}. Based on this discussion, it appears
that twisted factorizations with $e$-representation seem to be a good
choice. As the off-diagonal entries of all the matrices stay the same, they
only need to be stored once and are reused during the entire
computation. 

\paragraph{Random perturbations.} 
In Line~\ref{line:mrrr:perturb} of
Algorithm~\ref{alg:mrrr}, to break up tight clusters, the data of 
$M_{root}$, $\{x_1,...,x_{2n-1}\}$, is perturbed element-wise by small
random relative amounts: 
$\tilde{x}_i = x_i (1 + \xi_i)$ with $|\xi_i| \leq \xi$ for all $1 \leq i
\leq {2n-1}$. In practice, a value like $\xi = 8
\varepsilon$ is used. Although our data is in \binaryy, we can be quite
aggressive and adopt $\xi = \varepsilon_x$ or a small multiple
of it.\footnote{Two comments: (1) If we later choose to relax the requirements on
  the representations, we do not do so for the root representation. Commonly,
we take a definite factorization that defines all eigenpairs to high relative
accuracy; (2) A reader might wonder if we loose the ability to attain the more
demanding bound on the residual. First, it is not the case in practice and,
second, it is irrelevant in context of dense eigenproblems.} 
For $y = 2x$, about half of the digits in each entry of the representation are chosen
randomly and with high probability, eigenvalues do not agree to
many more than $\lceil - \log_{10}{\varepsilon_x} \rceil$ digits.
This has two major effects: First, together with the changes in
$gaptol$ (see below), the probability to encounter large values for
$d_{max}$ (say 4 or larger) becomes extremely low. 
Second, it becomes easier to
find suitable shifts such that the resulting representation satisfies the
requirements of relative robustness and conditional element growth. 
The positive impact of small 
$d_{max}$ on the accuracy is apparent from \eqref{simpleorthobound}
and \eqref{simpleresbound}. Furthermore, as 
discussed below, due to limiting $d_{max}$, the computation can be
reorganized for efficiency. Although it might look innocent, the more
aggressive random perturbations lead to much improved robustness: A detailed
discussion can be found in~\cite{glued,Dhillon:Diss}.

\paragraph{Classification of the eigenvalues.} Due to the importance of the
$gaptol$-parameter, adjusting it to our requirements is key to the success
of our approach. The parameter influences nearly all stages of the
algorithm; most importantly, the classification of eigenvalues into
well-separated and clustered. As already discussed, the choice of {\it
  gaptol} is restricted by the loss of orthogonality that we are willing to
accept; in practice, the value is often
chosen to be $10^{-3}$~\cite{Dhillon:2004:MRRR}.\footnote{For instance,
  LAPACK's {\tt DSTEMR} uses $10^{-3}$ and {\tt SSTEMR} uses $3 \cdot
  10^{-3}$.} As we merely require 
orthogonality of $\varepsilon_x \sqrt{n}$, we usually accept
more than three orders of magnitude loss of orthogonality. Both terms in
\eqref{simpleorthobound} (and the in practice observed orthogonality)
grow as $n 
\varepsilon_y / gaptol$. 
Consequently, we might select any
 value satisfying
\begin{equation}
 \min\left\{ 10^{-3}, \frac{\varepsilon_y \sqrt{n}}{\varepsilon_x} \right\} \leq gaptol \leq
 10^{-3} \,,
\label{eq:gaptolinterval}
\end{equation}
 for $gaptol$, where the $10^{-3}$ terms are derived from practice and might be altered slightly.
 Note that $gaptol$ can become as small
 as $10^{-9} \sqrt{n}$ in the single/double case and $10^{-18} \sqrt{n}$ in the
 double/quadruple one. If we restrict the analysis to matrices with size $n \leq 10^6$, we can
 choose a constant $gaptol$ as small as $10^{-6}$ and $10^{-15}$ 
 respectively for the single/double and double/quadruple cases.

\begin{figure}[thb]
   \centering
   \subfigure[Double precision arithmetic.]{
     \includegraphics[width=.47\textwidth]{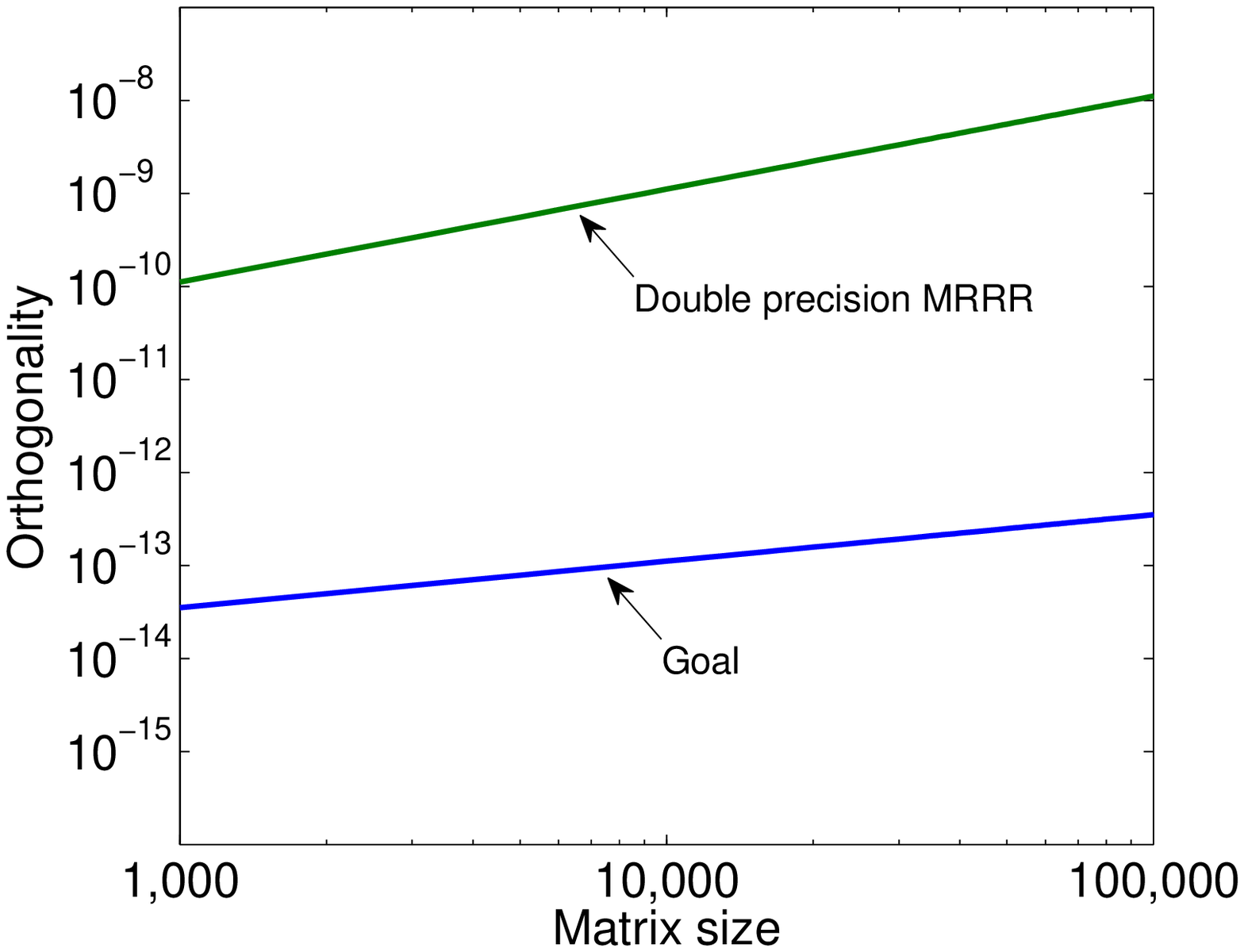}
     \label{fig:adjustinggaptola}
   } \subfigure[Quadruple precision arithmetic.]{
     \includegraphics[width=.47\textwidth]{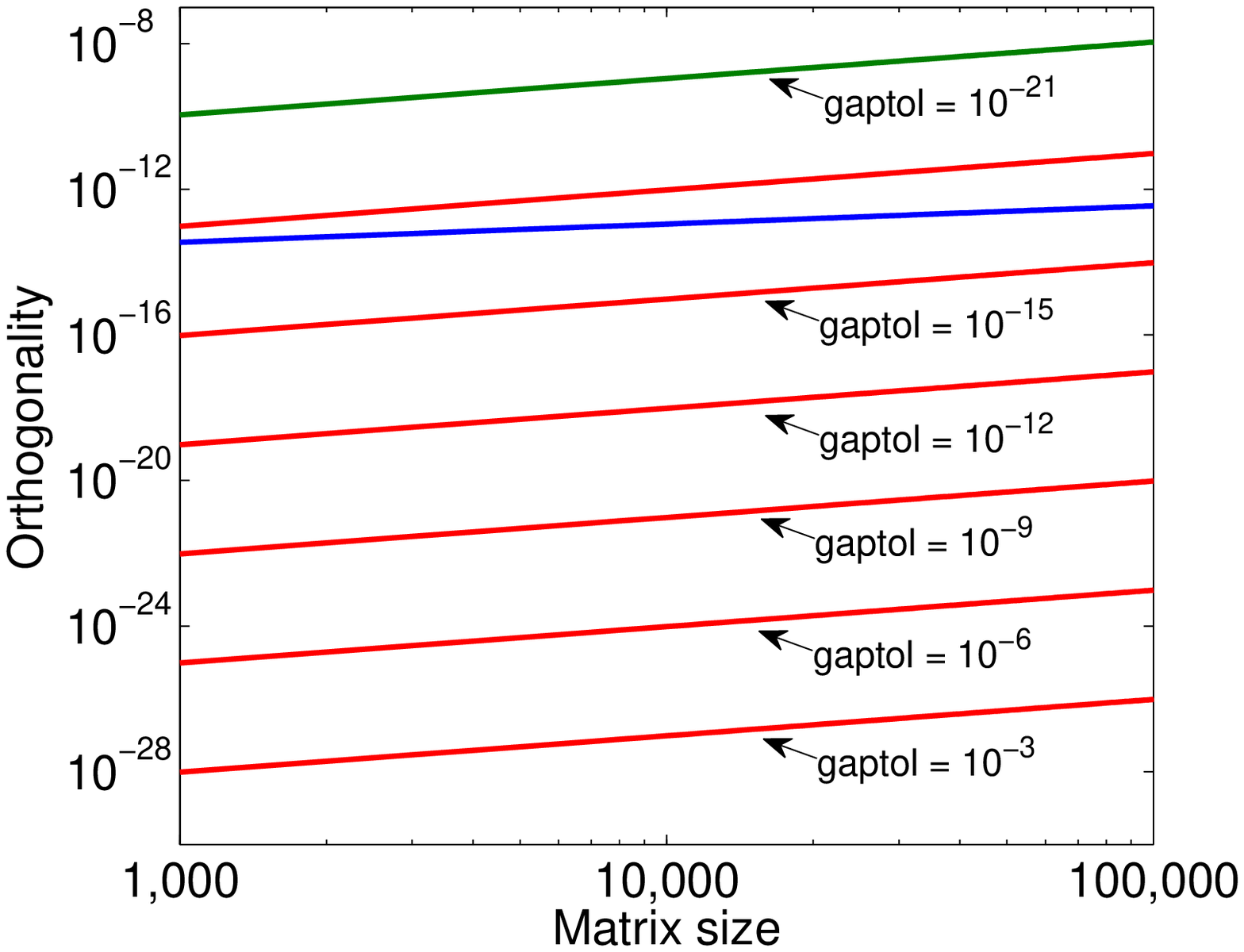}
     \label{fig:adjustinggaptolb}
   }
   \caption{
     Selection of the important parameter $gaptol$. By using higher precision
     arithmetic we gain freedom in the choice of the parameter. 
   }
   \label{fig:adjustinggaptol}
\end{figure}

\begin{figure}[thb]
   \centering
   \includegraphics[width=.70\textwidth]{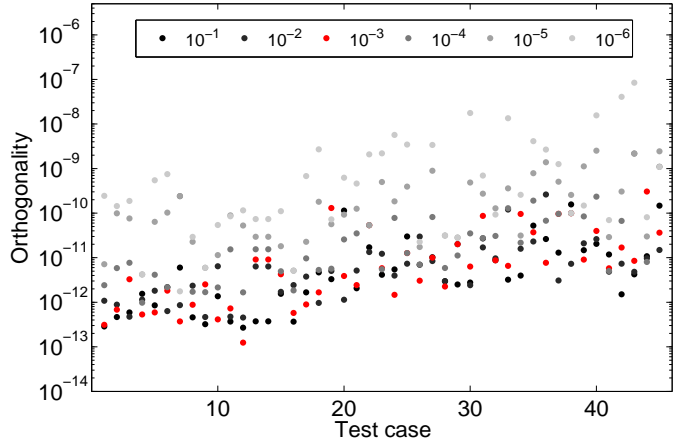}
   \caption{
     Orthogonality of {\tt mr3smp} (without mixed precision) for various values of $gaptol$. The
     experiment was performed on test set {\sc Application}, which is
     detailed in Appendix~\ref{appendix:matrices}. 
   }
   \label{fig:simplegaptola}
\end{figure}

We use again the double/quadruple case to illustrate the choice of
$gaptol$. With $gaptol = 10^{-3}$, Fig.~\ref{fig:adjustinggaptola} shows a
practical upper bound on the orthogonality of MRRR in double precision arithmetic, $1000 n
\varepsilon$. 
Depending on the specific goal, which for instance growths as
$\varepsilon \sqrt{n}$, we aim at improving the orthogonality. If the goal is set as in
Fig.~\ref{fig:adjustinggaptola}, a
reader looking at \eqref{simpleorthobound} might wonder whether it is possible to simply increase $gaptol$ until
the accuracy goal is met. Unfortunately, as Fig.~\ref{fig:simplegaptola}
illustrates, it is not possible to achieve better accuracy by
simply selecting a larger value for $gaptol$.
Even with $gaptol = 10^{-1}$, no significant accuracy improvement is observed. On the
contrary, such a large value of $gaptol$ negatively impacts performance and
reliability. If $gaptol$ is too large, it can become impossible to break
clusters and a code might fail completely. For this reason, $gaptol < 1$
is required and $gaptol \ll 1$ is desired. By Fig.~\ref{fig:simplegaptola},
even with $gaptol$ being several orders of magnitude smaller than $10^{-3}$,
the orthogonality is sufficient for {\em single} precision outputs. 
It is this observation that we exploit in our mixed precision MRRR. 
For double precision input/output,
instead of performing the entire computation in double precision arithmetic, we use
quadruple arithmetic. By \eqref{eq:gaptolinterval}, we gain freedom in
choosing $gaptol$. We might chose a constant value of $gaptol =
10^{-15}$ to fulfill our goal, as depicted in
Fig.~\ref{fig:adjustinggaptolb}. 
Alternatively, we can let $gaptol$ be a function of matrix 
size $n$ as in the lower bound of \eqref{eq:gaptolinterval}. However, if we do not
decrease $gaptol$ to the smallest possible value, by
\eqref{simpleorthobound} and \eqref{simpleresbound}, we can
instead relax other parameters ($k_{rr}$,$k_{rs}$) that influence accuracy.

 Returning to the general case of \binaryx/\binaryy, with any choice of $gaptol$ complying
 \eqref{eq:gaptolinterval}, accuracy to the desired level is guaranteed, and
 there is room to choose the specific value of $gaptol$, as well as other
 parameters, to optimize performance or 
 parallelism. In particular, by generally reducing the clustering of the
 eigenvalues, the smallest possible value of $gaptol$ provides the greatest parallelism.
 As we have done in Section~\ref{sec:remaininglimitations}, to quantify this
 statement, for any matrix, we define {\it clustering} $\rho \in [1/n, 1]$
 formally as the size of the largest cluster divided by the matrix
 size. 
 The two main advantages in decreasing $\rho$ were also discussed in
 Section~\ref{sec:remaininglimitations}: the work of processing the
 largest cluster introduces of $\order{\rho n^2}$ flops is reduced and the
 potential parallelism is increased.  
 A conservative estimate of the parallelism of a problem is provided by
 $\rho^{-1}$ (for instance, $\rho
 = 1/n$ implies that the problem is embarrassingly
 parallel\footnote{Communication of the initial eigenvalue approximation is
   still necessary, but the rest of the communication is removed.}). 
 Matrices with high clustering pose difficulties as they introduce load balancing issues and 
 communication, which can considerably reduce the parallel
 scalability~\cite{Vomel:2010:ScaLAPACKsMRRR,VoemelRefinedTree2007tr}. 
 Even if we do not desire to guarantee improved accuracy, we can use mixed precisions to 
 enhance parallelism. In this case, the $\sqrt{n}$-dependence on the lower
 bound for the value of $gaptol$ would be removed and the bound could be
 loosened by another three orders of magnitude; that is, we might choose a value of
 $10^{-12}$ and $10^{-21}$ for the single/double and double/quadruple cases,
 respectively.\footnote{If we select values $10^{-9}$ and $10^{-18}$, we still
 improve the bounds by three orders of magnitude, which is sufficient for many
 practical purposes; see also Fig.~\ref{fig:adjustinggaptolb}.} Most results
would still be improved, as the 
relative gaps of the eigenvalues are often larger than $gaptol$. 
In addition, we expect that {\it almost all} computations become embarrassingly parallel. 

 As an example, Table~\ref{tab:clustering}
 shows the clustering for double precision Hermite 
 type\footnote{For information on test
   matrices, see Appendix~\ref{appendix:matrices}.} test matrices of 
 various sizes with four distinct classification criteria:\footnote{Criterion I is used in LAPACK~\cite{Dhillon:DesignMRRR} and in results of {\tt
    mr3smp} in~\cite{mr3smp}, which usually uses II. Criterion II is used in
  ScaLAPACK~\cite{Vomel:2010:ScaLAPACKsMRRR} and
  Elemental~\cite{EleMRRR}. In massively parallel computing environments,
  criteria III and IV can (and should) additionally complemented with the
  splitting based on absolute gaps; see also~\cite{mixedtr}.} 
(I) $gaptol = 10^{-3}$, (II) $gaptol = 10^{-3}$, combined with splitting based on
the absolute gap as proposed in~\cite{VoemelRefinedTree2007tr} to enhance
parallelism, (III) $gaptol = 10^{-10}$, and (IV) $gaptol = 10^{-15}$.
As with this example, experience shows that, thanks to a reduced value of
$gaptol$ as in criteria III or IV, many problems 
become embarrassingly parallel {\em and} guarantee improved accuracy. 
In case $\rho = 1/n$, $d_{max}$ is zero, which not only benefits accuracy by
\eqref{simpleorthobound} and \eqref{simpleresbound}, but also has a 
more dramatic effect: {\it the danger of 
not finding representations that satisfy the requirements is entirely
removed.} This follows from the fact that a satisfactory root representation can 
always be found (e.g., by making $T - \mu I$ definite) and no other
representation needs to be computed. 
Even in cases with $d_{max} > 0$, the
number of times Line~\ref{line:mrrr:shifting} of Algorithm~\ref{alg:mrrr}
needs to be executed is often considerably reduced.\footnote{For example,
  consider the experiment in Appendix~\ref{sec:firstexperimentsmixed}
 and \ref{sec:moreexperimentsmixed}.} 
\begin{table}[htb]
\begin{center}
\small
\begin{tabular}{c@{\quad\quad\quad}cccc}
\toprule
Criterion &   \multicolumn{4}{c}{Matrix size} \\
               & $2{,}500$ & $5{,}000$ & $10{,}000$ & $20{,}000$ \\
\midrule
I  &  0.70  &         0.86   &  0.93  & 0.97 \\
II &  0.57   &        0.73   &  0.73  & 0.73 \\
III &  4.00e{-4}  &   2.00e{-4}   &  1.00e{-4}  & 5.00e{-5}  \\
IV &  4.00e{-4}  &   2.00e{-4}   &  1.00e{-4}  & 5.00e{-5} \\
\bottomrule\noalign{\smallskip}
\end{tabular} 
\end{center}
\caption{The $gaptol$-parameter effect on clustering $\rho \in [1/n,1]$.  
}
\label{tab:clustering}
\end{table}

On the downside, selecting a smaller $gaptol$ can result in more work in the
initial approximation and later refinements\footnote{For instance, if bisection is used to obtain
  initial approximations to the eigenvalues.} -- in both cases, eigenvalues must be
approximated to relative accuracy of about $gaptol$; as a result, optimal
performance is often not achieved 
for the smallest possible value of $gaptol$. 
Moreover, as we discuss below, if one is willing to limit
the choice of $gaptol$, the approximation of eigenvalues can be
done (almost) entirely in $x$-bit arithmetic.\footnote{For the refinement of extreme
  eigenvalues prior to selecting shifts, we still need to resort to
  $y$-bit arithmetic.} If the $y$-bit arithmetic is significantly slower than
the $x$-bit one, it might be best to
take advantage of the latter. And, as we see below as
well, if not the smallest possible value is chosen for $gaptol$, the
requirements the intermediate 
representations must fulfill are relaxed.  

Another corollary of adjusting $gaptol$ is slightly hidden: in
Line~\ref{line:mrrr:shifting} of Algorithm~\ref{alg:mrrr}, we gain more freedom
in selecting $\tau$ such
that, at the next iteration, the cluster is broken apart. For instance, when choosing
$\tau$ close to one end of the cluster, we are able to ``back off'' further
away than usual from the end of the cluster in cases where we did not find a
representation satisfying the requirements in a previous attempt (see
Algorithm~\ref{alg:spectrumshift} in Section~\ref{sec:closerlookeigvecs}). 

We cannot overemphasize the positive effects an adjusted {\em gaptol} has on
robustness and parallel scalability. In particular, in parallel
computing environments, the smallest value for $gaptol$ can significantly
improve the parallel scalability. Since many problems become embarrassingly
parallel, the danger of failing is removed entirely.

\paragraph{Arithmetic used to approximate eigenvalues.} 
In Lines~\ref{line:mrrr:initialeigvals} and \ref{line:mrrr:refine} of
Algorithm~\ref{alg:mrrr}, eigenvalues are respectively computed and refined
to a specified relative accuracy. 
In both cases, we are given a
representation, which we call $M_y$ henceforth, and an index set
$\mathcal{I}$ that indicates the eigenvalues that need to be approximated. 
When the $y$-bit arithmetic is much slower than the $x$-bit one (say a
factor 10 or more), the use of the latter is preferred: 
One creates a temporary copy of $M_y$ in \binaryx\ -- called $M_x$ henceforth
-- that is used for the eigenvalue computation in $x$-bit arithmetic. The creation of $M_x$
corresponds to an element-wise relative perturbation of $M_y$ bounded by
$\varepsilon_x$. By the relative robustness of $M_y$, 
\begin{equation}
\abs{\lambda_i[M_x] - \lambda_i[M_y]} \leq
k_{rr} n \varepsilon_x \abs{\lambda_i[M_y]} \,.
\end{equation}
For instance, bisection can be used to compute eigenvalue approximations
$\hat{\lambda}_i[M_x]$ to high relative accuracy, after which $M_x$ is discarded. As
casting the result back to \binaryy\ causes no additional error, it is $\hat{\lambda}_i[M_y] =
\hat{\lambda}_i[M_x]$ and
\begin{equation*}
\abs{\hat{\lambda}_i[M_y] - \lambda_i[M_x]} \leq
k_{bi} n \varepsilon_x \abs{\lambda_i[M_x]} \,,
\end{equation*}
where $k_{bi}$ is a constant given by the bisection
method.\footnote{In practice, $k_{bi}$ is bounded by a small multiple of
  $k_{rr}$, meaning the eigenvalues are computed to the accuracy granted by
  the representation.} 
To first order, by the
triangle inequality, it holds
\begin{equation}
\abs{\hat{\lambda}_i[M_y] - \lambda_i[M_y]} \leq \left( k_{rr} + k_{bi}\right) n
\varepsilon_x \abs{\lambda_i[M_y]} \,.
\label{eq:accuracyeigvalinxarithmetic}
\end{equation}
Provided $\left( k_{rr} + k_{bi}\right) n \varepsilon_x \lesssim
gaptol$, $x$-bit arithmetic can be used to approximate the
eigenvalues. Thus, an additional constraint on both the size $n$ and
$gaptol$ arises:
Given $gaptol$, we must limit the matrix size up to which we can do the
computation purely in $x$-bit arithmetic. Similarly, for a given matrix size, we need to adjust
the lower bound on $gaptol$ in \eqref{eq:gaptolinterval}. As an example, if
say $k_{rr} \leq 10$, $k_{bi} \leq 10$, $n \leq 10^6$, and $\varepsilon_x =
\varepsilon_d = 2^{-53}$, it is required that that $gaptol \gtrsim 10^{-10}$. 
When resorting to $x$-bit arithmetic or if $gaptol$ is chosen too small, one might
respectively verify or refine the result of the $x$-bit eigenvalue computation using
$y$-bit arithmetic without significant costs.\footnote{If the first
  requirement in Definition~\ref{def:RRR} is 
  removed, we can still make use of $x$-bit arithmetic although
  \eqref{eq:accuracyeigvalinxarithmetic} might
  not always be satisfied anymore.} 

\paragraph{Requirements on the representations.}
As long as $k_{elg} n\varepsilon_y \ll \varepsilon_x \sqrt{n}$, by \eqref{simpleresbound},
the residual with respect the $M_{root}$ is mainly influenced by the local
residual. In our mixed precision approach, without loss of accuracy, it
is possible to allow for  
\begin{equation}
k_{elg}\leq \max \left\{10, \frac{\varepsilon_x}{\varepsilon_y \sqrt{n}}
\right\} \,,
\label{eq:newkelgbound}
\end{equation}
where we assumed 10 was the original value of $k_{elg}$. As a result, the
requirement on the conditional element growth is considerably relaxed. 
For instance, in the single/double and double/quadruple cases,
assuming $n \leq 10^6$, bounds on $k_{elg}$ of about $10^6$ and $10^{15}$ are
sufficient, respectively. 
If $gaptol$ is
not chosen as small as possible, the bound on $k_{rr}$ can be loosened in a
similar fashion:
\begin{equation}
k_{rr}\leq \max \left\{10, \frac{\varepsilon_x}{\varepsilon_y
    \sqrt{n}} \cdot gaptol \right\} \,.
\label{eq:newkrrbound}
\end{equation}
As an example, in the double/quadruple case, assuming $n \leq 10^6$ and
$gaptol$ set to $10^{-10}$, $k_{rr} \leq 10^5$ would be sufficient to ensure
accuracy. 

\paragraph{Rayleigh quotient iteration.}
Our willingness to lose orthogonality up to a certain level, which is
noticeable in the lower bound on $gaptol$, 
is also reflected in stopping criterion for RQI, which is given by \eqref{localresbound}. 
As $n \varepsilon_y / gaptol \leq \varepsilon_x 
 \sqrt{n}$, we can stop the RQI when
\begin{equation}
\norm{\bar{r}^{(local)}} \leq k_{rs}  \cdot
gap\left(\hat{\lambda}_i[M]\right) \, \varepsilon_x \sqrt{n} \,,
\end{equation}
where $k_{rs} $ is $\order{1}$. In practice, we take $k_{rs} \approx 1$ or
even $k_{rs} \approx 1/\sqrt{n}$.
As a consequence, the iteration is stopped
earlier on and overall work reduced. 

As a side note: In the rare cases where RQI fails to converge (or as a general alternative to
RQI), we commonly resort to bisection to approximate the eigenvalue
$\lambda_i$ and then use only one step of RQI (with or without
applying the correction term). In the worst case, we require the eigenvalue to be
approximated to high relative accuracy, $|\hat{\lambda}_i -
\lambda_i| = \order{n \varepsilon_y
  |\lambda_i|}$~\cite{Dhillon:2004:Ortvecs}. With mixed precision, we 
can relax the condition to $|\hat{\lambda}_i - \lambda_i| =
\order{\varepsilon_x \sqrt{n} |\lambda_i| \, gaptol}$, which is less
restrictive if $gaptol$ is not chosen as small as
possible.\footnote{The implied constants being the same and given by the
  requirement of a regular solver based on $y$-bit arithmetic. In a similar way, we
  could say that the Rayleigh quotient correction does not
  improve the eigenvalue essentially anymore 
  if $|\gamma_r|/\norm{\hat{z}_i} = \order{\varepsilon_x |\hat{\lambda}_i|
    \, gaptol / \sqrt{n}}$, instead of $|\gamma_r|/\norm{\hat{z}_i} =
  \order{\varepsilon_y |\hat{\lambda}_i|}$. We never employed such a
  change as it will hardly have any effect on the computation time.} If
$relgap(\hat{\lambda}_i) \gg gaptol$, the restriction on the accuracy of the
approximated eigenvalue can be lifted even further~\cite{Willems:Diss}. 

\paragraph{Traversal of the representation tree.} Thanks to the random
perturbation of the root representation and a properly adjusted
$gaptol$-parameter, we rarely expect to see large values for $d_{max}$. For all
practical purposes, in the case of $y = 2x$, we may assume $d_{max} \leq
2$. As a result,
the computation can be rearranged, as discussed
in~\cite{Willems:framework} and summarized in the following: To bound the memory
consumption, a breath-first strategy such as in Algorithm~\ref{alg:mrrr} is used; see for
instance in~\cite{Dhillon:DesignMRRR,mr3smp}. This means that, at
any level of the representation tree, all singletons are processed before the
clusters. A depth-first strategy would instead process entire clusters,
with the
only disadvantage that meanwhile up to $d_{max}$ representations need to 
be kept in memory. If $d_{max}$ is limited as in our case, the depth-first
strategy can be used without disadvantage. In fact, a depth-first strategy
brings two advantages: ($i$) 
copying representations to and from the eigenvector matrix is avoided entirely (see
the next section on the benefit for the mixed precision MRRR) and ($ii$) if
at some point in the computation no
suitable representation is found, there is the possibility of
backtracking, that is, we can process the cluster again by choosing different
shifts at a higher level of the representation tree. For these reasons, in the mixed
precision MRRR, a depth-first strategy is preferred. 

\subsection{Memory cost}
\label{sec:memcost}

We stress both input and output are in \binaryx\
format; only {\it internally} (i.e., hidden to a user) $y$-bit
arithmetic is used.
The memory management of an actual implementation of MRRR is affected by the
fact that output matrix $Z \in \mathbb{R}^{n \times k}$, containing
the desired eigenvectors, 
is commonly used as intermediate workspace. Since $Z$ is in \binaryx\ format, whenever $y >
x$, the workspace is not sufficient anymore for its customary use: For
each cluster, a
representation is stored in the corresponding columns of
$Z$~\cite{Dhillon:DesignMRRR,mr3smp}. 
As these representations consist of $2n-1$ \binaryy\ numbers, this approach
is generally not applicable. If we restrict to $y \leq 2x$, we can store the $2n-1$ 
\binaryy\ numbers whenever a cluster of size four and more is encountered. 
Thus, the computation
must be reorganized so that at least clusters containing less than four
eigenvalues are processed without storing any data in $Z$ temporarily. In
fact, using a depth-first strategy, we remove the need to use $Z$ as
temporary storage entirely. 
Furthermore, immediately after computing an eigenvector
in \binaryy, it is converted to \binaryx, written into $Z$, and
discarded. 
Consequently, while our approach slightly increases the memory usage, we
do not require much more memory: 
with $p$ denoting the level of
parallelism (i.e., number of threads or processes used), the mixed precision MRRR
still needs only $\mathcal{O}(pn)$ \binaryx\ floating point numbers extra workspace.

\section{Practical aspects}
\label{sec:implementation}

We have implemented the mixed precision MRRR for three cases:
{\it single/double}, {\it double/extended}, and {\it double/quadruple}. The
first solver accepts single precision input and produces single precision
output, but internally uses double precision. The
other two are for double precision input/output. The performance of the
solvers, compared with the traditional implementation, depends entirely on
the difference in speed between the two involved arithmetic. 
If the higher precision arithmetic is not much slower (say less
than a factor four), the approach is expected to work well, even for
sequential executions and relatively small matrices. If the higher
precision arithmetic is considerably slower, the mixed precision MRRR
might still perform well for large 
matrices or, due to increased parallelism, when executed on highly parallel systems. 
Our target application is the computation of a {\em subset of eigenpairs} of
{\it large-scale dense} Hermitian matrices. For such a scenario, we 
tolerate a slowdown of the tridiagonal eigensolver due to the use of mixed precisions
without affecting overall performance significantly~\cite{EleMRRR,mixedtr}.

\subsection{Implementations}
In Section~\ref{sec:mixed:experiments}, we present experimental results of our
implementations. All mixed precision implementations are based on {\tt mr3smp}, presented
in Chapter~\ref{chapter:parallel}, and use
$N$-representations of lower bidiagonal 
factorizations. 
As discussed in Algorithm~\ref{alg:initialeigvals} in
Section~\ref{sec:closerlook}, bisection is used 
for the initial eigenvalue computation if a small 
subset of $k$ eigenpairs is requested or if the
number of executing threads exceeds $12k/n$. If
all eigenpairs are requested and the number of threads is less than 12, the fast
sequential {\it dqds algorithm}~\cite{AccurateSVDandQDtrans,dqds99} is used
instead of bisection. As a consequence, speedups compared to the sequential
execution appear less than perfect even for an embarrassingly parallel
computation. 

As several design decisions can be made and the run time depends on both the
specific input and the architecture, optimizing a code for performance
is non-trivial. However,  
we can choose settings in a way that in general yields good, but not necessarily
optimal, performance. 
For instance, on a highly parallel machine one would pick a small value for $gaptol$ to
increase parallelism. For testing purposes, we disabled the
classification criterion based on the absolute gaps of the eigenvalues
proposed in~\cite{VoemelRefinedTree2007tr}, which might reduce clustering even
further (it has no consequences for our test cases shown in the next section). 

For now, we did not relax the requirements on the representations
according to \eqref{eq:newkelgbound} and \eqref{eq:newkrrbound}; we only
benefit from the possibility of 
doing so indirectly: As shown in Algorithm~\ref{alg:spectrumshift} in
Section~\ref{sec:closerlook}, if no suitable 
representation is found, a good candidate is chosen, which might fulfill the
{\it relaxed} requirements. 
In the following, we provide additional comments to all of the mixed
precision solvers individually.  

\paragraph{Single/double.} With widespread language and hardware support for
  double precision, the mixed precision MRRR is most easily implemented
  for the {\it single/double} case. In our test implementation, we fixed
  $gaptol$ to $10^{-5}$. When bisection is used, the initial approximation
  of eigenvalues is done to a relative accuracy of $10^{-2} \cdot gaptol$;
  the same tolerance is used for the later refinements. 
  We additionally altered the computation of compared with {\tt mr3smp}: Thanks to the 
  reduced clustering, the initial eigenvalue approximation often becomes the
  most expensive part of the computation. In order to achieve better load
  balancing in this stage, we reduced the granularity of the tasks and used
  dynamic scheduling of tasks.

  As on most machines the double precision arithmetic is not more than a factor two
  slower than the single precision one, we carry out {\it all} computations in
  the former. Data conversion is only necessary when reading the input and
  writing the output. As a result, compared with a double precision solver
  using a depth-first strategy,
  merely a number of convergence criteria and thresholds must be
  adjusted, and the RQI must be performed using a temporary vector that is, after convergence,
  written into the output eigenvector matrix. The mixed precision code
  closely resembles a conventional double precision implementation of MRRR.

\paragraph{Double/extended.} Many current architectures have hardware
  support for a 80-bit extended floating point format (see
  Table~\ref{tab:precisions} in Section~\ref{sec:mixed:asolver}). As the
  unit roundoff is only 
  about three orders of magnitude smaller than for double precision, we can
  improve the accuracy of MRRR by this amount. For matrices of moderate
  size, the accuracy becomes comparable to that of the best methods (see
  Appendix~\ref{sec:firstexperimentsmixed}). The
  main 
  advantage of the extended format is that, compared with double precision, its
  arithmetic comes without any or only a small loss in speed. On the
  downside, we cannot make any further adjustments in the algorithm to 
  positively effect its robustness and parallelism. 
  We do not include test results in the experimental section; however, we
  tested the approach and results can be found in Appendix~\ref{sec:firstexperimentsmixed}.  

\paragraph{Double/quadruple.} As quadruple precision arithmetic is not
  widely supported by today's processors, we had to resort to a rather slow
  software-simulated arithmetic. For this reason, we
  used double precision for the initial approximation and for the refinement
  of the eigenvalues. The necessary intermediate data conversions make the
  mixed precision approach slightly more complicated to implement than
  the {\it single/double} one. 
  We used the value $10^{-10}$ for $gaptol$ in our tests. Further details
  can be found in~\cite{mixedtr}.  

\subsection{Portability} 
The biggest problem of the mixed precision approach
is a potential lack of support for the involved data types. As single and
double precisions are supported by 
virtually all machines, languages, and compilers, the mixed precision
approach can be incorporated to any linear algebra library for single
precision input/output. However, for double precision input/output, we
need to resort to either 
extended or quadruple precision.  
Not all architectures, languages, and compilers support these formats. For
instance, the 80-bit floating point format is not supported by all
processors. Furthermore, while the FORTRAN {\tt REAL*10} data
type is a non-standard feature of the language and is not supported by all
compilers, a C/C++ code can use the standardized {\tt long double} data type
(introduced in ISO C99) that achieves the 
desired result on most architectures that support 80-bit arithmetic. 
For the use of quadruple precision, there are presently two major drawbacks:
($i$) it is usually not supported in hardware, which means that one has to resort to
a rather slow software-simulated arithmetic, and ($ii$) the support from compilers  
and languages is rather limited. 
While FORTRAN has a {\tt REAL*16}
data type, the quadruple precision data type in C/C++ is compiler-dependent:
  for instance, there exist the {\tt \rule{8pt}{0.5pt}float128} and {\tt
  \rule{4pt}{0.5pt}Quad} data types for the GNU and Intel
compilers, respectively. An external library implementing the
software arithmetic might be used for portability. In all cases, the
performance of quadruple arithmetic depends on its specific 
implementation. It is however likely that the hardware/software support for
quadruple precision will be improved in the near future.  

\subsection{Robustness}
The mixed precision MRRR provides improved robustness. 
To quantify the robustness of an eigensolver, as discussed in
Section~\ref{sec:objectives}, we propose the
following measure: For a given test set of matrices, {\sc TestSet}, the
robustness $\phi$ is expressed as 
\begin{equation}
  \phi(\mbox{\sc TestSet}) = 1 - \frac{\mbox{\sc NumFailures}}{|\mbox{\sc TestSet}|} \,
\end{equation}
where {\sc NumFailures} is the number of inputs for which the method
``fails''. At this point, we elaborate on what constitutes as
failure.\footnote{For some test cases, even {\tt 
  xSTEVX} or {\tt xSTEDC} fail to return correct
results~\cite{Dhillon98currentinverse,Willems:Diss}.
For measure $\phi$ to be meaningful, the {\sc TestSet} should be
large and include matrices that lead or led to failure
of different solvers.}

If the output does not comply with even the theoretical
error bounds (which includes failing to return a result at all), one or more assumptions
in the derivation of the bounds are not
satisfied. For MRRR, usually a representation that is not relative
robustness is accepted as an RRR. We therefore suggest that cases in which
at least one representation is accepted without passing the test for relative
robustness are considered failures. Furthermore, for any method, we might be more strict
and classify an execution as troublesome if the output exceeds {\it practical}
error bounds as well, which often signals problems earlier on. 

If we use such a strict standard to measure
robustness, we believe that use of mixed precisions might be an important
ingredient for MRRR to achieve robustness comparable to the most reliable solvers
(in particular, implementations of QR).
A number of improvements of MRRR's robustness are proposed in
\cite{Willems:Diss} (e.g., changing the form to represent intermediate tridiagonals, using
of the substructure of clusters to obtain more reliable envelope information,
and allowing shifts inside clusters\footnote{Several of these points were
  already suggested in \cite{Dhillon:Diss}.}). 
We remark that our implementations of the mixed precision MRRR do not implement
all of them, but since these measures are orthogonal to our approach, they
can and should be additionally adopted for maximal robustness. 
\section{Experimental results}
\label{sec:mixed:experiments}

All tests, in this section, were run on an multiprocessors system comprising four
eight-core {\it Intel Xeon X7550 Beckton} processors, with a nominal clock
speed of 2.0 GHz. 
Subsequently, we refer to this
machine as {\sc Beckton} (see Appendix~\ref{appendix:hardware}). 
We used LAPACK version
3.4.2 and linked the library with the vendor-tuned MKL BLAS version
12.1. In addition to the results for LAPACK's routines and our mixed
precision solvers, which we call {\tt mr3smp \hspace{-.4em}(mixed)} subsequently, we also
include results for {\tt 
mr3smp} without mixed precisions. 
All routines of this experiment were compiled with
Intel's compiler version 12.1 and optimization level {\tt -O3} enabled.
Although we present only results for computing {\it all}
eigenpairs (LAPACK's DC does not allow the computation of subsets), we
mention that {\it MRRR's strength and main application lies in the
  computation of subsets of 
eigenpairs}. 

For our tests, we used matrices of size ranging from $2{,}500$ to
$20{,}000$ (in steps of $2{,}500$) of six different types: uniform eigenvalue distribution,
geometric eigenvalue distribution, 1--2--1, Clement, Wilkinson, and
Hermite. The
dimension of the Wilkinson type matrices is $n+1$, as they are
only defined for odd sizes. Details on these matrix types can be found
in Appendix~\ref{appendix:matrices}. To help the exposition of the
results, in the accuracy plots, the matrices are sorted by type first and
then by size; vice versa, in the plots relative to timings, the matrices are
sorted by size first and then by type. Subsequently, we call the test set
{\sc Artificial}. 

We use a second test set, {\sc Application}, which consists
of 45 matrices arising in scientific applications. Most matrices, , listed in
Appendix~\ref{appendix:matrices}, are part of the publicly available {\sc
  Stetester} suite~\cite{Marques:2008} and range from $1{,}074$ to $8{,}012$ in size.

\subsection{Tridiagonal matrices}

\begin{figure}[thb]
   \centering
   \subfigure[Execution time: sequential.]{
     \includegraphics[width=.47\textwidth]{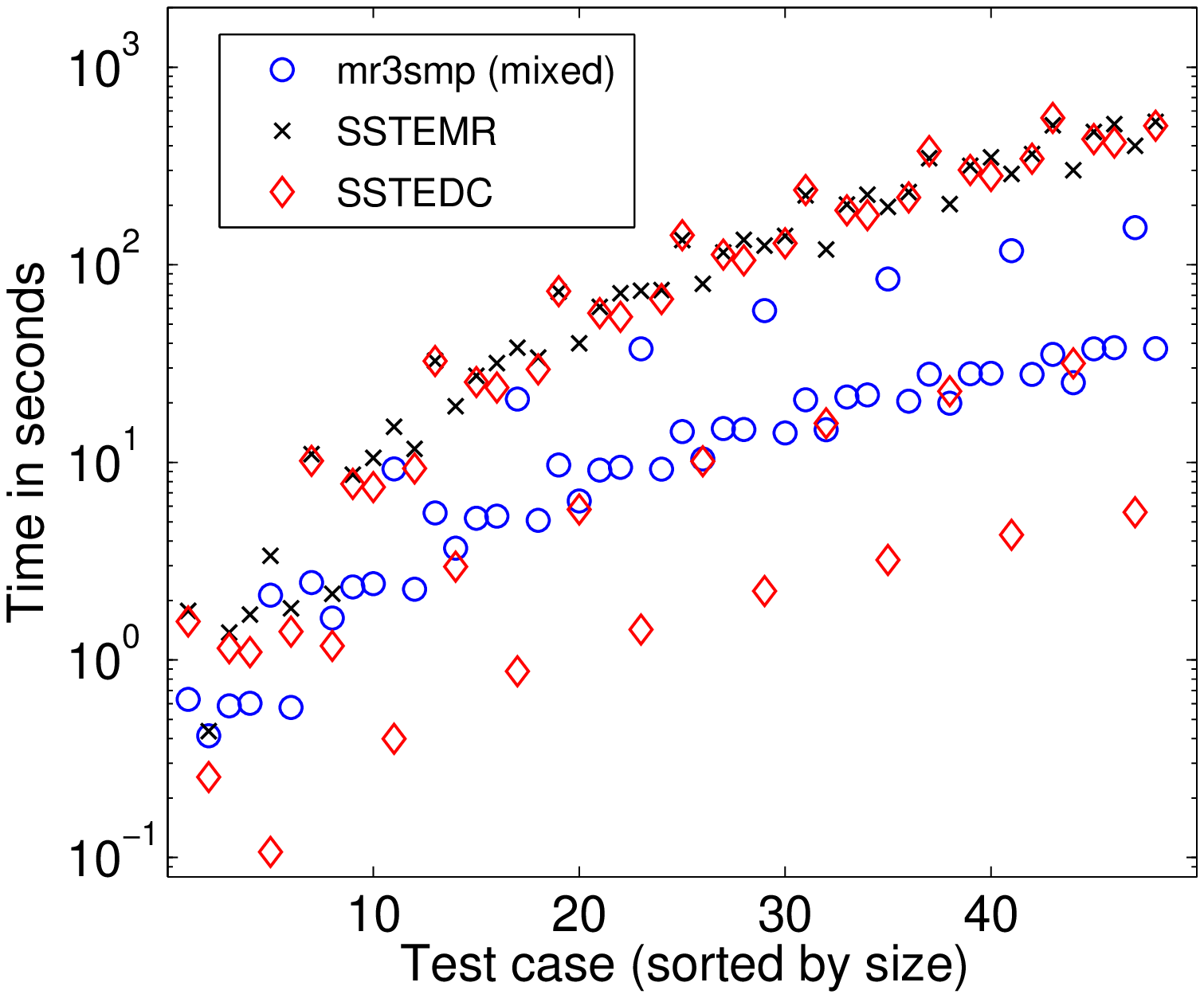}
     \label{fig:timeartifialsinglea}
   } \subfigure[Execution time: multi-threaded.]{
     \includegraphics[width=.47\textwidth]{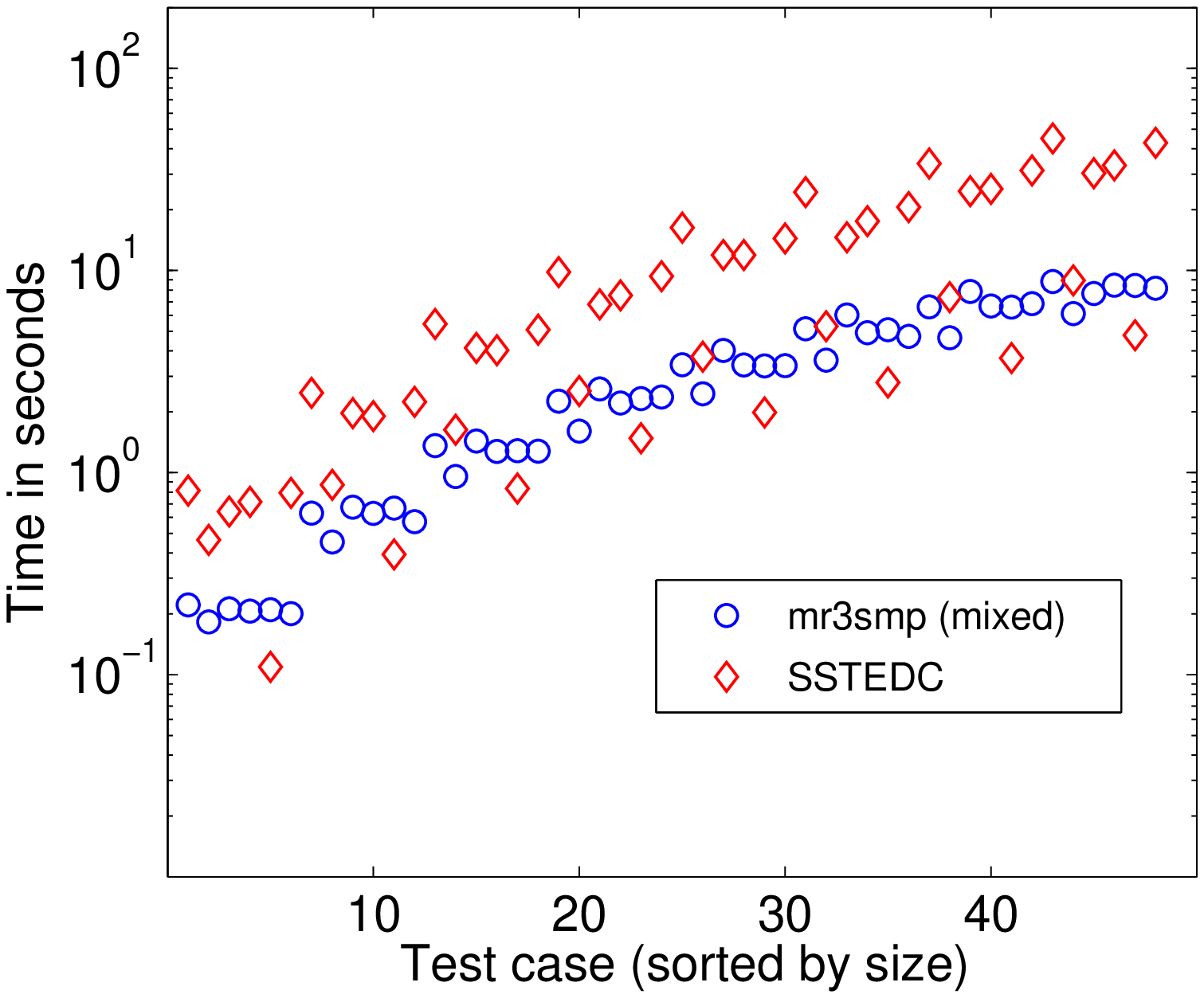}
     \label{fig:timeartifialsingleb}
   }
   \caption{
     Timings for test set {\sc Artificial} on {\sc Beckton}. The results of
     LAPACK's {\tt SSTEMR} (MRRR) and {\tt SSTEDC} (Divide \& Conquer) are
     used as a reference.
   }
   \label{fig:timeartifialsingle}
\end{figure}

\begin{figure}[thb]
   \centering
   \subfigure[Largest residual norm.]{
     \includegraphics[width=.47\textwidth]{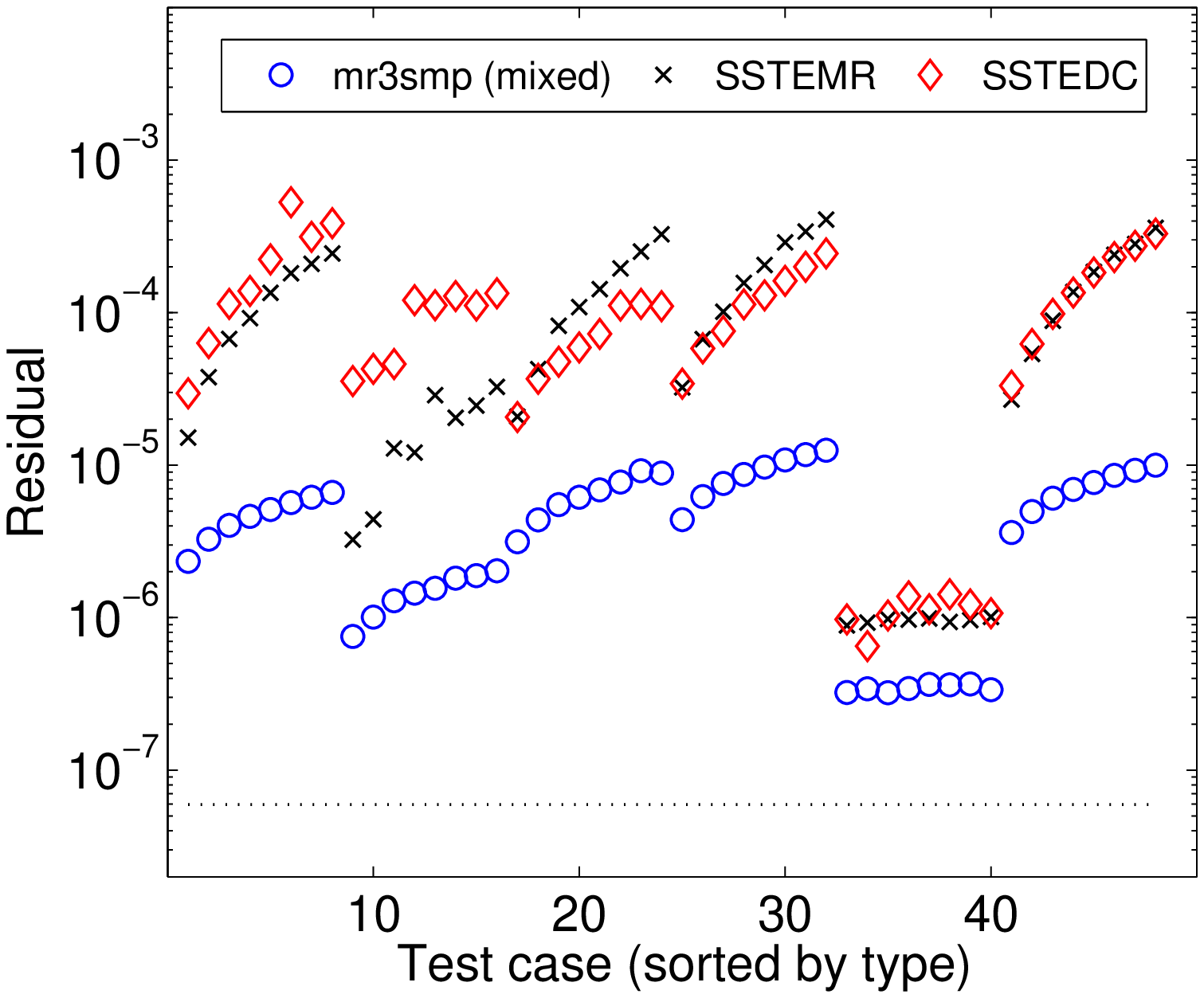}
     \label{fig:accartifialsinglec}
   } \subfigure[Orthogonality.]{
     \includegraphics[width=.47\textwidth]{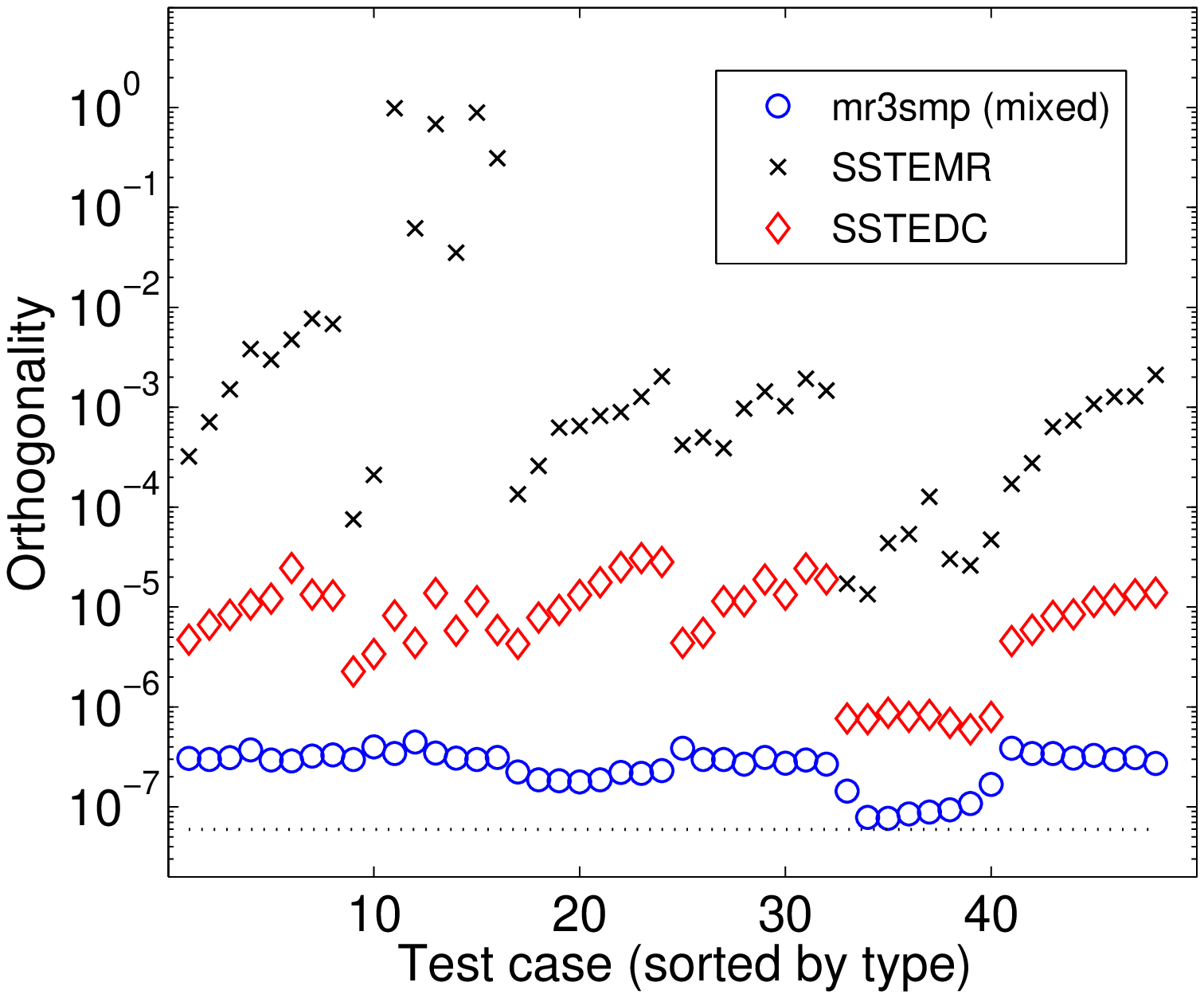}
     \label{fig:accartifialsingled}
   }
     
   \caption{
     Accuracy for test set {\sc Artificial}. The largest residual norm and the
orthogonality are measured as in~\eqref{def:defresortho2}. The results of
     LAPACK's {\tt SSTEMR} (MRRR) and {\tt SSTEDC} (Divide \& Conquer) are
     used as a reference. The dotted lines indicate unit roundoff $\varepsilon_s$.
   }
   \label{fig:accartifialsingle}
\end{figure}

For the {\sc Artificial} matrices in single precision,
Figs.~\ref{fig:timeartifialsingle} and \ref{fig:accartifialsingle} shows
timing and accuracy results, respectively.  
As a reference, we include
results for LAPACK's {\tt SSTEMR} (MRRR) and {\tt SSTEDC} (Divide \&
Conquer).\footnote{For all relevant LAPACK routine names, see
  Appendix~\ref{appendix:routinenames}.}  
Even in a sequential execution, Fig.~\ref{fig:timeartifialsinglea}, our
mixed precision solver {\tt mr3smp \hspace{-.4em}(mixed)} is up to an 
order of magnitude {\it
  faster} than LAPACK's {\tt SSTEMR}. 
For one type of matrices, {\tt SSTEDC} is considerably faster than for all
the others. These are the Wilkinson matrices, which represent a class of
matrices that allow for heavy deflation within the Divide \& Conquer
approach. For all other matrices, which do 
not allow such extensive deflation, our solver is usually {\it faster} than {\tt
  SSTEDC}. As seen in Fig.~\ref{fig:timeartifialsingleb}, in a
parallel execution with one thread/core, the performance gap for the
Wilkinson matrices almost entirely vanishes, while for the other matrices
{\tt mr3smp \hspace{-.4em}(mixed)} remains faster than {\tt SSTEDC}. As depicted in
Fig.~\ref{fig:accartifialsingle}, our routine is not only as accurate as
desired, but it is the most accurate. In particular, for the large matrices
with geometric eigenvalue distribution, {\tt SSTEMR} even fails to return
numerically orthogonal eigenvectors, while {\tt mr3smp
  \hspace{-.4em}(mixed)} returns accurate results.

\begin{figure}[t]
   \centering
   \subfigure[Single precision.]{
     \includegraphics[width=.47\textwidth]{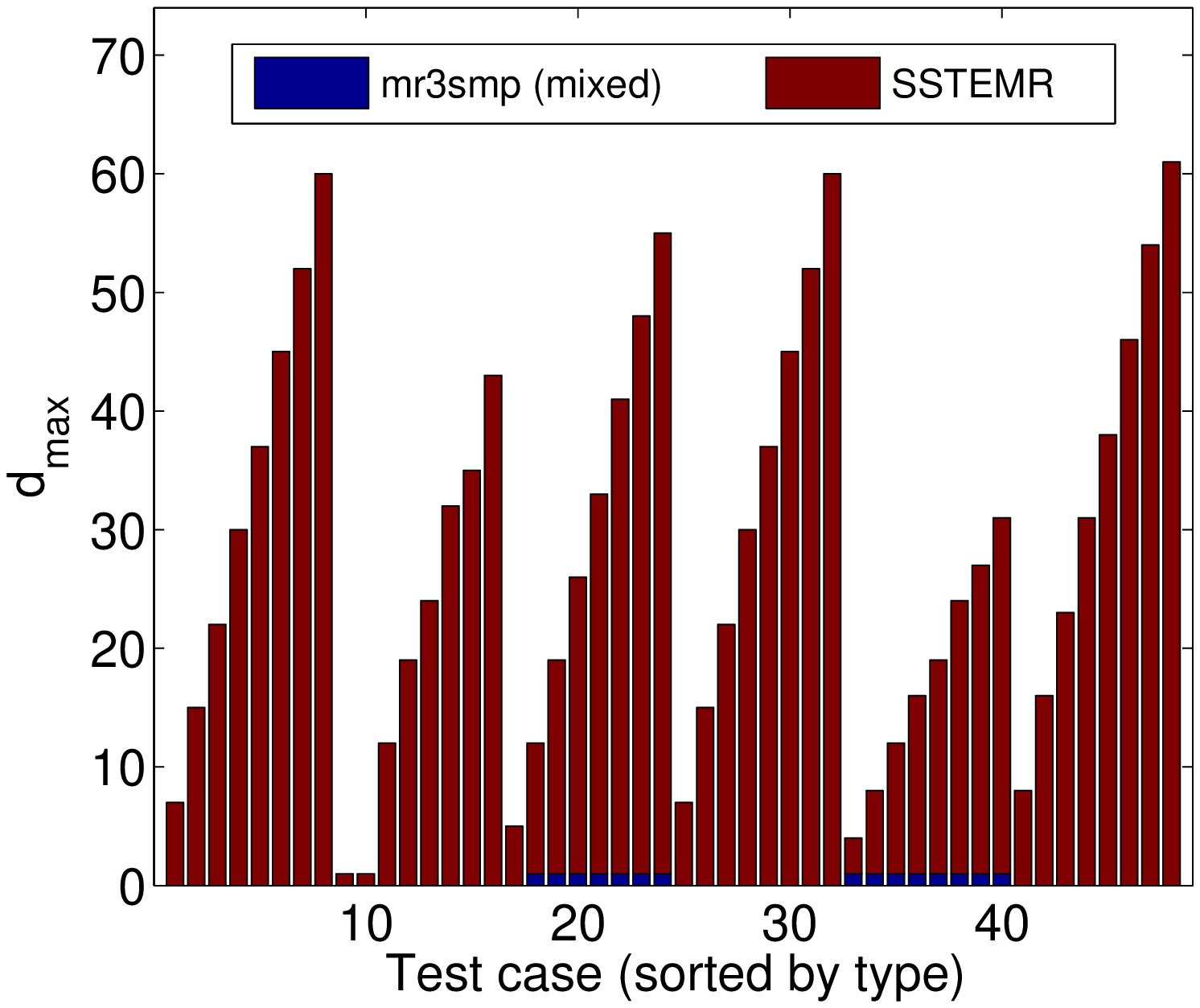}
     \label{fig:dmaxartifiala}
   } \subfigure[Double precision.]{
     \includegraphics[width=.47\textwidth]{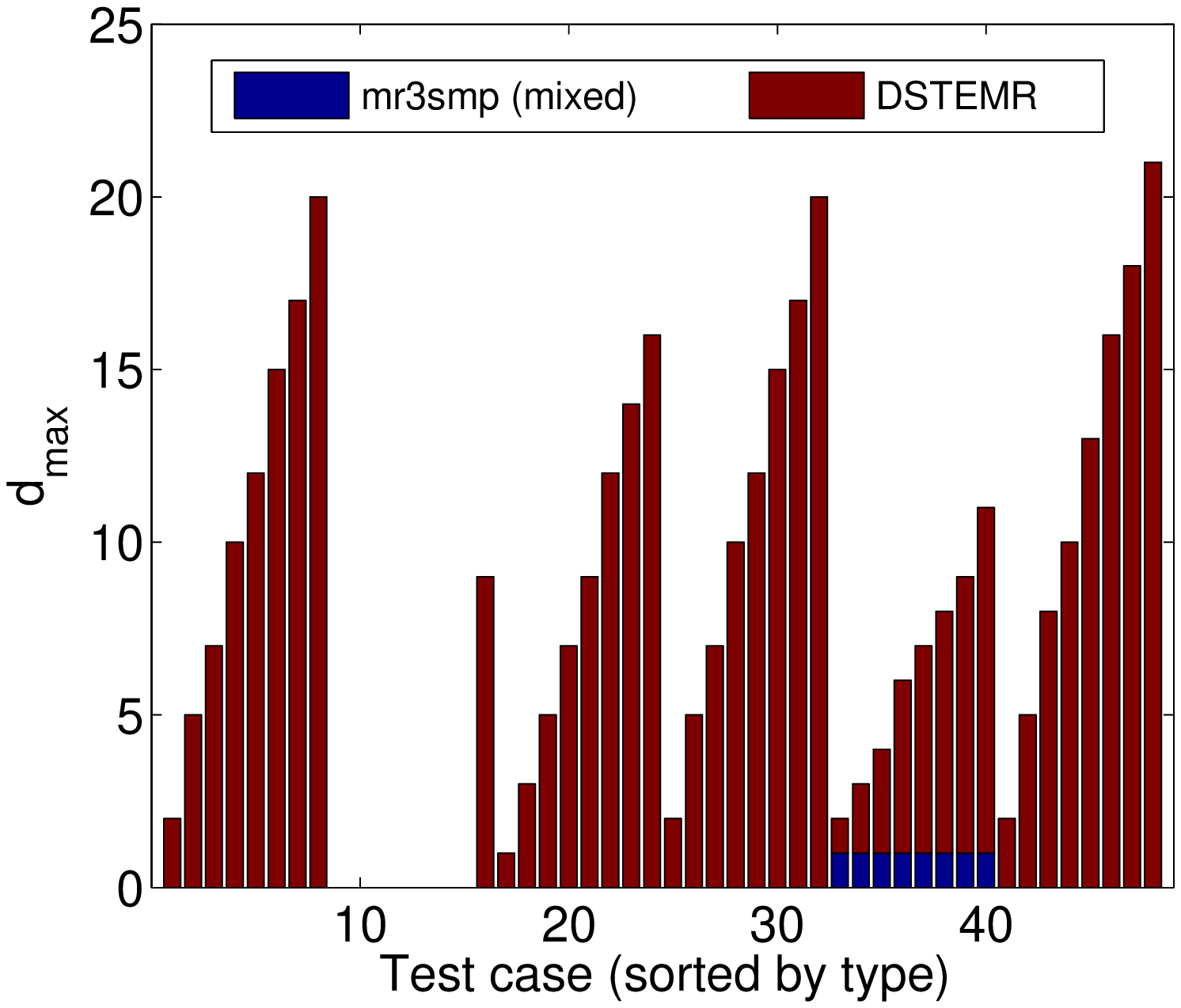}
     \label{fig:dmaxartifialb}
   }
   \caption{
     Maximal depth of the representation tree, $d_{max}$.
   }
   \label{fig:dmaxartifial}
\end{figure}

The faster execution of the mixed precision solver relative to {\tt SSTEMR} is explained by
Fig~\ref{fig:dmaxartifiala}, which shows the maximal depth of the
representation tree, $d_{max}$.
A standard MRRR using single precision requires
significant effort to construct a sequence of relative robust representations
from which the eigenpairs are computed. In contrast, using mixed precisions,
$d_{max}$ is limited to one -- for all but two matrix types, it is even zero. 

If we consider each computation in which MRRR accepts at least one
representation not passing the test for relative robustness as a failure,
for {\tt SSTEMR}, 38 out of the 48 test cases are problematic and
$\phi(\mbox{\sc Artificial}) \approx 0.21$. The number indicates that in
almost 80\% of the test cases, {\tt SSTEMR} 
might produce erroneous results. However, only in three out of the 38
problematic cases this is reflected in an orthogonality exceeding
$n\varepsilon/gaptol$. Consequently, an improved test for relative
robustness, such as proposed in \cite{Willems:Diss}, could probably reduce
the number of failures significantly. Even without altering the selection of
RRRs, {\tt mr3smp \hspace{-.4em}(mixed)} was able to find suitable representations and
$\phi(\mbox{\sc Artificial}) = 1$. Furthermore, the orthogonality is bounded
by $\varepsilon \sqrt{n}$ as desired. 
For single precision input/output
arguments, we obtain a solver that is more accurate {\it and} faster than the
original single precision solver. In addition,
it is more robust and more scalable. 

\begin{figure}[thb]
   \centering
   \subfigure[Execution time: sequential.]{
     \includegraphics[width=.47\textwidth]{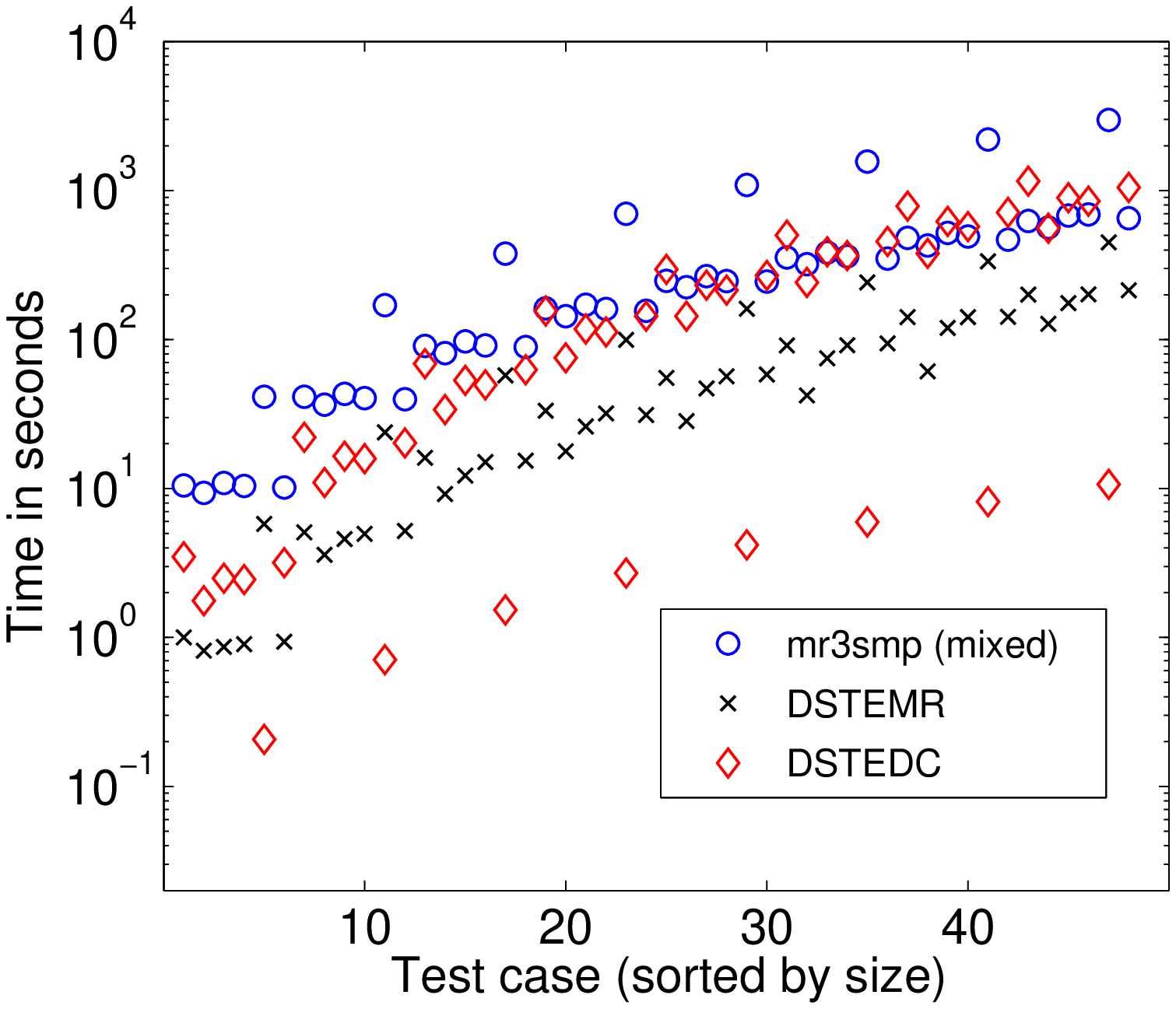}
     \label{fig:timeartifialdoublea}
   } \subfigure[Execution time: multi-threaded.]{
     \includegraphics[width=.47\textwidth]{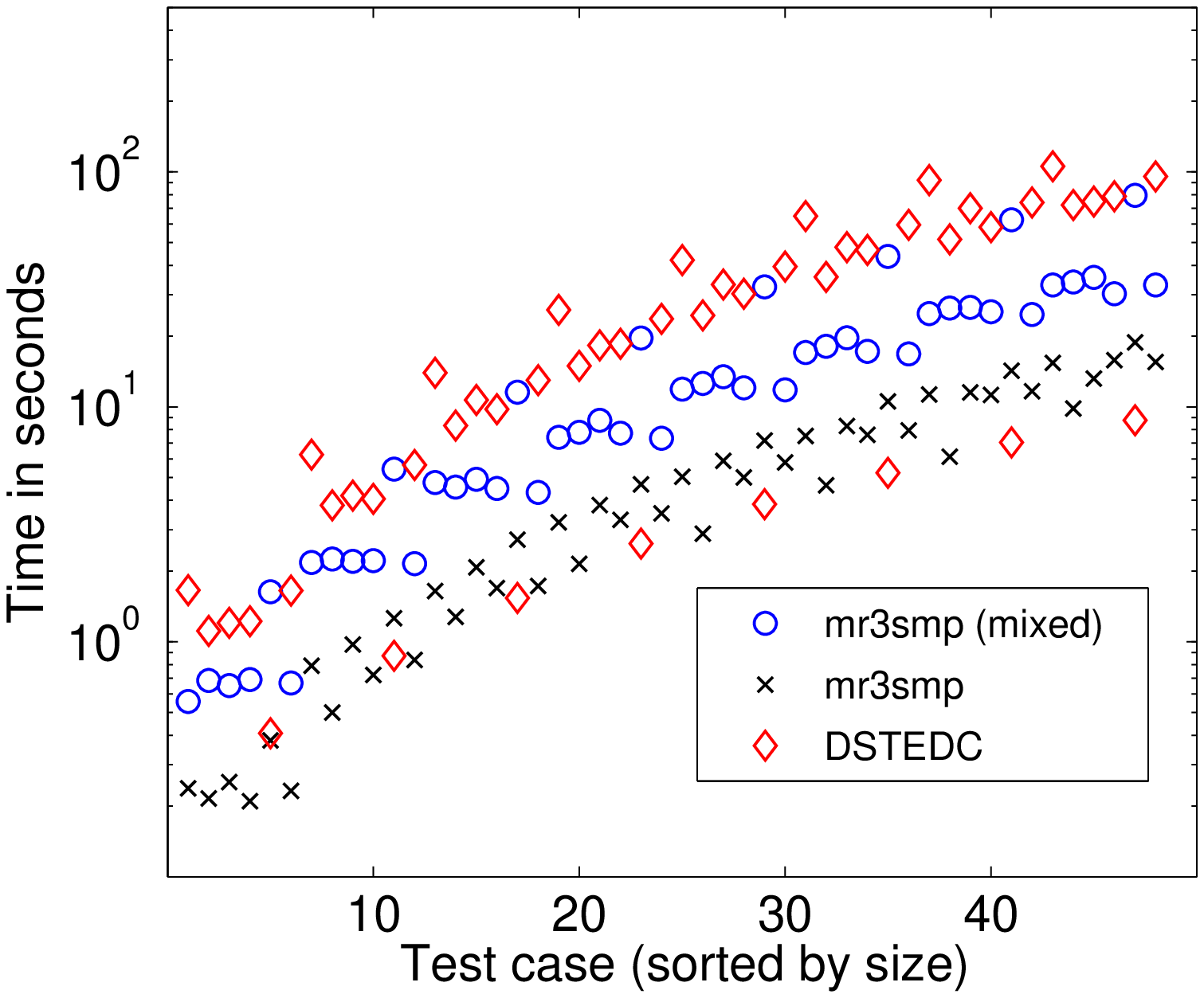}
     \label{fig:timeartifialdoubleb}
   }
   \caption{
     Timings for test set {\sc Artificial} on {\sc Beckton}. The results of
     LAPACK's {\tt DSTEMR} (MRRR) and {\tt DSTEDC} (Divide \& Conquer), as
     well as the multi-threaded {\tt mr3smp} as introduced Chapter \ref{chapter:parallel}, are
     used as a reference.  
   }
   \label{fig:timeartifialdouble}
\end{figure}

\begin{figure}[thb]
   \centering
   \subfigure[Largest residual norm.]{
     \includegraphics[width=.47\textwidth]{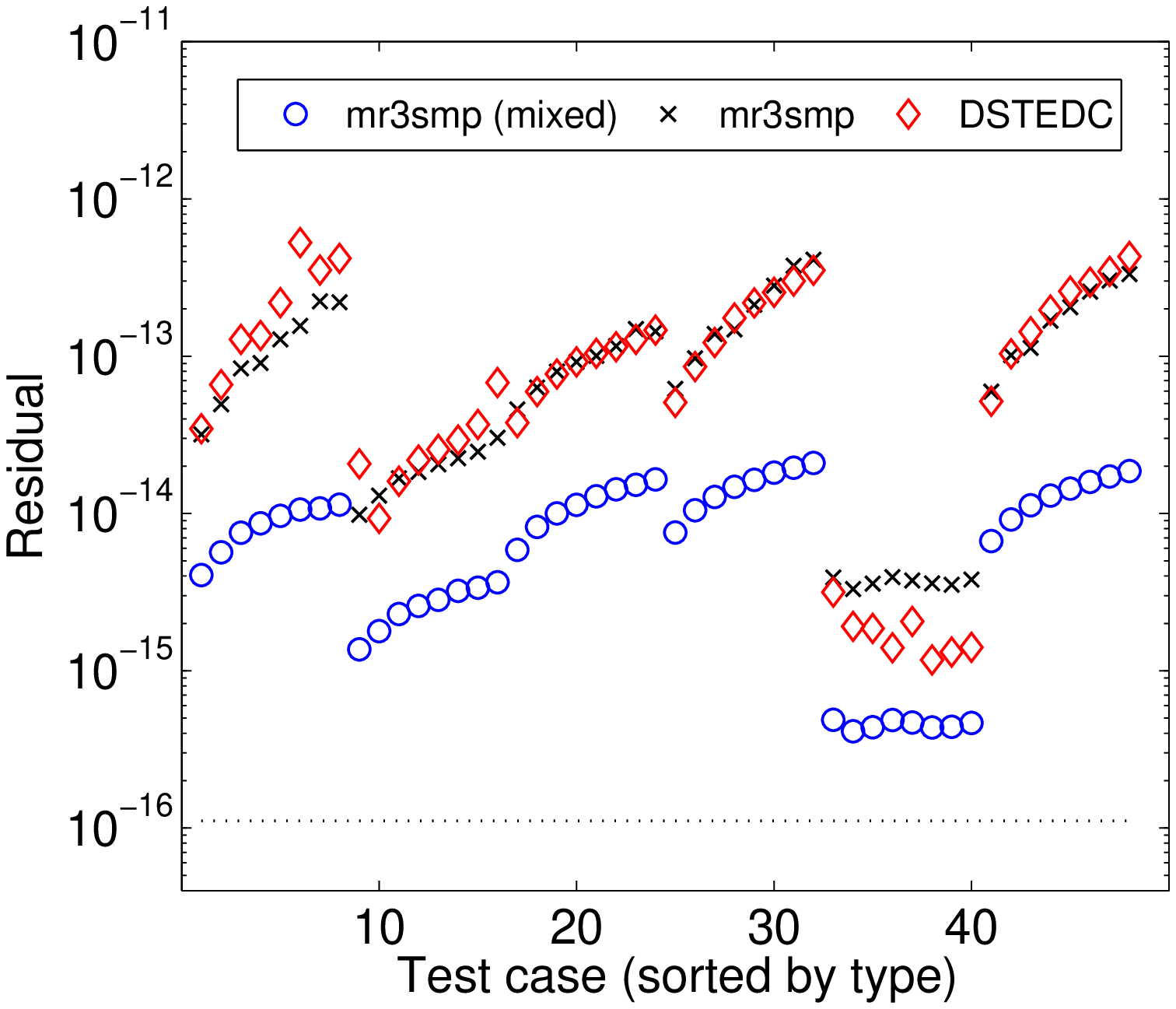}
     \label{fig:accartifialdoublec}
   } \subfigure[Orthogonality.]{
     \includegraphics[width=.47\textwidth]{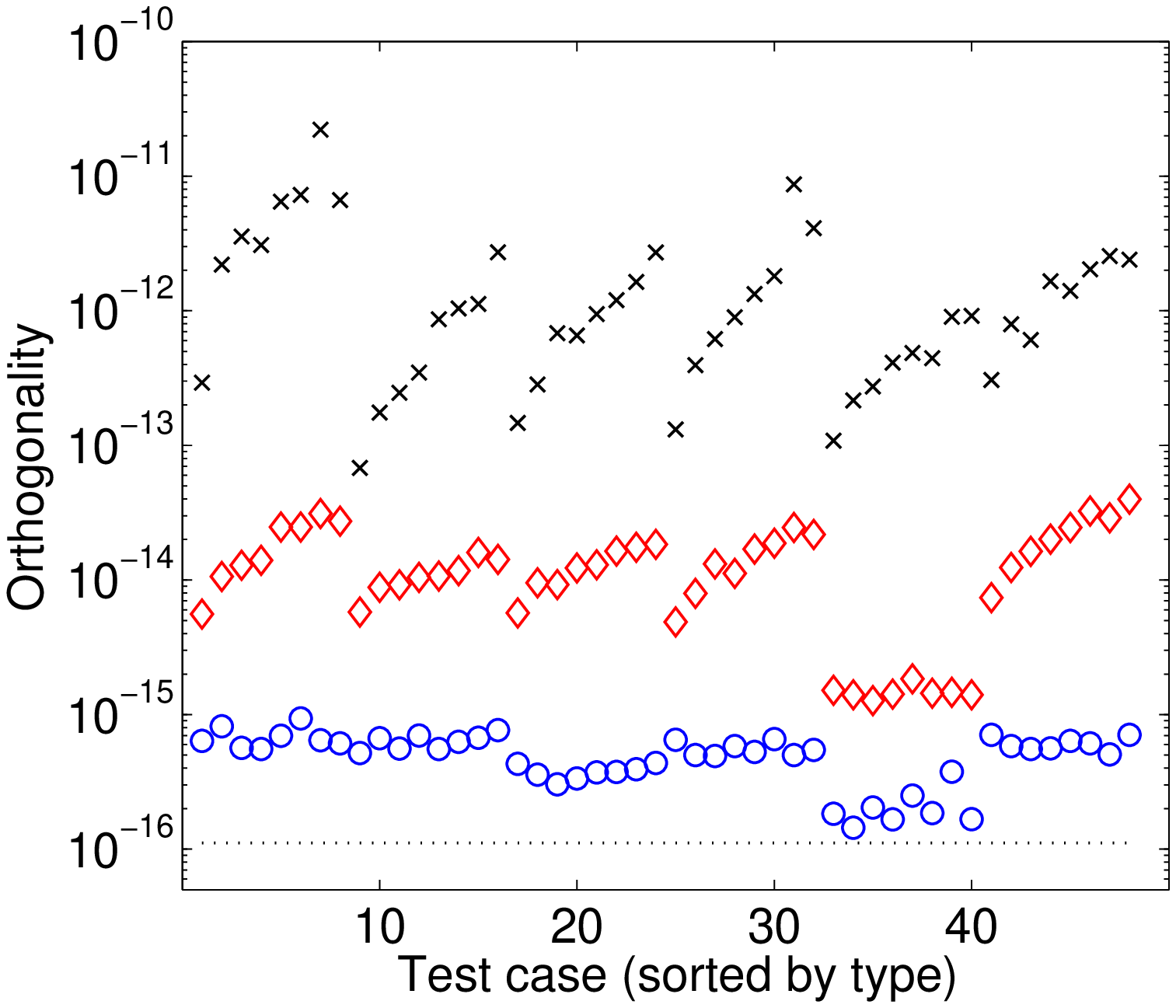}
     \label{fig:accartifialdoubled}
   }     
   \caption{
     Accuracy for test set {\sc Artificial}. The largest residual norm and the
orthogonality are measured as in~\eqref{def:defresortho2}. The results of {\tt mr3smp} 
    and  LAPACK's {\tt DSTEDC} (Divide \& Conquer) are
     used as a reference. In general, {\tt mr3smp} obtains accuracy equivalent to 
    LAPACK's {\tt DSTEMR} (MRRR). The dotted lines indicate unit roundoff $\varepsilon_d$.
   }
   \label{fig:accartifialdouble}
\end{figure}

We now turn our attention to double precision input/output, for which timings and
accuracy are presented in Figs.~\ref{fig:timeartifialdouble} and
\ref{fig:accartifialdouble}, respectively. 
We included the results for {\tt mr3smp} without mixed precisions, which 
in the sequential case is just a wrapper to LAPACK's {\tt DSTEMR}. In general,  {\tt
  mr3smp} obtains accuracy equivalent to 
 {\tt DSTEMR}.  

Figure~\ref{fig:timeartifialdoublea} shows timings for sequential
executions: {\tt mr3smp \hspace{-.4em}(mixed)} is slower than {\tt DSTEMR}, which is
not 
a surprise, as we make use of {\it software-simulated} quadruple precision
arithmetic. What might be a surprise is that even with the use of such slow
arithmetic, for large matrices, {\tt mr3smp \hspace{-.4em}(mixed)} is often as fast as {\tt
  DSTEDC}. As in the single precision case, only for matrices that allow
heavy deflation, {\tt
  DSTEDC} is considerably faster. As Fig.~\ref{fig:timeartifialdoubleb}
shows, for parallel executions, such performance difference reduces and is
expected to eventually vanish, see also Fig.~\ref{fig:timetrdeigb} in
Section~\ref{sec:studyscalapackissues}. For matrices 
that do not allow for extensive 
deflation, {\tt mr3smp \hspace{-.4em}(mixed)} is about a factor two faster
than {\tt DSTEDC}.  

The reason that {\tt mr3smp \hspace{-.4em}(mixed)} is, despite its use of
software-simulated arithmetic, not much slower than {\tt DSTEMR} is
depicted in Fig.~\ref{fig:dmaxartifialb}. While for {\tt DSTEMR} $d_{max}$
is as large as 21, for {\tt mr3smp \hspace{-.4em}(mixed)}, we
have $d_{max} \leq 1$. In fact, for
all but the Wilkinson type matrices, we have $d_{max}$ equals zero and as a
consequence: {\it no danger of failing to find suitable representations and
embarrassingly parallel computation}. To illustrate the difference
between the standard MRRR and the mixed precision variant, we report the
clustering $\rho \in [1/n,1]$ for various matrices in Table~\ref{tab:clusteringdouble}.
\begin{table}[t]
\centering
\footnotesize
\begin{tabular}{ l l@{\quad\quad} c c c c} \toprule
 Matrix   & Routine & \multicolumn{4}{c}{Matrix size}  \\ 
\cmidrule(r){3-6}
    & & 2{,}500             & 5{,}000         &  10{,}000 &  20{,}000 \\
\midrule  
 {\sc Uniform}   &  {\tt DSTEMR} & 0.60 & 0.80 & 0.90 & 0.95 \\
     & {\tt mr3smp (mixed)}             & 4.00e-4  & 2.00e-4 & 1.00e-4 & 5.00e-5  \\
\midrule
 {\sc Geometric}   &  {\tt DSTEMR} & 4.00e-4 & 2.00e-4 & 1.00e-4 & 0.87 \\
     & {\tt mr3smp (mixed)}             & 4.00e-4  & 2.00e-4 & 1.00e-4 & 5.00e-5  \\
\midrule
 {\sc 1--2--1}   &  {\tt DSTEMR} & 0.43 & 0.64 & 0.81 & 0.90 \\
     & {\tt mr3smp (mixed)}             & 4.00e-4  & 2.00e-4 & 1.00e-4 & 5.00e-5  \\
\midrule
 {\sc Clement}   &  {\tt DSTEMR} & 0.60 & 0.80 & 0.90 & 0.95 \\
     & {\tt mr3smp (mixed)}             & 4.00e-4  & 2.00e-4 & 1.00e-4 & 5.00e-5  \\
\midrule
 {\sc Wilkinson}   &  {\tt DSTEMR} & 0.20  & 0.60 & 0.80 & 0.90 \\
     & {\tt mr3smp (mixed)}             & 8.00e-4 & 4.00e-4 & 2.00e-4 & 1.00e-4 \\
     \bottomrule 
\end{tabular}
\caption{Clustering $\rho \in [1/n,1]$ for different types of test
  matrices. The results for Hermite type matrices were already presented in
  Table~\ref{tab:clustering} as criterion I 
  (\DSTEMR) and III ({\tt mr3smp} with mixed precision).  
}
\label{tab:clusteringdouble}
\end{table}
(Recall that the smaller $\rho$ the more natural parallelism is provided by the
problem.)  
\DSTEMR\ is confronted with significant
clustering. In contrast, in the mixed precision solver, for all but the
Wilkinson matrices, $\rho = 1/n$. For Wilkinson matrices, clustering $\rho$ was limited to
$2/n$, which still implies ample parallelism. 
The data suggest that our approach is especially well-suited for highly parallel 
systems. In particular, {\it solvers for distributed-memory systems should greatly
benefit from better load balancing and reduced communication.} 

In addition to enhanced parallelism, robustness is improved. For {\tt
  DSTEMR}, we have $\phi(\mbox{\sc Artificial}) \approx 0.40$ and,
consequently, the accuracy of {\tt DSTEMR} might have turned out problematic
for about 60\% of the inputs. However, no execution actually resulted in
insufficient accuracy. For {\tt mr3smp \hspace{-.4em}(mixed)}, we have
$\phi(\mbox{\sc Artificial}) = 1$, that is, all computed
representations passed the test for relative robustness. Furthermore, as
Fig.~\ref{fig:accartifialdouble} shows, {\tt mr3smp \hspace{-.4em}(mixed)} attains
excellent accuracy; both the residuals and the orthogonality are improved to
the desired level. In fact, the latter is several orders of magnitude better
than what can be expected by the standard MRRR. 


\subsection{Real symmetric dense matrices}
\label{section:densemixedexperiments}

For single precision inputs [Fig.~\ref{fig:timeartifialsingle}] or for
double precision inputs in a parallel 
setting [Fig.~\ref{fig:timeartifialdoubleb}], in terms of execution time,
our mixed precision tridiagonal
eigensolver is highly competitive with
Divide \& Conquer and the standard MRRR. 
Hence, when used in context of dense Hermitian
eigenproblems, the accuracy improvement of the tridiagonal stage carry over to the
dense problem without any performance penalty. 
Even for sequential executions with double precision inputs
[Fig.~\ref{fig:timeartifialdoublea}], the slowdown 
due to mixed precisions is often not dramatic. The reason is that, to compute $k$ 
eigenpairs, MRRR only requires $\order{kn}$ arithmetic operations,
while the reduction to tridiagonal form requires $\order{n^3}$
operations. Consequently, the time spent in the tridiagonal stage is
asymptotically negligible. 

\begin{figure}[thb]
   \centering
   \subfigure[Execution time: sequential.]{
      \includegraphics[width=.47\textwidth]{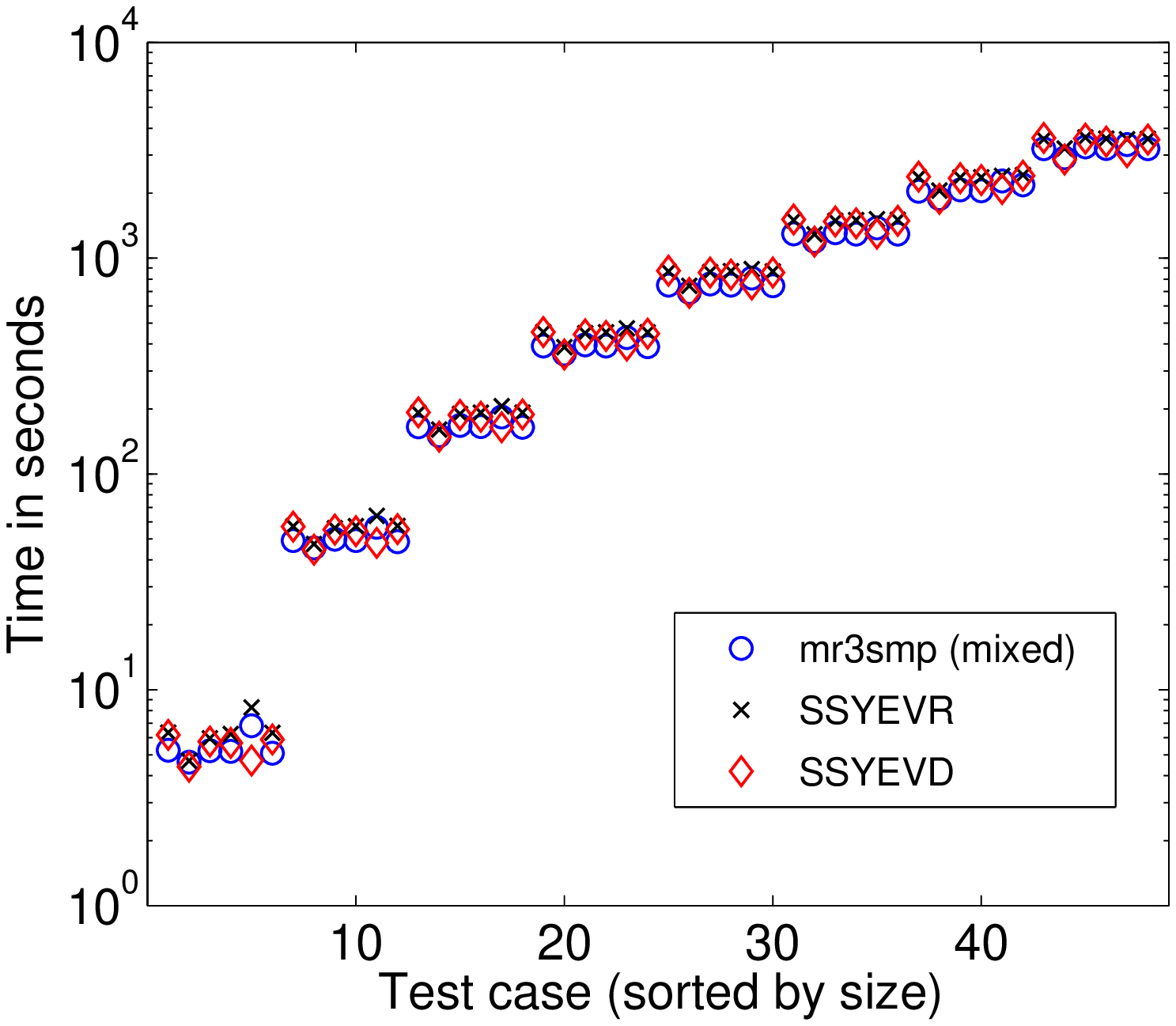}    
      \label{fig:timeartifialsingledensea}
    } \subfigure[Execution time: multi-threaded.]{
      \includegraphics[width=.47\textwidth]{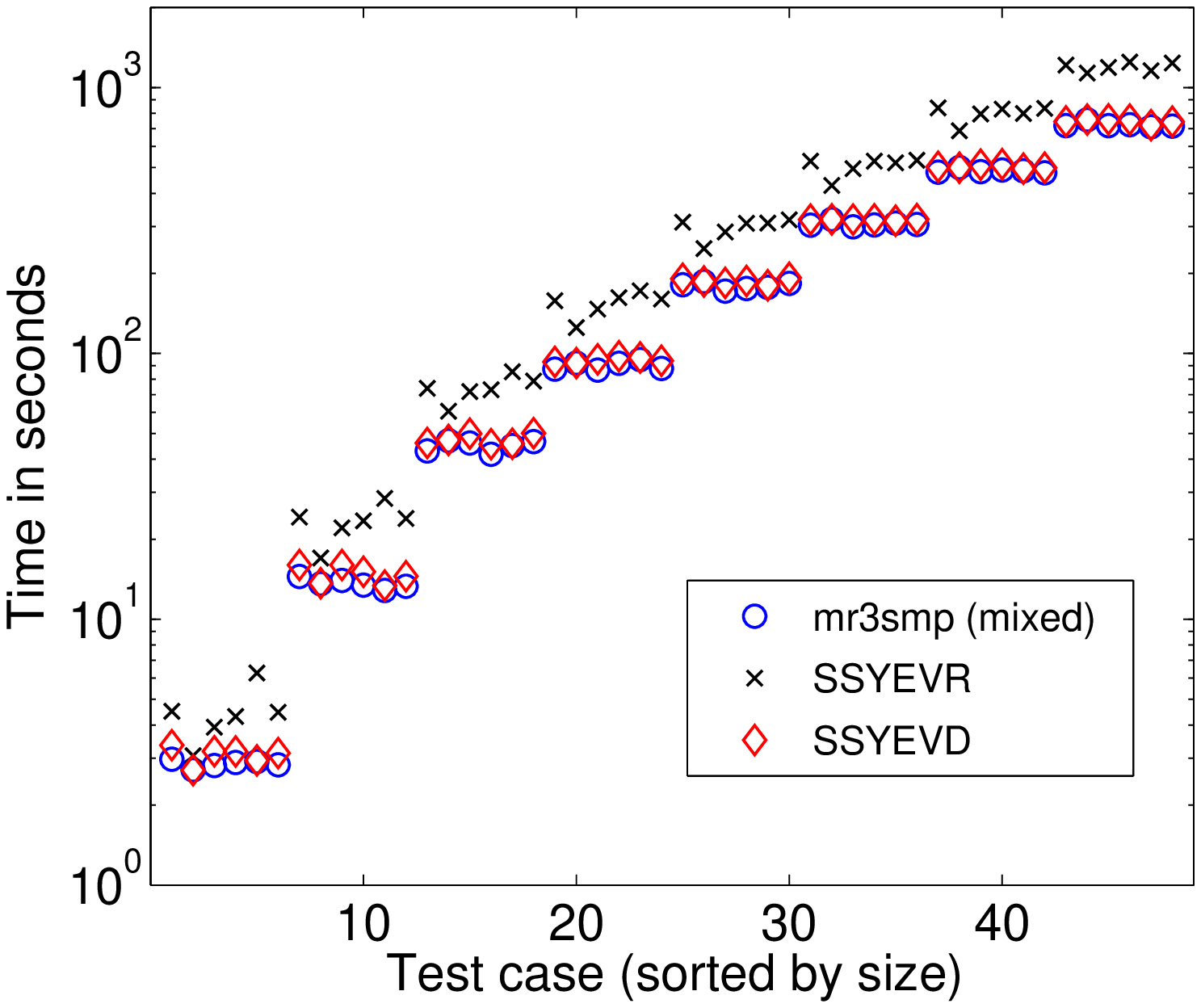}        
      \label{fig:timeartifialsingledenseb}
    }
   \caption{
     Timings for test set {\sc Artificial} on {\sc Beckton}. The results of
     LAPACK's {\tt SSYEVR} (MRRR) and {\tt SSYEVD} (Divide \& Conquer) are
     used as a reference. In a multi-threaded execution, {\tt SSYEVR} is
     slower than the other two routines, as it makes use of the sequential
     routine {\tt SSTEMR}.  
   }
   \label{fig:timeartifialsingledense}
\end{figure}

\begin{figure}[thb]
   \centering
    \subfigure[Largest residual norm.]{
      \includegraphics[width=.47\textwidth]{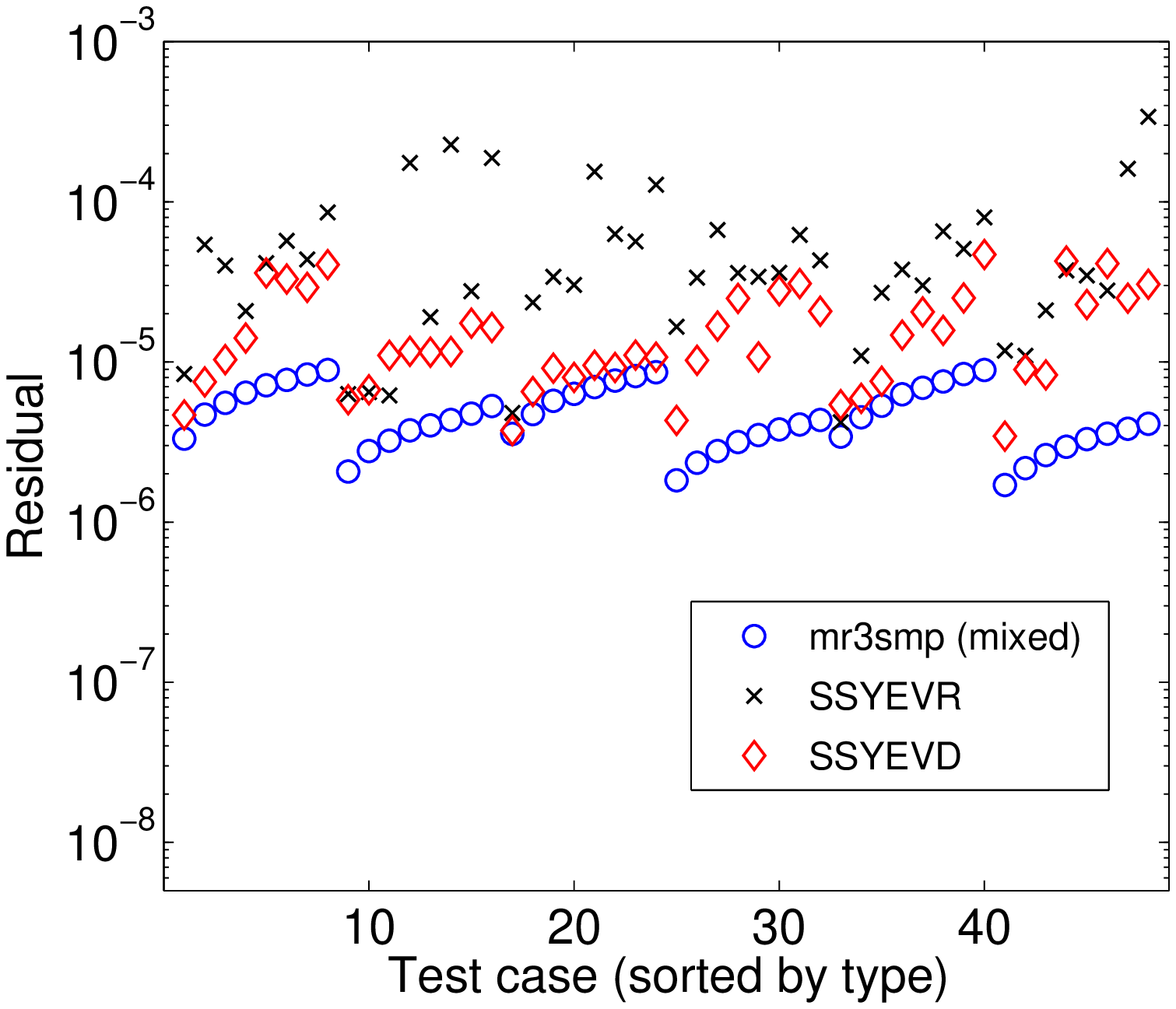}        
    } \subfigure[Orthogonality.]{
      \includegraphics[width=.47\textwidth]{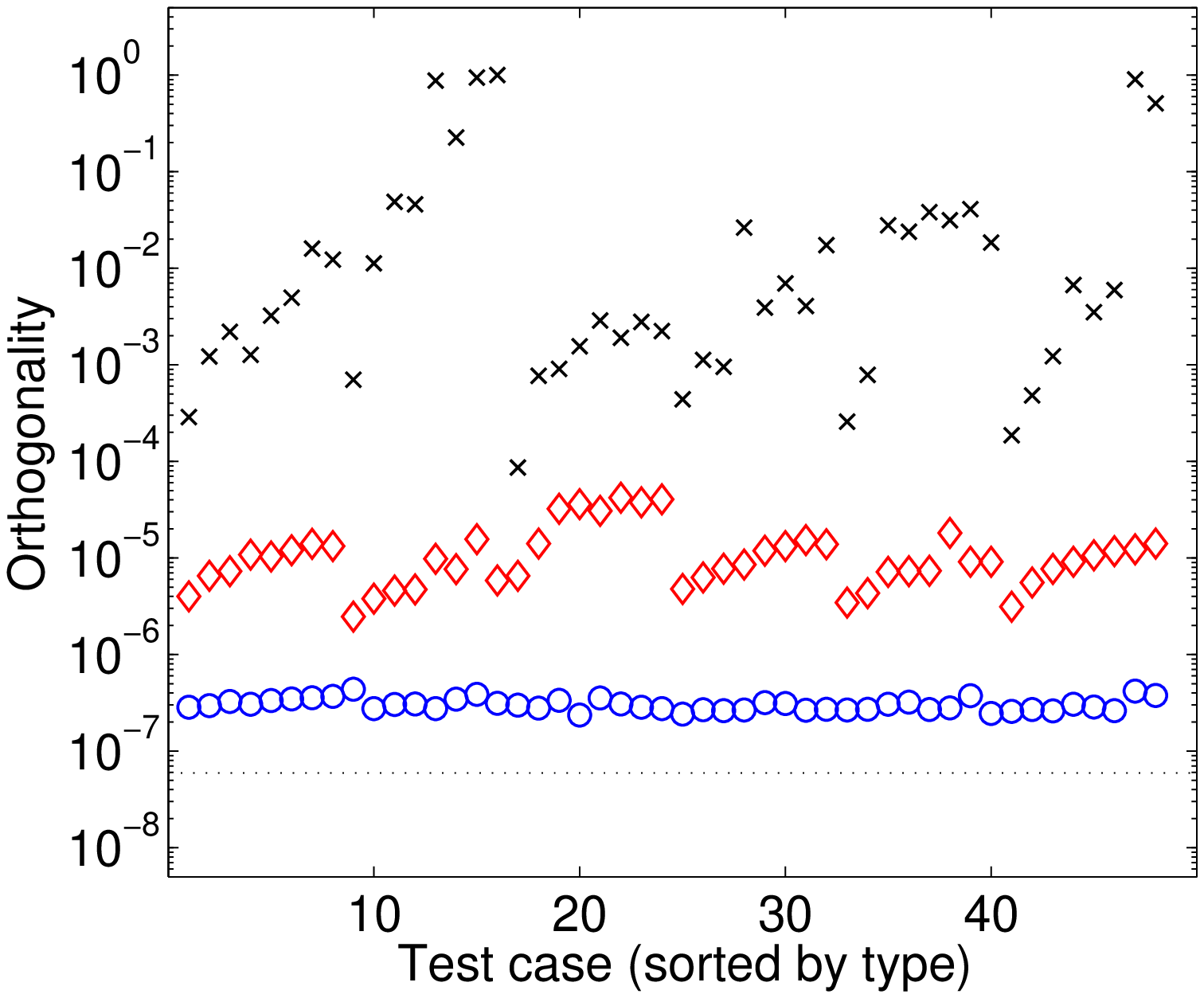}        
    }
   \caption{
     Accuracy for test set {\sc Artificial}. The largest residual norm and the
orthogonality are measured as in~\eqref{def:defresortho2}. The results of
     LAPACK's {\tt SSYEVR} (MRRR) and {\tt SSYEVD} (Divide \& Conquer) are
     used as a reference. The dotted line indicates unit roundoff $\varepsilon_s$.
   }
   \label{fig:accartifialsingledense}
\end{figure}

In Figs.~\ref{fig:timeartifialsingledense} and
\ref{fig:accartifialsingledense}, we present respectively timings and
accuracy for real symmetric matrices in single precision.  
The matrices are generated by applying random orthogonal similarity
transformations to the tridiagonal matrices of the previous experiments:
$A = QTQ^*$, with random orthogonal matrix $Q \in \Rnn$.\footnote{Similar to LAPACK's
auxiliary routine {\tt xLARGE}.} 
By Theorem~\ref{thm:gapthm}, the error in the eigenvalue is a lower bound on
the residual norm: $|\hat{\lambda} - \lambda| \leq \norm{A \hat{x} -
  \hat{\lambda} \hat{x}}$. In general, we can only hope to compute
eigenvalues with an $\order{n \varepsilon \norm{A}}$ error and therefore do
not expect improvements in the residuals. However, the
improvements in the orthogonality directly translate to the dense
eigenproblem. 
Figure~\ref{fig:accartifialsingledense} demonstrates that excellent
accuracy is achieved with the 
mixed precision approach. 
On top of that, as shown in Fig.~\ref{fig:timeartifialsingledense}, {\tt mr3smp \hspace{-.4em}(mixed)} is faster than LAPACK's {\tt 
  SSYEVR} --  sequentially and multi-threaded. In a multi-threaded execution,
it becomes apparent that the tridiagonal stage of {\tt SSYEVR}, {\tt SSTEMR}, is not
parallelized. However, the vast majority of time is spent in the reduction to
tridiagonal form and the backtransformation of the eigenvectors. For a
comparison of Figs.~\ref{fig:timeartifialsingledensea} and \ref{fig:timeartifialsingledenseb}, we remark that the reduction to tridiagonal form, {\tt
  SSYTRD}, does not scale and limits the speedup of a
parallel execution. A similar experiment using the {\sc Application}
matrices can be found in Appendix~\ref{sec:moreexperimentsmixed}. The
experiment underpins that generally good performance and accuracy can be
expected for single precision input/output.

Interestingly, if we allow an
orthogonality of up to $n\varepsilon/gaptol$ (or close to that), {\tt
  SSYEVR} cannot be used reliably for matrices as large as in our test set. 
The orthogonality bound is close to or even exceeds
one. Consequently, the applicability of MRRR is
limited. However, in certain situations, it might be beneficial 
to use single precision computations for such large matrices: the memory requirement and
execution time of a solver is reduced by a factor two. 
The mixed precision MRRR does not introduce such a tight restriction on the
matrix size. As shown in Fig.~\ref{fig:accartifialsingledense}, even for
large matrices, {\tt mr3smp \hspace{-.4em}(mixed)} delivers accuracy to a
level that might be expected. 

We now turn our attention again to double precision
input/output. 
\begin{figure}[t]
   \centering
   \subfigure[Execution time: sequential.]{
      \includegraphics[width=.47\textwidth]{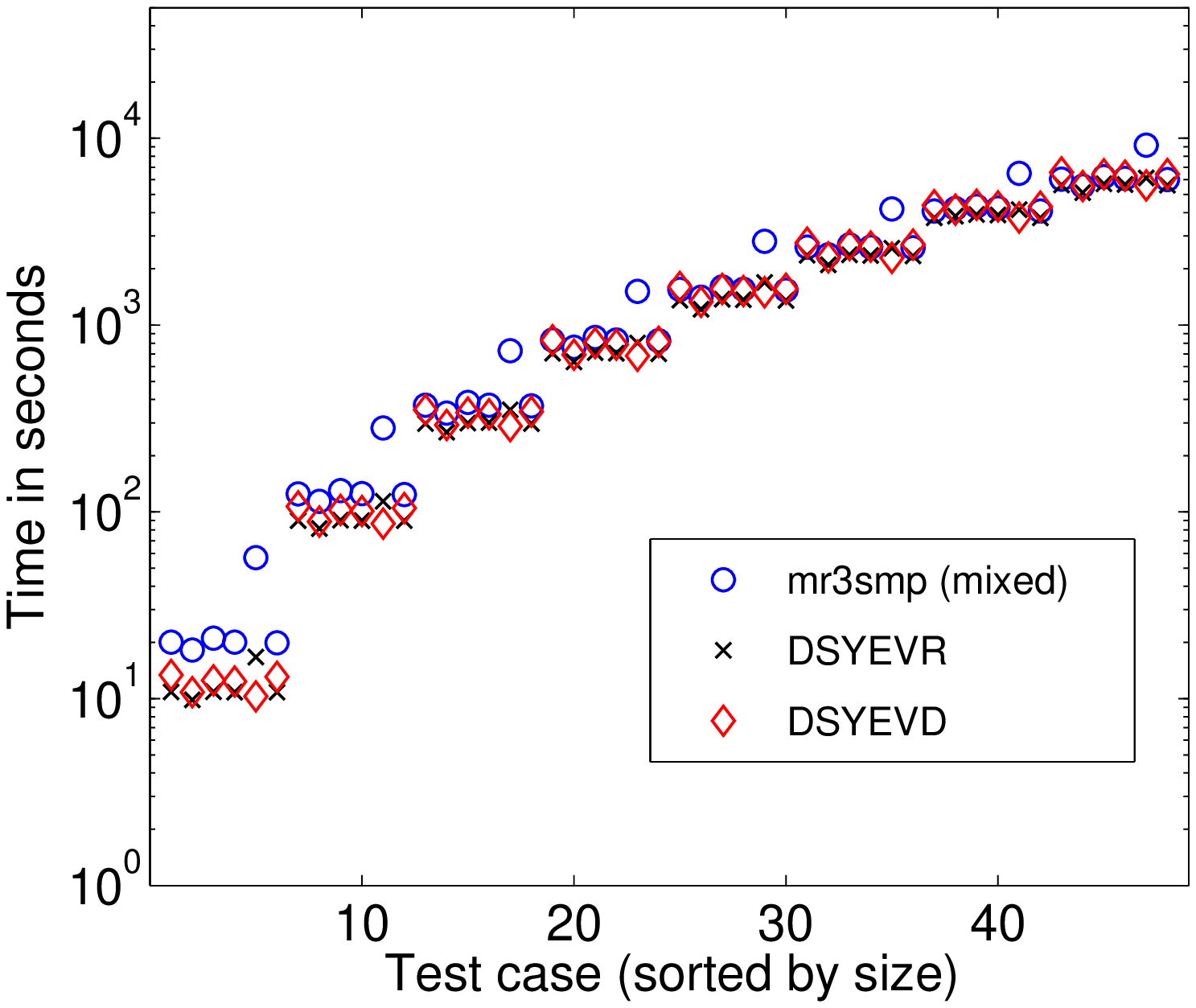}    
      \label{fig:timeartifialdoubledensea}
    } \subfigure[Execution time: multi-threaded.]{
      \includegraphics[width=.47\textwidth]{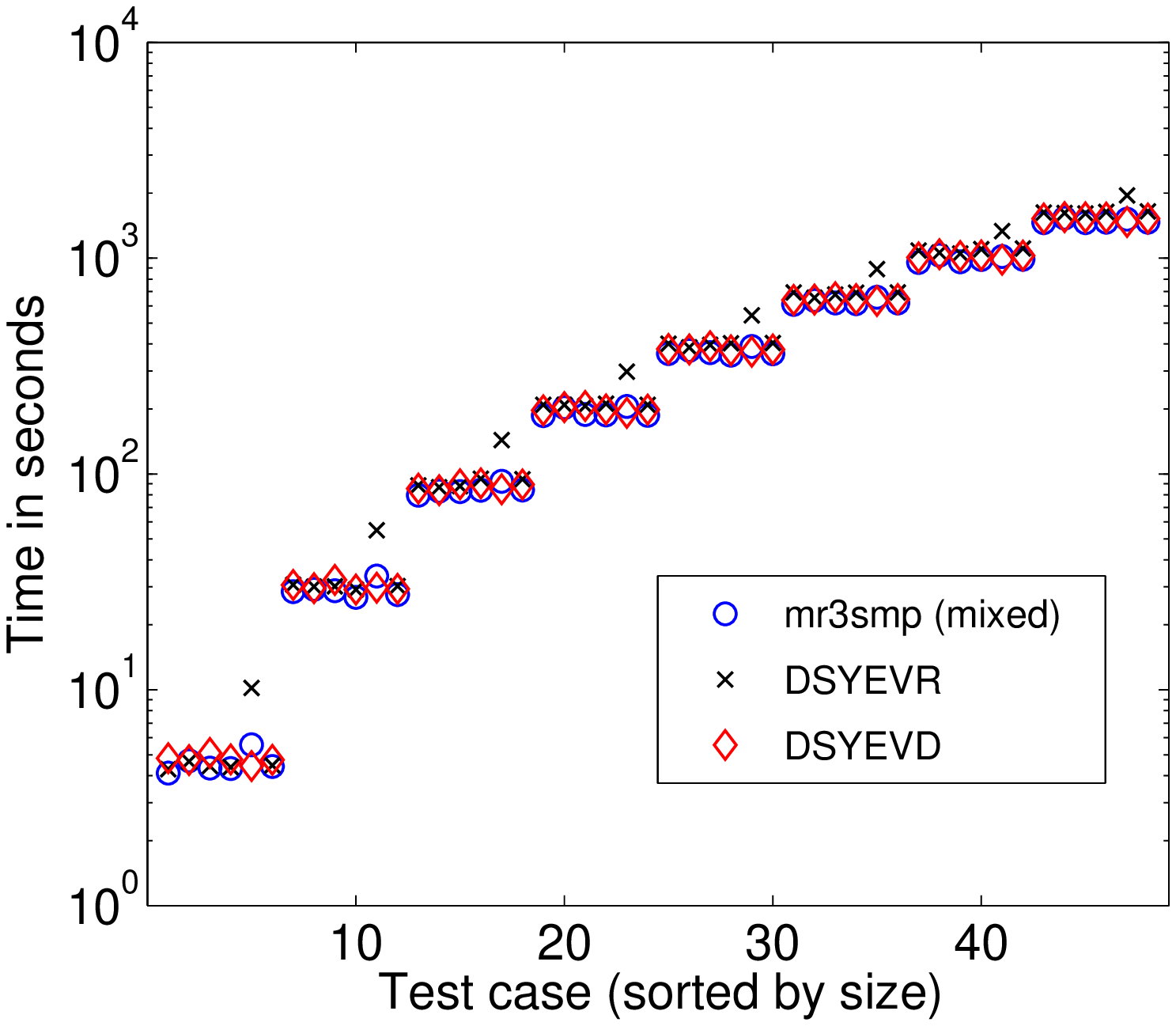}        
      \label{fig:timeartifialdoubledenseb}
    }
   \caption{
     Timings for test set {\sc Artificial} on {\sc Beckton}. The results of
     LAPACK's {\tt DSYEVR} (MRRR) and {\tt DSYEVD} (Divide \& Conquer) are
     used as a reference. 
   }
   \label{fig:timeartifialdoubledense}
\end{figure}
The timings for tridiagonal inputs in
Fig.~\ref{fig:timeartifialdoublea} indicate that, in sequential execution, for small matrix
sizes, our approach introduces overhead. However, as seen in
Fig.~\ref{fig:timeartifialdoubledensea}, for dense eigenproblems, the
execution time is effected less severely.  
In a parallel execution, Fig.~\ref{fig:timeartifialdoubledenseb}, {\tt
  mr3smp \hspace{-.4em}(mixed)} is competitive even for smaller matrices. In
all cases, the improved 
accuracy of the tridiagonal stage carries over to the dense
eigenproblem, as exemplified by Fig.~\ref{fig:accartifialdoubledense}.
\begin{figure}[thb]
   \centering
    \subfigure[Largest residual norm.]{
      \includegraphics[width=.47\textwidth]{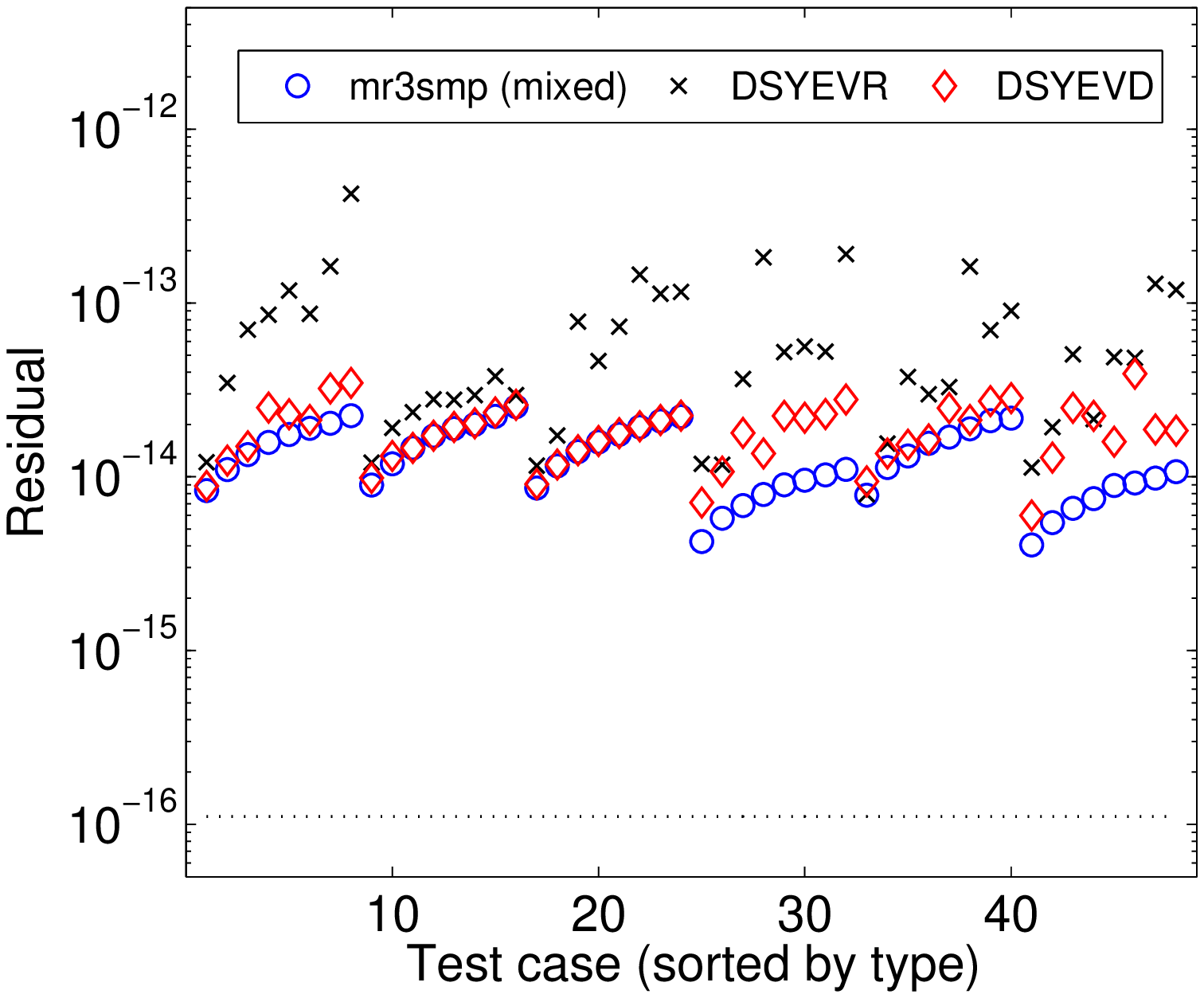}        
    } \subfigure[Orthogonality.]{
      \includegraphics[width=.47\textwidth]{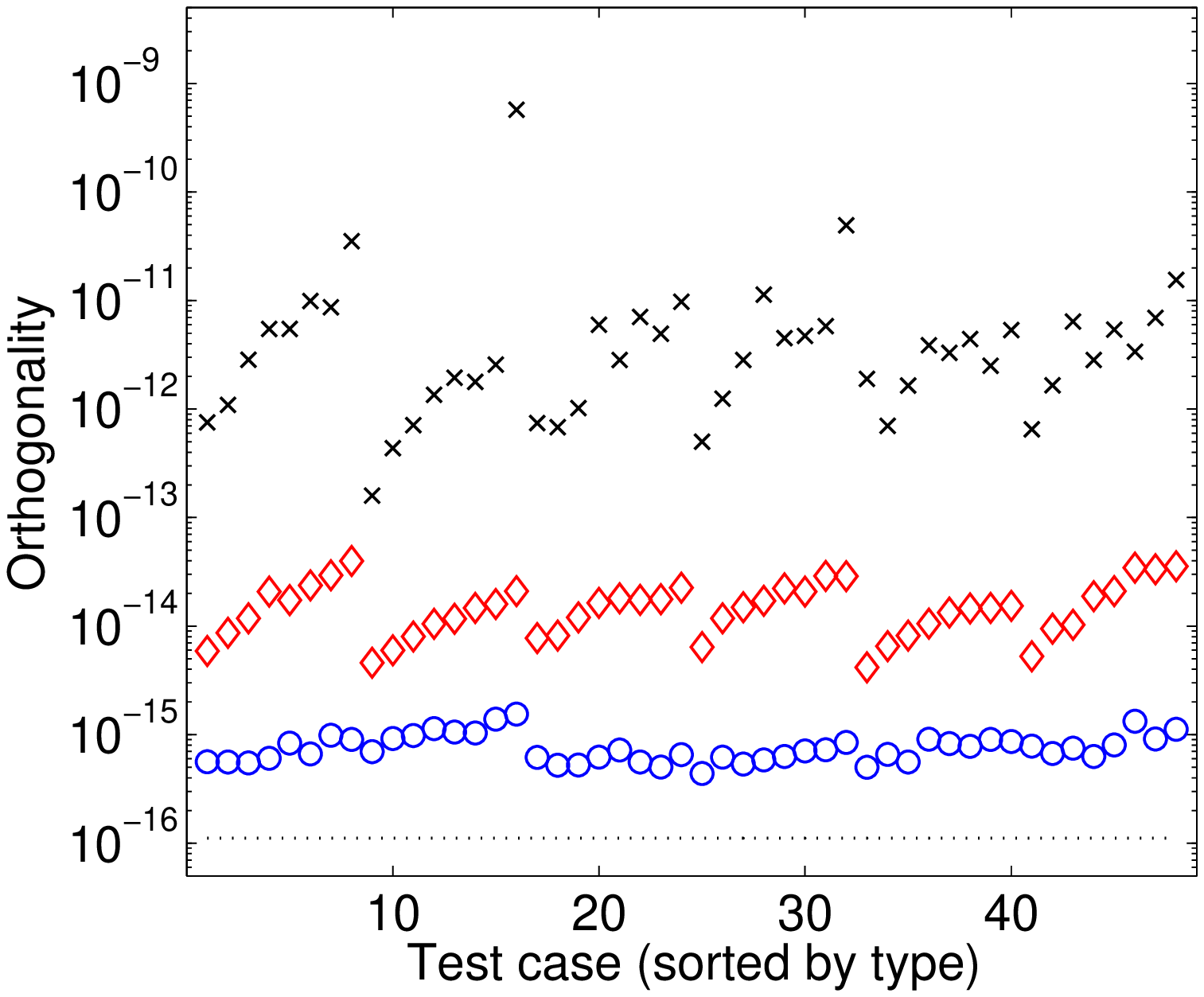}        
    }

   \caption{
     Accuracy for test set {\sc Artificial}. The largest residual norm and the
orthogonality are measured as in~\eqref{def:defresortho2}. The results of
     LAPACK's {\tt DSYEVR} (MRRR) and {\tt DSYEVD} (Divide \& Conquer) are
     used as a reference. The dotted lines indicate unit roundoff $\varepsilon_d$.
   }
   \label{fig:accartifialdoubledense}
\end{figure}

We also performed a similar experiment for the {\sc Application} matrices. 
The results are presented in Appendix~\ref{sec:moreexperimentsmixed} and
support two previously made statements: First, for small matrices, 
the sequential execution is slower than {\tt DSYEVR}, but the performance
gap reduces as the matrix size increases. Second, the accuracy improvements
are usually limited to the orthogonality; the residuals are often comparable
for all solvers. 


The above tests were limited to 
computing {\it all} eigenpairs of {\it real symmetric} matrices. If the matrices were
complex-valued and/or only a subset of eigenpairs were computed, the mixed
precision approach would work even better.\footnote{For complex-valued matrices and
for subset computations, experimental results can be found in
Appendix~\ref{sec:firstexperimentsmixed}. Results for a second test set can
be found in Appendix~\ref{sec:moreexperimentsmixed}.} As a result, using mixed
precisions, we obtain eigensolvers for large-scale Hermitian eigenproblems
that are {\it fast and accurate}. In particular, our solvers compete with
the fast Divide \& Conquer method when all eigenpairs are computed and are
faster when only a small subset of eigenpairs is desired. Furthermore, they
are accurate and promise to be highly scalable.

\chapter{Conclusions}

Recent developments in computer hardware dictate that programs need to make
efficient use of the ever growing parallelism  
in order to improve their performance.  We concentrated on how eigensolvers based on the
algorithm of Multiple Relatively Robust Representations efficiently exploit
modern parallel computers.  

For today's multi-core and future many-core architectures, we presented a
parallelization strategy,  MR$^3$-SMP, that breaks the computation into tasks to be
executed by multiple threads. 
The tasks are both created and scheduled dynamically. While a static division of work would
introduce smaller overheads, the dynamic approach is flexible and
produces remarkable workload balancing.  Our approach matches or outperforms
parallel routines designed for distributed-memory architectures as well as
all the eigensolvers in LAPACK and Intel's MKL. 

For massively parallel supercomputers, which are themselves built out of
multi-core processors, we created an eigensolver, PMRRR, that merges the
previously introduced task-based approach with a parallelization using
message-passing. 
With a proper scheduling of tasks and the use of non-blocking communication, we are able
to achieve better scalability than all the solvers included in
ScaLAPACK. Furthermore, experiments indicate that our solver is among 
the fastest tridiagonal eigensolvers available. 

PMRRR was integrated in the publicly available Elemental library for the
solution of standard and generalized dense Hermitian eigenproblems. 
A performance study
on two supercomputers at the Research Center J\"ulich, Germany, revealed
that Elemental's eigensolvers outperform the widely used ScaLAPACK
 -- sometimes significantly. For highly parallel executions, the tridiagonal
stage of ScaLAPACK's eigensolvers contributed considerably to the overall
run time. In contrast, Elemental's tridiagonal stage, which
makes use of PMRRR, scales well and, in all experiments, was negligible in terms of
execution time. 
As with the tridiagonal stage, for standard and generalized dense Hermitian
eigenproblems, much of Elemental's performance advantage
over ScaLAPACK comes from a better scalability in the various stages of the computation. 

Although fast, the accuracy of MRRR-based eigensolvers can be several orders
of magnitude worse compared with solvers based on the Divide \& Conquer or the
QR algorithms. 
Additionally, for solvers targeting distributed-memory systems, input matrices with highly
clustered eigenvalues introduce communication and lead to load imbalance.  
For such matrices, the parallel scalability is limited. 
Even worse, as every now and then a crucial assumption in the proof of MRRR's correctness
cannot be verified, the accuracy of the output is not always guaranteed. 

To address the limitations of MRRR, we introduced a mixed precision variant of the
algorithm. 
Our approach adopts a
new perspective: Given input/output arguments in a \binaryx\ floating
point format, internally to the algorithm, we use a higher precision \binaryy\ arithmetic to
obtain the desired accuracy. The use of mixed precisions provides us with
more freedom to choose important parameters of the algorithm. In particular,
while meeting more demanding accuracy goals, we 
reduce the operation count, increase robustness, and improve parallelism. 

Combining all the techniques presented in this thesis, eigensolvers based on
our mixed precision MRRR are not only as accurate as eigensolvers based on
the Divide \& Conquer or the QR algorithms, but -- in many circumstances --
are also faster or even faster than solvers using a conventional MRRR
implementation. Due to their superior scalability, such a statement is
particularly true for massively parallel computing environments.

\appendix

\chapter{A list of (Sca)LAPACK's Eigensolvers}
\label{appendix:routinenames}

\section{(Sca)LAPACK's symmetric tridiagonal eigensolvers}

As LAPACK is the de facto standard for dense linear algebra computations, we frequently show performance and
accuracy results for its implementations of various algorithms. In Table~\ref{tab:lapackroutinesSTEP}, we compile a list of currently available routines for the STEP within
LAPACK.\footnote{The list is based on version 3.4.2 of LAPACK.} 
The placeholder {\tt x} in the names stands for one of the following: {\tt S} single precision, {\tt
  C} single precision complex, {\tt D} double precision, or {\tt Z} double
precision complex. 

\begin{table}[htb]
\begin{center}
\footnotesize
\begin{tabular}{ l  l  c  c  c } \hline\noalign{\smallskip}
  Method & Routine       & Functionality & Subset & Reference \\ \hline\hline\noalign{\smallskip}
  Bisection     & {\tt xSTEBZ} & EW & Yes & \cite{Kahan:1966} \\
  Inverse Iteration     & {\tt xSTEIN} & EV & Yes & \cite{Jessup:1992:Invit,Dhillon98currentinverse} \\
  $\sqrt{\;\,}$-free QR Iteration & {\tt xSTERF} & EW & No &
  \cite{Ortega01011963,Reinsch1971,Parlett:1998:SEP} \\
  QR Iteration (positive def.)  & {\tt xPTEQR} & EW + EV & No &
  \cite{Demmel90accuratesingular,AccurateSVDandQDtrans} \\
  QR Iteration  & {\tt xSTEQR} & EW + EV & No & \cite{EISPACKqr} \\
  Divide and Conquer     & {\tt xSTEDC} & EW + EV & No & \cite{Rutter} \\
  MRRR     & {\tt xSTEMR} & EW + EV & Yes & \cite{Dhillon:DesignMRRR} \\[1mm]
  \hline 
\end{tabular}
\end{center}
\caption{LAPACK routines for the STEP. EW means eigenvalues only, EV means eigenvectors only, and EW + EV means eigenvalues and optionally eigenvectors. Additionally,
  {\tt xSTEGR} exist, which is just a wrapper to {\tt xSTEMR}.
}
\label{tab:lapackroutinesSTEP}
\end{table}

The routines are mainly used through four different expert routines: {\tt
  xSTEV}, {\tt xSTEVX}, {\tt xSTEVD}, and {\tt xSTEVR}. They work as
follows:
\begin{itemize}[noitemsep,nolistsep]
\item {\tt xSTEV} uses
QR Iteration. If only
eigenvalues are desired, the routine calls the square-root free variant {\tt
  xSTERF}  and, otherwise, it calls {\tt xSTEQR}. 
\item {\tt xSTEVX} uses bisection
  and inverse iteration. The routine calls {\tt xSTEBZ}, followed by {\tt xSTEIN}.
\item {\tt xSTEVD} uses Divide \& Conquer. If only eigenvalues are desired,
  the routine uses QR ({\tt xSTERF}) by default and, otherwise, calls {\tt xSTEDC}.
\item {\tt xSTEVR} uses MRRR. If all eigenvalues are requested, the routine
  uses QR ({\tt xSTERF}), and, if a subset of eigenvalues is required, it
  uses bisection ({\tt xSTEBZ}). If all eigenpairs are requested, the
  routine uses {\tt xSTEMR}, otherwise, it makes use of bisection and
  inverse iteration ({\tt xSTEBZ} and {\tt xSTEIN}).\footnote{Older version
    of LAPACK used MRRR for the subset case as well.}
\end{itemize}

ScaLAPACK contains a subset of the above methods.\footnote{The list is based on
  version 2.0.2 of ScaLAPACK.} By convention, the corresponding ScaLAPACK
routines have a preceding {\tt P} in their names, indicating the parallel
version. Table~\ref{tab:scalapackroutinesSTEP} gives an overview of the
available routines.

\begin{table}[htb]
\begin{center}
\footnotesize
\begin{tabular}{ l  l  c  c  c } \hline\noalign{\smallskip}
  Method & Routine       & Functionality & Subset & Reference \\ \hline\hline\noalign{\smallskip}
  Bisection     & {\tt PxSTEBZ} & EW & Yes & \cite{Demmel:bisec} \\
  Inverse Iteration     & {\tt PxSTEIN} & EV & Yes & \cite{Stanley94theperformance} \\
  Divide and Conquer     & {\tt PxSTEDC} & EW + EV & No & \cite{Tisseur:1999:PDC} \\
  \hline 
\end{tabular}
\end{center}
\caption{ScaLAPACK routines for the STEP. EW means eigenvalues only, EV means eigenvectors only, and EW + EV means eigenvalues and optionally eigenvectors. 
}
\label{tab:scalapackroutinesSTEP}
\end{table}

ScaLAPACK also implements the QR algorithm and MRRR (included in 2011), but they
are not encapsulated in a separate routine. They are however used for the
solution of the HEP. We use the name {\tt PxSTEMR} for the tridiagonal
MRRR.

\section{(Sca)LAPACK's Hermitian eigensolvers}

Table~\ref{tab:lapackroutinesHEP} lists all routines for the
standard HEP. Similar routines for packed storage and banded matrices exist, but
are not of any relevance in our discussion. All routines
  compute eigenvalues and (optionally) eigenvectors.

\begin{table}[htb]
\begin{center}
\footnotesize
\begin{tabular}{ l  l l c } \hline\noalign{\smallskip}
  Method & $\Cnn$ & $\Rnn$       &  Subset \\ \hline\hline\noalign{\smallskip}
  Bisection \& Inverse Iteration    & {\tt xHEEVX}  & {\tt xSYEVX}  & Yes \\
  QR Iteration & {\tt xHEEV}   & {\tt xSYEV} & No \\
  Divide-and-Conquer     & {\tt xHEEVD}  & {\tt xSYEVD} & No \\
  MRRR   & {\tt xHEEVR}  & {\tt xSYEVR} & Yes \\[1mm]
  \hline 
\end{tabular}
\end{center}
\caption{LAPACK routines for the HEP. 
}
\label{tab:lapackroutinesHEP}
\end{table}

LAPACK's routines are based on a direct reduction to tridiagonal
form and the aforementioned tridiagonal eigensolvers. The reduction is
performed by routines {\tt xHETRD} and {\tt 
  xSYTRD} for the complex-valued and real-valued case, respectively.
For all methods but QR, the backtransformation is implemented by routines
{\tt xUNMTR} and {\tt xORMTR} for the complex-valued and real-valued case,
respectively. For QR, the transformation matrix, implicitly given by
Householder reflectors, is built using routines {\tt xUNGTR} or
{\tt xORGTR}. The Givens rotations of the tridiagonal QR are applied to
this matrix. 
We give some of comments regarding the LAPACK routines:
\begin{itemize}[noitemsep,nolistsep]
\item {\tt xHEEVX}, {\tt xSYEVX}: Depending on the user, when all
  eigenvalues or all eigenpairs are desired, the routine uses QR. 
\item {\tt xHEEVD}, {\tt xSYEVD}: If only eigenvalues are desired, QR is
  used.
\item {\tt xHEEVR}, {\tt xSYEVR}: If only eigenvalues are desired, QR is
  used and, if only a subset of eigenvalues or eigenpairs is desired, BI is
  used. MRRR is only used to compute all eigenpairs.
\end{itemize}

ScaLAPACK offers the same functionality. Additionally to the regular reduction routines, {\tt PxHETRD} and
{\tt PxSYTRD}, ScaLAPACK offers {\tt PxHENTRD} and {\tt PxSYNTRD}, which are
optimized for square grids of processors and should be preferred. However, not all the ScaLAPACK routines make use of those 
routines for the reduction.  We give some of comments regarding the ScaLAPACK routines:
\begin{itemize}[noitemsep,nolistsep]
\item {\tt xHEEV}, {\tt xSYEV}: Make use of the (often) inferior reduction
  routines {\tt PxHETRD} and {\tt PxSYTRD}.
\item {\tt xHEEVX}, {\tt xSYEVX}: Orthogonality of the computed eigenvectors
  is only guaranteed if enough workspace is provided.
\item {\tt xHEEVD}, {\tt xSYEVD}: Make use of the (often) inferior reduction
  routines {\tt PxHETRD} and {\tt PxSYTRD}. Can only be used to compute all
  eigenvalues {\it and} eigenvectors.
\end{itemize}

\section{(Sca)LAPACK's generalized eigensolvers}

Table~\ref{tab:lapackroutinesGHEP} lists all available routines for the
full GHEP of type $Ax = \lambda B x$, $A Bx = \lambda x$, or $B Ax = \lambda 
x$. Similar routines for packed storage and banded matrices exist, but 
are not of any relevance in our discussion. All routines
 compute eigenvalues and (optionally) eigenvectors.

\begin{table}[htb]
\begin{center}
\footnotesize
\begin{tabular}{ l  l l c } \hline\noalign{\smallskip}
  Method & $\Cnn$ & $\Rnn$       &  Subset \\ \hline\hline\noalign{\smallskip}
  Bisection \& Inverse Iteration    & {\tt xHEGVX}  & {\tt xSYGVX}  & Yes \\
  QR Iteration    & {\tt xHEGV}  & {\tt xSYGV}  & No \\
  Divide and Conquer    & {\tt xHEGVD}  & {\tt xSYGVD}  & No \\
  \hline 
\end{tabular}
\end{center}
\caption{LAPACK routines for the GHEP.
}
\label{tab:lapackroutinesGHEP}
\end{table}

LAPACK's routines are based on a transformation to a HEP and
the corresponding routine for the HEP. The Cholesky
factor of $B$ is computed via routine {\tt xPOTRF} and the transformation to
standard form performed by routines {\tt xHEGST} and {\tt xSYGST} for
complex-valued and real-valued case, respectively. The final transformation
of the eigenvectors is done via routines {\tt xTRSM} for type 1 and 2 and
{\tt xTRMM} for type 3 of the problem. No routine based on MRRR exist, but
can be built by a user out of the existing components.

As shown is
Table~\ref{tab:scalapackroutinesGHEP}, ScaLAPACK offers only one routine. 
The routine compute eigenvalues and (optionally) eigenvectors and functions
as its LAPACK analog. 
Additionally to the standard
transformation routines to the HEP, {\tt xHEGST} and {\tt xSYGST}, ScaLAPACK
contains {\tt xHENGST} and {\tt xSYNGST}, which are
optimized for square grids of processors and should be
preferred. These routines are only optimized for the case that the
{\it lower} triangular part of the matrices is stored and referenced. 
Solvers based on QR, DC, and MRRR are not available, but
can be built by a user out of the existing components.

\begin{table}[htb]
\begin{center}
\footnotesize
\begin{tabular}{ l  l l c } \hline\noalign{\smallskip}
  Method & $\Cnn$ & $\Rnn$       &  Subset \\ \hline\hline\noalign{\smallskip}
  Bisection \& Inverse Iteration    & {\tt PxHEGVX}  & {\tt PxSYGVX}  & Yes \\
  \hline 
\end{tabular}
\end{center}
\caption{ScaLAPACK routines for the GHEP.
}
\label{tab:scalapackroutinesGHEP}
\end{table}

\chapter{Algorithms}
\label{appendix:somealgorithms}

\begin{figure*}
  \begin{minipage}[t]{.48\textwidth}
    \begin{algorithm}[H]
      \footnotesize
      {\bf Input:} Symmetric tridiagonal matrix $T \in \Rnn$ given by its diagonal
      $(\alpha_1,\ldots,\alpha_n)$ and off-diagonal $(\beta_1,\ldots,\beta_{n-1})$. \\
      {\bf Output:} Number of eigenvalues smaller than $\sigma$.
      
      \vspace{1mm}
      \algsetup{indent=2em}
      \begin{algorithmic}[1]
        \STATE $\mbox{count} := 0$
        \STATE $d := \alpha_1 - \sigma$
        \STATE $\mbox{count} := \mbox{count} + \mbox{SignBit}(d)$      
        \FOR{$i = 2, \ldots, n$}
        \STATE $d := (\alpha_i - \sigma) - \beta_{i-1}^2 / d$
        \STATE $\mbox{count} := \mbox{count} + \mbox{SignBit}(d)$
        \ENDFOR
        \RETURN $\mbox{count}$
      \end{algorithmic}
      \caption{NegCount}
      \label{alg:negcountT}
    \end{algorithm}
  \end{minipage}
  %
  \begin{minipage}[t]{0.03\textwidth}
    \begin{center}
    \makebox{\begin{minipage}{0.03\textwidth}
      \hspace{0.01in}
      \vspace{0.03in}
      \end{minipage}
     }
    \end{center}
  \end{minipage}
  \begin{minipage}[t]{.48\textwidth}
\begin{algorithm}[H]
 \footnotesize
    {\bf Input:} Symmetric tridiagonal matrix $LDL^* \in \Rnn$ given by the
    non-trivial entries $(d_1,\ldots,d_n)$ and $(\ell_1,\ldots,\ell_{n-1})$. \\
    {\bf Output:} Number of eigenvalues smaller than $\sigma$.
    
    \vspace{1mm}
    \algsetup{indent=2em}
    \begin{algorithmic}[1]
      \STATE $\mbox{count} := 0$
      \STATE $s := -\sigma$      
      \FOR{$i = 1, \ldots, n-1$}
      \STATE $d_{+} := d(i) + s$
      \IF{$|s| = \infty$ $\wedge$ $|d_{+}| = \infty$}
      \STATE $q := 1$
      \ELSE
      \STATE $q := s/d_{+}$
      \ENDIF
      \STATE $s := q \cdot d(i) \ell(i) \ell(i) - \sigma$
      \STATE $\mbox{count} := \mbox{count} + \mbox{SignBit}(d_{+})$
      \ENDFOR
      \STATE $d_{+} := d(n) + s$
      \STATE $\mbox{count} := \mbox{count} + \mbox{SignBit}(d_{+})$
      \RETURN $\mbox{count}$
    \end{algorithmic}
  \caption{NegCount}
  \label{alg:negcountLDL}
\end{algorithm}
  \end{minipage}
\end{figure*}

\begin{figure*}
  \begin{minipage}[t]{.48\textwidth}
\begin{algorithm}[H]
 \footnotesize
    {\bf Input:} Symmetric tridiagonal $T \in \mathbb{R}^{n
    \times n}$ given by its diagonal $(\alpha_1,\ldots,\alpha_n)$ and off-diagonal
  $(\beta_1,\ldots,\beta_{n-1})$. \\
    {\bf Output:} Gershgorin interval $[g_\ell, g_u]$.
    
    \vspace{1mm}

  \algsetup{indent=2em}
  \begin{algorithmic}[1]
    \STATE $g_\ell := \alpha_1 - \beta_1$ 
    \STATE $g_u   := \alpha_1 + \beta_1$ 
    \FOR{$i = 2$ to $n-1$}
    \STATE $g_\ell := \min( g_\ell, \alpha_i - |\beta_{i-1}| - |\beta_i| )$ 
    \STATE $g_u   := \max( g_u, \alpha_i + |\beta_{i-1}| + |\beta_i| )$ 
    \ENDFOR
    \STATE $g_\ell := \min( g_\ell, \alpha_n - |\beta_{n-1}|)$ 
    \STATE $g_u   := \max( g_u, \alpha_n + |\beta_{n-1}|)$ 
    \STATE $bnorm := \max(|g_\ell|, |g_u|) $
    \STATE $g_\ell := g_\ell - (2n + 10) \cdot bnorm \cdot \varepsilon$ 
    \STATE $g_u := g_u + (2n + 10) \cdot bnorm \cdot \varepsilon$ 
    \RETURN $[g_\ell, g_u]$
  \end{algorithmic}
  \caption{\ Gershgorin}
  \label{alg:gershgorin}
\end{algorithm}
  \end{minipage}
\end{figure*}

{\sc Remarks on Algorithm~\ref{alg:negcountT}:} (1) Previous scaling to ensure $|\beta_i| \geq
\sqrt{\omega}$ is assumed, where $\omega$ denotes the underflow threshold; (2)
$\mbox{SignBit}(d) = (d < 0)$, i.e., it is equals one whenever $d < 0$ and 
equals zero whenever $d > 0$; in particular, $\mbox{SignBit}(-0) = 1$ and
$\mbox{SignBit}(+0) = 0$; (3) Even if $d = 0$ in some iteration, the count
will be produced correctly~\cite{Demmel97appliednumerical}; (4) It is
usually beneficial to precompute and reuse quantities $\beta_i^2$, which can
serve as an input instead of the off-diagonals; (5)
See~\cite{Demmel:bisec,Demmel97appliednumerical} for an error analysis and further details.  

{\sc Remarks on Algorithm~\ref{alg:negcountLDL}:} (1) It is can be beneficial to precompute
quantities $d(i) \ell(i) \ell(i)$; (2) The floating point exception handling
technique discussed in~\cite{Marques:2006:BIF} can accelerate the
computations.

{\sc Remarks on Algorithm~\ref{alg:gershgorin}:} See~\cite{Demmel:bisec} for further details.

\chapter{Hardware}
\label{appendix:hardware}

In this section, we collect information regarding of our
experiments. We specify hardware, compilers, compiler
flags, and external libraries.  

\vspace{2mm}
{\sc Dunnington}: Refers to an SMP system comprising four
six-core {\it Intel Xeon X7460 Dunnington} processors, running at a
frequency of 2.66 GHz. Each core possesses 32 KB data and 32 KB instruction
cache; two cores share a common 3 MB L2 cache and all six cores of a processor share a
common 16 MB L3 cache.  
For all our experiments, routines were compiled with
version~$11.1$ of the Intel compilers \textit{icc} and \textit{ifort}, with optimization level
three enabled. LAPACK routines were linked with MKL BLAS version
10.2.   

\vspace{2mm}
{\sc Beckton}: Refers to an SMP system comprising four
eight-core {\it Intel Xeon X7550 Beckton} processors, with a nominal clock
speed of 2.0 GHz. Each processor is equipped 18 MB L3 cache; each core
is equipped 256KB L2 cache as well as 32KB L1 data cache. For all our
experiments, routines were compiled with version~$12.1$ of the Intel
compilers \textit{icc} and \textit{ifort}, with optimization level three enabled.
LAPACK routines were from version 3.4.2 and linked to
the vendor-tuned MKL BLAS version 12.1.

\vspace{2mm}
{\sc Westmere}: Refers to an SMP system comprising four
ten-core {\it Intel Xeon E7-4850 Westmere} processors, with a nominal clock
speed of 2.0 GHz. Each processor is equipped 18 MB L3 cache; each core
is equipped 256KB L2 cache as well as 32KB L1 data cache. For all our
experiments, routines were compiled with version~$12.1$ of the Intel
compilers \textit{icc} and \textit{ifort}, with optimization level three enabled.
LAPACK routines were from version 3.3.0 and linked to
the vendor-tuned MKL BLAS version 12.1.

\vspace{2mm}
{\sc Juropa}: The machine is installed at the Research Center J\"ulich,
Germany. It consists of $2{,}208$ nodes, each comprising two Intel
{\it Xeon X5570 Nehalem} quad-core processors running at 2.93 GHz
with 24 GB of memory. The nodes are connected by an {\it Infiniband QDR}
network with a fat-tree topology.
All tested routines were compiled using the {\it Intel compilers}
version~11.1, with the flag {\tt -O3}, and linked to the {\it
  ParTec's ParaStation MPI} library (version~5.0.23 and, when support for
  multithreading was needed,  5.0.24)

\vspace{2mm}
{\sc Jugene}: The machine is installed at the Research Center J\"ulich,
Germany. It consists of $73{,}728$ nodes, each of which is equipped with 2 GB of
memory and a quad-core PowerPC 450 processor running at 850 MHz.  
All routines were compiled using the {\it IBM XL} compilers (ver.~9.0) in
combination with the vendor tuned 
{\it IBM} MPI library.

\chapter{Test Matrices}
\label{appendix:matrices}

In this section, we collect a number of frequently used test matrices. More
information on these matrices can be found in~\cite{Marques:2008}.

\begin{itemize}
\item{\sc Uniform}: Eigenvalue distribution $\lambda_k = \varepsilon + (k-1)(1 -
  \varepsilon)/(n-1), \enspace k=1,\ldots,n$.

\item{\sc Geometric}: Eigenvalue distribution $\lambda_k = \varepsilon^{(n-k)/(n-1)}, \enspace k=1,\ldots,n$.

\item{\sc 1--2--1}: Contains ones on the subdiagonals
and twos on the diagonal; its eigenvalues are $\lambda_k = 2-2 \cos(
     \pi k / (n + 1)), \enspace k=1,\ldots,n$.

\item{\sc Clement}: Has zeros on its diagonal; the off-diagonal
  elements are given by $\beta_k = \sqrt{k(n-k)}, \enspace k=1,\ldots,n-1$.
   Its eigenvalues are the integers $\pm n, \pm n-2, \ldots$ with the
   smallest (in magnitude) eigenvalue $\pm 1$ for even $n$ and $0$ for odd $n$.

\item{\sc Wilkinson}: The off-diagonals are ones and the 
diagonal equals the vector $(m, m-1, \dots, 1, 0, 1, \dots, m)$, with $m =
(n-1)/2$ and odd size $n$.  The matrices ``strongly favor Divide
and Conquer over the MRRR algorithm. [...] It can be verified
that Divide and Conquer deflates all but a small number of eigenvalues (the
number depends on the precision and the deflation
threshold)''~\cite{Marques:2008}. Almost all eigenvalues come in
increasingly close pairs. 

\item{\sc Legendre}: Has zeros on its diagonal; its the off-diagonal
  elements are given by $\beta_k = k/\sqrt{(2k-1)(2k+1)}, \enspace k=2,\ldots,n$

\item{\sc Laguerre}: Its off-diagonal elements are $(2,3,\ldots,n-1)$ and
  its diagonal elements $(3,5,7,\ldots,2n-1,2n+1)$.

\item{\sc Hermite}: Has zeros on its diagonal; its the off-diagonal
  elements are given by $\beta_k = \sqrt{k}, , \enspace k=1,\ldots,n-1$.
\end{itemize}

\noindent We further used a test set of application matrices in our experiments. 

\begin{itemize}
\item{\sc Application}:{\tt \,T\_bcsstkm08\_1, T\_bcsstkm09\_1, T\_bcsstkm10\_1,
  T\_1138\_bus, T\_bcsstkm06\_3, T\_bcsstkm07\_3, T\_bcsstkm11\_1,
  T\_bcsstkm12\_1, Fann02, T\_nasa1824, T\_nasa1824\_1, T\_plat1919,
  T\_bcsstkm13\_1, bcsstkm13, Fann03, T\_nasa2146, T\_nasa2146\_1,
  T\_bcsstkm10\_2, T\_nasa2910, T\_nasa2910\_1, T\_bcsstkm11\_2,
  T\_bcsstkm12\_2, T\_bcsstkm10\_3, T\_nasa1824\_2, T\_bcsstkm13\_2,
  T\_sts4098, T\_sts4098\_1, Juelich4237k1b, Juelich4289k2b, T\_nasa2146\_2,
  T\_bcsstkm10\_4, T\_bcsstkm11\_3, T\_bcsstkm12\_3, T\_nasa4704,
  T\_nasa4704\_1, T\_nasa1824\_3, T\_nasa2910\_2, T\_bcsstkm11\_4,
  T\_bcsstkm12\_4, T\_bcsstkm13\_3, T\_Alemdar, T\_Alemdar\_1,
  T\_nasa2146\_3, input7923, T\_bcsstkm13\_4}

\end{itemize}

\chapter{Elemental on Jugene}
\label{appendix:resultsjugene}

We verify the prior results obtained on {\sc Juropa} for
a different architecture -- namely, the {\it BlueGene/P} installation {\sc
  Jugene}.

We used a square processor grid $P_r = P_c$ whenever possible
and $P_r = 2 P_c$ otherwise.\footnote{As noted in
    Section~\ref{sec:elemrrr:experiments}, $P_c \approx P_r$ or the 
  largest square grid possible should be
  preferred. These choices do
  not affect the qualitative behavior of our performance results.} 
Similarly, ScaLAPACK (ver.~1.8) in 
conjunction 
with the vendor-tuned BLAS included in the ESSL library (ver.~4.3) was used
throughout. In contrast to the {\sc Juropa}
experiments, we concentrate on the weak scalability of the {\it
  symmetric-definite} generalized eigenproblem. Therefore ScaLAPACK's DC
timings correspond to the  
sequence of routines {\tt PDPOTRF}--{\tt PDSYNGST}--{\tt PDSYNTRD}--{\tt PDSTEDC}--{\tt
  PDORMTR}--{\tt PDTRSM}. 
Accordingly, ScaLAPACK's MRRR corresponds to the same sequence of routines
with {\tt PDSTEDC} replaced by {\tt PDSTEMR}.\footnote{As {\tt
    PDSTEMR} is not contained in ScaLAPACK, it corresponds to the sequence
  {\tt PDPOTRF}--{\tt PDSYNGST}--{\tt PDSYEVR}--{\tt PDTRSM}.} 
In both cases, a block size of
48 was found to be nearly optimal 
and used in all experiments. As explained in
Section~\ref{sec:elemrrr:experiments}, we avoided the use of the routines
{\tt PDSYGST} and {\tt PDSYTRD} for the reduction to standard and tridiagonal
form, respectively.
For EleMRRR's timings we used Elemental (ver.~0.66), which integrates
PMRRR (ver.~0.6). A block size of 96 was identified as nearly optimal and
used for all experiments.\footnote{
The block size for matrix vector products, which does not have a significant
influence on the performance, was fixed to 64 in all cases.}

\begin{figure}[thb]
   \centering
   \subfigure[Execution time.]{
     \includegraphics[width=.47\textwidth]{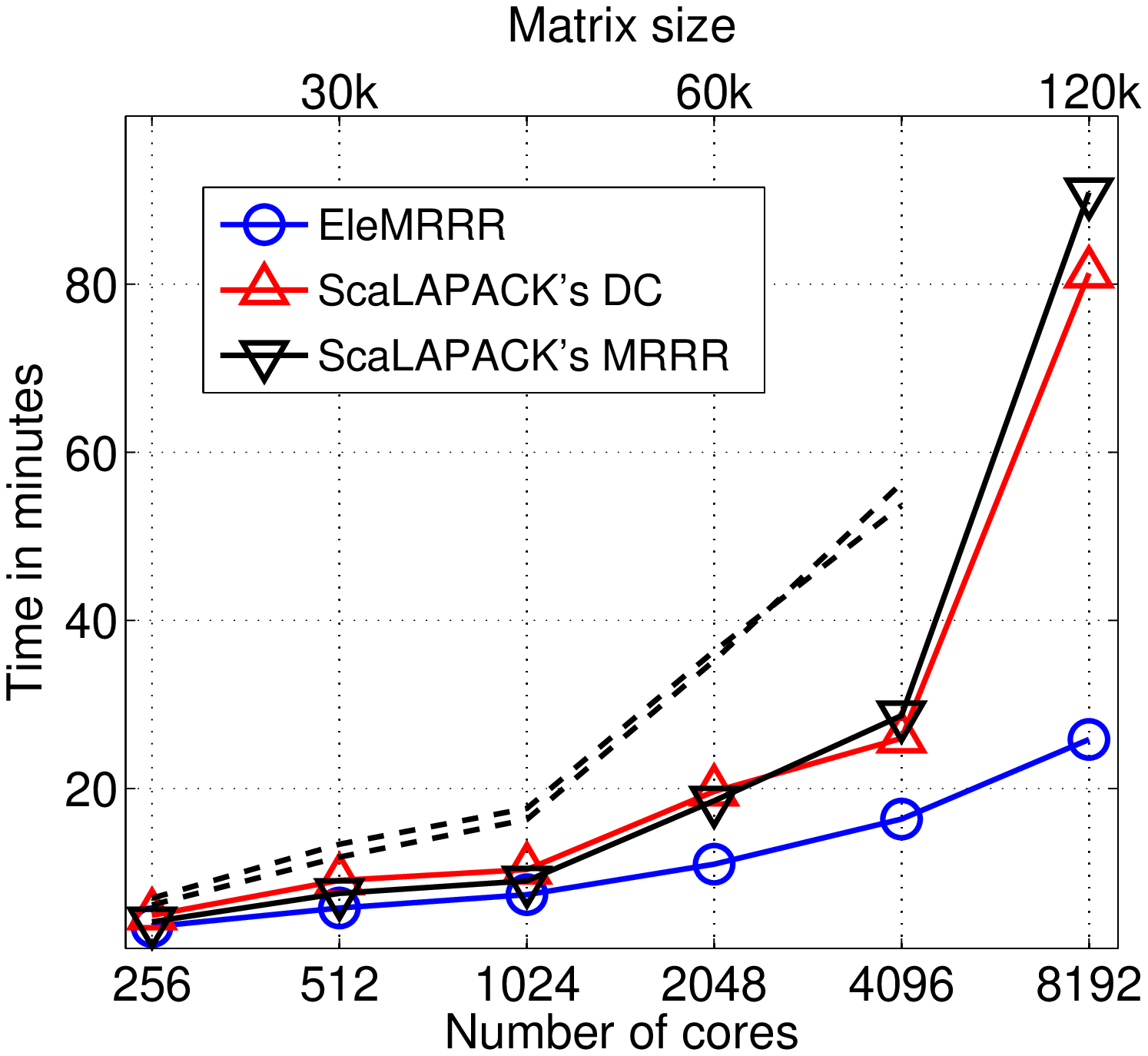}
     \label{fig:bgp:weaka}
   } \subfigure[Breakdown of time by stages.]{
     \includegraphics[width=.47\textwidth]{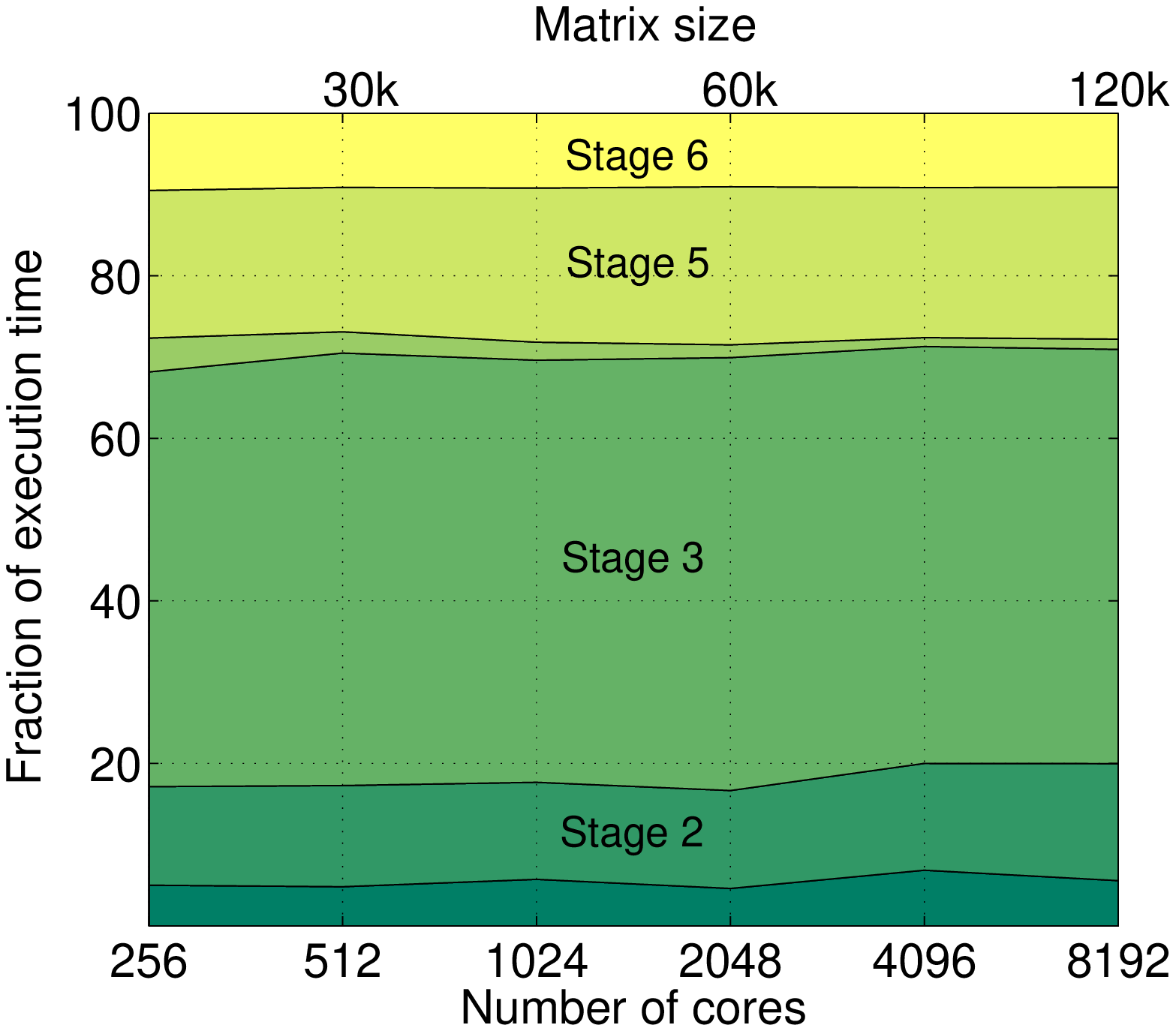}
     \label{fig:bgp:weakb}
   }
   \caption{
     Weak scalability for computing all eigenpairs of $Ax = \lambda B
     x$. 
     The dashed lines refer to ScaLAPACK's solvers when the matrices $A$ and
     $B$ are stored in the upper triangular part; in this scenario, the 
     non-square routines for the reductions are used. 
   } 
   \label{fig:bgp:weak}
\end{figure}

In the left panel of Fig.~\ref{fig:bgp:weak} we present EleMRRR's timings
for the computation of all 
eigenpairs of the generalized problem in the form of $Ax = \lambda Bx$. While
the size of the test matrices ranges from 
$21{,}214$ to $120{,}000$, the number of cores increases from $256$ to
$8{,}192$ ($64$ to $2{,}048$ nodes). In the right panel, the execution time is broken down
into the six stages of the generalized eigenproblem.

Both graphs show a similar behavior to the experiments performed on
{\sc Juropa}. In all experiments EleMRRR outperforms both of the
ScaLAPACK's solvers. 
Most importantly, ScaLAPACK again suffers from a breakdown in scalability. 

When analyzing the results of Fig.~\ref{fig:bgp:weak} for
ScaLAPACK's solver we the following three observations: (1) 
The scalability issues can be attributed mainly to
Stages 1, 2, and 4. In particular, for the largest experiment
ScaLAPACK's reduction to standard form (28 minutes) and MRRR
(25 minutes) each exceed the time that EleMRRR spends for the entire
problem. 
(2) 
Although ScaLAPACK's tridiagonal eigensolver is usually not very time
consuming, for highly parallel systems it might become the bottleneck.
As an example, in our experiment on 8{,}192 cores the tridiagonal
problem accounts for 54\% (MRRR) and 33\% (DC) of the total execution time
of the standard eigenproblem.  
(3) Due to better
  scalability, DC becomes faster than ScaLAPACK's MRRR.

We conclude with four comments regarding EleMRRR’s behavior:
 (1)
While stage 4 of ScaLAPACK's MRRR and DC take up to 
28\% and 20\% of the  total
execution time, respectively, PMRRR accounts for less than 4\%. In
particular, for the largest problem 25 minutes were spent in
ScaLAPACK's MRRR, whereas PMRRR required only 20 seconds. 
In all experiments
PMRRR's execution time was negligible. 
(2) The
timings corresponding to the standard eigenproblem account for 70\%--74\%  
 of the generalized problem's execution time.
(3) The part which is roughly proportional to the
fraction of desired eigenpairs 
makes up 26\%--32\% of both the generalized and
standard problem. 
(4) All of the six stages
scale equally well and no computational bottleneck emerges.

\chapter[Mixed Precision MRRR: Experiments I]{First Tests of Mixed Precision MRRR}
\label{sec:firstexperimentsmixed}

In this section, we report on the first performance and accuracy results
that we obtained for the mixed precision MRRR. All experiments were performed
{\it sequentially} on {\sc Westmere}.\footnote{In~\cite{mixedtr} we
  mistakenly wrote that we used {\sc Beckton}.} 
For the mixed precision
results, we used the same reduction and backtransformation routines as
LAPACK.   
For the extended precision results, we used version 4.7 of the
GNU compilers. 

Since for single precision input/output the mixed precision MRRR usually
is faster than the conventional MRRR, we concentrate on double precision
input/output. In this case, the mixed precision approach uses either
extended or quadruple precision. We refer to these cases as {\tt
  mr3smp-extended} and {\tt mr3smp-quad}, respectively. 
The use of quadruple is more 
critical as it shows that the approach is applicable in many
circumstances, even when the higher 
precision arithmetic used in the tridiagonal stage is much slower than
arithmetic in the input/output format.

We confine ourselves to experiments on a small set of
application matrices as listed in
Table~\ref{tab:applmatrices}, coming from quantum chemistry and structural
mechanics problems.\footnote{These matrices are stored in tridiagonal
  form. In order to create real and complex dense matrices, we generated a
  random Householder matrix $H = I - \tau v v^*$ and applied the similarity
transformation $H T H$ to the tridiagonal matrix $T$.}
As the
performance depends on the spectra of the input matrices, the
platform of the experiment, and the implementation of
the quadruple arithmetic, we cannot draw final
conclusions from these limited test.
However, the orthogonality improvements are
quite general and are observed for a much larger test set
originating from~\cite{Marques:2008}. We do not report on residuals as {\it the
  largest residual norms are generally comparable for all solvers}.

\begin{table}[htb]
  \small
  \begin{center}
\begin{tabular}[htb]{c@{\quad}c@{\quad}c@{\quad}c@{\quad}}
\hline\noalign{\smallskip}
Matrix & Size & Application & Reference \\
\hline\noalign{\smallskip}

$A$   & $2{,}053$ &  Chemistry & ZSM-5 in~\cite{Fann:1995} and Fann3 in~\cite{Marques:2008}  \\ 
$B$   & $4{,}289$ &  Chemistry & Originating from~\cite{fleur}     \\ 
$C$   & $4{,}704$ &  Mechanics & T\_nasa4704 in~\cite{Marques:2008} \\ 
$D$   & $7{,}923$ &  Mechanics & See
\cite{Bientinesi:2005:PMR3} for information \\ 
$E$   &$12{,}387$ &  Mechanics & See \cite{Bientinesi:2005:PMR3} for information \\ 
$F$   &$13{,}786$ &  Mechanics & See
\cite{Bientinesi:2005:PMR3} for information \\ 
$G$   &$16{,}023$ &  Mechanics & See
\cite{Bientinesi:2005:PMR3} for information \\ 
\hline\noalign{\smallskip}
\end{tabular}
  \end{center}
  \vspace{-0.5cm}
  \caption{A set of test matrices.}
  \label{tab:applmatrices}
\end{table}

To better display the effects of the use of mixed precisions, the
performance results are simplified in the following sense: As the routines
for the reduction to tridiagonal form and the backtransformation of all
solvers are the same, we used for these stages the {\it minimum}
execution time of all runs for all solvers. 
In this way, the cost of the mixed precision approach becomes more visible and we do not
have to resort to statistical metrics for the timings. We point out that
especially for the subset tests, the run time of the tridiagonal stage for larger matrices
is often smaller than the fluctuations in the timings for the reduction to
tridiagonal form.

\section{Real symmetric matrices}

\paragraph{Computing all eigenpairs.}

\begin{figure}[thb]
   \centering
\includegraphics[scale=.38]{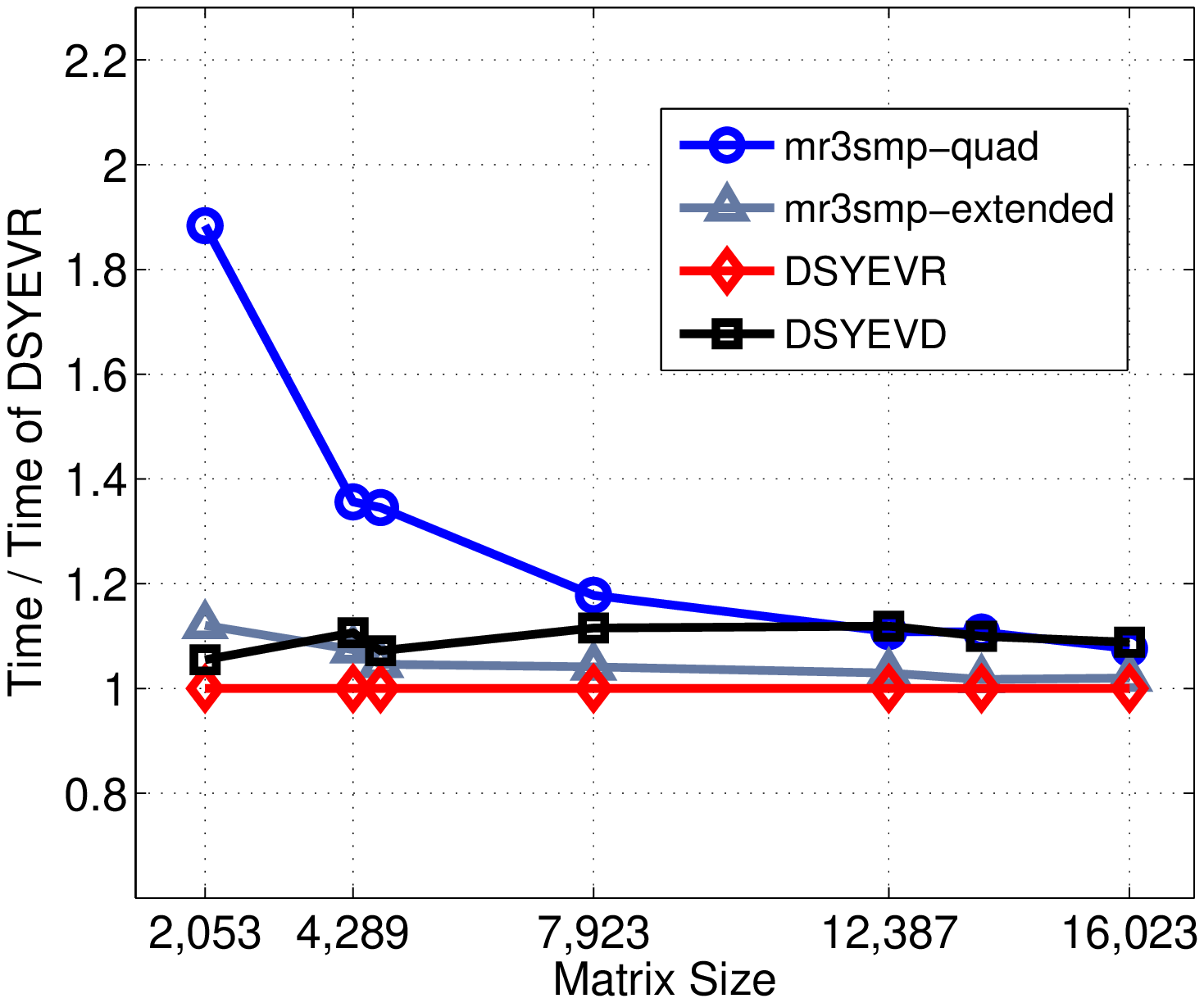}
\includegraphics[scale=.38]{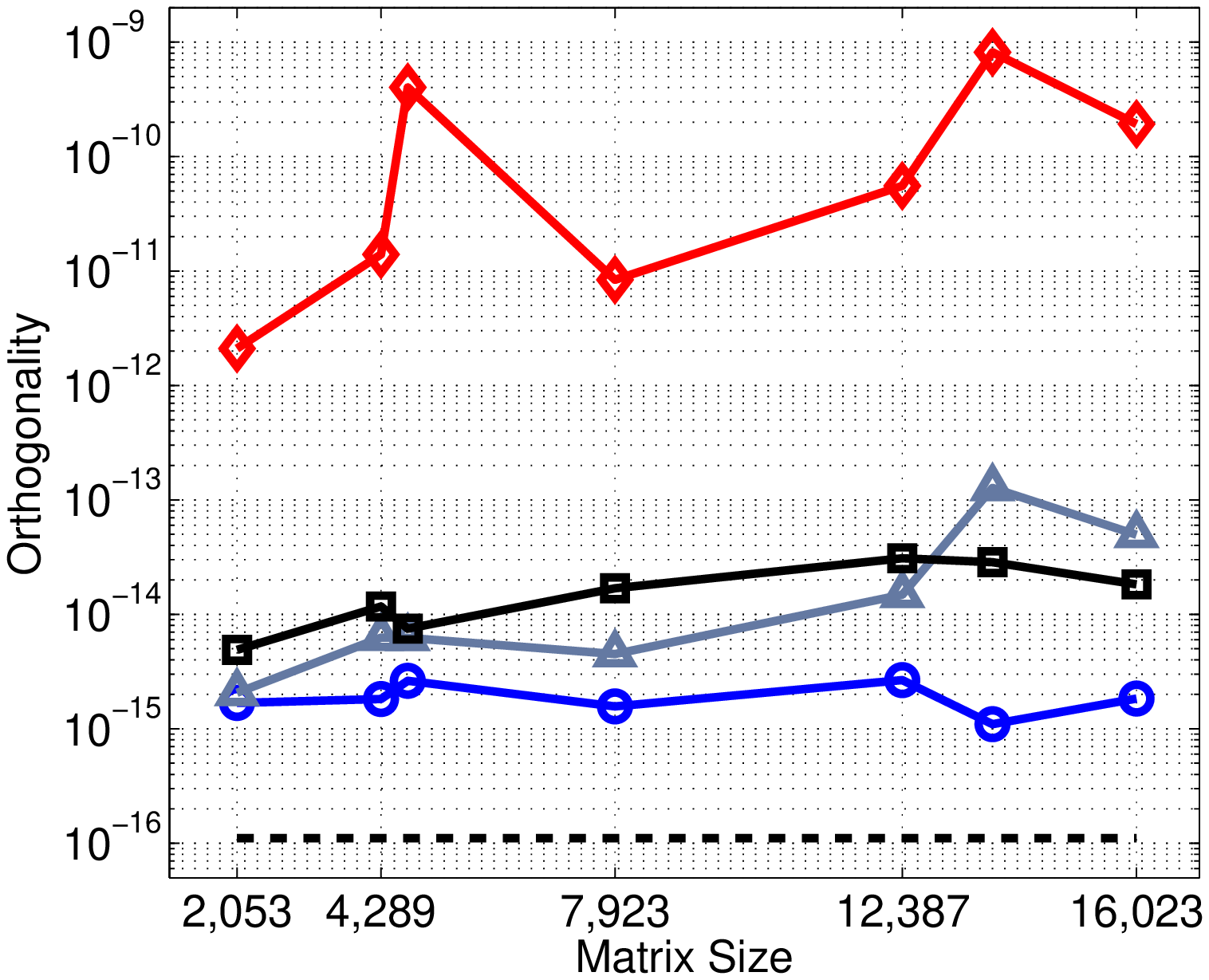}
   \caption{
     Computation of all eigenpairs. {\it Left:} Execution time relative to
     routine {\tt DSYEVR}. {\it Right:}
     Orthogonality. 
   }
   \label{fig:dblquadapplsymall}
\end{figure}

Figure~\ref{fig:dblquadapplsymall} refers to the computation of all 
eigenpairs. We report on the execution time of the mixed precision routines
relative to LAPACK's MRRR ({\tt DSYEVR}) and the obtained orthogonality. As
a reference, results for 
LAPACK's DC ({\tt DSYEVD}) are 
included. 
The orthogonality is improved using extended and
quadruple precision. The left plot shows the  
performance penalty that we pay for the improvements. In particular, for
larger matrices, the additional cost of the mixed precision approach becomes
negligible, making it extremely attractive for large-scale problems. For
example, for test matrices $E$, $F$, and $G$, our solver 
{\tt mr3smp-quad} is as fast as {\tt DSYEVD}, although it uses software-simulated arithmetic, while achieving better
orthogonality. Since the quadruple arithmetic is currently much slower than 
the double one, {\tt mr3smp-quad} carries a performance penalty for small
matrices. 
In our case, for matrices with $n <
2{,000}$, one must expect an increase in the total execution time of a factor
larger than two. 
The situation is similar to the one
reported in~\cite{Baboulin20092526} for the mixed precision iterative refinement of the
solution of linear systems of equations, where the mixed precision
approach comes with a performance penalty for small matrices. 

\begin{figure}[thb]
   \centering
\includegraphics[scale=.38]{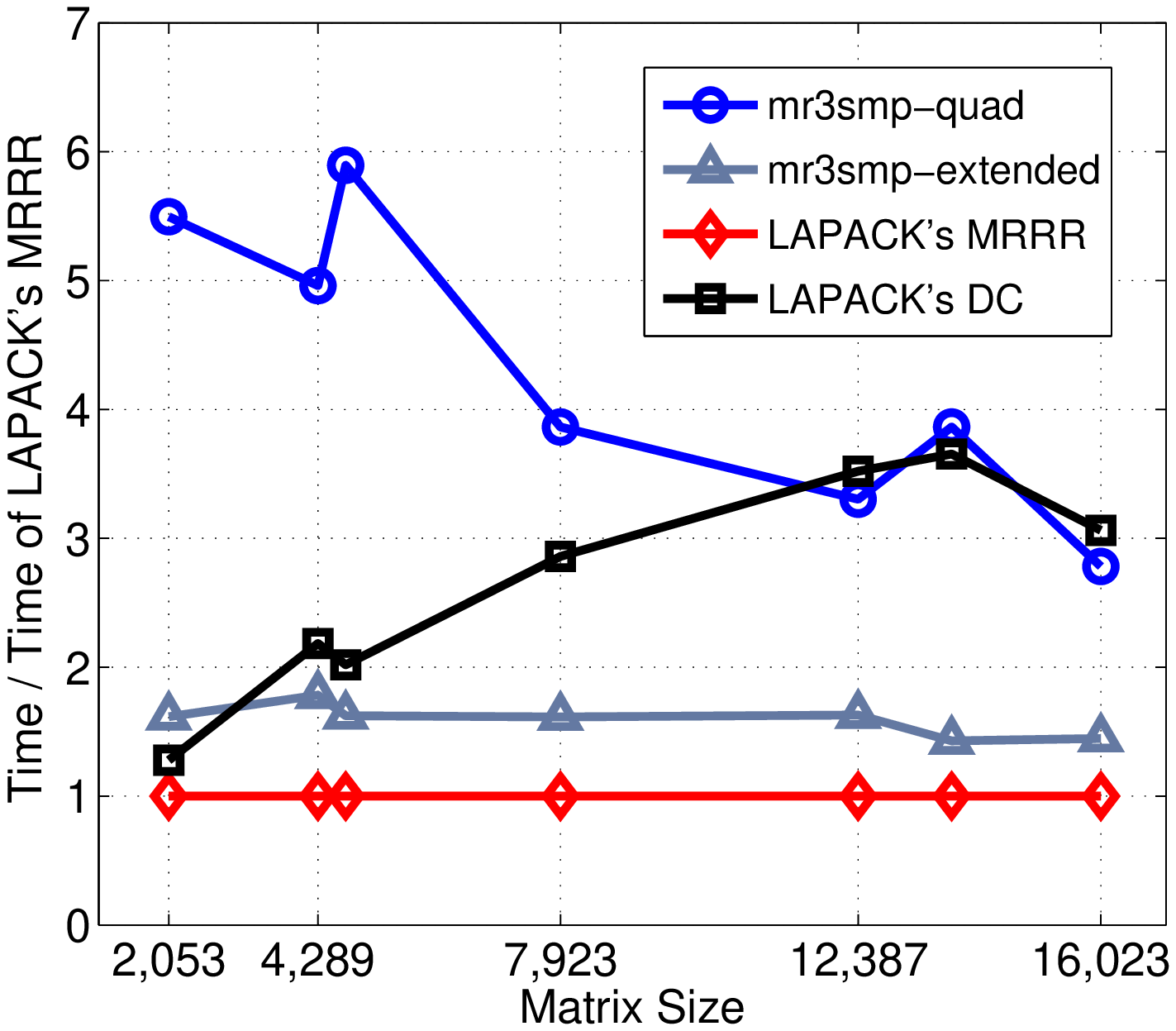}
\includegraphics[scale=.38]{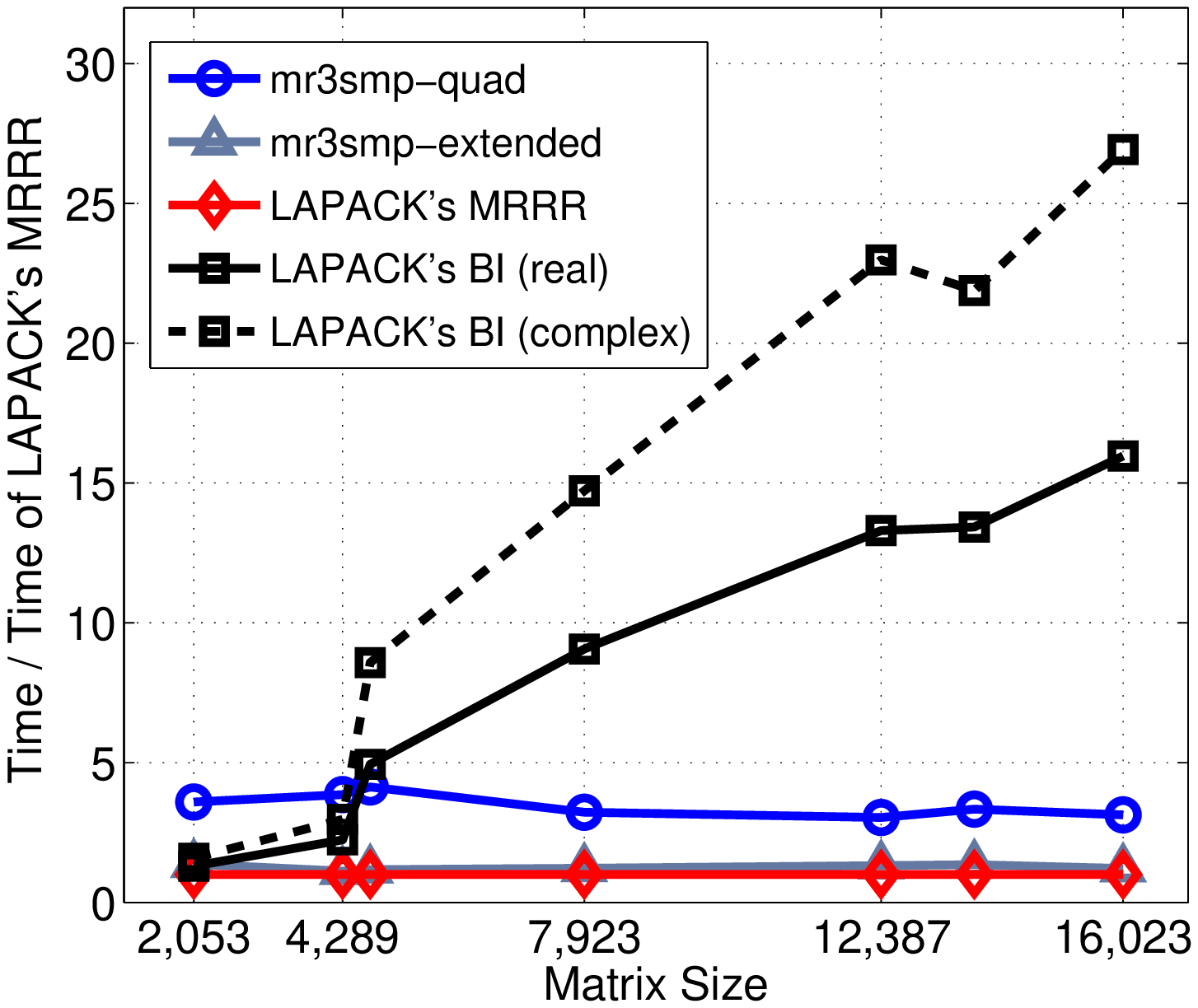}
   \caption{
     Execution time of the tridiagonal stage relative to
     LAPACK's MRRR. {\it Left:}  Computation
     of all eigenpairs. {\it  
       Right:} Computation of 20\% of the
     eigenpairs corresponding to the smallest eigenvalues. 
   }
   \label{fig:dblquadapplsubtridiag}
\end{figure}

In contrast to {\tt mr3smp-quad}, the use of extended precision does not
significantly increase the execution time even for smaller matrices, while still
improving the orthogonality. As the reason for different
performance is solely due to the tridiagonal eigensolver, in the
left panel of Fig.~\ref{fig:dblquadapplsubtridiag} we show the execution
time of the tridiagonal eigensolvers relative to LAPACK's MRRR ({\tt DSTEMR}). 

We remark that, although the mixed precision approach slows down the
tridiagonal stage compared to {\tt DSTEMR} (at least with the current support
for quadruple precision arithmetic, see
Fig.~\ref{fig:dblquadapplsubtridiag}), it has two features compensating this
disadvantage: 
the approach increases robustness and parallel scalability of 
the code. 
To underpin these statements, in
Table~\ref{tab:depthreptree}, for the computation of all eigenpairs, we present the recursion depth $d_{max}$, the maximal
cluster size and the number of times Line~\ref{line:mrrr:shifting} in Algorithm
\ref{alg:mrrr} (the only possible source of failure) is executed.

\begin{table}[htb]
  \small
  \begin{center}
\begin{tabular}[htb]{l@{\quad}l@{\quad}c@{\quad}c@{\quad}c@{\quad}c@{\quad}c@{\quad}c@{\quad}c@{\quad}}
\hline\noalign{\smallskip}
 & & $A$ & $B$ & $C$ & $D$ & $E$ & $F$ & $G$ \\
\hline\noalign{\smallskip}

max.~depth &  {\tt DSTEMR} &  $2$ & $2$ & $2$ & $4$ & $5$ & $2$ & $5$   \\ 
           &  {\tt mr3smp-quad} &  $0$ & $0$ & $0$ & $0$ & $0$ & $0$ & $0$   \\
\hline\noalign{\smallskip} 
largest cluster & {\tt DSTEMR}  & $324$ & $1{,}027$ & $10$ & $5{,}011$ & $8{,}871$ & $1{,}369$ & $14{,}647$    \\ 
                & {\tt mr3smp-quad}         & $1$ & $1$ &  $1$ &  $1$ & $1$ & $1$ & $1$    \\ 
\hline\noalign{\smallskip} 
new RRR & {\tt DSTEMR} & $311$ & $638$ & $1{,}043$ & $1{,}089$ & $1{,}487$ & $1{,}798$ & $1{,}825$   \\ 
        & {\tt mr3smp-quad} & $0$ & $0$ & $0$ & $0$ & $0$ & $0$ & $0$    \\ 
\hline\noalign{\smallskip}
\end{tabular}
  \end{center}
  \vspace{-0.5cm}
  \caption{Recursion depth, largest encountered cluster, and number of times
    an RRR for a cluster needs to be computed by executing
    Line~\ref{line:mrrr:shifting} in Algorithm~\ref{alg:mrrr} for {\tt DSTEMR} and
    {\tt mr3smp-quad}.} 
  \label{tab:depthreptree}
\end{table}
In all cases, {\tt mr3smp-quad} computes the eigenpairs directly from the
root representation. Since this representation can be made definite, no danger of
element growth in its computation exist (thus, the RRR can be found). Such a
danger occurs in Line~\ref{line:mrrr:shifting}, where a 
new RRR for each cluster needs to be computed. By executing
Line~\ref{line:mrrr:shifting} only a few times -- often no times at all -- the 
danger of not finding a 
proper RRR is reduced and robustness increased.\footnote{Besides
  the fact that less RRRs need to be found, additionally, the restriction of
what constitutes an RRR might be relaxed.} 
Since our approach is independent of the actual form of the RRRs, it is possible to
additionally use twisted or blocked factorizations as proposed
in~\cite{Willems:twisted,Willems:blocked}. 

The mixed precision MRRR is especially appealing in the context of
distributed-memory systems.
The fact that all eigenpairs in our experiment are computed directly from the
root representation implies that the execution is truly {\it embarrassingly
parallel}.
That MRRR is embarrassingly parallel was already announced -- somewhat
prematurely -- with its introduction~\cite{Dhillon:Diss}. Only later,
parallel versions of
MRRR~\cite{Bientinesi:2005:PMR3,Vomel:2010:ScaLAPACKsMRRR}
found that ``the eigenvector computation in 
MRRR is only embarrassingly parallel if the root representation
consists of singletons''~\cite{VoemelRefinedTree2007tr} and that
otherwise ``load imbalance can occur and hamper the overall
performance''~\cite{Vomel:2010:ScaLAPACKsMRRR}.  

While one can expect limited clustering of eigenvalues for
application matrices arising from dense inputs, it is not always the case
that the recursion depth is zero. Experiments on all the tridiagonal matrices
provided explicitly by the {\sc Stetester}~\cite{Marques:2008} -- a
total of 176 matrices ranging in size from 3 to 24{,}873 -- showed that the
largest residual norm and worst case orthogonality were given by
respectively $1.5 
\cdot 10^{-14}$ and $1.2 \cdot 10^{-15}$ and $d_{max} \leq 2$. In fact, only four
artificially constructed matrices, glued Wilkinson
matrices~\cite{glued}, had clusters within clusters.  
In most cases, with the settings of our experiments, the
clustering was very benign or even no clustering was observed. 
For example,
the largest matrix in the test set, {\it Bennighof\_24873}, had only a single
cluster of size 37. 
Furthermore, it is also possible to significantly lower
the $gaptol$ parameter, say to $10^{-16}$, and reduce clustering
even more. 
For such small values of $gaptol$,
in the approximation and refinement of eigenvalues
we need to resort to quadruple precision,
which so far we avoided for performance reasons, see
Chapter~\ref{chapter:mixed}.  

Our results suggest that even better results can be expected for parallel
executions. The MRRR algorithm was already reasonably scalable, and the mixed precision
approach additionally improves scalability -- often making the computation 
truly embarrassingly parallel.  


\paragraph{Computing a subset of eigenpairs.}

\begin{figure}[thb]
   \centering
\includegraphics[scale=.38]{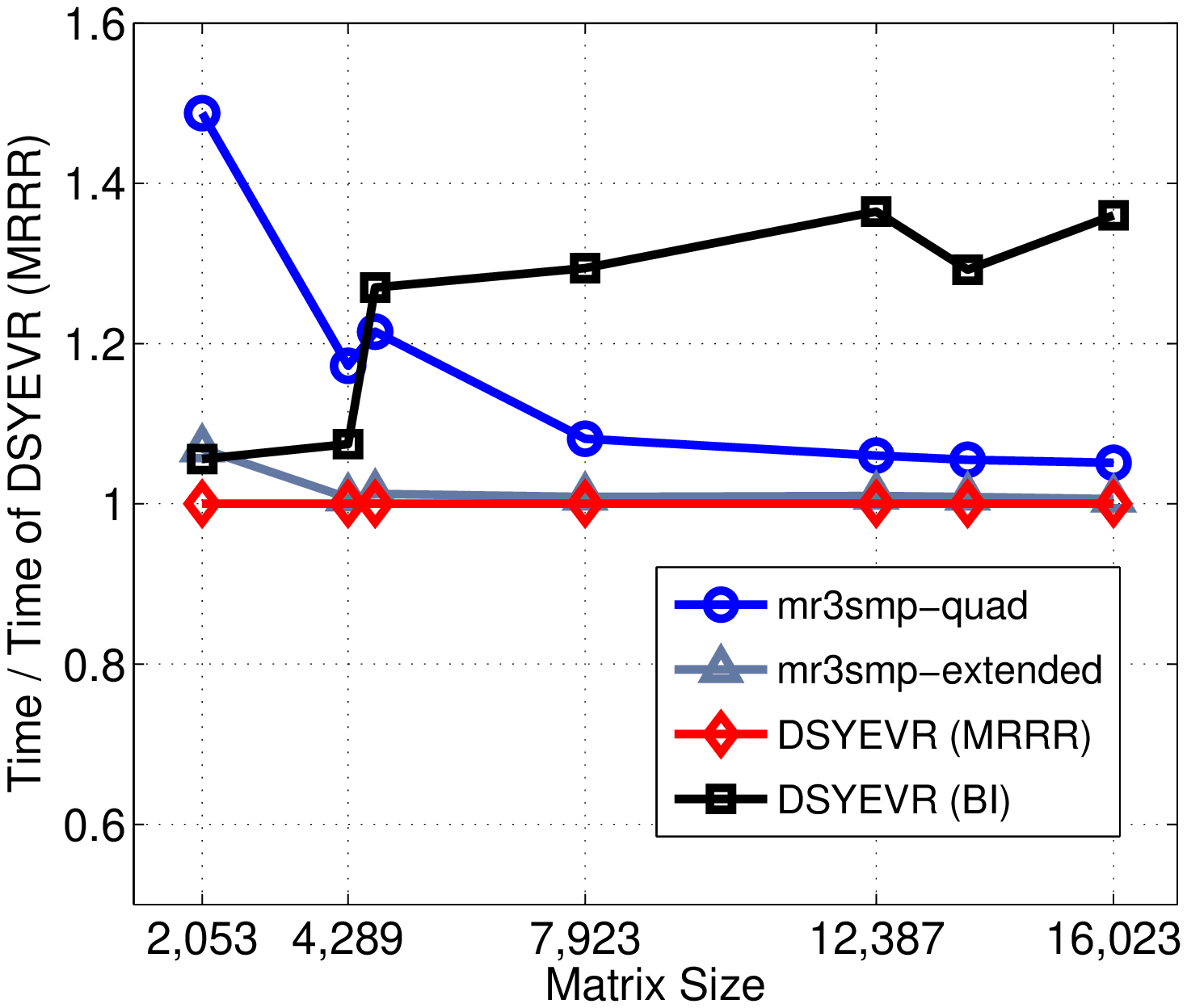}
\includegraphics[scale=.38]{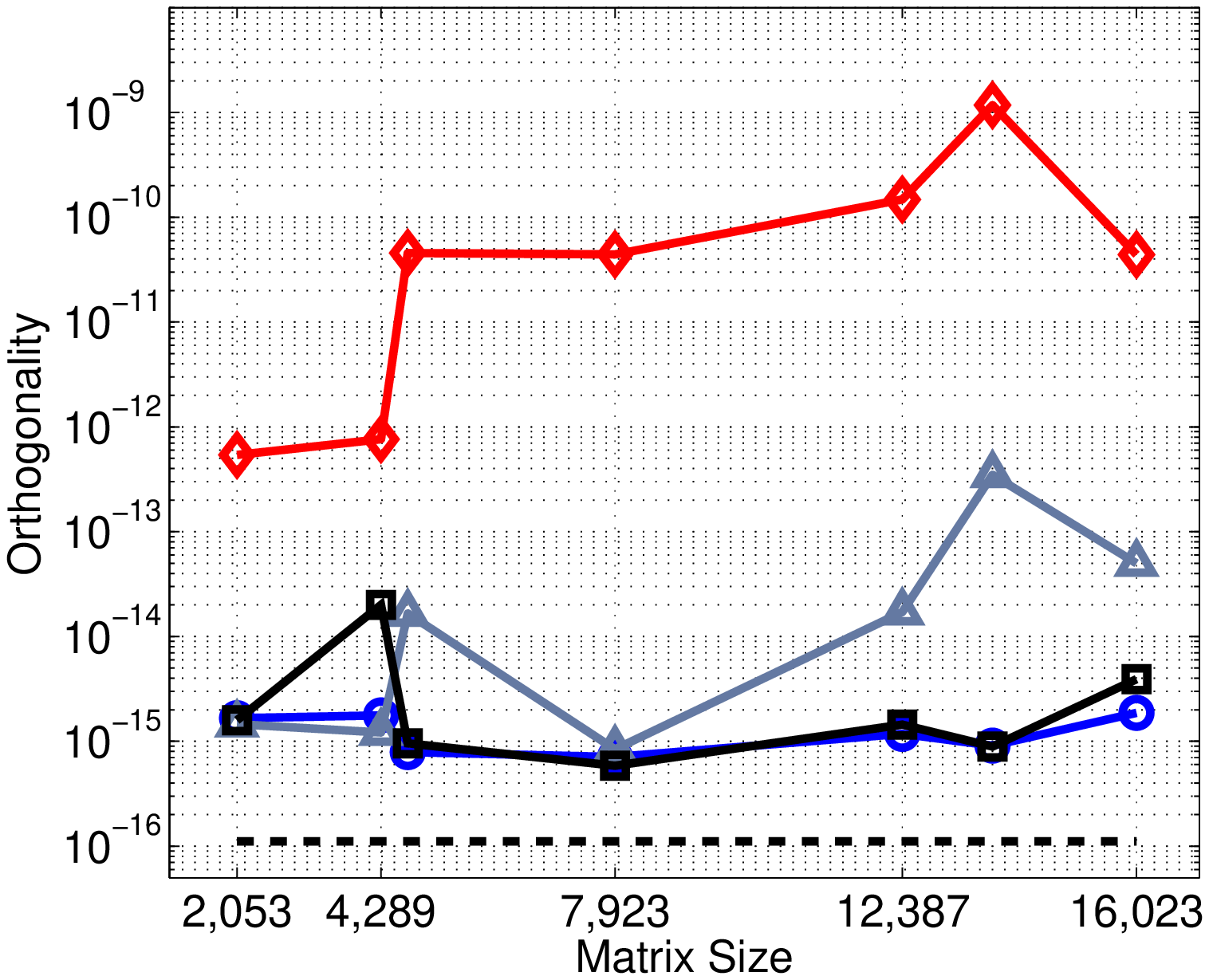}
   \caption{
     Computation of 20\% of the eigenpairs corresponding to the smallest
     eigenvalues. 
     {\it Left:} Execution time relative to routine {\tt DSYEVR} that is
     forced to use MRRR. 
     {\it Right:} Orthogonality.
   }
   \label{fig:dblquadapplsymsub}
\end{figure}

The situation is more favorable when only a subset of eigenpairs needs to be computed. 
As {\tt DSYEVD} does not allow subset computations at reduced cost, a user
can resort to either BI or MRRR. 
The capabilities of BI are accessible via LAPACK's routine {\tt DSYEVX}. 
Recently, the routine {\tt
  DSYEVR} was edited, so that it uses BI instead of MRRR in the
subset case. We therefore refer to {\tt 'DSYEVR (BI)'} when we use BI and {\tt
  'DSYEVR (MRRR)'} when we force the use of MRRR
instead.\footnote{In all experiments, we used BI with default parameters.}
In Fig.~\ref{fig:dblquadapplsymsub}, we report the  
execution time relative to LAPACK's MRRR for computing 20\% of
the eigenpairs associated with the smallest eigenvalues and the
corresponding orthogonality. 

BI and {\tt mr3smp-quad} achieved the best orthogonality. 
At the moment, for small matrices ($n \ll 2{,}000$), the use quadruple might be
be too expensive. In this case, the mixed precision routine can easily be
run in double precision only or BI can be used for accuracy and
performance. As support for quadruple precision improves, the overhead will
further decrease or completely vanish. The use of extended precision comes
almost without any performance penalty. Unfortunately, for larger matrices, the orthogonality might still
be inferior to other methods and no increased robustness and parallelism is observed. 
To illustrate the source of the differing run times, the right panel of Fig.~\ref{fig:dblquadapplsubtridiag} presents
the execution time of the tridiagonal eigensolver relative to LAPACK's
MRRR. As expected, due to explicit
orthogonalization via the Gram-Schmidt procedure, BI potentially becomes
considerably slower than MRRR.

\section{Complex Hermitian matrices}

\begin{figure}[thb]
   \centering
\includegraphics[scale=.38]{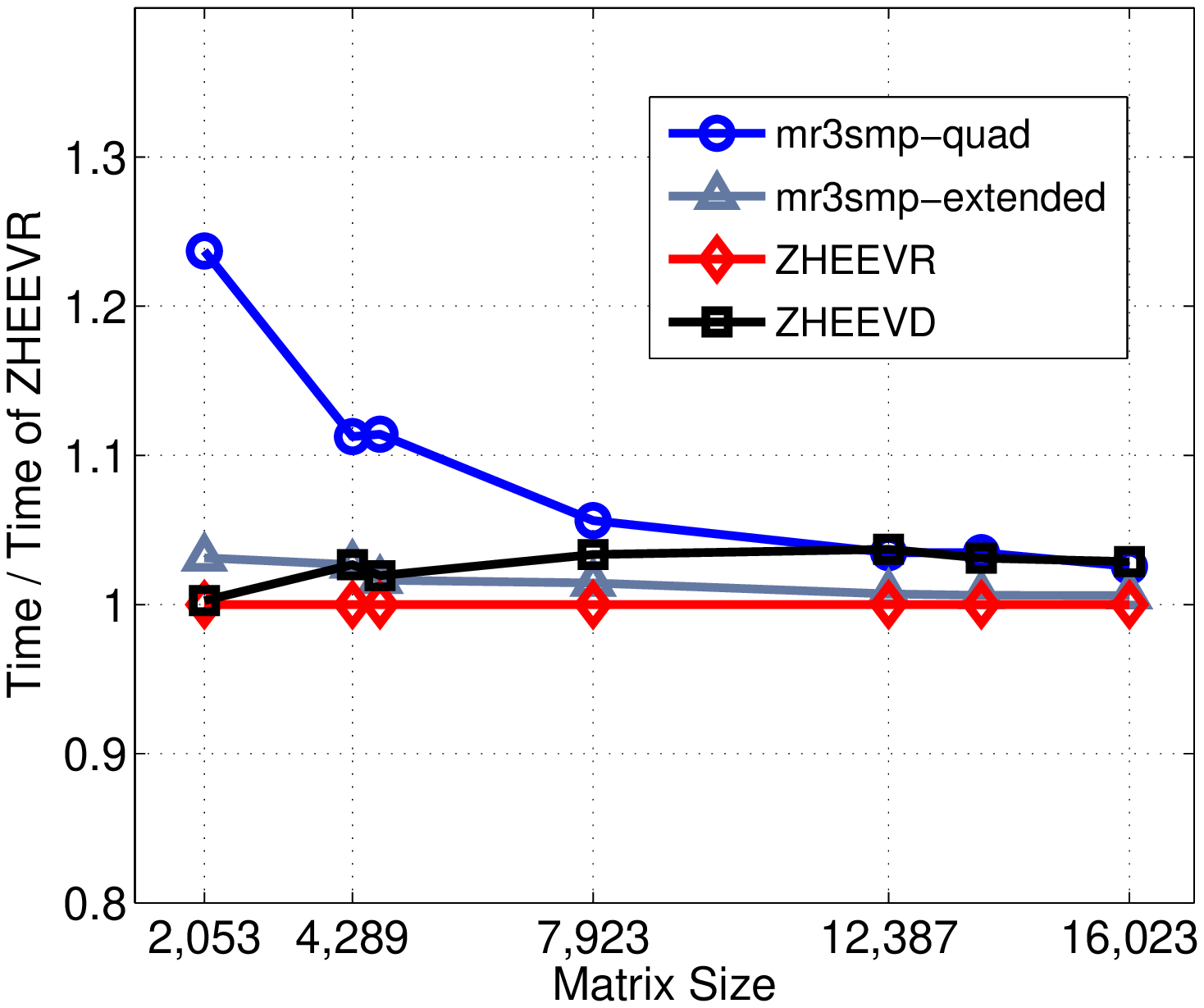}
\includegraphics[scale=.38]{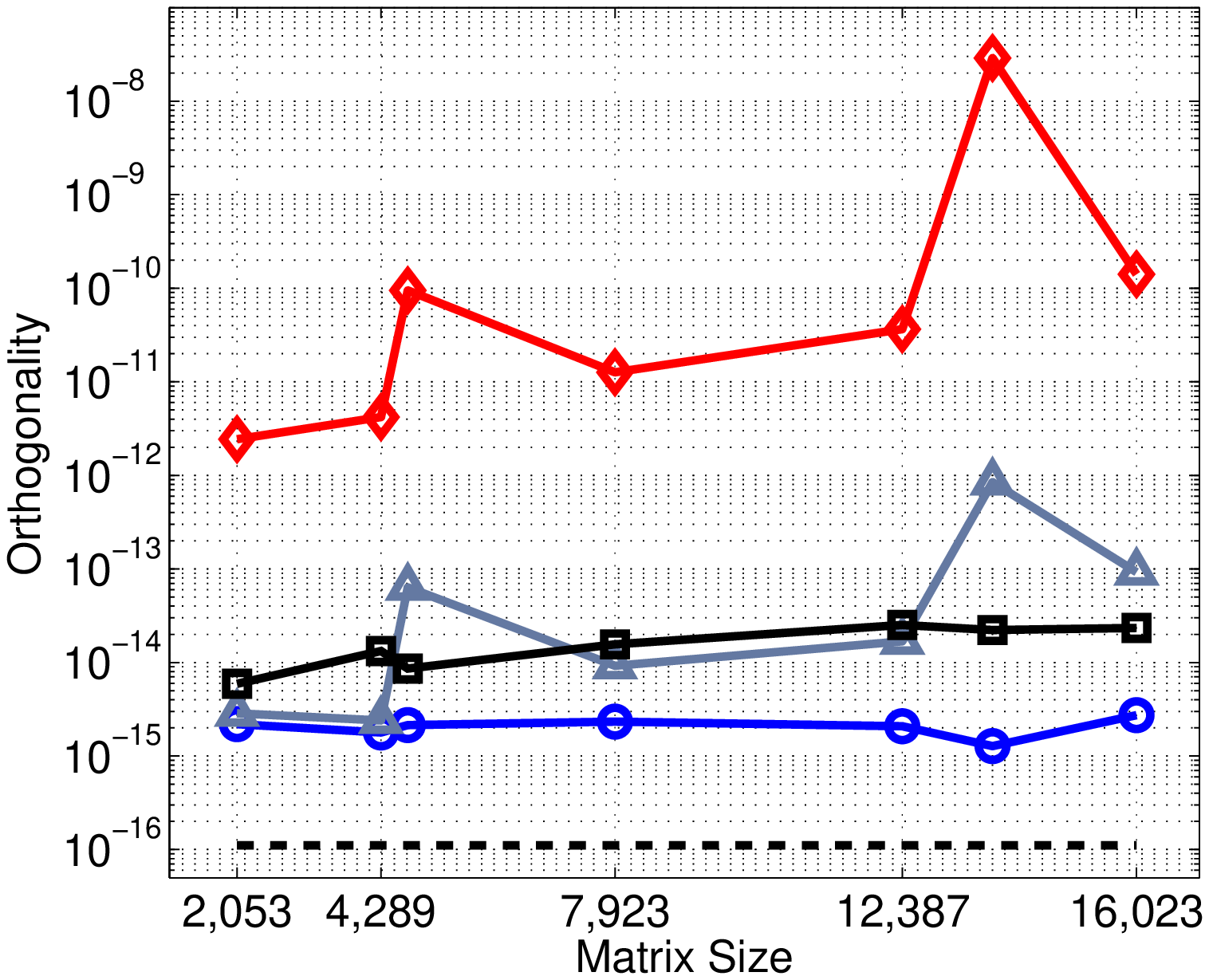}
   \caption{
     Computation of all eigenpairs. {\it Left:} Execution time relative to
     routine {\tt ZHEEVR}. {\it Right:} Orthogonality.
   }
   \label{fig:dblquadapplhermall}
\end{figure}

\paragraph{Computing all eigenpairs.}
In Fig.~\ref{fig:dblquadapplhermall}, we show results for computing all
eigenpairs. The left and right panel
display the execution time of all solvers relative
to LAPACK's MRRR ({\tt ZHEEVR}) and the orthogonality, respectively.
As predicted, the extra cost due to the higher precision becomes relatively
smaller for 
complex-valued input compared to real-valued input -- compare
Figs.~\ref{fig:dblquadapplsymall} and \ref{fig:dblquadapplhermall}. 
Similarly, if
the mixed precision solver is used for the generalized eigenproblem based on
Cholesky-Wilkinson algorithm (see Section~\ref{sec:GHEPsixstages}), the
approach increases the execution only 
marginally even for relatively small problems.  

We remark that the timing plots might be misleading and suggest
that the cost of the mixed precision approach is high. Indeed, for
test matrix $A$, using quadruple precision increased the run time by about
23\% relative to {\tt ZHEEVR}. This means in absolute time that the mixed
precision approach required about 27 seconds and {\tt ZHEEVR} only 22
seconds. For larger matrices the absolute execution time
increases as $n^3$ and the performance gap between mixed precision approach
and pure double precision solver vanishes. Such a scenario is observed with
test matrix $F$, for which 
we obtain an orthogonality of $1.3
\cdot 10^{-15}$ with {\tt mr3smp-quad} compared to $2.9 \cdot 10^{-8}$ with
{\tt ZHEEVR}, while spending only about 4\% 
more in the total execution time.

\begin{figure}[thb]
   \centering
\includegraphics[scale=.38]{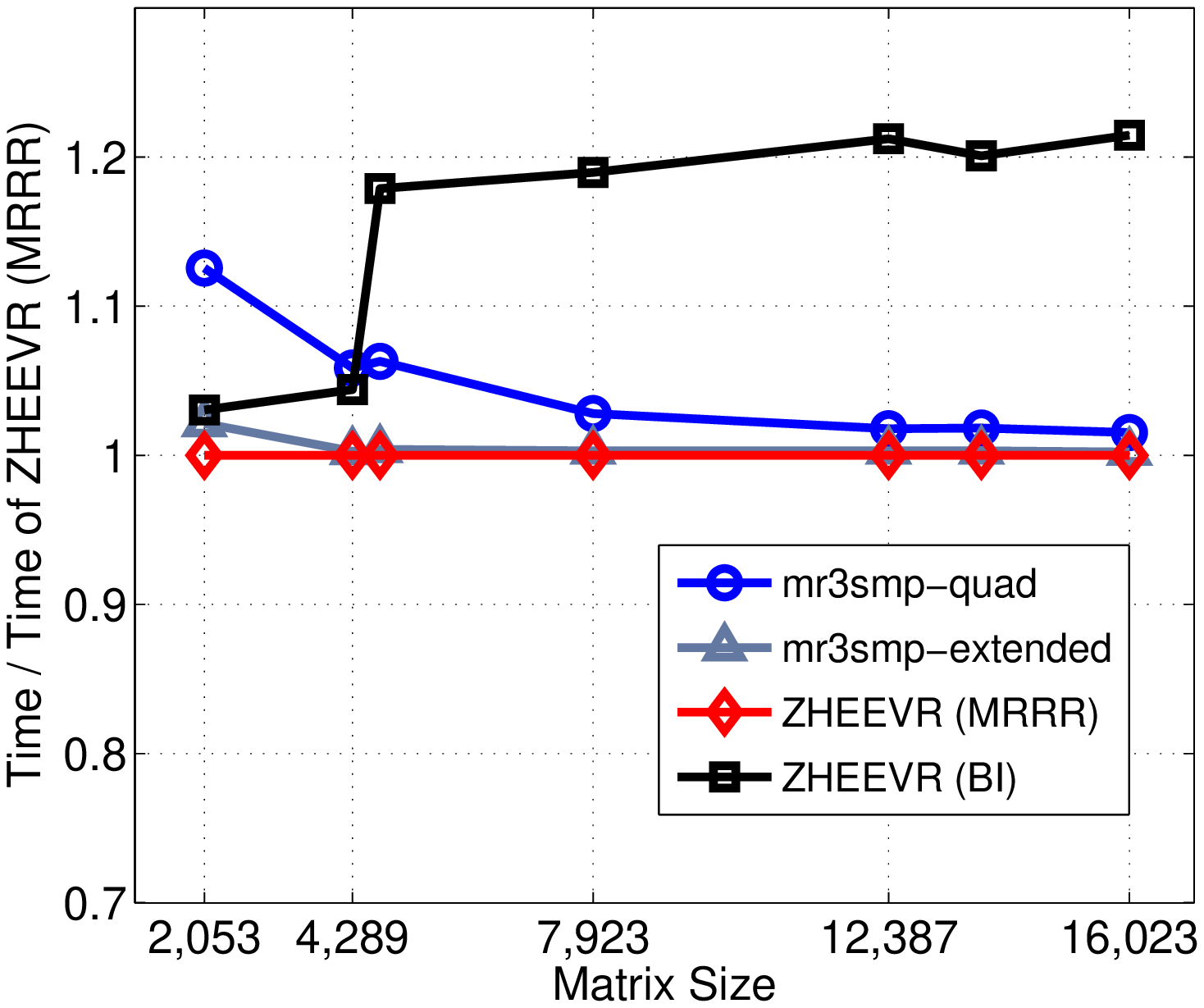}
\includegraphics[scale=.38]{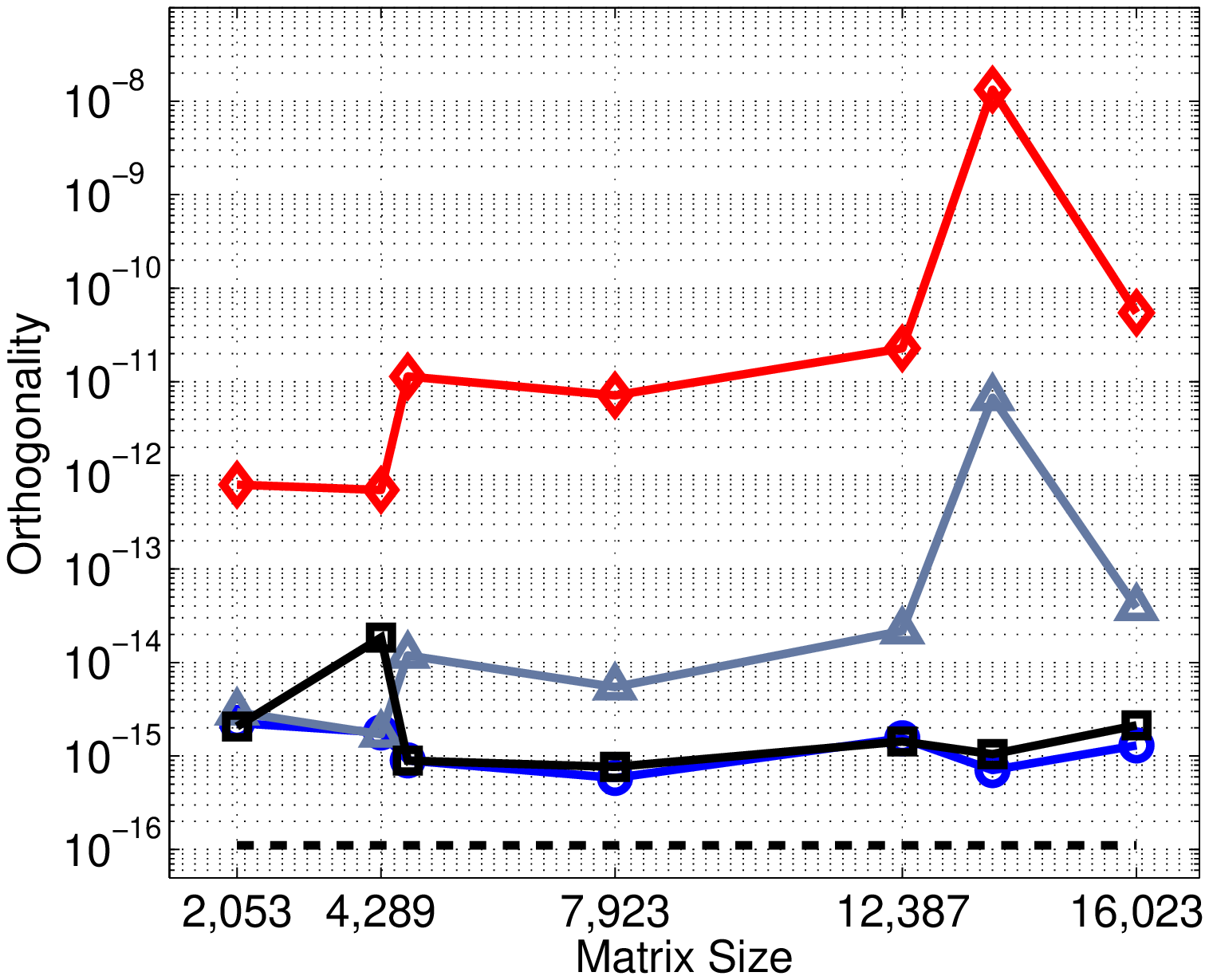}
   \caption{
     Computation of 20\% of the eigenpairs corresponding to the smallest
     eigenvalues are computed. 
     {\it Left:} Execution time relative to routine {\tt ZHEEVR} (when
     forced to use MRRR).  
     {\it Right:} Orthogonality.
   }
   \label{fig:dblquadapplhermsub}
\end{figure}

\paragraph{Computing a subset of eigenpairs.}
We compute 20\% of the eigenpairs associated
with the smallest eigenvalues. The execution
time relative to LAPACK's MRRR and the corresponding orthogonality are
displayed in Fig.~\ref{fig:dblquadapplhermsub}.
The extra cost due to the use of higher precision in {\tt
  mr3smp-extended} or {\tt 
  mr3smp-quad} were quite small. Even using quadruple precision, the run
time only increased by maximally 13\%. 
In a similar experiment -- computing 10\% of the eigenpairs corresponding to
the smallest eigenvalues -- the extra cost for {\tt mr3smp-quad}
was less than 6\% compared to LAPACK's MRRR. 
Such a penalty in the
execution time is already below the fluctuations observed in repeated
experiments.
While {\tt mr3smp-extended} is faster than {\tt mr3smp-quad} for
smaller problems, it cannot quite
deliver the same orthogonality.
 
The relative timings of the tridiagonal eigensolvers are depicted in the
right panel of Fig.~\ref{fig:dblquadapplsubtridiag}. Interestingly, BI is
almost by a factor two slower than in the real-valued case. The reason is that the Gram-Schmidt
orthogonalization, a memory bandwidth-limited operation, is 
performed on complex data (although all imaginary parts of the involved
vectors are zero).

\section{Summary}

The mixed precision MRRR obtains much improved orthogonality -- possibly at
the cost of some performance. 
The performance penalty depends on the difference in speed between the precision of
input/output and the higher precision used in
the tridiagonal stage. In the single/double case, the mixed precision
approach does not introduce a penalty and usually leads
faster executions; in the double/quadruple case, the mixed precision
approach does introduce a penalty for small matrices.  
For larger matrices, the
additional cost becomes negligible as the tridiagonal stage has a lower
complexity than the other two stages of the standard Hermitian
eigenproblem. 
In the future, with improved support for
quadruple -- through (partial) hardware support or advances in the
algorithms for software simulation -- the additional cost of
the mixed precision 
approach vanishes. The use of a hardware supported 80-bit extended floating
point format provides an alternative. In
this case, the execution time is hardly affected, but
it cannot guarantee the same orthogonality. In addition to
improving the orthogonality, our approach 
increases both robustness and scalability of the solver.  
For this reasons, the mixed precision approach
is ideal for large-scale distributed-memory solvers.  

\chapter[Mixed Precision MRRR: Experiments II]{Mixed Precision MRRR on {\sc Application} Matrices}
\label{sec:moreexperimentsmixed}

We use test set {\sc Application}, detailed in
Appendix~\ref{appendix:matrices}, to augment our previous results for test
set {\sc Artificial}. Most {\sc Application} matrices 
are part of the publicly available {\sc Stetester} suite~\cite{Marques:2008}
and range from $1{,}074$ to $8{,}012$ in size.

\begin{figure}[thb]
   \centering
   \subfigure[Execution time: sequential.]{
     \includegraphics[width=.47\textwidth]{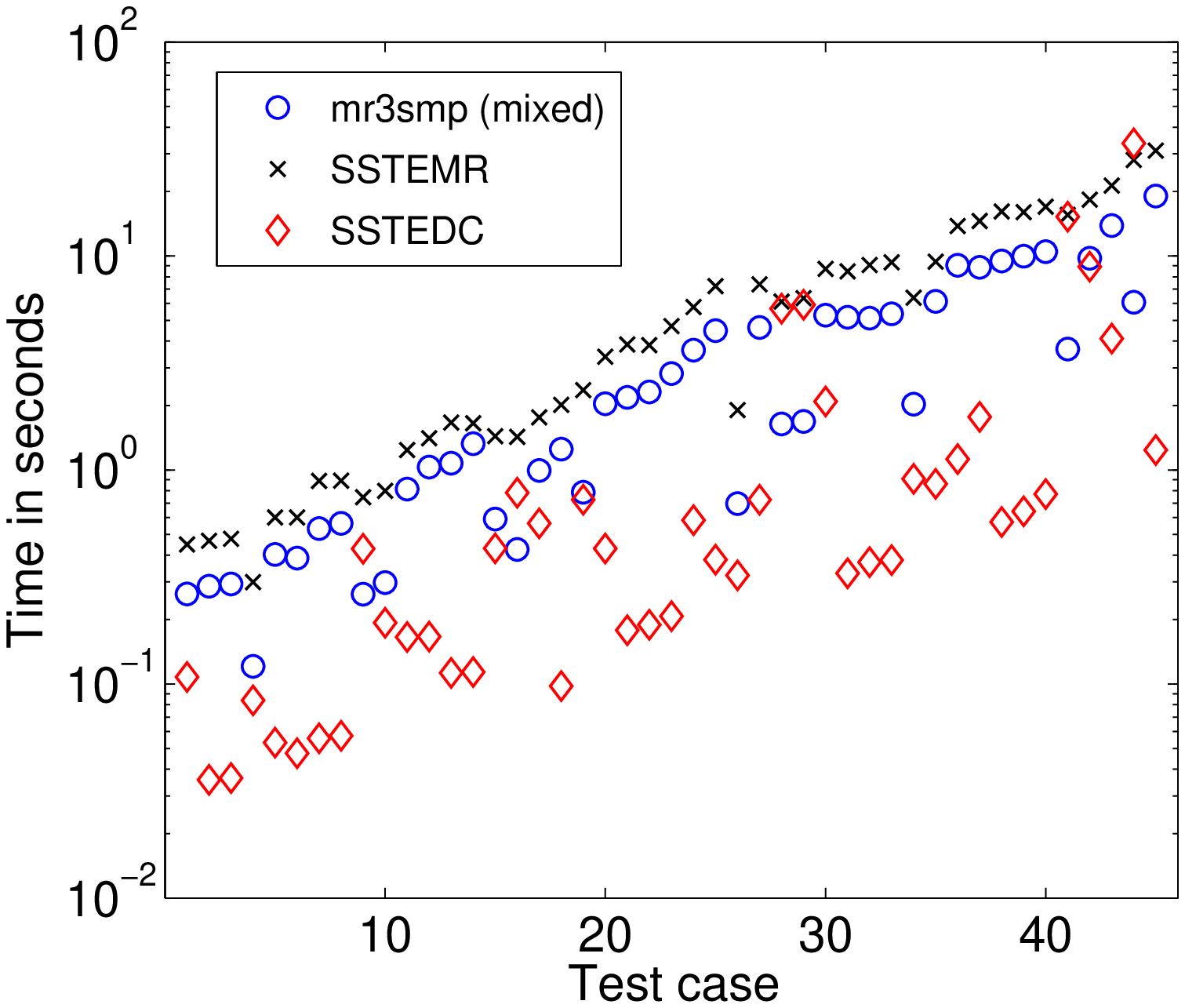}
   } \subfigure[Execution time: multi-threaded.]{
     \includegraphics[width=.47\textwidth]{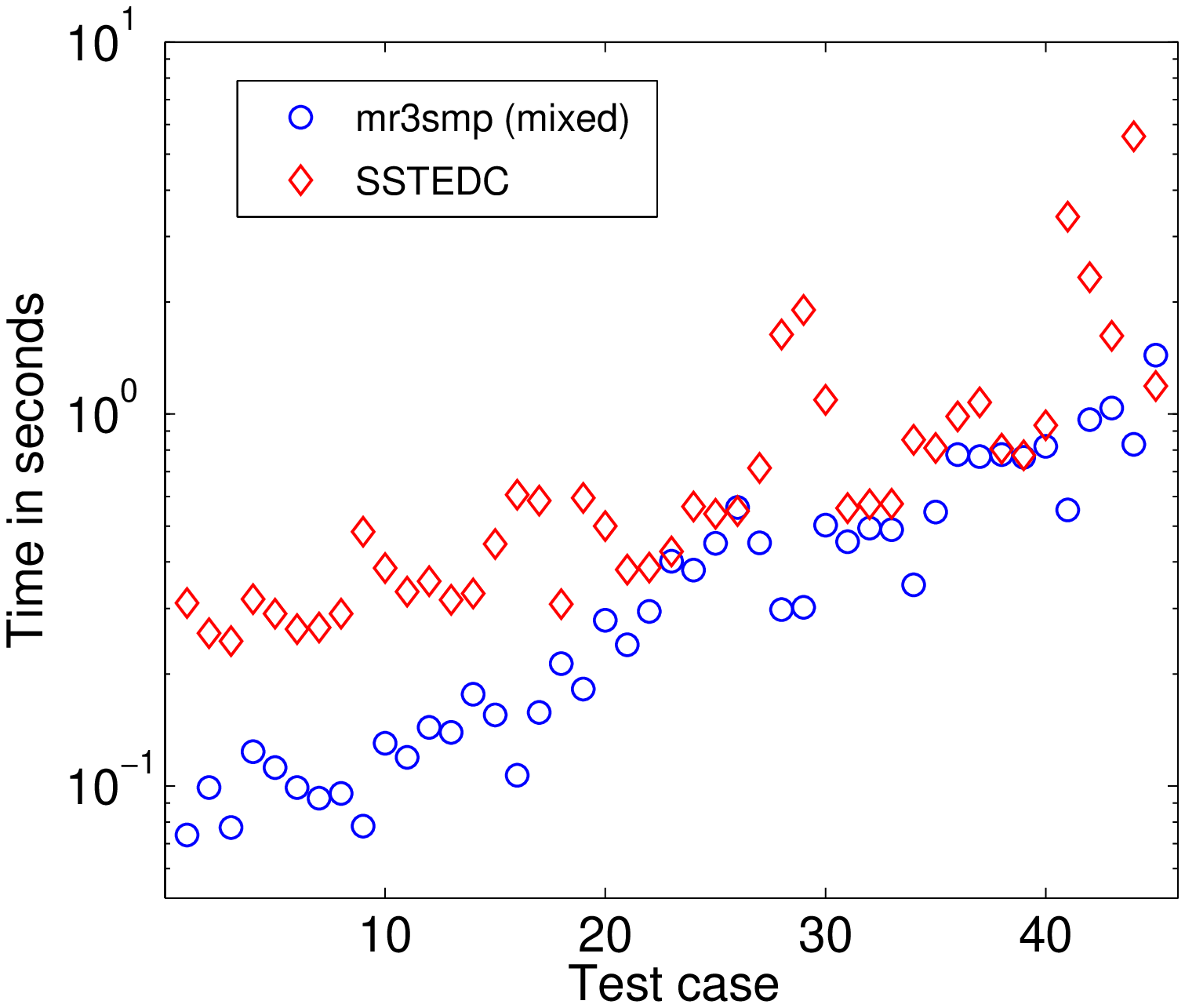}
   }
   \caption{
     Timings for test set {\sc Application} on {\sc Beckton}. The single
     precision input matrices $T \in \Rnn$ are tridiagonal.
   }
   \label{fig:timestetestersingle}
\end{figure}

\begin{figure}[thb]
   \centering
   \subfigure[Largest residual norm.]{
     \includegraphics[width=.47\textwidth]{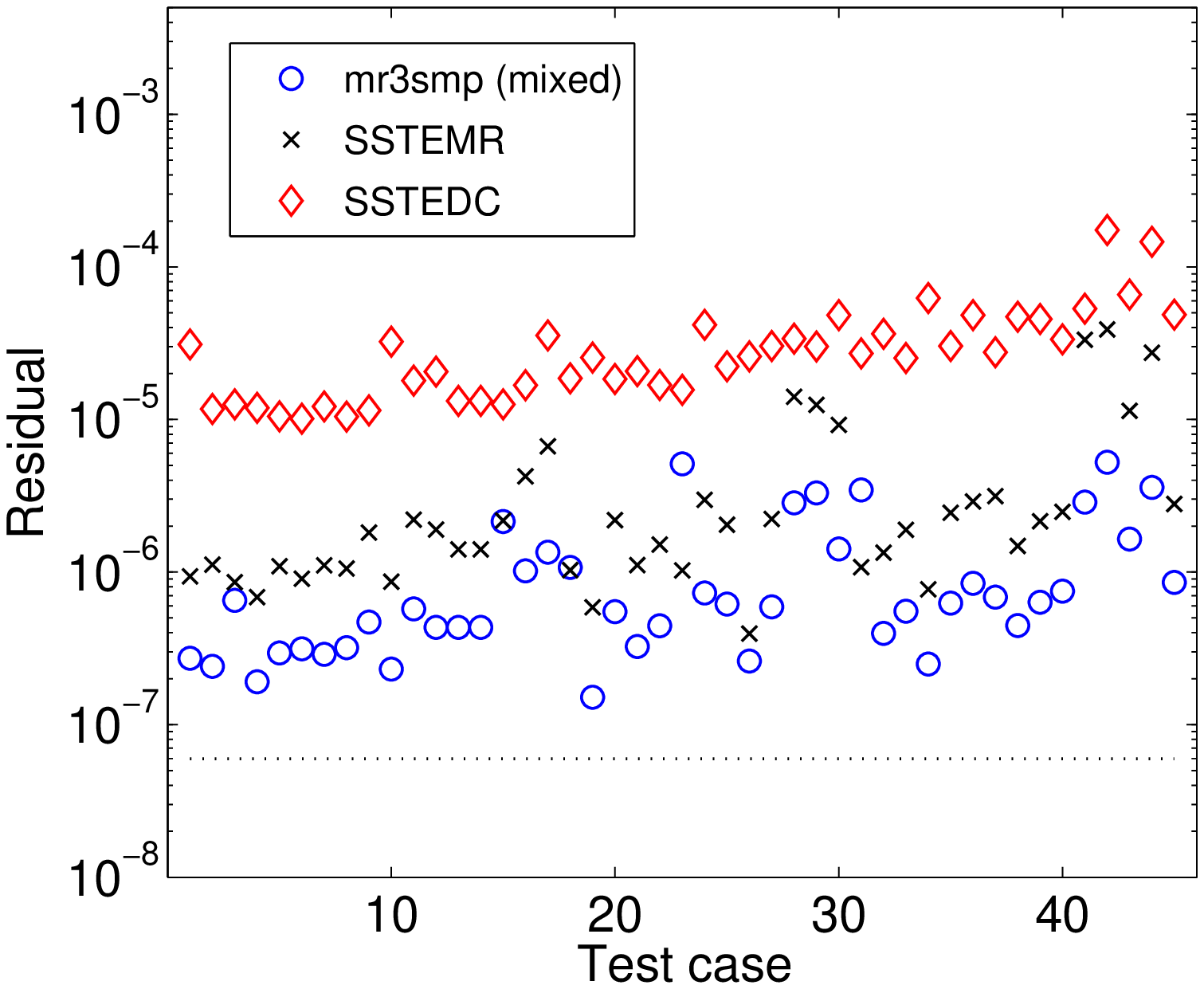}
   } \subfigure[Orthogonality.]{
     \includegraphics[width=.47\textwidth]{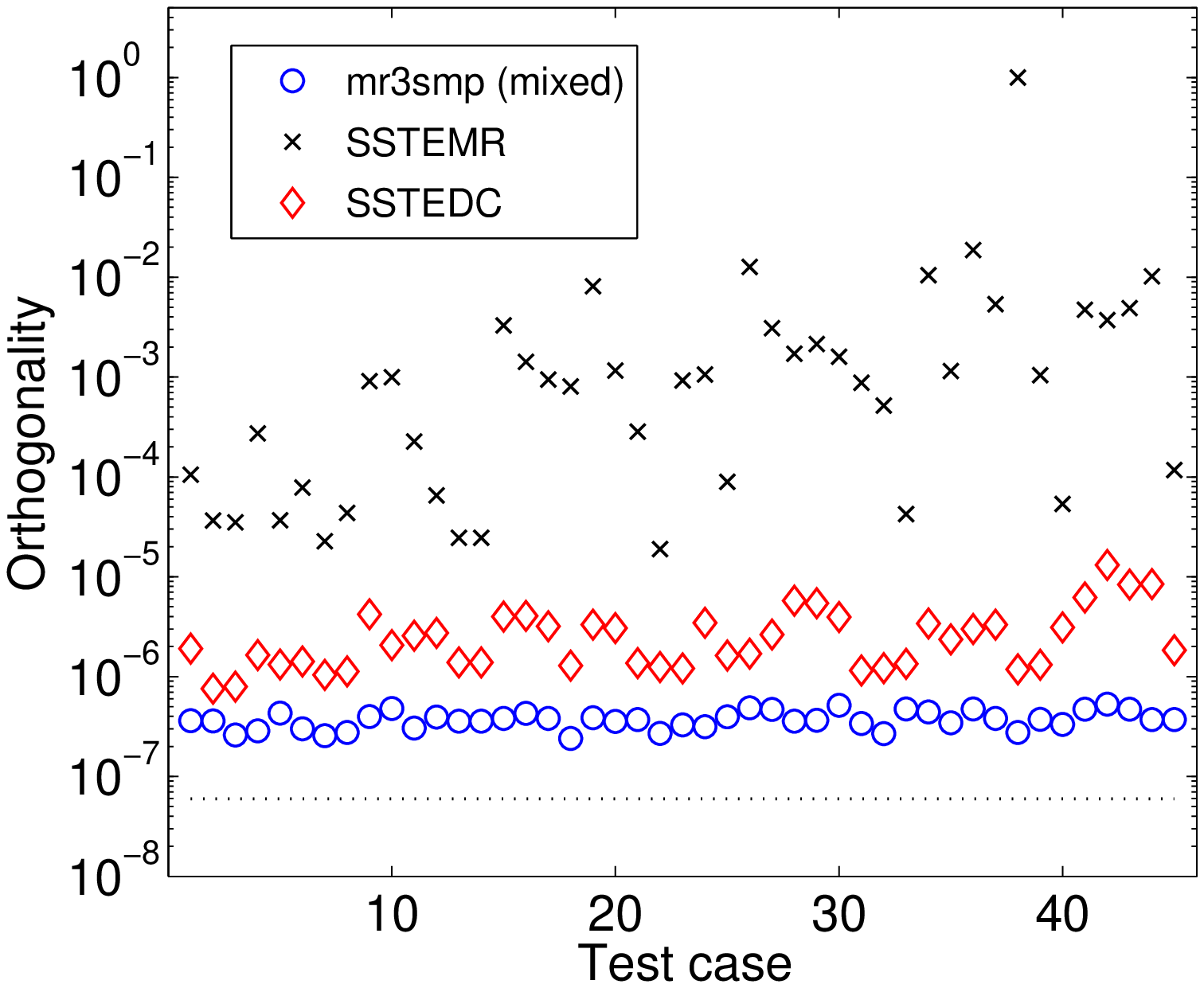}
   }     
   \caption{
     Accuracy for test set {\sc Application}. The single precision input
     matrices $T \in \Rnn$ are tridiagonal.
   }
   \label{fig:accstetestersingle}
\end{figure}

In Figs.~\ref{fig:timestetestersingle} and \ref{fig:accstetestersingle}, we
present accuracy and timings for the single precision matrices in
tridiagonal form. 
The tests correspond to Figs.~\ref{fig:timeartifialsingle} and
\ref{fig:accartifialsingle} in Section~\ref{sec:mixed:experiments}.
As matrices are quite small for the resources, we hardly
see any speedup through multi-threading for the smallest
matrices. Nonetheless, the mixed precision MRRR is highly competitive with
DC, both in terms of performance and accuracy.

\FloatBarrier

\begin{figure}[thb]
   \centering
   \subfigure[Execution time: sequential.]{
     \includegraphics[width=.47\textwidth]{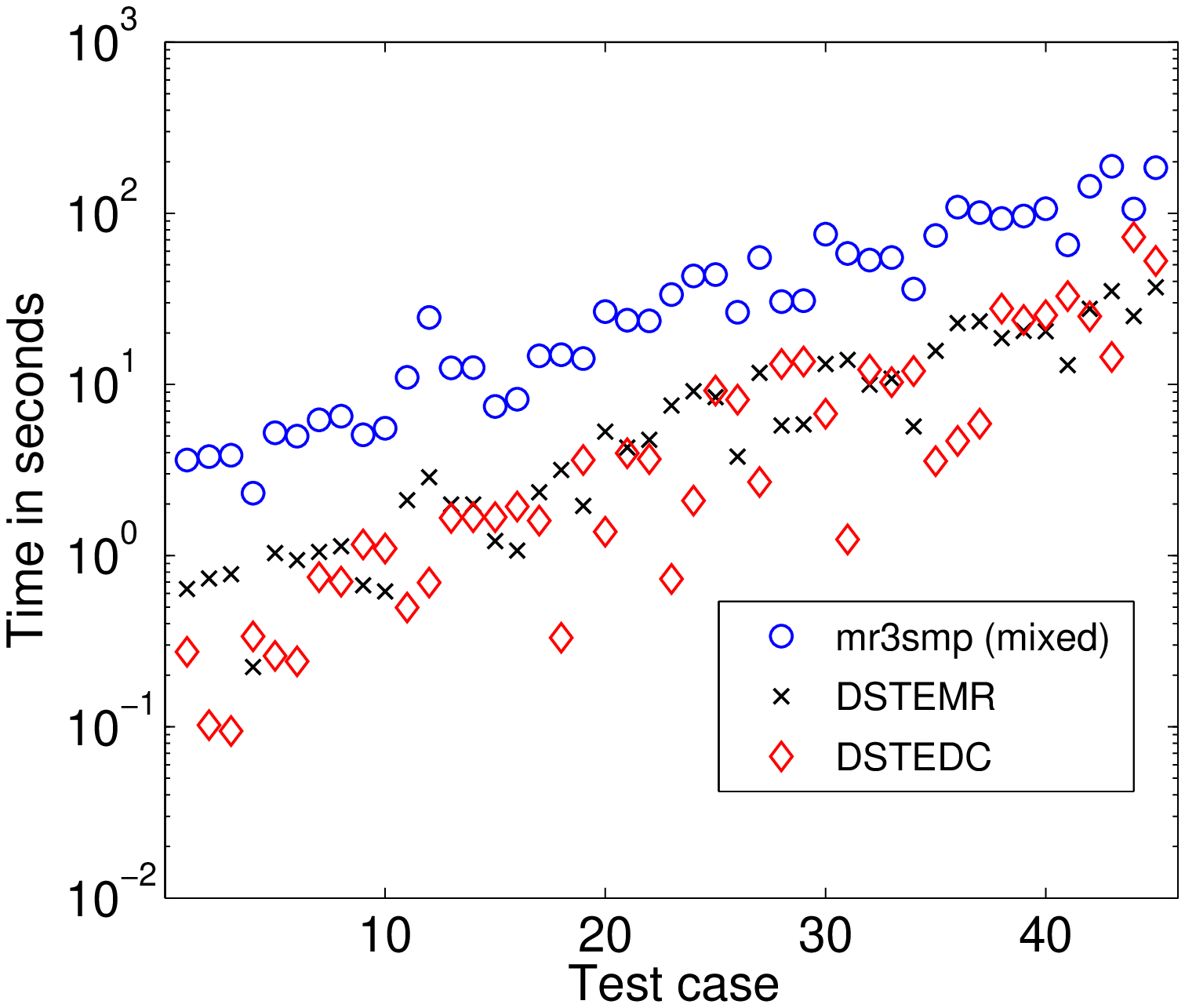}
     \label{fig:accstetesterdoubleca}
   } \subfigure[Execution time: multi-threaded.]{
     \includegraphics[width=.47\textwidth]{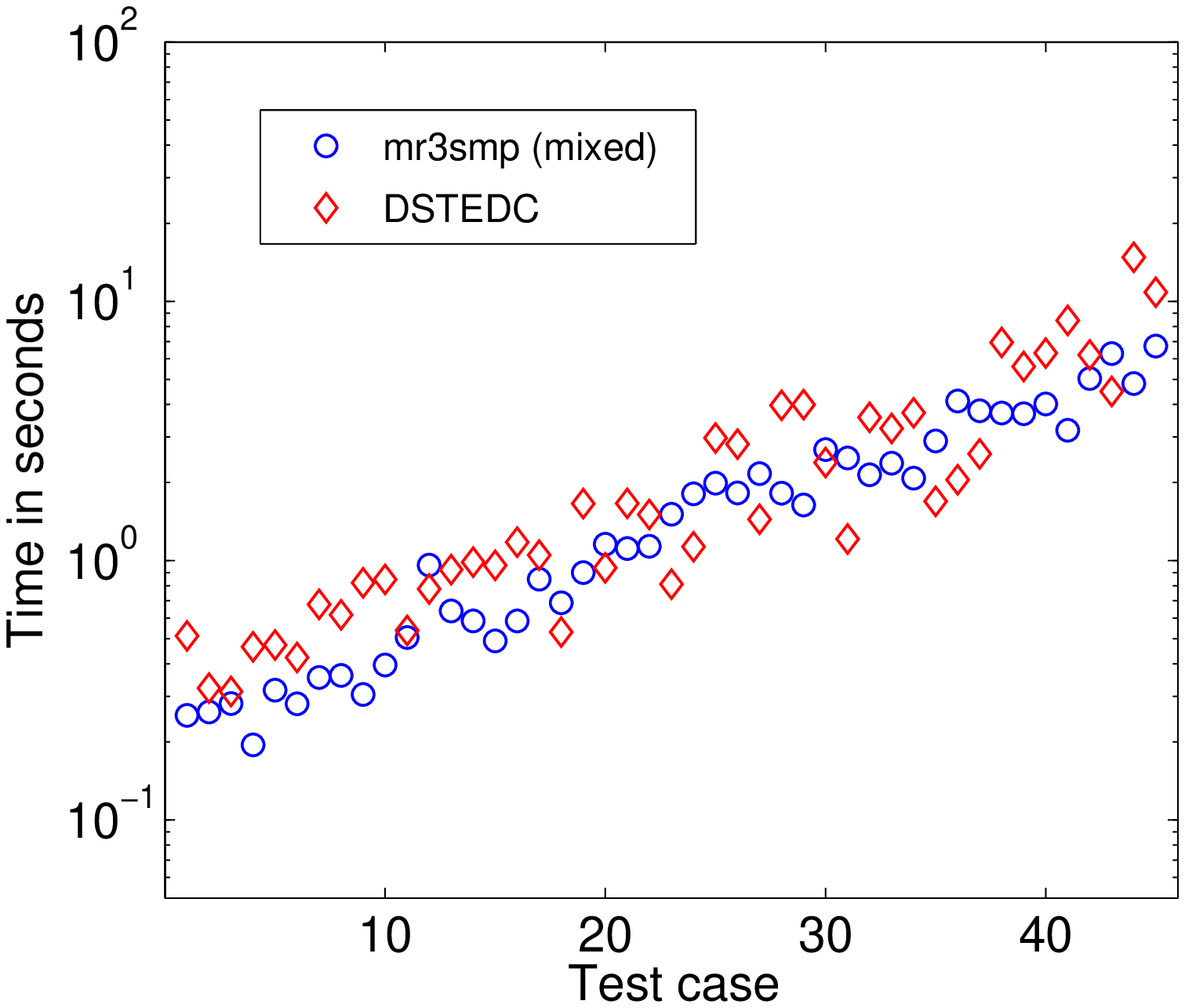}
     \label{fig:accstetesterdoubleb}
   }     
   \caption{
     Timings for test set {\sc Application} on {\sc Beckton}. The double
     precision input 
     matrices $T \in \Rnn$ are tridiagonal.
   }
   \label{fig:timestetesterdouble}
\end{figure}

\begin{figure}[thb]
   \centering
   \subfigure[Largest residual norm.]{
     \includegraphics[width=.47\textwidth]{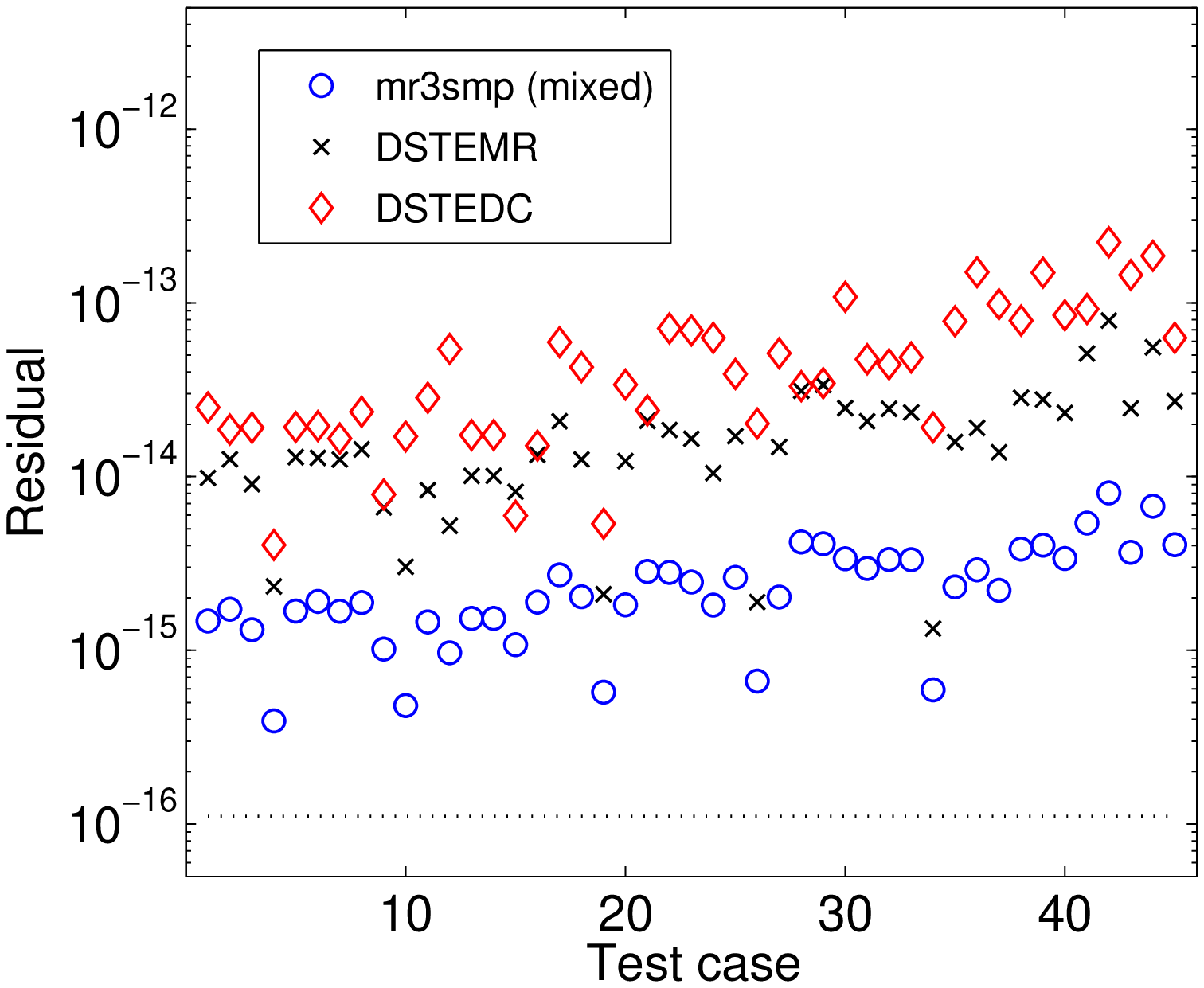}
     \label{fig:accstetesterdoublec}
   } \subfigure[Orthogonality.]{
     \includegraphics[width=.47\textwidth]{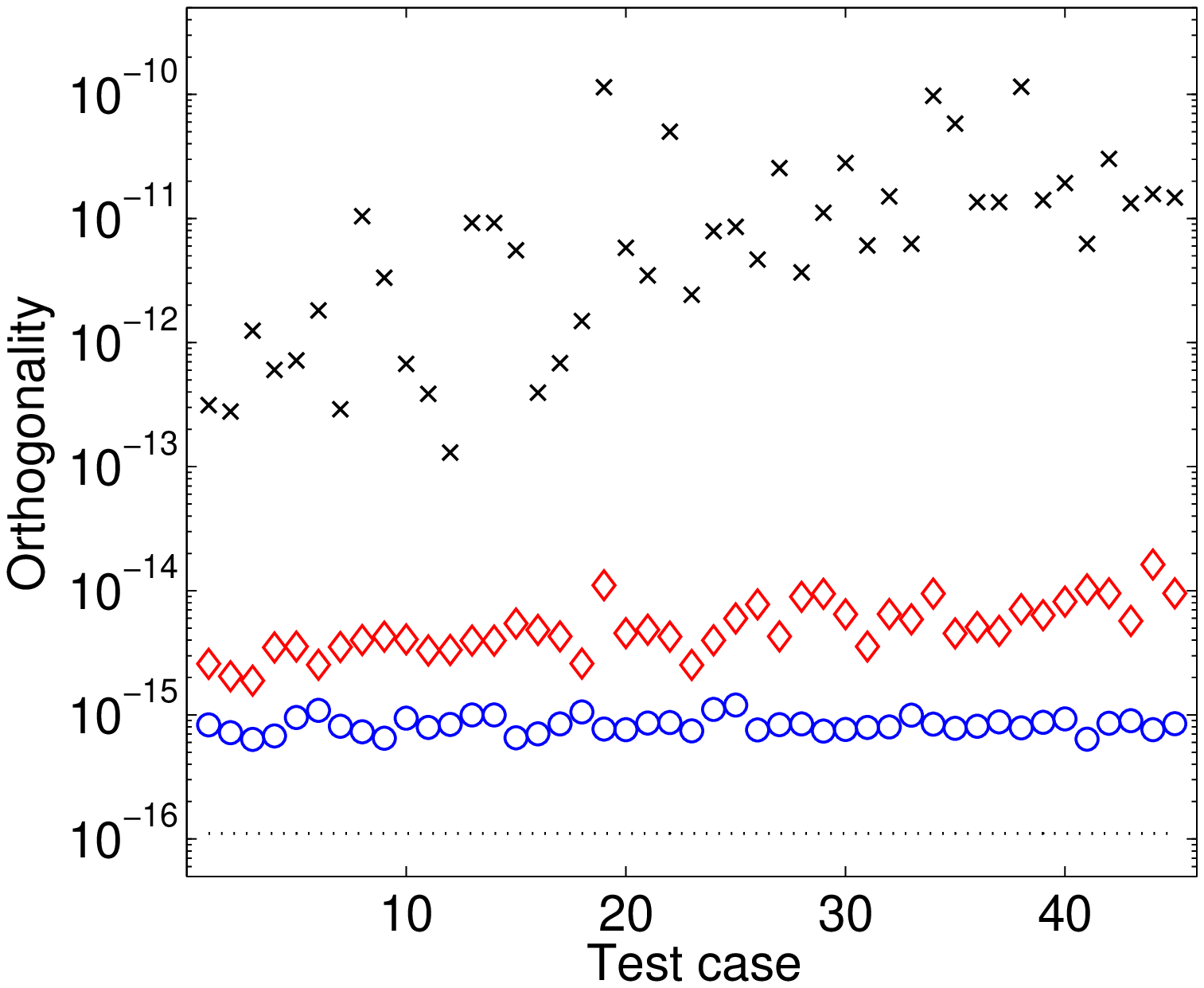}
     \label{fig:accstetesterdoubled}
   }
   \caption{
     Accuracy for test set {\sc Application}. The double precision input
     matrices $T \in \Rnn$ are tridiagonal.
   }
   \label{fig:accstetesterdouble}
\end{figure}

Corresponding to Figs.~\ref{fig:timeartifialdouble} and
\ref{fig:accartifialdouble} in Section~\ref{sec:mixed:experiments}, we
performed the experiment on test set {\sc Application} in double
precision. The results are presented in Figs.~\ref{fig:timestetesterdouble}
and \ref{fig:accstetesterdouble}.
In sequential executions, as the matrices are smaller than for test set {\sc
  Artificial} and we use a rather slow software-simulated quadruple
precision arithmetic, the mixed precision solver is considerably slower than
both \DSTEDC\ and \DSTEMR. Such a performance penalty vanishes for parallel
executions. 

Robustness, measured by $\phi(\mbox{\sc Application})$, is increased: 
while for {\tt DSTEMR} $\phi \approx 0.02$, we have $\phi \approx 0.71$ for the mixed
precision MRRR, even without taking the relaxed requirements of
\eqref{eq:newkelgbound} and \eqref{eq:newkrrbound} into account. 
By only very conservatively relaxing the requirements on what constitutes
an RRR, we already achieve $\phi \approx 0.98$; for all matrices, only a
single representation is accepted without passing the test for being an
RRR.\footnote{This has to be compared to the $14{,}356$ problematic cases
  for {\tt DSTEMR}.} 
We suspect that, by properly adjusting 
the test for relative robustness, we can easily achieve $\phi(\mbox{\sc
  Application}) = 1$ and thereby guarantee accuracy for all inputs.\footnote{Additionally, 
all the measures to improve MRRR's robustness proposed in
\cite{Willems:Diss} can and should be implemented for maximal robustness.}
These numbers support our believe that the use of mixed precisions might be an important
ingredient for MRRR to achieve robustness comparable to the most reliable solvers.

\begin{figure}[thb]
   \centering
   \subfigure[Single precision.]{
     \includegraphics[width=.47\textwidth]{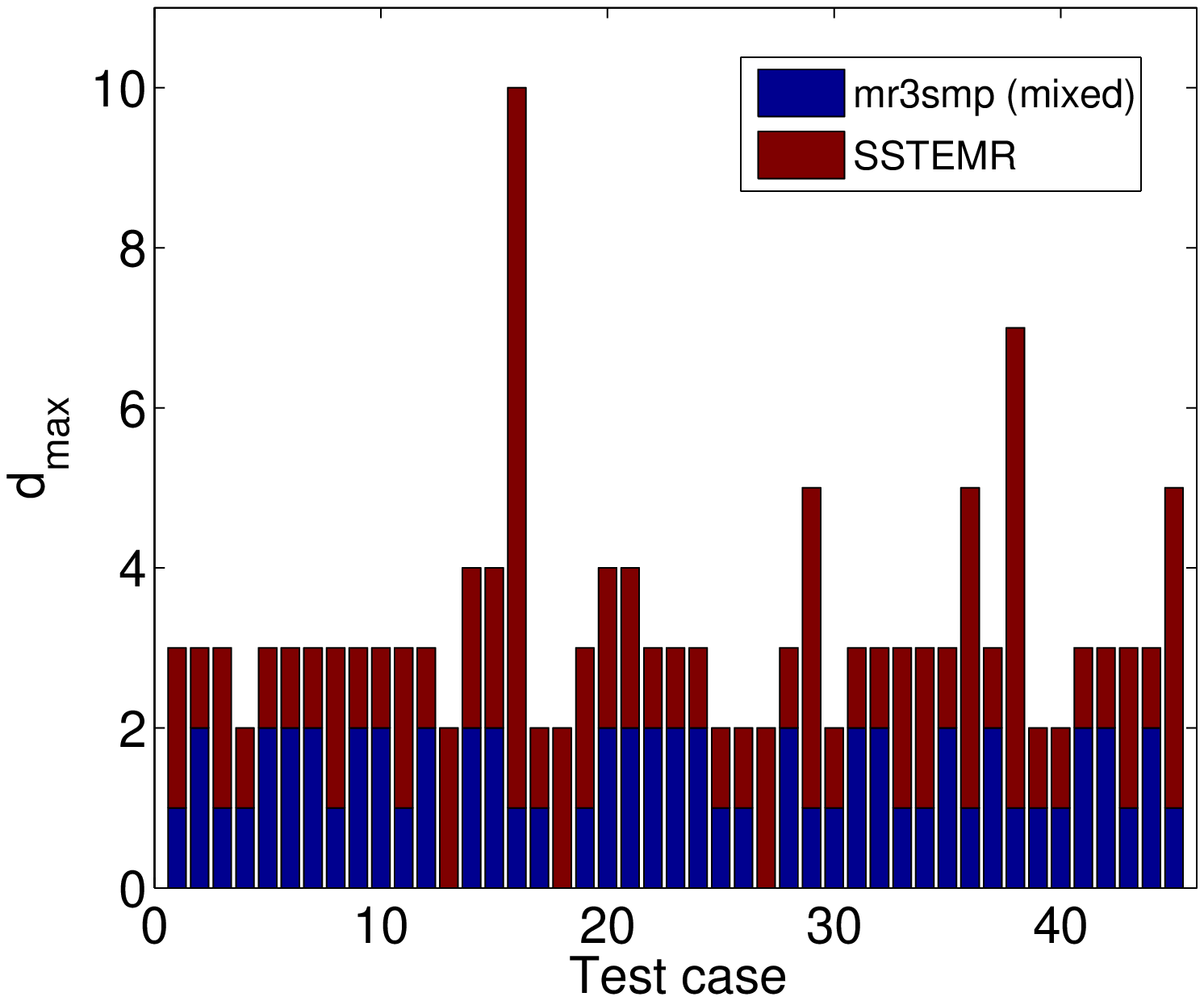}
   } \subfigure[Double precision.]{
     \includegraphics[width=.47\textwidth]{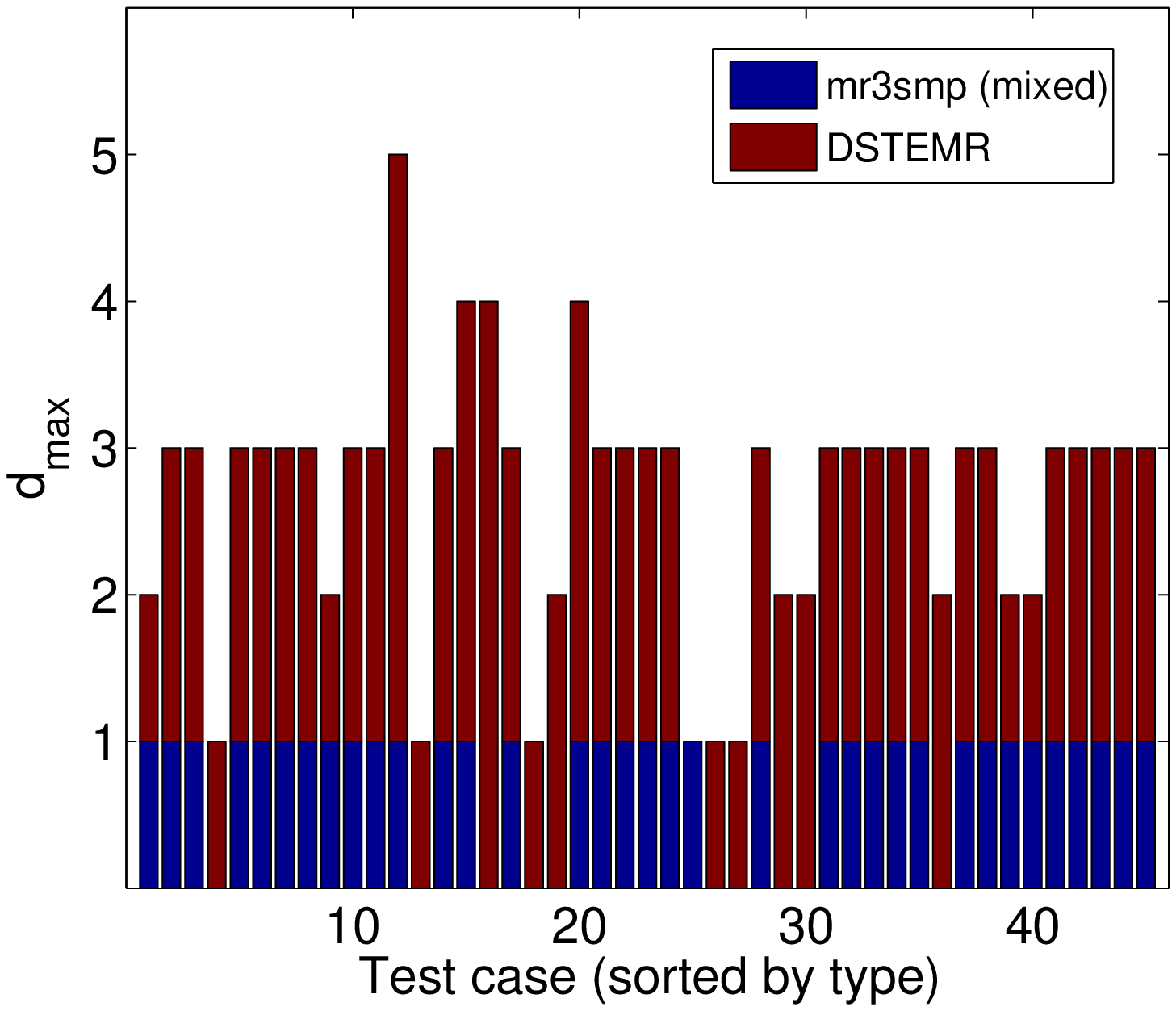}
   }     
   \caption{
     Maximal depth of the representation tree, $d_{max}$.
   }
   \label{fig:dmaxstetester}
\end{figure}

Similar to Fig.~\ref{fig:dmaxartifial} in
Section~\ref{sec:mixed:experiments}, we show in Fig.~\ref{fig:dmaxstetester}
the maximal depth of the representation tree, $d_{max}$. 
In contrast to test set {\sc
  Artificial}, $d_{max}$ is limited to small values for both \SSTEMR\
and \DSTEMR;  the {\sc Application} matrices are smaller and have
less clustering of eigenvalues. For the mixed precision MRRR,
$d_{max}$ is limited to two in the single/double case and to one in the
double/quadruple case.

\FloatBarrier

\begin{figure}[thb]
   \centering
   \subfigure[Execution time: sequential.]{
      \includegraphics[width=.47\textwidth]{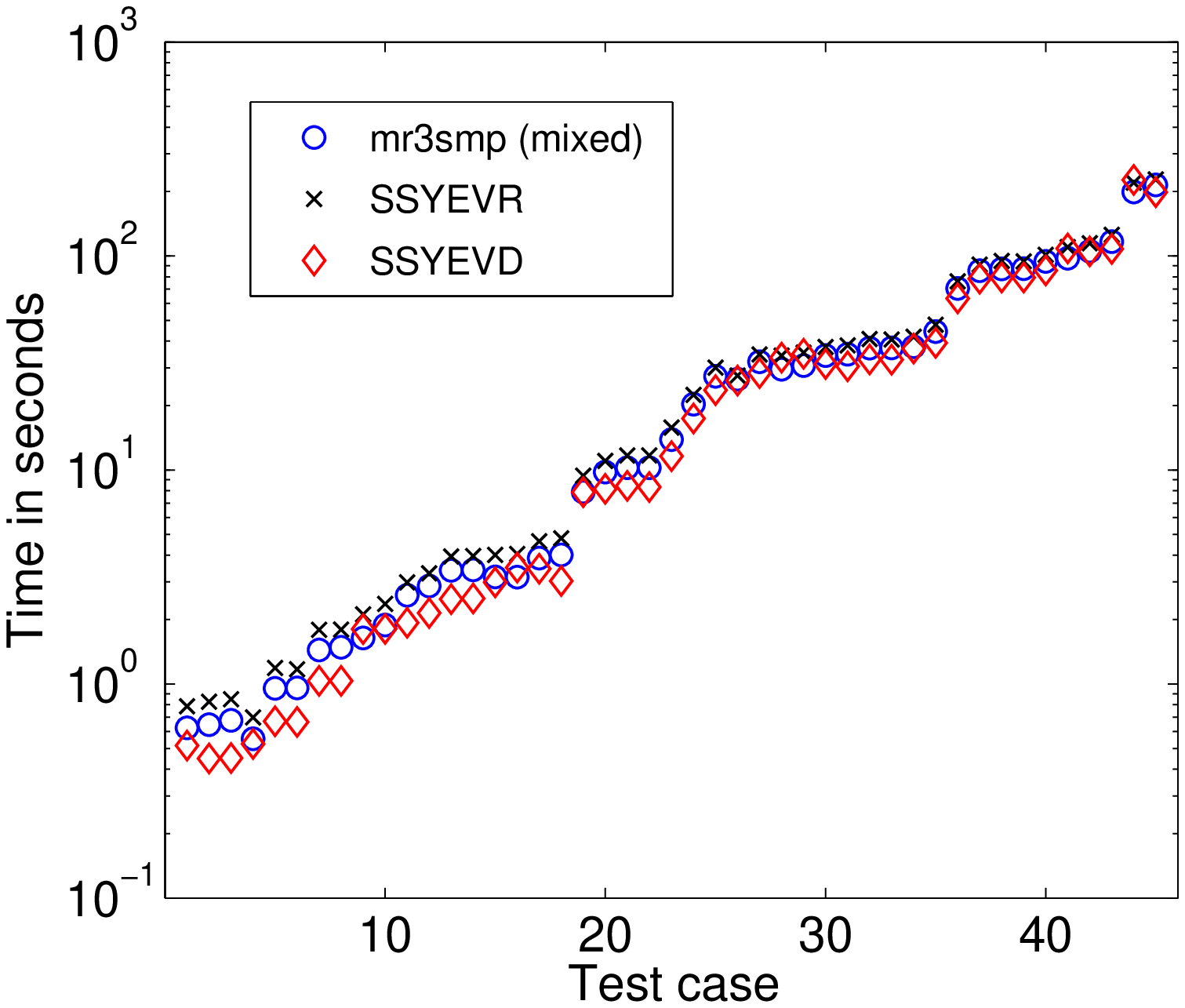}    
    } \subfigure[Execution time: multi-threaded.]{
      \includegraphics[width=.47\textwidth]{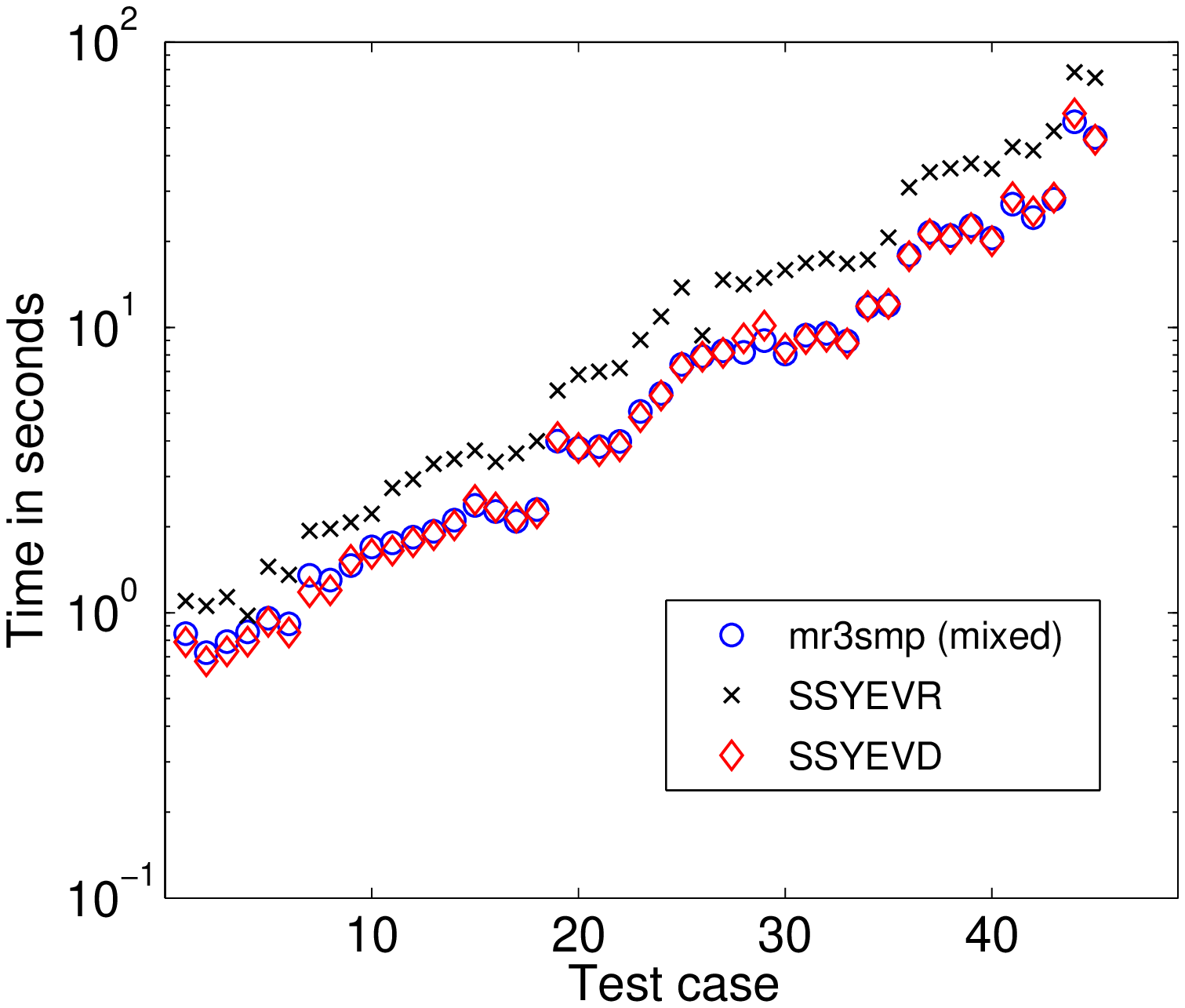}        
    }
   \caption{
     Timings for test set {\sc Application} on {\sc Beckton}. The single
     precision input
     matrices $A \in \Rnn$ are dense.
   }
   \label{fig:timestestersingledense}
\end{figure}

\begin{figure}[thb]
   \centering
    \subfigure[Largest residual norm.]{
      \includegraphics[width=.47\textwidth]{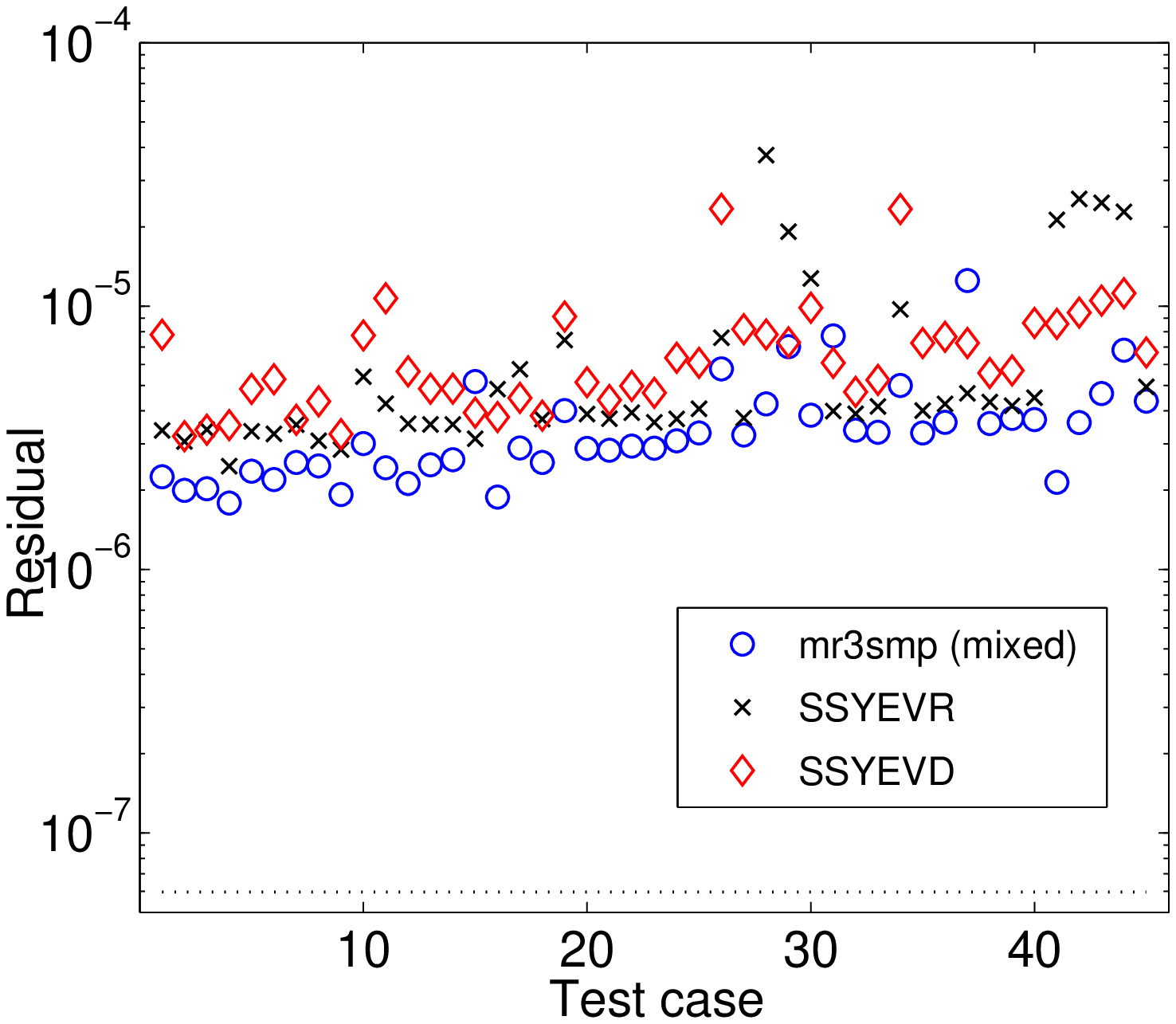}        
    } \subfigure[Orthogonality.]{
      \includegraphics[width=.47\textwidth]{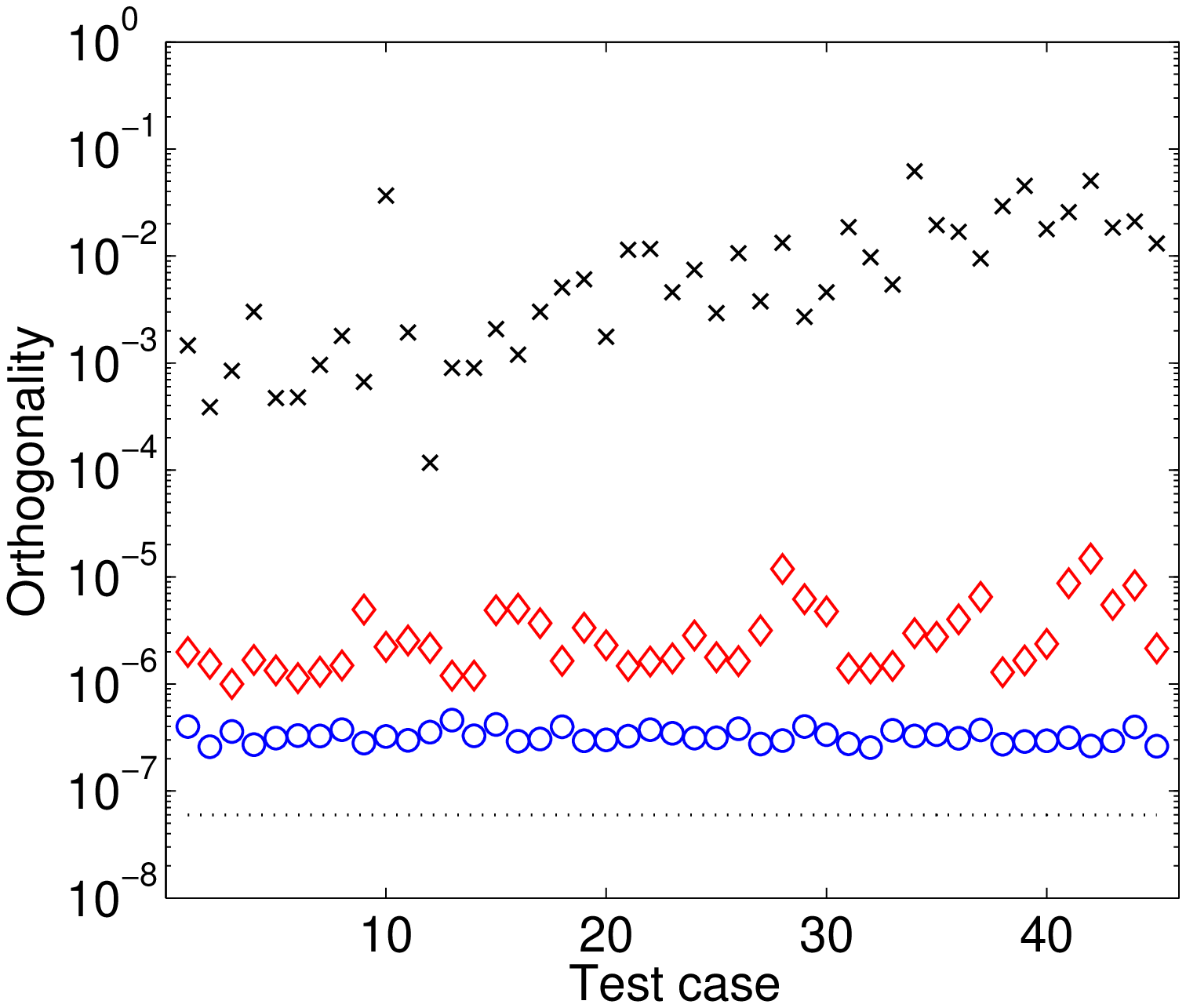}        
    }
   \caption{
     Accuracy for test set {\sc Application}. The single precision input
     matrices $A \in \Rnn$ are dense.
   }
   \label{fig:accstestersingledense}
\end{figure}

For single precision symmetric dense inputs $A \in \Rnn$, we show results in
Figs.~\ref{fig:timestestersingledense}
and \ref{fig:accstestersingledense}. 
The experiment corresponds to
Figs.~\ref{fig:timeartifialsingledense} and \ref{fig:accartifialsingledense}
in Section~\ref{sec:mixed:experiments}. The matrices are generated by
applying random orthogonal similarity transformations to the tridiagonal
matrices of the previous experiments: $A = QTQ^*$, with random orthogonal
matrix $Q \in \Rnn$. 
The execution time is dominated by the reduction to tridiagonal form and
the backtransformation of the eigenvectors. As the first stage, {\tt SSYTRD}, does not
scale well, so does the overall problem. Similar results are also obtained
for executions with a smaller number of threads. 

\FloatBarrier

\begin{figure}[thb]
   \centering
   \subfigure[Execution time: sequential.]{
      \includegraphics[width=.47\textwidth]{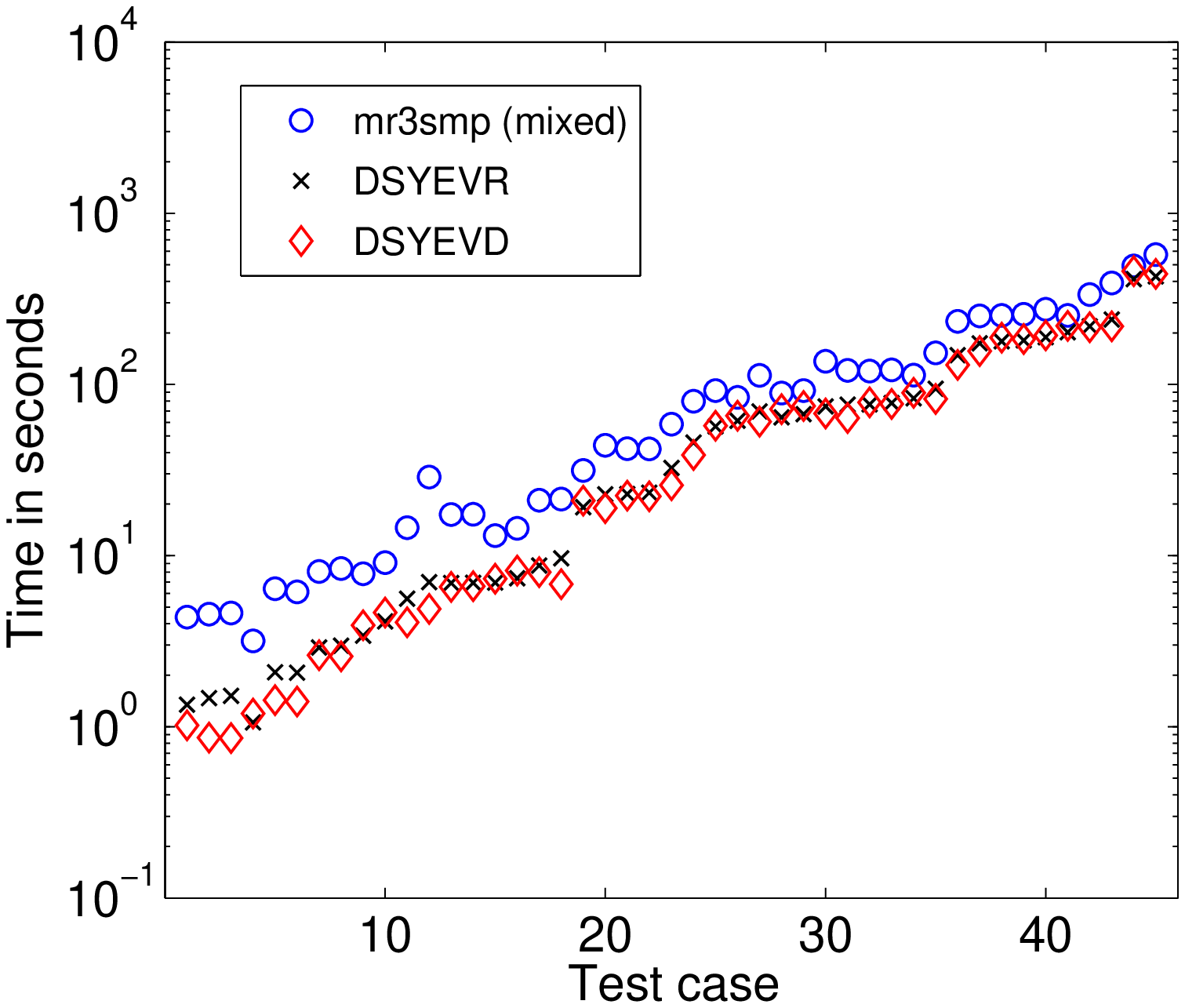}    
      \label{fig:timestesterdoubledensea}
    } \subfigure[Execution time: multi-threaded.]{
      \includegraphics[width=.47\textwidth]{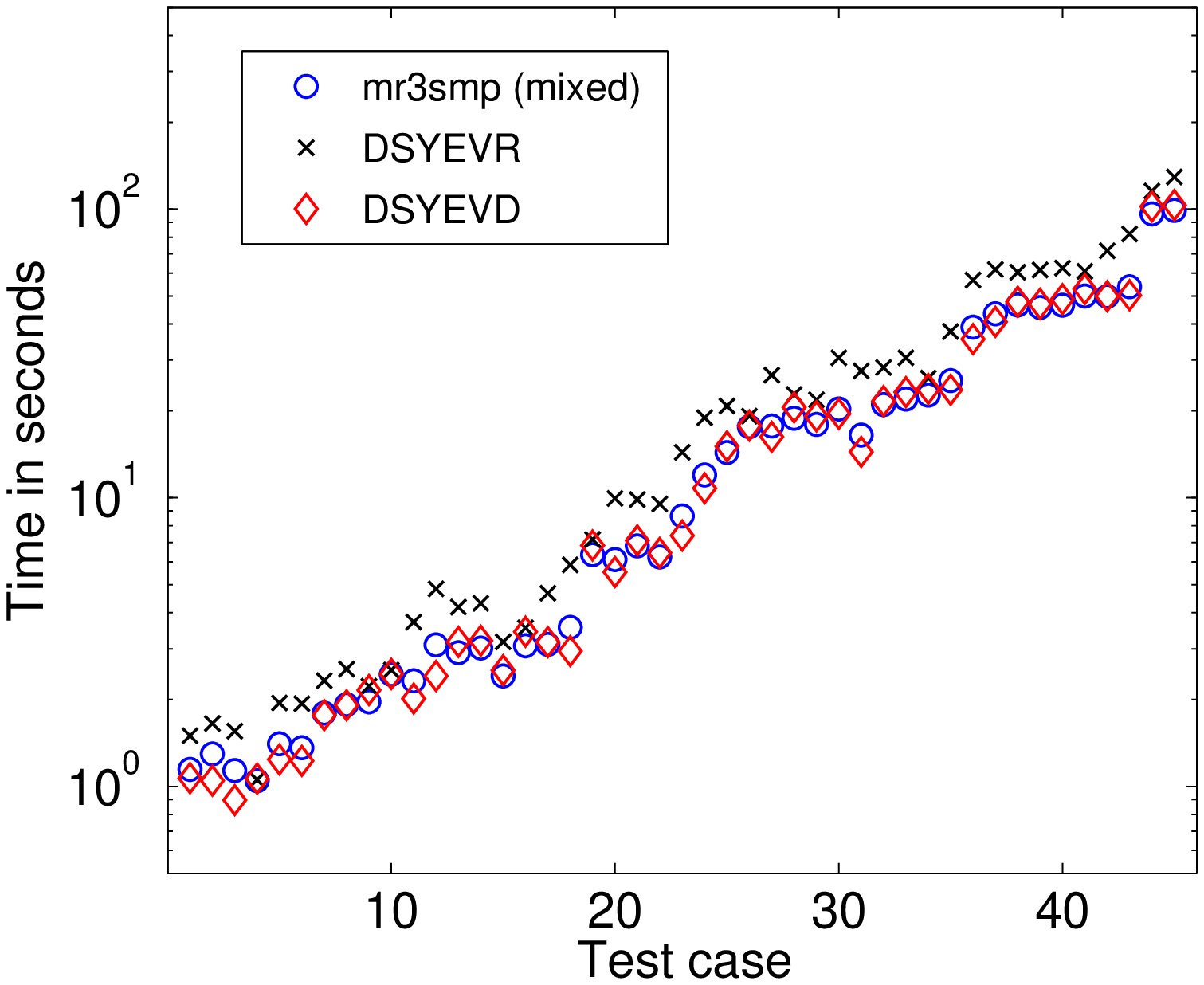}        
      \label{fig:timestesterdoubledenseb}
    }
   \caption{
     Timings for test set {\sc Application} on {\sc Beckton}. The double
     precision input matrices $A \in \Rnn$ are dense.
   }
   \label{fig:timestesterdoubledense}
\end{figure}

\begin{figure}[thb]
   \centering
    \subfigure[Largest residual norm.]{
      \includegraphics[width=.47\textwidth]{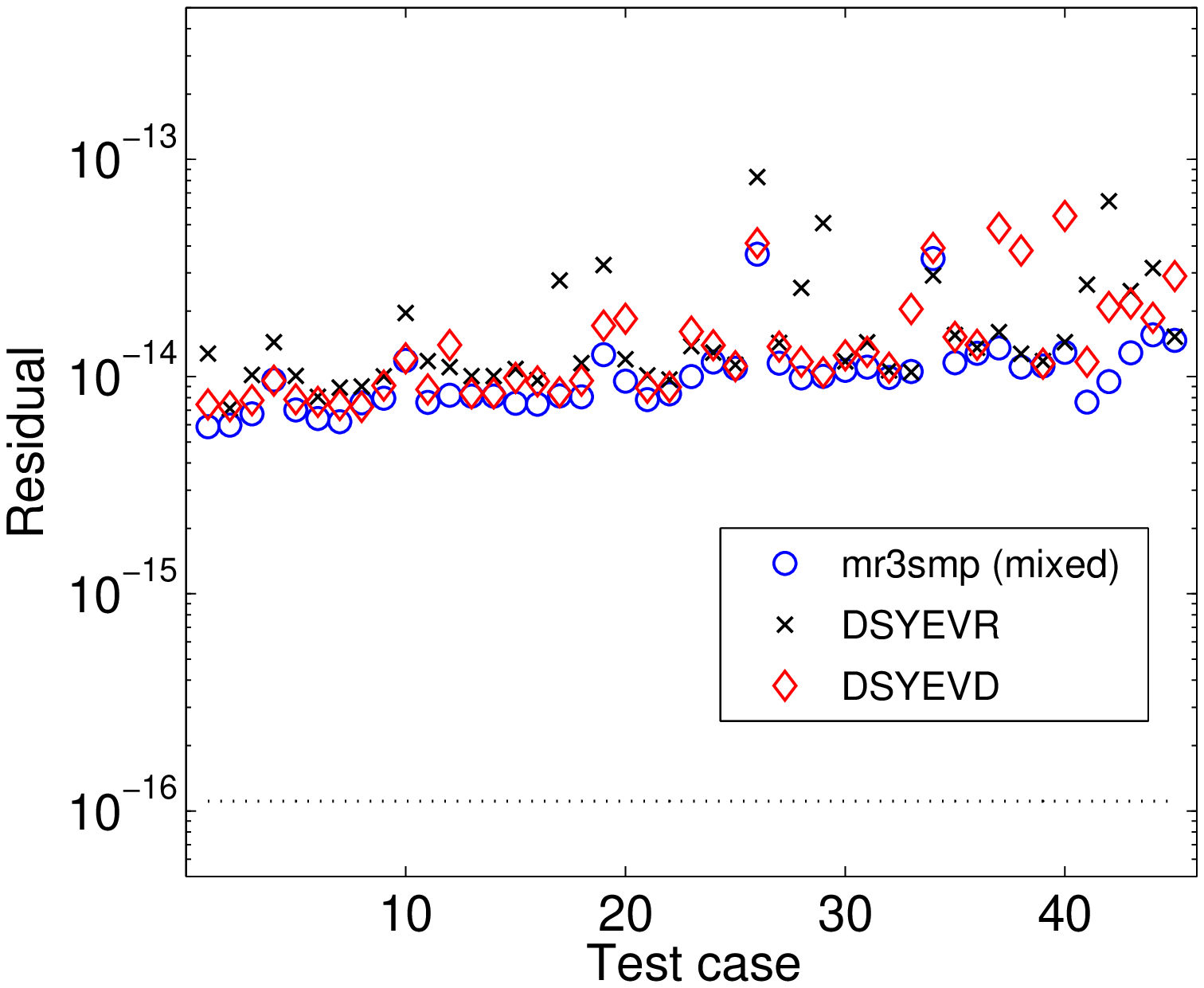}        
      \label{fig:accstesterdoubledensea}
    } \subfigure[Orthogonality.]{
      \includegraphics[width=.47\textwidth]{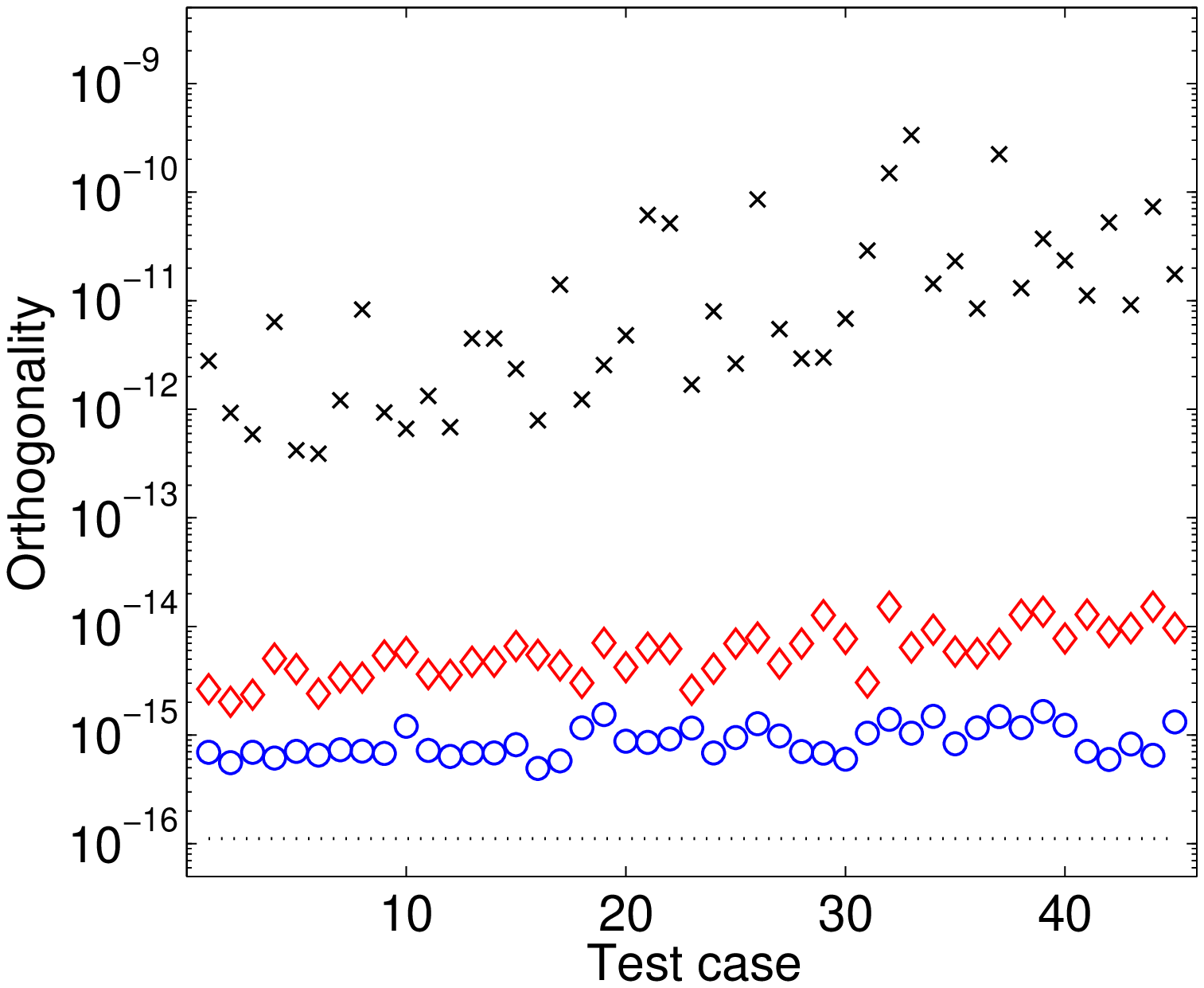}        
      \label{fig:accstesterdoubledenseb}
    }
   \caption{
     Accuracy for test set {\sc Application}. The double precision input
     matrices $A \in \Rnn$ are dense.
   }
   \label{fig:accstesterdoubledense}
\end{figure}

For double precision real symmetric dense inputs $A \in \Rnn$, we show results in
Figs.~\ref{fig:timestesterdoubledense}
and \ref{fig:accstesterdoubledense}. The experiment corresponds to
Figs.~\ref{fig:timeartifialdoubledense} and \ref{fig:accartifialdoubledense}
in Section~\ref{sec:mixed:experiments}. 
For small matrices, 
the sequential execution is slower than {\tt DSYEVR}, but the performance
gap reduces as the matrix size increases. 
The accuracy improvements are limited to the orthogonality; the residuals
are often comparable for all solvers. 
If input matrices become complex-valued and/or only a subset of eigenpairs
needs to be computed, a possible overhead due to the use of mixed precisions
is reduced as the reduction to tridiagonal form would carry even more
weight relative to the tridiagonal stage.

{\small
\bibliographystyle{acm}
\bibliography{thesis}
}

\end{document}